\documentclass[12pt]{book}
\usepackage[T1]{fontenc}
\usepackage[utf8]{inputenc}

\usepackage{these}
\usepackage[nottoc]{tocbibind}
\french
\usepackage{amsmath,amssymb,amsfonts,amsthm}
\usepackage{stmaryrd,wasysym}
\usepackage{calc}
\usepackage{empheq}
\usepackage{graphicx}
\usepackage[all]{xy}
\usepackage[dvips]{hyperref}

\PrerenderUnicode{Ã©}
\hypersetup{%
  pdftitle={Théories de jauge en géométrie non commutative et généralisation du modèle de Born-Infeld},%
  pdfauthor={Emmanuel Sérié},
  pdfsubject={Thèse de l'Université Paris 6 soutenue le 20 septembre 2005},
  pdfkeywords={géométrie non commutative, géométrie différentielle, modèle de Born-Infeld, symétries, théorie des champs, théories de jauge.},
  pdfproducer={pdflatex},
  bookmarksopen=false,
}

\usepackage{macromath}
\usepackage{figures}
\usepackage{mesthm}

\etablissement{
 \raisebox{-30pt}[0pt][0pt]{
    \hspace*{-15pt}\makebox[0pt][l]{
      \includegraphics{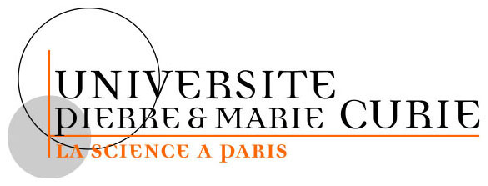}
    }}
}

\universite{Université Paris 6} 
\discipline{physique théorique}
\auteur{Emmanuel Sérié} 
\adresseelectronique{emmanuel.serie@th.u-psud.fr}

\titre{Théories de jauge en géométrie non commutative et généralisation du modèle de Born-Infeld}  
\directeur{Richard Kerner} 
\datesoutenance{20 septembre 2005}

\jury{%
  Jean& Iliopoulos & président\\%
&&\\    
  Richard& Kerner & directeur\\%
  Thierry& Masson & co-directeur  \\%
&&\\    
  Robert& Coquereaux  &rapporteur\\%
  Michel& Rausch de Traubenberg \ \  & rapporteur \\%
&&\\
  Alain& Comtet & examinateur \\%
&&\\
  Michel & Dubois-Violette &examinateur invité   %
}

\begin{document}
\version{}
\pagedegarde

\pagestyle{empty}
\vfill
\begin{center}
  \textbf{\large Résumé}
\end{center}
Les algèbres d'endomorphismes peuvent remplacer la notion de fibré principal. Dans ce cadre algébrique, les théories de jauge sont reformulées et généralisées, unifiant ainsi connexions ordinaires et champs de Higgs. Un modèle de ``Maxwell non commutatif'' est construit pour des fibrés non triviaux nécessitant le développement de la notion de structure Riemannienne. Les techniques de la géométrie non commutative  utiles à l'étude des algèbres associatives sont présentées et une nouvelle méthode permettant d'obtenir le morphisme de Chern-Weil usuel est développée. Ensuite, les résultats d'une étude sur les connexions non commutatives généralisent ceux connus sur les fibrés symétriques; une extension de l'ansatz de Witten est énoncée. Enfin, une action est proposée pour généraliser le modèle de Born-Infeld à des connexions non commutatives. Les Lagrangiens obtenus sont non polynomiaux et on étudie l'existence de solutions de type solitonique sur quelques exemples explicites.
\english

\vfill
\begin{center}
  \textbf{\large Abstract}
\end{center}
Endomorphisms algebras can replace the concept of principal fiber bundle. Gauge theories are reformulated within this algebraic framework and further generalized to unify ordinary connections and Higgs fields. A ``noncommutative Maxwell'' model is built starting from non trivial fiber bundles thus requiring the development of the notion of Riemannian structure. The tools involved in the study of associative algebras are presented and an algebraic method to characterize the usual Chern-Weil morphism is proposed. Then, current results on symmetric fiber bundles are generalized to noncommutative connections and an extension of the Witten ansatz is given. Finally, a generalization of the Born-Infeld action for noncommutative connections is proposed. The corresponding Lagrangians are non-polynomial and the existence of solitonic solutions is shown on several examples.
\french
\vfill

\pagestyle{empty}
\chapter*{Remerciements}
\addcontentsline{toc}{chapter}{\protect\numberline{}Remerciements}
\chaptermark{Remerciements} 

Ma thèse s'est déroulée au Laboratoire de Physique Théorique des Liquides de l'Université Paris VI et au Laboratoire de Physique Théorique de l'Université Paris XI à Orsay.
Je remercie donc Bertrand Guillot, Dominique Schiff et Henk Hilhorst de m'avoir accueilli dans leur laboratoire qu'ils dirigent ou ont dirigé afin que je puisse y effectuer mon travail de thèse.
Mon séjour y a été des plus agréables et j'ai pu y bénéficier d'une ambiance stimulante.

Ce double statut est la résultante du fait que Richard Kerner et Thierry Masson ont tous deux accepté de diriger mes travaux de recherche durant ces trois années.
Je dois dire que cette situation tout à fait exceptionnelle m'a permis de bénéficier d'un encadrement de très grande qualité et je tiens à leur exprimer ma plus sincère reconnaissance et toute ma sympathie.
J'ai beaucoup apprécié Richard pour son enthousiasme scientifique et Thierry pour son attention et son souci de rigueur permanent.
Je les remercie pour l'expérience et les savoirs qu'ils ont su partager avec moi et pour m'avoir initié au vaste domaine qu'est la physique mathématique.

Je tiens à remercier Robert Coquereaux et Michel Rausch de Traubenberg pour avoir accepté de juger cette thèse, pour m'avoir accordé leur temps et leur attention, ainsi que pour les remarques qu'ils ont pu me faire et qui m'ont permis de clarifier certains points de ce manuscrit.
Je tiens également à remercier Alain Comtet, Michel Dubois-Violette et Jean Iliopoulos pour avoir accepté d'être membres du jury et d'avoir porté une attention particulière à mon travail.

Je tiens à remercier toutes les personnes que j'ai rencontrées et qui m'ont permis durant mon parcours d'avancer dans ma perception de certains problèmes:
Michel Dubois-Violette, 
Yvon Georgelin,
John Madore,
Vincent Rivasseau,
Jean-Christophe Wallet
et bien d'autres encore, enseignants et chercheurs, dont la liste serait trop longue à énumérer ici.
Je remercie également
Marco Maceda,
Todor Popov et
Fabien Vignes-Tourneret
qui ont accompagné mon destin de thésard et avec qui des échanges fructueux ont eu lieu.
Merci également à tous les thésards et postdocs que j'ai rencontrés au LPT et au LPTL au cours de ces trois années, qui participent grandement aux souvenirs que je conserve de cette période.

Je remercie Philippe Boucaud, Olivier Brand-Foissac, Michel Quaggeto et Jean-Pierre Leroy pour leur dévouement et gentillesse face aux problèmes informatiques que j'ai pu rencontrer. Je remercie également Patricia Flad pour son aide dans mes recherches bibliographiques ainsi que les secrétaires de ces laboratoires pour leur gentillesse. Je tiens aussi à remercier Gérard Hoffeurt pour son aide précieuse lors de l'impression de ma thèse.

Je remercie ma compagne, Aurélie, pour son soutien affectif et moral et pour avoir relu sans relâche les prototypes successifs de mon manuscrit.
Je remercie également ma mère, Dominique, d'avoir fait une dernière relecture attentive.
Si il y a moins de vingt fautes par page, c'est grâce à elles!

Enfin, merci à ma mère et mon père, Philippe, pour m'avoir toujours soutenu dans ce que j'entreprenais, mes soeurs, Jeanne et Léa, mon frère Julien pour avoir rythmé ces années, mes grands-parents Yvonne, Jacqueline et Jean, toute ma famille et celle d'Aurélie. Bien sûr je n'oublie pas mes amis qui m'ont encouragé et supporté, en particulier Bertrand, Christophe, Faustine, Géraldine, Mathieu, Sébastien et Stéphane.

J'adresse un grand merci à tout le monde et à toutes les personnes que j'aurais pu oublier.\\

\cleardoublepage%
\pagestyle{myheadings}


\tableofcontents
\chapter*{Introduction}
\addcontentsline{toc}{chapter}{\protect\numberline{}Introduction}
\chaptermark{Introduction} 

Je me suis intéressé dans cette thèse à la formulation des théories de jauge dans le cadre de la géométrie non commutative et à la généralisation du modèle de Born Infeld pour des champs de jauge non abéliens.
Les théories de jauge constituent un langage mathématique dans lequel sont formulées les interactions fondamentales en physique.

Je me propose de faire un rappel, dans cette introduction, sur la manière dont les théories de jauge ont été introduites en physique, ce qui mettra en évidence leurs principales caractéristiques.
Cela permettra également au lecteur de cette thèse de mieux en comprendre les motivations.
J'expliquerai ensuite en quoi la géométrie non commutative semble être un outil approprié pour formuler les théories de jauge et je ferai également un bref rappel historique sur son apparition en mathématiques et en physique.
Enfin, j'expliquerai mon intérêt pour la théorie de Born-Infeld et ses généralisations et donnerai un plan détaillé de cette thèse.

\section*{Pourquoi les théories de jauge?}

Le terme  ``jauge''  fut introduit pour la première fois par Hermann Weyl en 1919 dans une tentative d'unifier l'électromagnétisme et  la gravitation.
Cette terminologie fut empruntée à celle des  tablettes de jauge utilisées comme étalons de longueur dans les ateliers d'usinage.
Ainsi, dans la théorie de Weyl, la jauge est une référence de mesure permettant d'étalonner l'échelle qui va servir à mesurer une quantité physique. Les quantités physiques, ou observables, sont supposées être invariantes sous des transformations locales d'échelle (ou de jauge).
L'invariance de jauge, telle qu'elle fut introduite par Weyl, était directement inspirée\footnote{Il se questionna en 1919 suite au succès remporté par la théorie d'Einstein: ``Si les effets d'un champ gravitationnel peuvent être décrits par une connexion exprimant l'orientation relative entre des référentiels locaux de l'espace-temps, d'autres forces de la nature telles que l'électromagnétisme peuvent-elles être associées aussi à des connexions similaires?''}
de la théorie des connexions linéaires utilisée par Albert Einstein dans sa théorie de la relativité générale
 et avait donc, dès sa première formulation, un statut géométrique.

Malheureusement, pour diverses raisons, cette tentative d'unification échoua. 
Mais par la suite, lors de l'apparition de la mécanique quantique ondulatoire développée par Schrödinger en 1926,
Weyl, avec Vladimir Fock et Fritz London, travailla à l'élaboration d'un nouveau type de jauge en passant d'une jauge de type facteur d'échelle à une jauge complexe de type changement de phase.

Rétrospectivement, on pourrait considérer que c'est en fait James Clerk Maxwell qui introduisit la première théorie de jauge en formulant les lois de l'électromagnétisme telles que nous les connaissons aujourd'hui.
Dans la théorie de Maxwell, la symétrie de jauge, associée à un groupe de structure $U(1)$, fut pendant longtemps considérée comme une simple liberté intervenant dans le choix des outils mathématiques pour décrire une même réalité physique.
L'invariance locale des équations de Maxwell ne fut donc réellement exploitée qu'à partir des travaux de Weyl\footnote{En faisant encore une fois un parallèle avec la théorie de la relativité, on peut remarquer que les transformations de Lorentz, laissant également les équations de Maxwell invariantes, étaient aussi considérées par Poincaré et Lorentz, au début du 20ème siècle, comme une simple commodité mathématique permettant de décrire une réalité  physique plus compliquée.  C'est Einstein qui, le premier,  proposa de donner au groupe des transformations de Lorentz un statut fondamental. C'est, ensuite, en voulant rendre ces transformations locales qu'il fut amené à considérer la notion de connexion linéaire reliant différents repères. Il utilisa pour cela la notion de connexion de Levi-Cività qu'il interpréta comme décrivant les effets du champ gravitationnel.}.
Dans sa deuxième proposition, il identifia ce groupe $U(1)$ à l'invariance sous transformations de phase de la fonction d'onde d'une particule chargée en mécanique quantique.
Le simple fait de demander que les équations du mouvement soient invariantes sous transformations de jauge permit alors de comprendre l'origine du couplage entre le champ électromagnétique et la fonction d'onde d'une particule chargée.
Cette symétrie $U(1)$ prit alors un statut fondamental en devenant associée à la conservation locale de la charge électrique.

Ce modèle a ensuite mené à celui de l'électrodynamique quantique développé par Feynman, Schwinger et Tomonaga, décrivant l'interaction entre photons et particules chargées.
Ce modèle est exprimé au sein d'un formalisme visant à rendre compatible la mécanique quantique et la relativité restreinte, appelée théorie quantique des champs (TQC).
Dans cette théorie, les particules élémentaires sont associées à des représentations irréductibles du groupe de Poincaré dont les représentations sont indexées par deux nombres quantiques qui sont la masse et le spin (ou hélicité dans le cas des états de masse nulle) et la dynamique de ces états est gouvernée par ce qu'on appelle un champ quantique\footnote{Un champ quantique est une distribution à valeurs dans les opérateurs sur un espace de Hilbert.}.

Dans la théorie de l'électrodynamique quantique, qui fut le premier modèle de théorie quantique des champs, les photons sont des représentations d'hélicité $1$ et de masse nulle du groupe de Poincaré et sont décrits par le champ quantique de Maxwell; les particules chargées correspondent à des représentations de spin $1/2$ et de masse positive ou nulle et sont décrites par le champ quantique de Dirac.
Le couplage de ces deux champs est obtenu en demandant que le champ de Dirac soit également dans une représentation du groupe $U(1)$ associé à l'invariance de jauge des équations de Maxwell. 
Cette demande impose alors d'introduire un couplage minimal entre le champ de Dirac et le champ de Maxwell.
De même nous avons appris, du célèbre (second) théorème d'Emmy Noether, que l'on pouvait associer à  l'invariance sous des transformations locales de symétrie des quantités conservées appelées  \textit{courants de Noether} (resp. \textit{identités de Noether}) auxquels correspond l'algèbre des courants (resp. \textit{identités de Ward}) en TQC.
Appliquées à la symétrie $U(1)$, les identités de Noether expriment la loi de conservation locale de la charge.
Ainsi, dans ce modèle, le groupe de structure $U(1)$ apparaît comme étant un \textit{groupe de symétrie interne}.

Par la suite, on put mettre en évidence d'autres types d'interactions en observant la conservation de certains nombres quantiques dans les réactions nucléaires.
Ainsi, en 1954, C.N.~Yang et R.~Mills introduisirent une jauge non abélienne généralisant de manière directe la théorie de Maxwell.
Il remplacèrent le groupe $U(1)$ par un groupe non abélien $SU(2)$ et la connexion de Maxwell,  prescrivant aux phases de la fonction d'onde la relation qui existe entre elles en chaque point par une connexion à valeurs dans une algèbre non abélienne.
Bien que la notion de connexion fut présente dès l'introduction des théories de jauge par Hermann Weyl, ce modèle permit de mettre en évidence une généralisation de la théorie des connexions, utilisée en relativité générale pour les repères linéaires, à une théorie des connexions pour un groupe de Lie compact arbitraire assimilé en physique à un groupe de symétries internes.
Il fut reconnu par la suite que cette théorie des connexions correspond en fait à la théorie des connexions sur les fibrés principaux développée quelques années auparavant par les mathématiciens Elie Cartan et Charles Ehresmann.

Bien que le modèle de Yang et Mills ne put être retenu pour décrire les interactions entre nucléons étant donné son incapacité à reproduire des interactions nucléaires de courte portée (reconnues par Yukawa en 1935 comme se traduisant par l'échange de particules massives), l'idée de pouvoir décrire les interactions fondamentales à l'aide d'un principe d'invariance de jauge séduisit fortement les physiciens et est devenue par la suite un véritable guide pour établir de nouvelles théories.
En effet, c'est dans la théorie des interactions faibles que resurgit le modèle de Yang et Mills du fait des difficultés rencontrées dans la théorie de Fermi et la théorie V-A. L'idée fut donnée par S.L.~Glashow et J.~Schwinger qui constatèrent l'analogie possible entre la symétrie d'isospin constituant la base de la théorie de Yang-Mills et la symétrie observée entre leptons.
Cependant, le principal obstacle à la formulation d'une théorie de jauge des interactions faibles fut encore une fois l'obligation d'avoir des interactions de courte portée.

C'est finalement le mécanisme de brisure spontanée de symétrie proposé par Peter Higgs en 1964 qui débloqua la situation et permit à Steven Weinberg (1967) et Abdus Salam (1968) de développer (de manière indépendante) une théorie unifiée des interactions électromagnétiques et faibles, dont Gerard 't~Hooft et Martinus Veltman prouvèrent en 1971 qu'elle était renormalisable. 
Ce modèle a été construit avec le groupe de structure $SU(2)\times U(1)$ et a prédit l'existence des bosons intermédiaires $W^{\pm}$ et $Z^{0}$.
Le fait de devoir décrire en même temps l'interaction faible et les interactions électromagnétiques était, lui, motivé par la nécessité de rendre compte, dans un cadre cohérent, de la violation de la parité dans laquelle sont impliquées des particules chargées telles que l'électron.
Dans ce modèle, l'électron est partenaire du neutrino et ils doivent tous deux être de masse nulle dans une formulation invariante de jauge.
La présence d'un angle de mélange, appelé angle de Weinberg,  permet alors que le mécanisme de Higgs n'attribue une masse qu'à l'électron et aux bosons intermédiaires.
Cet angle de mélange permet également de caractériser le sous groupe $U(1)$ correspondant à l'électromagnétisme, partie de la symétrie de jauge n'étant pas brisée.

Parallèlement, dans les années 1960-70, fut élaborée la théorie de la Chromodynamique Quantique. C'est une théorie de jauge construite sur le modèle de Yang et Mills avec un groupe de symétrie interne $SU(3)$ qui décrit les \textit{interactions fortes} entre quarks et gluons.
Cette théorie est dotée de deux propriétés physiques essentielles qui sont la liberté asymptotique et le confinement des quarks dans les hadrons.
Cette dernière propriété est à l'heure actuelle toujours non démontrée analytiquement mais elle semble être vérifiée par les expériences et lors de simulations numériques.

Par la suite, le modèle des interactions électro-faible et le modèle de la Chromodynamique Quantique furent affinés.
Jusqu'à la fin du  XXème siècle, une succession de prédictions et découvertes de nouvelles particules  mena à l'élaboration du \textit{modèle standard} décrivant l'ensemble des particules élémentaires connues à ce jour et les trois interactions fondamentales (interaction électro-faible et forte).
Ce modèle est une théorie de jauge avec groupe de structure $SU(2)\times U(1)\times SU(3)$.
Toutes les particules de ce modèle ont été observées ou mises en évidence à l'exception du boson de Higgs qui reste la pièce manquante.
Le mécanisme de brisure de symétrie de Higgs y apparaît comme fondamental et est rendu responsable de la masse de toutes les particules élémentaires.

Ce que nous pouvons conclure de cette analyse des théories de jauge est que les lois dynamiques de la nature semblent indépendantes de la jauge choisie, même si celle-ci est locale. Le mot jauge faisant référence à un type de mesures physiques donné.

\section*{Pourquoi la géométrie non commutative?}

D'une certaine manière, on peut considérer que la géométrie non commutative est née avec la mécanique quantique.
En effet, dans sa première formulation, qui fut celle de Heisenberg, la mécanique quantique apparaît sous une forme comparable à la mécanique Hamiltonienne où les coordonnées de l'espace des phases sont remplacées par des opérateurs qui ne commutent pas entre eux.
Dirac lui-même sembla être fasciné par cette idée et suggéra la possibilité d'interpréter la mécanique quantique dans un formalisme géométrique non commutatif.
Enfin, les travaux de von Neumann sur la mécanique quantique furent à l'origine du domaine des mathématiques que nous appelons aujourd'hui les \textit{algèbres d'opérateurs}. La théorie des algèbres de von Neumann (terme introduit par Dixmier plus tard) peut d'un certain point de vue être considérée comme une théorie de la mesure non commutative.

L'idée de remplacer un espace géométrique par une algèbre fut concrétisée assez tôt en mathématiques par la \textit{géométrie algébrique} où les points correspondent aux idéaux maximaux d'algèbres de polynômes.
Les travaux de Gelfand et Naimark dans les années 1940 firent un pas de plus en établissant un pont entre la topologie et les algèbres: ils montrèrent que les $C^{*}$-algèbres fournissent une théorie des espaces topologiques non commutatifs dans le sens où la catégorie des $C^{*}$-algèbres commutatives est équivalente à la catégorie des espaces topologiques.
Ce lien permit alors de développer des techniques analogues en topologie et en algèbre d'opérateurs et un certain nombre de notions géométriques trouvèrent leurs équivalents algébriques.
Par exemple, les fibrés vectoriels de rang fini au dessus d'un espace topologique correspondent aux modules projectifs de type fini sur les $C^{*}$-algèbres.
Cette identification permet alors de formuler la $K$-théorie sur les espaces topologiques (groupe de Grothendieck formé à partir d'une certaine classe d'équivalence sur les fibrés vectoriels) en des termes algébriques et  permet ainsi de formuler une $K$-théorie ``topologique'' pour les $C^{*}$-algèbres.
En physique, l'idée que les coordonnées de l'espace temps puissent ne pas commuter fut émise par Heisenberg en 1930 dans l'idée que cela puisse résoudre le problème des divergences ultraviolettes en théorie quantique des champs.
Cette idée fut ensuite reprise et concrétisée dans un article de Snyder en 1947.
Enfin, le terme de ``géométrie non commutative'' (GNC) fut introduit par Alain Connes dans les années 1980 comme étant un programme visant à généraliser différents concepts empruntés à la géométrie ordinaire en des concepts équivalents pour des algèbres non commutatives et en particulier les concepts venant de la géométrie différentielle.
Il montra qu'il est possible de généraliser un certain nombre de notions. 
Ainsi, l'homologie de de~Rham peut être remplacée par la cohomologie cyclique; le caractère de Chern vu comme un morphisme entre la $K$-théorie et la cohomologie de de~Rham se généralise en un morphisme entre la $K$-théorie des $C^{*}$-algèbres  et l'homologie cyclique; les théorèmes d'indice ont également leurs équivalents \dots\
La géométrie non commutative représente aujourd'hui un ensemble de techniques sur les algèbres d'opérateurs permettant de traiter des problèmes mathématiques très variés tels les représentations de groupes ou encore l'étude d'espace, considérés comme pathologiques en géométrie ordinaire (feuilletage, fractales,  etc\dots).

Du point de vue de la physique, la géométrie non commutative constitue un cadre mathématique dans lequel un certain nombre de concepts physiques peuvent être exprimés et parfois unifiés.
La remarque effectuée par  Dirac sur l'analogie entre commutateur en mécanique quantique et crochet de Poisson en mécanique Hamiltonienne peut par exemple être concrétisée par l'introduction d'une structure symplectique non commutative.
Ainsi, il fut montré par Michel Dubois-Violette~\cite{dubois-violette:rapallo:90} que le commutateur en mécanique quantique peut se comprendre dans ce cadre comme un crochet de Poisson. 
En 1985, M.  Dubois-Violette, Richard Kerner et John Madore montrèrent~\cite{dubois-violette:89,dubois-violette:89:II,dubois-violette:90,dubois-violette:90:II,dubois-violette:91} que les théories de jauge formulées sur l'algèbre des fonctions à valeurs matricielles possèdent de manière naturelle un mécanisme de brisure de symétrie de jauge analogue à celui proposé par Higgs.  
Ils utilisèrent pour cela le \textit{calcul différentiel basé sur les dérivations}~\cite{dubois-violette:88}.
Ce modèle fut ensuite généralisé par Robert Coquereaux~\cite{coqu:90,coquereaux:93} et A. Connes et J. Lott~\cite{conn-lott:90} à d'autres types de calculs différentiels.
En particuliers, il fut montré~\cite{cham-conn:96,connes:96} qu'il est possible d'exprimer le Lagrangien du modèle standard dans un cadre totalement algébrique en utilisant un calcul différentiel construit à partir d'un \textit{triplet spectral} composé d'un opérateur de Dirac, d'un espace de Hilbert et d'une algèbre.
Cette approche permet également d'incorporer les spineurs et la notion de métrique de manière naturelle~\cite{Conn:94} et de formuler un principe de moindre action~\cite{cham-conn:96}.
Il a également été fait usage des techniques de la géométrie non commutative par Jean Bellissard afin de décrire certains systèmes de physique statistique tels que les quasi-cristaux ou encore l'effet Hall quantique.
Enfin, il fut mis en évidence par Dirk Kreimer et Alain Connes que la structure du groupe de renormalisation perturbatif peut se comprendre en termes d'algèbres de Hopf (généralisation de la notion de groupe en géométrie non commutative).

Je voudrais maintenant donner quelques motivations supplémentaires pour l'utilisation de la géométrie non commutative en physique.
Il est bien connu qu'il est difficile de passer de la mécanique quantique à la mécanique classique.
Réciproquement, il ne semble pas exister de technique pour passer de manière rigoureuse et canonique d'une théorie classique à une théorie quantique; l'exemple en est la théorie de la relativité générale ou bien même la mécanique analytique.
La géométrie non commutative fournit un langage dans lequel ces deux types de théories peuvent être formulées.
En effet,  la géométrie non commutative nous apprend que les points de la mécanique classique ou de la géométrie ordinaire peuvent être considérés comme des idéaux d'algèbres commutatives.
D'autre part, aussi bien pour la mécanique quantique d'Heisenberg que pour la théorie quantique des champs, la notion d'algèbre d'opérateurs est essentielle comme cela fut mis en évidence par von Neumann.
Nous pourrions qualifier le passage de la mécanique classique à la mécanique quantique (1ère quantification) ou le passage de la théorie classique des champs à la théorie quantique des champs (2ème quantification) de transferts de niveau de réalité.
En effet, dans le premier passage, les objets fondamentaux que sont les points sont remplacés par des fonctions exprimant des relations entre points. 
De même, les relations entre fonctions peuvent être décrites à l'aide d'opérateurs comme ceux introduits dans la mécanique matricielle d'Heisenberg ou encore ceux de la théorie quantique des champs.
Ce type de passage consiste donc à considérer les relations entre objets en tant qu'objets d'un type nouveau et correspond à ce que nous pourrions appeler un transfert de niveau de réalité ou de niveau de relations.
Le fait remarquable est que les différentes structures intervenant aussi bien dans les théories classiques que dans les théories quantiques semblent pouvoir s'interpréter de manière naturelle dans un langage commun qu'est celui de la géométrie non commutative et nous pouvons espérer que le fait d'avoir un cadre commun pour décrire ces objets puisse permettre un jour de mieux comprendre les liens qui peuvent exister entre eux.

Je me suis intéressé dans cette thèse à la formulation des théories de jauge en tant que théories classiques des champs dans un langage algébrique.
Ainsi, nous verrons que les fibrés principaux, habituellement utilisés pour décrire les théories de Yang-Mills, peuvent être remplacés par des algèbres d'endomorphismes.
Ce type d'algèbres généralise de manière directe les algèbres de fonctions à valeurs matricielles introduites par Dubois-Violette, Kerner et Madore.
De manière générale, une algèbre d'endomorphismes sera considérée comme étant l'algèbre des sections du fibré des endomorphismes associé à un fibré vectoriel.
L'étude de ces algèbres s'avère être intéressante du point de vue de la géométrie non commutative dans le sens où un certain nombre de notions utilisées en géométrie différentielle dans le cadre des fibrés principaux  peuvent être traduites dans un langage algébrique.
Du point de vue de la physique, le groupe de jauge\footnote{groupe des transformations locales de symétrie.} semble jouer un rôle plus fondamental que le groupe de structure\footnote{groupe des transformations ``rigides'' de symétrie que nous avons appelé \textit{groupe de symétrie interne} jusqu'à présent. C'est en général un groupe de Lie compact de dimension finie.} du fait que c'est son action qui génère les couplages minimaux entre champs de jauge et champs de matière.
Ainsi, l'invariance d'une théorie vis à vis du groupe de jauge peut être associée au caractère local des interactions et est reliée à des lois de conservation locales.
Le groupe de jauge intervient également de manière essentielle en théorie quantique des champs (algèbres de courants, symétrie B.R.S.T., etc \dots).
Ainsi, nous pouvons considérer ce groupe comme étant le ``vrai'' groupe de symétrie.
Je voudrais montrer que de ce point de vue, il semble plus naturel d'utiliser des algèbres d'endomorphismes pour modéliser les théories de jauge plutôt que des fibrés principaux.
Afin d'illustrer ce propos, nous pouvons faire une brève description de quelques notions pouvant être introduites dans ces deux cadres:
\begin{itemize}
\item 
  Un fibré principal est généralement construit à partir d'un groupe de   structure $G$.  
  Le groupe de jauge apparaît alors comme le groupe des   automorphismes verticaux de ce fibré.  
  L'algèbre de Lie du groupe de jauge   correspond aux champs de vecteurs verticaux.  
  Grâce aux connexions sur ce fibré, on peut alors construire un morphisme qui associe à   tout polynôme invariant sur $\kg$, l'algèbre de Lie de $G$, une classe   caractéristique qui est un élément de la cohomologie de de~Rham de la   variété de base.  
  Ce morphisme est appelé le morphisme de Chern-Weil. 
\item  
  Pour une algèbre d'endomorphismes $A$, le groupe de jauge correspond au   groupe des automorphismes intérieurs de l'algèbre et est donc décrit   directement par les éléments de l'algèbre.
  L'algèbre de Lie du groupe de   jauge correspond aux dérivations intérieures de l'algèbre $A$.  
  Le morphisme de   Chern-Weil peut se construire directement en considérant les éléments   invariants d'un certain module sur l'algèbre de Lie des dérivations de $A$. 
\end{itemize}
On voit ainsi que l'algèbre des endomorphismes donne une description plus directe des objets qui nous intéressent en physique.
Un des buts de cette thèse est de montrer comment peut s'effectuer ce changement de point de vue entre fibrés principaux et algèbres d'endomorphismes.

On verra également que le fait de formuler les théories de jauge dans le cadre des algèbres d'endomorphismes permet de considérer des généralisations des théories de jauge usuelles et d'explorer de nouveaux mécanismes de brisure de symétrie.

\section*{Pourquoi la théorie de Born-Infeld ?}
Dans la dernière partie de cette thèse, je décrirai certaines théories des champs non linéaires généralisant la théorie de Born-Infeld à des champs de jauge non abéliens possédant ou ne possédant pas de mécanisme de brisure de symétrie.

Ces généralisations sont principalement motivées  par les théories de cordes.
Cette théorie est une théorie conforme bidimensionnelle dont les champs sont les coordonnées de plongement de la feuille d'univers (bidimensionnelle) d'objets étendus appelés cordes.
Le groupe de Poincaré devient alors le groupe de symétrie interne des coordonnées de plongement.
L'invariance de jauge, c'est à dire l'invariance sous difféomorphismes de ces coordonnées, est retrouvée au niveau des théories effectives de cordes.
Ceci rend alors possible de décrire d'une certaine façon la gravitation dans un formalisme quantique cohérent.
Les symétries de jauge, telles que nous les avons décrites auparavant, semblent émerger d'une toute autre manière.
En effet, cette théorie possède des solutions de type solitons associées aux modes de masse nulle de la théorie.
Ces solutions peuvent être décrites par des objets étendus, appelés $D$-branes, correspondant à des conditions au bord de type Dirichlet imposées aux cordes.
Les champs de jauge, tels que nous les connaissons, apparaissent alors comme décrivant les fluctuations quantiques de ces solutions. 
Ils peuvent être reliés par $T$-dualité aux coordonnées transverses de ces $D$-branes.
Sous certaines approximations, la dynamique de ces objets peut être décrite par un élément de volume généralisé correspondant à l'action de Born-Infeld introduite dans les années 1920 pour l'électromagnétisme couplé à la gravitation\footnote{Un rappel sur l'histoire de la théorie de Born-Infeld sera fait dans le chapitre \ref{chap:theorie-de-born}.}.
Ces solutions possèdent des propriétés particulières (solutions BPS super-symétriques, \textit{etc}) et il est ainsi possible de considérer des configurations où $N$ $D$-branes se superposent.
 La dynamique de ces configurations est alors décrite par des champs de jauge non abéliens associés au groupe de structure $U(N)$, le facteur $N$ correspondant aux facteurs de Chan-Paton associés à chaque extrémité des cordes.
Une des questions ouvertes en théorie des cordes est de savoir quel type d'action effective pour ces champs de jauge non abéliens vient remplacer l'action de Born-Infeld.
Il est proposé dans cette thèse certaines généralisations de l'action de Born-Infeld pour des champs de jauge non abéliens considérés comme décrivant une connexion non commutative sur un module libre de rang $1$. 
Une telle généralisation semble naturelle dans le cadre de la géométrie non commutative des fonctions à valeurs matricielle dans laquelle l'action de ``Maxwell'' non commutative pour une connexion non commutative particulière correspond à l'action de Yang-Mills.

Il est intéressant de noter qu'un certain type de géométrie non commutative émerge de manière naturelle en théorie des cordes.
Cette non commutativité se manifeste par la présence d'un champ de fond qui est une $2$-forme, souvent notée $B$, et vient s'ajouter à la métrique dans l'action de Born-Infeld.
 La correspondance découverte par Seiberg et Witten~\cite{seib-witt:99} établit que la dynamique des $D$-branes dans un tel champ de fond peut être décrite à l'aide d'une théorie des champs non commutative de type Moyal, où la $2$-forme $B$ exprime la non commutativité des coordonnées d'une $D$-brane.

Il n'est pas dans le but de cette thèse d'étudier ce type de géométrie non commutative, mais l'on peut cependant noter que les techniques développées pour les algèbres d'endomorphismes pourraient être utiles dans le contexte de la théorie des cordes.
En effet, il fut reconnut \cite{minasian:97,witten:98,bouwknegt:00} que la charge d'une $D$-brane, associée à la $2$-forme $B$, correspond à un certain élément de K-théorie ``twistée''. Cet invariant correspond en fait à une classe de Dixmier-Douady qui est une classe de cohomologie de de~Rham de degré $3$ (construite ici à partir de la $2$-forme $B$) et qui fut introduite par Dixmier et Douady en 1963~\cite{dixm-doua:63} afin de classifier les champs continus de $C^{*}$-algèbres.
Ce type d'algèbres généralisent dans certaines situations les algèbres d'endomorphismes qui seront étudiées dans cette thèse.

\section*{Plan de la thèse et description des chapitres}

Dans la première partie de cette thèse, je fais la description de certaines méthodes générales de la géométrie non commutative.
J'ai rédigé ce chapitre dans l'idée de faire un exposé pédagogique sur des techniques de base que sont l'homologie de Hochschild et l'homologie cyclique. 
J'ai également voulu faire une synthèse des différents points de vue adoptés dans la littérature sur la notion de calcul différentiel universel.
Enfin, je présente le calcul différentiel basé sur les dérivations qui nous sera utile pour la formulation des théories de jauge en géométrie non commutative.

Dans la deuxième partie, je montre comment le concept de connexion utilisé en géométrie ordinaire peut être généralisé à des structures non commutatives.
Je fais également une présentation des algèbres d'endomorphismes introduites dans~\cite{dubois-violette:98} et tente d'éclaircir leurs liens avec les fibrés principaux.
Cet exposé reflète une partie de mes travaux effectués dans \cite{masson-serie:04}.
A la fin de ce chapitre, je présente un travail encore non publié, susceptible de développements ultérieurs, sur la manière de construire le morphisme de Chern-Weil à partir d'une algèbre d'endomorphismes.
Cette construction repose sur l'adaptation d'une méthode développée par Lecomte~\cite{lecomte:85} dans le cadre des algèbres de Lie.
Cette construction permet de jeter un nouveau regard sur les classes caractéristiques.

La partie 3 correspond aux travaux que j'ai effectués dans~\cite{masson-serie:04} et traite des réductions de fibrés principaux et d'algèbres d'endomorphismes.
Il est fait une synthèse des différents travaux connus traitant de réductions sur les fibrés principaux et de la caractérisation des connexions ordinaires invariantes.
Il est ensuite montré comment ces résultats peuvent être généralisés dans le cadre des algèbres d'endomorphismes pour caractériser les connexions non commutatives invariantes sous l'action d'un groupe de Lie compact.
En particulier, cela illustrera le fait que des résultats obtenus sur les fibrés principaux peuvent s'exprimer en terme d'algèbres d'endomorphismes et que l'utilisation d'opérations algébriques (par exemple la dérivée de Lie par rapport à une dérivation intérieure) simplifie certaines considérations.
Il est développé deux exemples, dont un qui permet de généraliser l'ansatz de Witten utilisé dans les théories de Yang-Mills pour caractériser les connexions $SU(2)$ à symétrie sphérique.

Dans la quatrième partie, je présente des modèles de théories de jauge pouvant être construits dans le cadre des algèbres d'endomorphismes.
Dans un premier temps, je fais un rappel sur les modèles construits avec des algèbres d'endomorphismes correspondant à la situation des fibrés triviaux.
Cela correspond essentiellement à la description du modèle de Dubois-Violette, Kerner, Madore~\cite{dubois-violette:90:II} et ses raffinements exposés dans~\cite{dubois-violette:japan99}.
Cela nous permettra de mettre en évidence le mécanisme de brisure de symétrie présent dans ce type de théories.
Dans un second temps, je présente des travaux originaux non encore publiés.
Je montre comment le concept de structure Riemannienne peut être introduit sur l'algèbre des endomorphismes et j'essaie de clarifier la définition donnée dans~\cite{masson:99}.
Cela permet également d'approfondir, dans le cas particulier des algèbres d'endomorphismes, la notion de structure Riemannienne introduite dans \cite{dubo-mich:95}.
Je fais ensuite une présentation originale des théories de Kaluza-Klein où il est fait usage de l'algèbre des endomorphismes plutôt que du fibré principal habituellement utilisé.
Enfin, je construit l'action de ``Maxwell'' non commutative généralisant le modèle de Dubois-Violette, Kerner, Madore pour des algèbres non triviales.
Il est montré comment le mécanisme de brisure de symétrie est modifié dans ce cas. 

Dans la  cinquième partie, il est fait une synthèse de travaux originaux publiés~\cite{serie:03,serie:04} durant la thèse ainsi qu'un rappel historique sur la théorie de Born-Infeld.
Il est proposé une généralisation de l'action de Born-Infeld pour des champs de jauge non abéliens ainsi qu'une comparaison avec d'autres généralisations possibles.
Cette action est ensuite étendue aux connexions non commutatives introduites dans le chapitre 2.
Enfin, il est fait une étude numérique de solutions, tout d'abord dans le cas de l'action de Born-Infeld pour des champs de jauge non abéliens avec groupe $SU(2)$, permettant ainsi d'obtenir des solutions du même type que celles obtenues par Bartnik et McKinnon~\cite{bartnik:88} en relativité générale  (solutions du type sphaleron utilisées pour décrire certains types de trous noirs), puis dans le cas de connexions non commutatives ne comportant qu'une partie de type  champ scalaire.
Enfin, je présente une étude sur le couplage d'un champ scalaire à une métrique de Friedmann-Robertson-Walker en relativité générale.


\chapter{Géométrie non commutative}

Pour pouvoir formuler les théories de jauge dans un cadre algébrique, il est nécessaire de faire quelques rappels sur les principaux outils utilisés en géométrie non commutative.
Ainsi, le but de ce chapitre est de clarifier la notion de calcul différentiel sur une algèbre.

Tout d'abord, différentes notions de calcul différentiel universel seront données.
Nous montrerons, ensuite, comment ces notions interviennent en (co)homologie de Hochschild et en (co)homologie cyclique qui sont des outils essentiels de la géométrie non commutative.
Enfin, nous aborderons le calcul différentiel basé sur les dérivations.

\section{Algèbres d'opérateurs}
Nous allons donner quelques définitions de base sur les différents types d'algèbres que nous utiliserons par la suite et énoncer le théorème de Gelfand-Naimark qui fait le pont entre topologie et algèbre.
Ce théorème établit une équivalence entre la catégorie des espaces topologiques et la catégorie des $C^{*}$-algèbres commutatives.
Cette correspondance est essentielle pour comprendre comment différentes notions de géométrie peuvent être traduites en des notions algébriques équivalentes puis être généralisées à des algèbres non commutatives.
L'ensemble des méthodes algébriques ainsi obtenues contribue à  former ce qu'on appelle communément  la géométrie non commutative.

\subsection{\texorpdfstring{$C^{*}$-Algèbres}{C*-Algèbres}}
\label{sec:star-algèbres}

\begin{defn}
  Une $C^{*}$-algèbre est une algèbre de Banach involutive $\cal{A}$ telle que pour tout $a,b \in \cal{A}$, on ait:
  \begin{align}
    &    \norm{ab} \leq \norm{a}\norm{b}\\
    &    \norm{a^{*}a}=\norm{a}^{2}\ .
  \end{align}
\end{defn}

\begin{thm}[Gelfand-Neimark]
  Toute $C^{*}$-algèbre avec unité est isomorphe à une algèbre de fonctions $C(X)$, où $X$ est un espace de Hausdorff compact.
\end{thm}
Cet isomorphisme se construit de la manière suivante: soit $A$ une $C^{*}$-algèbre commutative.
Alors, $X=\Delta(A)$,  l'espace des caractères de $A$ (l'ensemble des homomorphismes de $A $ vers $\gC$) est un espace de Hausdorff localement compact.
L'application $A \to C(X) : a \mapsto [\hat{a}: \omega \mapsto \omega(a)]$, appelée transformation de Gelfand et définie de manière plus générale pour les algèbres de Banach, est un isomorphisme lorsque $A$ est une $C^{*}$-algèbre. 

Ce théorème peut s'étendre aux algèbres sans unité et nous pouvons l'énoncer de la manière suivante:
\begin{thm}[Gelfand-Neimark II]
  La catégorie dont les objets sont les $C^{*}$-algèbres commutatives et les morphismes les $*$~-~homomorphismes est duale de la catégorie dont les objets sont les espaces de Hausdorff localement compacts et les morphismes les applications continues propres.
\end{thm}

\subsection{Algèbres topologiques localement convexes}
Nous avons vu que la catégorie des $C^{*}$-algèbres est à la géométrie non commutative ce qu'est la topologie à la géométrie ordinaire.
De même, les algèbres localement convexes généralisent les $C^{*}$-algèbres et semblent être les objets naturels à étudier dans ce que l'on pourrait appeler la ``géométrie différentielle non commutative''.
En particulier, la catégorie des algèbres localement convexes contient les algèbres de Fréchet obtenues du calcul différentiel sur les variétés (par exemple, l'algèbre des fonctions $C^{\infty}$ sur une variété). Elle contient également des algèbres définies de manière purement algébrique comme par exemple les algèbres ayant une base dénombrable.
Beaucoup d'exemples d'algèbres non commutatives associées à des structures différentiables sont également dans cette catégorie, comme les algèbres d'opérateurs (pseudo-)différentiels, les algèbres construites à partir de formes différentielles, les déformations d'algèbres de fonctions différentiables, etc $\dots$

Les outils développés dans ce cadre sont essentiellement la K-théorie topologique et sa version bivariante qui a été développée pour les $C^{*}$-algèbres par Kasparov~\cite{kasp:80,kasp:83} et pour les algèbres topologiques localement convexes par Cuntz~\cite{cuntz}, ainsi que l'homologie cyclique qui fut découverte par Connes~\cite{Conn:85}.
Comme nous le verrons dans la section suivante, l'homologie cyclique peut être définie dans un cadre purement algébrique.
Ces outils permettent de généraliser des constructions obtenues en géométrie ordinaire telles que les théorèmes de l'index ou encore la notion de caractère de Chern.

Les algèbres que nous allons rencontrer par la suite seront essentiellement des algèbres topologiques localement convexes pouvant être obtenues à partir d'algèbres des fonctions $C^{\infty}$ sur une variété.
Ces algèbres sont obtenues à l'aide de familles de seminormes et en voici une définition:
\begin{defn}
  Une algèbre localement convexe est une algèbre $\cA$ sur $\gC$ équipée d'une topologie localement convexe pour laquelle la multiplication est continue.
De fait, la topologie sur $\cA$ est définie par une famille $\cP$ de seminormes, telle que pour tout $p \in \cP$, il existe $q \in \cP$ tel que:
  \begin{align}
    &    p(xy)\leq q(x)q(y) &
    & \forall x,y \in \cA \ .
  \end{align}
\end{defn}

Nous allons maintenant passer à l'étude des algèbres associatives.
Nous ne supposerons pas que les algèbres sont munies d'une norme ou d'une topologie, bien que cela soit le cas dans la plupart des situations pratiques que nous allons rencontrer.
Les techniques que nous allons développer peuvent s'adapter dans la plupart des cas aux $C^{*}$-algèbres et aux algèbres topologiques localement convexes.

\section{Calculs différentiels}
\label{sec:calculs-différentiel}
\subsection{Calcul différentiel universel}
\label{sec:calc-diff-univ}

 \begin{defn}
  Une {\bf  algèbre différentielle graduée}  $(\cA^{*},\delta)$ est une algèbre graduée, $\cA^{*}=\bigoplus_{n \geq 0} \cA^{n}$ munie d'un produit:
  \begin{equation}
    \begin{aligned}
      \cA^{n} \times \cA^{m} &\to \cA^{n+m}\\
      (a,b) &\mapsto ab
    \end{aligned}
  \end{equation}
  et d'un  morphisme $\delta: \cA^{n} \to \cA^{n+1}$ de degré $+1$ satisfaisant la \textit{règle de Leibnitz graduée}:
    \begin{equation}
      \begin{aligned}
        &   \delta(ab)=(\delta a) b +(-1)^{n}a (\delta b) &
        &\forall a \in \cA^{n}, b \in  \cA^{m} \\
      \end{aligned}
    \end{equation}
    et satisfaisant $\delta^{2}=0$.\\
    Lorsqu'il n'y aura pas d'ambiguïté, nous noterons $\cA$ ou $\cA^{*}$ au lieu de $(\cA^{*},\delta)$.
    En particulier,  $\cA^{0}$ est une algèbre et nous appellerons parfois  $(\cA^{*},\delta)$ un calcul différentiel sur ${\cA^{0}}$ ou tout simplement un calcul différentiel lorsqu'il n'y aura pas d'ambiguïté sur l'algèbre $\cA^{0}$.
  \end{defn}
  
 Soit $A$ une algèbre associative sur un corps $k$ (on prendra $\gC$ dans la plupart des cas).
 Nous pouvons associer à cette algèbre un {\bf calcul différentiel universel} $(\Omega(A),d)$ défini par la propriété universelle suivante:  \begin{prop}
   Tout morphisme $\psi~:~A~\to~C^{0}$ de $A$ dans une algèbre différentielle graduée $(C^{*},\delta)$ s'étend en un unique morphisme d'algèbres différentielles graduées $\bar{\psi}: (\Omega(A),d) \to (C,\delta)$.  
\label{def:calc-diff-univ}
\end{prop}
L'existence de l'algèbre différentielle graduée $\Omega(A)$ se montre facilement en donnant une construction explicite.
Nous pouvons par exemple considérer l'algèbre différentielle graduée libre\footnote{c'est-à-dire n'ayant d'autres relations constitutives que celles d'une algèbre différentielle graduée.} générée par les symboles  $a \in \Omega^{0}=A$. 
On peut vérifier qu'elle est solution au problème universel~\ref{def:calc-diff-univ} et est donc isomorphe à $\Omega(A)$.
En effet, les espaces  $\Omega^{n}(A)$ sont donc générés  par les combinaisons linéaires d'éléments de la forme $x_{0}dx_{1}\dots dx_{n}$ et $dx_{1}\dots dx_{n}$, avec $x_{i} \in A$. 
Alors pour tout morphisme $\psi : A \to C^{0}$ de $A$ dans une algèbre différentielle graduée $(C,\delta)$, on définit le morphisme $\bar{\psi}$ de $(\Omega(A),d)$ dans $(C,\delta)$ par:
\begin{align}
  \bar{\psi}( a_{0} da_{1}\dots da_{n} + db_{1} \dots db_{n} ) &= \psi(a_{0}) \delta(\psi(a_{1})) \dots \delta(\psi(a_{n})) +  \delta(\psi(b_{1})) \dots \delta(\psi(b_{n}))
\end{align}
pour $a_{i}, b_{i} \in A$.

Nous pouvons également donner une construction explicite de cette algèbre différentielle graduée à partir de $A$.
Pour cela, considérons le complexe défini par:
\begin{equation}
\Omega^{n}(A) =
\begin{cases}
  A & \text{pour $n=0$} \\ 
  \tilde{A}\otimes A^{\otimes n}  \simeq  A^{n+1} \oplus A^{n}  & \text{pour $n \geq 1$} 
\end{cases} \ ,
\label{Omega}
\end{equation}
où $\tilde{A}=  A \oplus k$ est l'extension de $A$ par $k$.
Ainsi, on a $\tilde{A}\otimes A^{\otimes n} \simeq A^{n+1} \oplus (k \otimes A^{n}) \simeq  A^{n+1} \oplus A^{n}$ et $\Omega(A)=\oplus_{n=0}^{\infty} (A^{n+1} \oplus A^{n})$.

Afin d'obtenir une algèbre différentielle graduée, on définit une structure de $A$-bi\-module, un produit et une différentielle.
La différentielle $d: \Omega^{n}(A) \to \Omega^{n+1}(A)$ est donnée par :
\begin{equation}
  \begin{cases}
    da = 1 \otimes a & n=0 \\
    d((a_{0} + \lambda_{0} 1)\otimes a_{1} \otimes \dots  \otimes a_{n})= 1 \otimes a_{0} \otimes \dots \otimes a_{n} & n \geq 1
  \end{cases} \ .
\end{equation}
Ainsi, par construction, on obtient $d^{2}=0$.
La structure de $A$-module à gauche est donnée par l'action:
\begin{align}
  a ( \tilde{a_{0}} \otimes a_{1} \otimes \dots \otimes a_{n})=  (a\tilde{a_{0}}) \otimes a_{1} \otimes \dots \otimes a_{n} \ .
\end{align}
La structure de $A$-module à droite est donnée par l'action:
\begin{align}
  ( \tilde{a_{0}} \otimes a_{1} \otimes \dots \otimes a_{n}) a = \sum_{i=0}^{n} (-1)^{n-i} \tilde{a_{0}} \otimes a_{1} \otimes \dots \otimes a_{i}a_{i+1} \otimes \dots \otimes a \ ,
\label{eq:action-droite}
\end{align}
où on pose $a_{n+1}=a$. 
On vérifie que l'on a bien $(\omega a)b =\omega (a b)$ pour tout $\omega \in \Omega^{n}(A)$ et $a,b \in A$.
On voit que cette action à droite sert à reproduire la règle de Leibnitz graduée et que les différents termes de la somme dans (\ref{eq:action-droite}) correspondent à diverses ``intégrations'' par partie. 
Cette action à droite s'étend  en une action unitaire de $\tilde{A}$ et nous permet de définir le produit:
\begin{equation}
\begin{aligned}
  \Omega^{n}(A) \times  \Omega^{m}(A) &\to \Omega^{m+n}(A)  \\
  (\omega , \tilde{a_{0}} \otimes a_{1} \otimes \dots \otimes a_{m}) &\mapsto  \omega \tilde{a_{0}} \otimes a_{1} \otimes \dots \otimes a_{m} 
\end{aligned} \ .
\end{equation}
Ce produit est associatif et vérifie la règle de Leibnitz graduée $d(\omega a)= d\omega \cdot  a + (-1)^{n} \omega \cdot da$, pour tout $\omega \in \Omega^{n}(A)$ et $a \in A$.
Ainsi, on a:
\begin{align}
  \tilde{a_{0}}da_{1}\dots da_{n} =   \tilde{a_{0}} \otimes a_{1}\dots \otimes a_{n} \ ,
\end{align}
ce qui montre que  $\Omega(A)$ est générée par $A$ en tant qu'algèbre différentielle graduée.

On doit maintenant vérifier la propriété universelle~\ref{sec:calc-diff-univ}.\\
Prenons un morphisme $\psi : A \to C^{0}$ de $A$ dans une algèbre différentielle graduée $(C^{*},\delta)$, on définit alors le morphisme $\bar{\psi}$ de $(\Omega(A),d)$ dans $(C,\delta)$ par:
\begin{align}
  \bar{\psi}( \tilde{a_{0}} da_{1}\dots da_{n} ) &= \psi(a_{0}) \delta(\psi(a_{1})) \dots \delta(\psi(a_{n})) + \lambda_{0} \delta(\psi(a_{1})) \dots \delta(\psi(a_{n})) \ ,
\end{align}
pour $a_{i} \in A$,$\tilde{a_{0}} \in \tilde{A}$ et $ \tilde{a_{0}} = a_{0} \oplus \lambda_{0}$.
Ainsi, l'algèbre différentielle graduée que nous venons de définir est isomorphe à $\Omega(A)$.
L'isomorphisme avec l'algèbre différentielle graduée libre engendrée par $A$ est donné, au niveau des espaces vectoriels, par: $ a_{0}da_{1}\dots da_{n} \oplus  da_{1}\dots da_{n} \cong \tilde{a_{0}} \otimes a_{1}\dots \otimes a_{n} \in \Omega^{n}(A)$, où $ \tilde{a_{0}} = a_{0} \oplus 1$.

Cette définition de $\Omega(A)$ fournit une représentation du calcul universel utile dans certains contextes, comme pour les calculs de l'homologie de Hochschild et de l'homologie cyclique.

\subsubsection{Version catégorielle}
Afin de pouvoir établir une notion de calcul différentiel universel pour d'autres types d'algèbres, il est utile et intéressant de formuler la propriété universelle de $\Omega(A)$ dans le langage des catégories.

On considère la catégorie des algèbres différentielles graduées ayant pour objets les algèbres différentielles graduées et pour flèches les morphismes d'algèbres différentielles graduées. 
Notons $\cO$ le foncteur d'oubli qui à toute algèbre différentielle graduée $(C^{*},\delta)$ associe l'algèbre $C^{0}$ en degré $0$.
Alors, dans la terminologie de Mac~Lane~\cite{maclane}, à toute algèbre $A$, on peut associer une flèche universelle  $\langle \Omega(A),u(A)\rangle $ de A vers $\cO$, que l'on appellera calcul différentiel universel. 
Le couple $\langle \Omega(A),u(A)\rangle $, où $\Omega(A)$ est une algèbre différentielle graduée et $u(A): A \to \Omega^{0}(A)=\cO(\Omega(A))$, est une flèche universelle de A vers $\cO$, c'est-à-dire que pour tout couple $\langle C,\rho\rangle $ avec $C$ une algèbre différentielle graduée et $\rho$ un morphisme  d'algèbre de $A$ dans $C^{0}=\cO(C)$, il existe un unique morphisme d'algèbres différentielles graduées $\bar{\rho}: \Omega(A)\to C$ tel que $\cO(\bar{\rho}) \circ u(A) = \rho$.
En d'autres termes, toute flèche de $A$ vers $\cO$ se factorise de manière unique par la flèche $u(A)$ comme dans le diagramme commutatif suivant:
\begin{align}
  \xymatrix{
    \Omega(A) \ar^-{\bar{\rho}}@{-->}[d] &  A \ar^-{u(A)}[r] \ar^-{\rho}[d] & \Omega^{0}(A) \ar^-{\cO \bar{\rho}}@{-->}[dl] \\
    C & C^{0}&
} \ .
\end{align}
Une manière de voir que la flèche $\langle \Omega(A),u(A)\rangle $ est unique à isomorphisme est de la considérer comme un objet initial dans la catégorie comma\footnote{La catégorie comma $(A \downarrow \cO)$ est la catégorie des objets ``$\cO$-sous'' $A$, \textit{i.e.} des couples $\langle f,C \rangle$ avec $C$ une algèbre différentielle graduée et $f$ un morphisme d'algèbre $f:A\to \cO C=C^{0}$. Les flèches $h:\langle f,C \rangle \to \langle f',C' \rangle$ sont données par les morphismes d'algèbres différentielles graduées pour lesquels $f'=\cO h \circ f$. Je renvoie le lecteur à~\cite{maclane} pour une approche plus systématique des catégories.}%
 $(A \downarrow \cO)$.

\begin{rem}
  Dans notre situation, $u(A)$ est la flèche identité.
\end{rem}

Cette caractérisation nous permet de définir, de manière plus générale, la notion de calcul différentiel universel associé à une catégorie d'algèbres différentielles graduées et un foncteur d'oubli vers une catégorie d'algèbres.
La définition d'un calcul différentiel universel reste alors formellement inchangée à la définition précédente mais nous devons prendre garde au fait que les morphismes et les objets satisfont des propriétés supplémentaires.
Par exemple, on peut faire une telle construction pour les algèbres différentielles graduées avec unité, ou bien commutatives avec unité, ou encore centrales \dots\ 
 Nous verrons plusieurs exemples dans les chapitres suivants.
Pour les algèbres différentielles graduées commutatives avec unité, on peut montrer que l'on obtient ainsi le calcul différentiel de Kähler (version algébrique du calcul de de~Rham).
Le cas des  algèbres différentielles graduées avec unité est étudié dans la section suivante et celui des algèbres différentielles graduées centrales dans la section~\ref{sec:bimodules-centraux}.

\begin{rem}
  La propriété universelle de $\Omega(A)$ fait de $\Omega$ un foncteur adjoint à droite au foncteur d'oubli $\cO$.
  Bien que l'on puisse définir la notion de calcul différentiel universel dans bien des situations, il faut s'assurer à chaque fois de l'existence de ce foncteur.
  Dans la plupart des cas, pour une algèbre $A$ dans une certaine catégorie d'algèbres, il suffira de considérer pour $\Omega(A)$ l'algèbre différentielle graduée libre générée par $A$ dans la catégorie d'algèbre différentielle graduée correspondante.
\end{rem}

\subsection{Calcul différentiel universel avec unité}

Lorsque $A$ possède une unité, il est naturel de considérer la catégorie des algèbres différentielles graduées avec unité dont les morphismes sont les morphismes d'algèbres différentielles préservant l'unité.
En particulier, pour une algèbre différentielle graduée $(C,\delta)$ avec unité,  $\delta 1=0$.
Ainsi, on a un foncteur d'oubli $\cO$  qui va de la catégorie des algèbres différentielles graduées avec unité dans la catégorie des algèbres associatives avec unité.
Le calcul différentiel universel avec unité est défini comme étant la flèche universelle de $A$ vers $\cO$.

On construit $\Ou(A)$ en considérant l'algèbre différentielle graduée avec unité libre générée par les symboles  $a \in \Omega^{0}=A$.
En tant que complexe de $A$-modules à gauche, $\Ou(A)$  est isomorphe  au complexe gradué $\bigoplus_{n\geq 0} A\otimes \bar{A}^{\otimes n}$, c'est-à-dire:
\begin{equation}
  \Omega^{n}(A) \simeq
  \begin{cases}
    A & \text{pour $n=0$} \\ 
    A\otimes \bar{A}^{\otimes n}  & \text{pour $n \geq 1$} 
  \end{cases} \ ,
\end{equation}
où $\bar{A}= A/k$.
L'isomorphisme est donné par $a_{0} da_{1}\dots da_{n}\mapsto a_{0} \otimes \bar{a_{1}} \dots \otimes \bar{a_{n}} $.
La propriété universelle est vérifiée immédiatement de manière analogue à celle de $\Omega(A)$.

Contrairement à la situation précédente, il existe une représentation de $\Ou(A)$ dans l'algèbre tensorielle sur $k$ (le corps de base de l'algèbre $A$), $TA$.
En effet, l'application
\begin{equation}
\begin{aligned}
  i:  &A\otimes A &
&\to A \otimes A \\
& x\otimes y  &
&\mapsto x (1\otimes y -y \otimes 1)
\end{aligned}
\end{equation}
a pour noyau $A\otimes k 1$ et est donc factorisée par la projection $\pi:A\otimes A  \to A\otimes \bar{A} $ comme dans le diagramme commutatif suivant:
\begin{equation}
\begin{aligned}
  \xymatrix{
A\otimes A \ar^-{\pi}[d] \ar^-{i}[r] & A \otimes A\\
A\otimes \bar{A} \ar@{-->}_{\bar{i}}[ur]&
} \ .
\end{aligned}
\end{equation}
Le produit dans l'algèbre, $\mu: A\otimes A\to A$, a un noyau généré par les éléments de la forme $1\otimes a -a \otimes 1$ et on a donc la suite exacte:
\begin{align}
  \xymatrix@R=2pt@M=6pt{
    0 \ar[r] &\Ou^{1}(A) \simeq A\otimes \bar{A} \ar^-{\bar{i}}[r] & A  \otimes A \ar^-{\mu}[r] & A \ar[r] & 0 \ .
  }
\label{eq:suite-exacte-omega}
\end{align}
Ainsi, on a l'identification,  $\Ou^{1}(A) \simeq \Im \bar{i} \simeq \Ker \mu$,  et comme   $\Ou(A)$  est générée par $A$, on a $\Ou^{n}(A)\simeq \Ou^{1}(A) \otimes_{A} \dots \otimes_{A} \Ou^{1}(A) \subset T^{n}A$ et le calcul différentiel universel $\Ou(A)$ est un sous-complexe du complexe gradué $TA$.
Ce complexe est étudié en détail dans~\cite{masson:95}.

\begin{rem}
  Nous savons qu'à toute algèbre $A$ nous pouvons associer l'algèbre $\tilde{A}=A \oplus k $ avec unité. 
  On pourrait donc dire qu'il est toujours possible de se ramener au calcul différentiel universel avec unité. 
  En effet, le complexe $\Ou(\tilde{A})$ coïncide avec $\Omega(A)$ en degré $n\geq1$, mais en degré $0$, nous avons $\Omega^{0}(A)=A$, tandis que $\Ou(\tilde{A})=\tilde{A}$. 
  De ce fait, $\Omega(A)$ est appelée complexe réduit de $\Ou(\tilde{A})$.
  Selon le contexte, on aura plutôt intérêt à travailler avec l'un ou l'autre de ces calculs différentiels. 
  
  Par exemple, dans le cadre de l'homologie cyclique, il est plus facile de travailler avec l'algèbre différentielle graduée $\Omega(A)$ (\textit{c.f.} section~\ref{sec:homologie-cyclique}). 
  
  D'autre part, on voit que $\Ou(A)$ admet une représentation comme sous algèbre de l'algèbre tensorielle $T(A)$ du fait qu'elle a une unité et que dans cette représentation, la structure de bimodule est plus simple. 
  Cette représentation a également d'autres avantages dans certains cas, comme nous le verrons plus loin.
\end{rem}

\begin{rem}
  L'algèbre $\Ou(A)$ étant générée par $A$, elle est nécessairement un quotient de $\Omega(A)$.
  On peut exhiber ce quotient  dans la  suite exacte courte suivante:
  \begin{align}
    \xymatrix{
      0 \ar[r] & M^{*} \ar[r] &\Omega(A) \ar[r] & \Ou(A) \ar[r] &  0 \ ,
  }
\end{align}
où $M^{*}$ est l'idéal engendré par les éléments de la forme $a_{0} da_{1} \dots d1 \dots da_{n}$, $da_{1} \dots d1 \dots da_{n}$ et $1da_{1} \dots  da_{n} -  da_{1} \dots  da_{n} $. 
\end{rem}

\section{Homologies et cohomologies}
\subsection{Homologie de Hochschild}

\label{sec:cohom-de-hochsch}
L'homologie de Hochschild  est définie de manière générale pour des algèbres associatives avec unité et un bimodule sur cette algèbre. 
Dans certains cas, nous expliquerons comment passer aux algèbres sans unité en se servant des propriétés fonctorielles de l'homologie de Hochschild.

L'homologie de Hochschild est caractérisée par une collection de  bifoncteurs notés $H_{n}$ qui à toute algèbre $A$ avec unité et tout $A$-bimodule $M$ associent des groupes abéliens $H_{n}(A,M)$.
Du point de vue de l'algèbre homologique~\cite{vermani}, $H$ est tout simplement défini par $H_{n}(A,M)= \Tor^{A^{e}}_{n}(A,M)$, où $A^{e}=A\otimes A^{op}$ est l'algèbre enveloppante et $A^{op}$ est  l'algèbre opposée%
\footnote{L'algèbre opposée de $A$ est l'algèbre générée par les éléments de $A$ avec le produit opposé $(a,b)\mapsto ba$.}
de $A$.
Le foncteur $\Tor$ est le foncteur dérivé du foncteur produit tensoriel $(\cdot \otimes\cdot)$.
Je renvoie à l'appendice~\ref{cha:algebre-homologique} pour plus de détails sur la construction de ce foncteur.
Ce point de vue, bien que formel, permet dans certains cas de calculer l'homologie de Hochschild grâce à des résolutions projectives (voir appendice~\ref{cha:algebre-homologique}) de modules et fournit donc une méthode de calcul efficace.

Nous allons tout d'abord donner une construction heuristique de l'homologie de Hochschild, correspondant à la construction historique, puis nous verrons qu'elle correspond bien à la définition précédente.

Ensuite, nous étudierons un cas particulier qui est celui où le bimodule $M$ est l'algèbre $A$ elle-même.
Ce point de vue est intéressant car il permet d'établir des liens avec les calculs différentiels universels introduits dans la section précédente et fournit une approche naturelle à l'homologie cyclique.  

\begin{defn}
Soit $A$ une algèbre avec unité et $M$ un $A$-bimodule.
L'homologie de Hochschild est l'homologie du complexe:
\begin{equation}
  \xymatrix{
    C_{\bullet}(A,M): \cdots  \ar^-{b}[r] &M \otimes A^{\otimes n}  \ar^-{b}[r]  & \cdots \ar^-{b}[r] &  M\otimes A \ar^-{b}[r]  & M \ .
}
\end{equation}
On a $\displaystyle C_{\bullet}(A,M)~=~\bigoplus_{n=0}^{\infty}C_{n}(A,M)$, avec
  \begin{equation}
    C_{n}(A,M)=
    \begin{cases}
      M &  \text{pour $n=0$} \\ 
      M \otimes A^{\otimes n}   & \text{pour $n \geq 1$} 
    \end{cases} \ .
  \end{equation}
L'opérateur $b: C_{n}(A,M) \to C_{n-1}(A,M)$, appelé opérateur de Hochschild, est défini de la manière suivante:
\begin{multline}
  b(m\otimes a_{1} \otimes \cdots \otimes a_{n})=  
  m a_{1} \otimes \cdots \otimes a_{n} 
  + \sum_{j=1}^{n-1}(-1)^{j} m \otimes a_{1} \otimes \cdots \otimes a_{i} a_{i+1} \otimes \cdots\otimes a_{n} \\
  + (-1)^{n} a_{n} m \otimes a_{1} \otimes \cdots \otimes a_{n-1} \ .
\label{eq:b-Hochschild}\end{multline}
\end{defn}

Afin de voir que l'on a bien $b^{2}=0$, il est utile d'introduire les opérateurs de face $b_{i} : C_{n} \to C_{n-1}$:
\begin{equation}
  b_{i}(m\otimes a_{1} \otimes \cdots \otimes a_{n})=
  \begin{cases}
    m a_{1} \otimes \cdots \otimes a_{n}  & \text{pour $i = 0$}\\
    m\otimes a_{1} \otimes \cdots \otimes a_{i} a_{i+1} \otimes \cdots\otimes a_{n} & \text{pour $0< i < n$}\\
    a_{n}m \otimes a_{1} \otimes \cdots \otimes a_{n-1}      &   \text{pour $i = n$}
  \end{cases} \ .
\end{equation}
Ainsi, $b=\sum_{i=0}^{n} (-1)^{i} b_{i}$ et on vérifie aisément que pour $j<i$,  $b_{j}b_{i}=b_{i-1}b_{j}$. 
On a donc:
\begin{equation}
  \begin{aligned}
    b^{2}&=\left(\sum_{j=0}^{n-1} (-1)^{j} b_{j}\right)\left(\sum_{i=0}^{n} (-1)^{i} b_{i}\right)\\
    &=\sum_{0\leq j<i \leq n} (-1)^{i+j}b_{j}b_{i} + \sum_{0 \leq i \leq j \leq n-1} (-1)^{i+j}b_{j}b_{i}\\
    &=\sum_{0\leq j\leq i-1 \leq n-1} (-1)^{i+j}b_{i-1}b_{j}+ \sum_{0\leq i<j \leq n-1} (-1)^{i+j} b_{j}b_{i}\\
    &=0 \ .
  \end{aligned}
\end{equation}

\begin{rem}
  Le fait de pouvoir décomposer l'opérateur de bord $b$ en la somme d'opérateurs de face, fait de $C_{\bullet}$ un complexe {\it présimplicial} (\textit{c.f.} ~\cite{Loda:92}).
\end{rem}

\subsubsection{\texorpdfstring{Interprétation en terme du foncteur $\Tor$}{Interprétation en terme du foncteur Tor}}
\label{sec:interpretation-tor}
Nous allons voir que l'homologie que l'on vient de définir est bien $\Tor^{A^{e}}_{n}(A,M)$ (voir appendice~\ref{cha:algebre-homologique} pour la définition du foncteur $\Tor$).
On introduit tout d'abord une résolution projective de $A$ en tant que $A^{e}$-module. 
Prenons pour cela le complexe suivant:
\begin{equation}
  \xymatrix{
    C_{\bullet}^{bar}(A): \cdots  \ar^-{b'}[r] & A^{\otimes n+1}  \ar^-{b'}[r]  & A^{\otimes n}  \ar^-{b'}[r] &\cdots \ar^-{b'}[r] & A\otimes A 
  }
\label{eq:Cbar}
\end{equation}
que l'on appelle souvent complexe \textbf{Bar}.

Si on considère $A^{e}$ comme un $A$-bimodule alors $C_{\bullet}^{bar}(A) \simeq C_{\bullet}(A, A^{e})$ et l'opération $b'$ sur $C_{\bullet}^{bar}(A) $ est définie à partir de l'opérateur de Hochschild $b$ sur  $C_{\bullet}(A, A^{e})$.
On a donc automatiquement $b'^{2}=0$.
On identifie $C_{n}(A,A^{e})$ et $C_{n}^{bar}(A)$ de la manière suivante:
\begin{equation}
  \begin{aligned}
    A^{e}\otimes A^{\otimes n}  &\to A^{\otimes n+2}\\
    (a\otimes b^{op})\otimes a_{1}\otimes \cdots \otimes a_{n} &\mapsto b \otimes a_{1}\otimes \cdots \otimes a_{n} \otimes a \ .
  \end{aligned}
\end{equation}

Les espaces  $C_{n}(A, A^{e})=A^{e}\otimes A^{\otimes n}$ sont munis de la structure de $A^{e}$-module à droite par le produit: $(a^{e}\otimes a_{1}\otimes \cdots \otimes a_{n})b^{e}=a^{e} b^{e} \otimes a_{1}\otimes \cdots \otimes a_{n}$.
On montre~\cite{Loda:92} que ce sont de plus des modules projectifs.

L'opérateur $b'$ s'exprime en fonction des opérateurs de face $b_{i}$ de la manière suivante:
\begin{align}
  b' &: A^{\otimes n+2} \to A^{\otimes n+1} ,&
  b'&=\sum_{i=0}^{n} (-1)^{i} b_{i}
\end{align}
et se prolonge de manière naturelle en l'application produit $\mu =b': A\otimes A\to A$.
Cette application définit ainsi une augmentation du complexe  $C_{\bullet}^{bar}(A)$.

Afin de voir que l'on a bien une résolution de $A$ en tant que $A^{e}$-module, on doit vérifier que l'homologie de $\xymatrix{C_{\bullet}^{bar} \ar^-{\mu}[r]& A}$ est nulle.
On introduit pour cela l'homotopie:
\begin{align}
  s:& A^{\otimes n+1} \to A^{\otimes n+2} ,&
  & s(a_{0} \otimes \cdots \otimes a_{n}) = 1\otimes a_{0} \otimes \cdots \otimes a_{n}
\label{eq:homotopie}
\end{align}
qui vérifie $b_{i}s=s b_{i-1}$ pour $1 \leq i \leq n$ et $b_{0}s=id$.
Ainsi, pour tout $n\geq0$, on a: 
\begin{equation}
\begin{aligned}
  b's+sb' &= \sum_{i=0}^{n+1}(-1)^{i} b_{i}s + \sum_{i=0}^{n} (-1)^{i} s b_{i} \\
  &=id +  \sum_{i=1}^{n+1}(-1)^{i} s b_{i-1} + \sum_{i=0}^{n} (-1)^{i} s b_{i} \\
  &= id \ .
\end{aligned}
\end{equation}
Ainsi, les applications $id$ et $0$ sont homotopes et  l'homologie du complexe  est nulle, ce qui montre que $\xymatrix{C_{\bullet}^{bar} \ar^-{\mu}[r]& A}$ est bien une résolution projective de $A$ en tant que $A^{e}$-module.

Maintenant, prenons un $A$-bimodule $M$.
C'est un $A^{e}$-module à gauche et le complexe nous permettant de calculer $\Tor^{A^{e}}_{n}(A,M)$ est donné en degré $n$ par $A^{\otimes n+2} \otimes _{A^{e}} M \simeq (A^{e}\otimes A^{\otimes n} )\otimes_{A^{e}} M \simeq M \otimes A^{\otimes n} $.
Pour les applications de bord correspondantes, on a les relations suivantes: $b' \otimes 1 \simeq b \otimes 1\simeq b$, ce qui montre%
\footnote{Il y a ici un abus de langage car nous utilisons la même notation pour l'opérateur de Hochschild sur $C_{n}(A, A^{e})=A^{e}\otimes A^{n}$ et $C_{n}(A, M)=M \otimes A^{n}$.} 
immédiatement que $H(A,M) \simeq \Tor^{A^{e}}_{n}(A,M)$.

\subsubsection{Homologie de Hochschild à valeurs dans l'algèbre}
\label{sec:HH}

L'homologie de Hochschild a été définie pour une algèbre avec unité et un bimodule $M$ sur cette algèbre.
Une situation particulièrement intéressante est celle où le bimodule est l'algèbre elle-même, c'est-à-dire $M=A$.
Dans ce cas, on note $H(A,A)=HH(A)$.
On vérifie que $HH_{n}(A)$ est un foncteur de la catégorie des algèbres sur $k$ associatives avec unité vers la catégorie des $k$-modules, respectant  le produit, {\it i.e.} $HH_{n}(A\times A')=HH_{n}(A) \times HH_{n}(A')$. 
Sa définition peut ainsi être étendue aux algèbres sur $k$ quelconques en posant :
\begin{align}
  HH_{n}(A) = \coker(H(k)\to H(\tilde{A})) \ .
\end{align}

On montre~\cite{Loda:92} que l'homologie de Hochschild  $HH(A)$, pour une algèbre $A$ quelconque est donnée par l'homologie du complexe total associée au bicomplexe $CC^{(2)}_{**}(A)$ défini de la manière suivante:
\begin{align}
  \begin{aligned}
    \xymatrix{
      \vdots \ar_-{b}[d] &
      \vdots  \ar_-{-b'}[d] \\
      A\otimes A^{\otimes n} \ar_-{b}[d] &
      A^{\otimes n+1} \ar_-{1-\lambda}[l] \ar_-{-b'}[d] \\
      A\otimes A^{\otimes n-1} \ar_-{b}[d] &
      A^{\otimes n}  \ar_-{1-\lambda}[l] \ar_-{-b'}[d] \\
      \vdots \ar_-{b}[d] &
      \vdots  \ar_-{-b'}[d] \\
      A\otimes A \ar_-{b}[d]&
      A^{\otimes 2}  \ar_-{1-\lambda}[l]  \ar_-{-b'}[d] \\
      A & \ar_-{1-\lambda}[l] A }
  \end{aligned} \hspace{60pt}\ . 
  \label{eq:HH}
\end{align}
Les opérateurs $b$ et $b'$ sont ceux rencontrés précédemment. L'opérateur $\lambda$ est défini de la manière suivante:
\begin{equation}
  \begin{aligned}
    \lambda &: &
    A^{\otimes n+1} &\longrightarrow A^{\otimes n+1} \\
    &&
    a_{0}\otimes \cdots \otimes a_{n} &\longmapsto  (-1)^{n} a_{n}\otimes a_{0}\otimes \cdots \otimes a_{n-1} 
\end{aligned}\ .
\end{equation}
Chaque carré du bicomplexe forme un diagramme anticommutatif, c'est-à-dire que l'on a:
\begin{equation}
  b (1-\lambda) =(1-\lambda)b' \ .
\end{equation}
\begin{rem}
  Cette formule peut être directement vérifiée à partir de la décomposition simpliciale de $b$ et $b'$ et en utilisant les relations: $b_{i} \lambda =-\lambda b_{i-1} $ pour $1\leq i \leq n$ et $b_{0} \lambda =(-1)^{n} b_{n}$.
\end{rem}

\begin{rem}
  Lorsque $A$ est une algèbre avec unité, la deuxième colonne a une homologie triviale (cf~\ref{sec:interpretation-tor}) et l'homologie du bicomplexe est l'homologie du complexe formée par la première colonne qui n'est autre que le complexe de Hochschild introduit dans la section~\ref{sec:cohom-de-hochsch}.
\end{rem}

\subsubsection{Lien avec le calcul différentiel universel}

Le complexe total associé au bicomplexe~(\ref{eq:HH}) est:
\begin{align}
  \xymatrix{
    \cdots  \ar^-{\partial}[r] &
    ( A\otimes A^{\otimes n}) \oplus  (k\otimes A^{\otimes n}) \ar^-{\partial}[r] &
    \cdots \ar^-{\partial}[r] &
    (A\otimes A) \oplus (k\otimes A) \ar^-{\partial}[r] &
    A  
  }
  \label{HH-tot}
\end{align}
où l'opérateur de bord $\partial$ est donné par la matrice:
\begin{equation}
  \partial= \left( 
    \begin{matrix}
      b & 1-\lambda\\
      0 & -b'
    \end{matrix}
  \right) \ .
\end{equation}
On reconnaît immédiatement le complexe associé au calcul différentiel universel  $\Omega(A)$ défini dans l'équation~(\ref{Omega}) en utilisant les isomorphismes $( A\otimes A^{\otimes n}) \oplus  (k\otimes A^{\otimes n}) \simeq \tilde{A}\otimes A^{\otimes n} \simeq \Omega^{n}(A)$.
En tant que complexe gradué, $\Omega(A)$ est le complexe réduit $\bar{C}(\tilde{A},\tilde{A})_{red}$ et on vérifie que l'opérateur de bord $\partial$ correspond à l'opérateur de Hochschild $b$ défini sur $\bar{C}(\tilde{A},\tilde{A})$ par la formule~(\ref{eq:b-Hochschild}).
Le complexe~(\ref{HH-tot}) est donc isomorphe au complexe suivant:
\begin{equation}
  \begin{aligned}
    \xymatrix{
      \cdots  \ar^-{b}[r] &
      \Omega^{n}(A) \ar^-{b}[r] &
      \cdots \ar^-{b}[r] &
      \Omega^{1}(A) \ar^-{b}[r] &
      \Omega^{0}(A)=A
    }
  \end{aligned} \ .
  \label{eq:HH-tot-Omega}
\end{equation}

Grâce à la structure d'algèbre différentielle de $\Omega(A)$, on peut définir l'opérateur $b$ de manière plus directe en posant:
\begin{align}
  &  \begin{aligned}
    b(\omega dx)&=(-1)^{deg \omega}[\omega, x]\\
    b(dx)&=0 \\
    b(x)&=0
  \end{aligned} &
  & \forall x \in A, \forall \omega\in \Omega(A) 
\label{eq:b-karoubi}
\end{align}
et $b$ est alors appelé opérateur de Karoubi.
Nous pouvons vérifier que ces deux définitions sont bien les mêmes.
Si l'on écrit  $\Omega^{n}(A)$ comme $\tilde{A}\otimes A^{\otimes n}$, pour $n>1$, d'après~(\ref{eq:b-karoubi}), on a:
\begin{multline}
  b(\tilde{a_{0}}\otimes a_{1} \otimes \cdots \otimes a_{n})=  
  \tilde{a_{0}} a_{1} \otimes \cdots \otimes a_{n} 
  + \sum_{j=1}^{n-1}(-1)^{j} \tilde{a_{0}} \otimes a_{1} \otimes \cdots \otimes a_{i} a_{i+1} \otimes \cdots\otimes a_{n} \\
  + (-1)^{n} a_{n} \tilde{a_{0}} \otimes a_{1} \otimes \cdots \otimes a_{n-1}
\end{multline}
et en degré 1, $b\left((a_{0}+\lambda_{0}1)\otimes a_{1}\right)=[a_{0},a_{1}]$.
Ainsi, $b$ correspond bien à l'opérateur de Hochschild usuel.

\subsection{Cohomologie de Hochschild}
Comme nous allons le voir, on peut considérer la cohomologie de Hochschild comme la version duale de l'homologie de Hochschild.
Les définitions et principaux résultats sont analogues à ceux rencontrés en homologie de Hochschild.
Cependant, ce sont bien des notions différentes et n'ayant pas la même interprétation. 
Ceci vient du fait que la cohomologie de Hochschild est construite à partir du foncteur  $Hom(\cdot, \cdot)$, alors que l'homologie de Hochschild est construite à partir du foncteur $(\cdot \otimes \cdot)$.

\begin{defn}
Soit $A$ une algèbre avec unité et $M$ un $A$-bimodule.
Les groupes de cohomologie de Hochschild $H^{n}(A,M)$ de $A$ à coefficients dans $M$ sont définis comme étant les groupes $\Ext^{n}_{A^{e}}(A,M)$, où $\Ext$ est le foncteur dérivé de $Hom_{A^{e}}(\cdot,\cdot)$.
\end{defn}

Donnons maintenant différentes manières de calculer ces groupes.
Nous avons vu que le complexe $C^{bar}(A)$ (\ref{eq:Cbar}) est une résolution projective de $A$ en tant que $A^{e}$ bimodule. Ainsi, par définition, on a:
\begin{align}
  H^{n}(A,M) &= H_{n}(Hom_{A^{e}}(C^{bar},M))
\end{align}
Soit $\phi$ une cochaîne dans $Hom_{A^{e}}(C^{bar}_{n},M)$.
L'application de cobord $\beta'$  est donnée par :
\begin{align}
  \beta'(\phi)&= (-1)^{n+1} \phi \circ b' \ .
\end{align}
De même que nous avions un isomorphisme de complexes 
$A^{\otimes n+2} \otimes _{A^{e}} M \simeq (A^{e}\otimes A^{\otimes n} )\otimes_{A^{e}} M \simeq M \otimes A^{\otimes n}$,
nous avons maintenant:
\begin{align}
  Hom_{A^{e}}(A^{\otimes n+2}, M) \simeq Hom_{A^{e}}(A^{e}\otimes A^{\otimes n} , M) \simeq Hom(A^{\otimes n},M) \ .
\end{align}
Ainsi, à toute cochaîne $\phi $ dans $Hom_{A^{e}}(C^{bar}_{n},M)= Hom_{A^{e}}(A^{\otimes n+2},M)$, on peut associer une application $f: A^{\otimes n}\to M$ définie de la manière suivante:
\begin{align}
  \phi(a_{0},a_{1},\cdots,a_{n+1})= a_{0} f(a_{1},\cdots,a_{n})a_{n+1} \ .
\end{align}
L'application $f$ peut être donnée explicitement en utilisant l'isomorphisme de $A^{e}$-modules à droite $A^{\otimes n+2}\simeq A^{e} \otimes A^{\otimes n}$.
On associe à la cochaîne $\phi$ un homomorphisme de $A^{e}$-modules à droite $\tilde{\phi}:A^{e} \otimes A^{\otimes n}\to M $ défini par la formule: 
\begin{align}
  \tilde{\phi}((a_{n+1}\otimes a_{0}^{op}) \otimes a_{1}\otimes \cdots \otimes  a_{n}) = \phi(a_{0},a_{1},\cdots,a_{n+1}) \  .
\end{align}
Si on pose $f(a_{1},\cdots,a_{n})=\tilde{\phi}((1\otimes 1^{op}) \otimes a_{1}\otimes \cdots \otimes a_{n})$, on a alors:
\begin{equation}
  \begin{aligned}
    \phi(a_{0},a_{1},\cdots,a_{n+1})%
    &=  \tilde{\phi}((a_{n+1}\otimes a_{0}^{op}) \otimes a_{1}\otimes \cdots \otimes a_{n}) \\
    &=  \tilde{\phi} \biggl(\bigl((1\otimes 1^{op}) \otimes a_{1}\otimes \cdots \otimes a_{n}\bigr) \cdot (a_{n+1}\otimes a_{0}^{op}) \biggr)\\
    &= \tilde{\phi} \biggl((1\otimes 1^{op}) \otimes a_{1}\otimes \cdots \otimes a_{n}\biggr)  \cdot (a_{n+1}\otimes a_{0}^{op}) \\
    &=  f(a_{1},\cdots,a_{n})   \cdot (a_{n+1}\otimes a_{0}^{op}) \\
    &= a_{0} f(a_{1},\cdots,a_{n})a_{n+1} \ .
  \end{aligned}
\end{equation}
On notera  $C^{n}(A,M)$ le complexe $Hom(A^{\otimes n},M)$ par la suite.
Voyons comment l'application de cobord $\beta'$ se répercute sur l'application $f$.
On a:
\begin{equation}
\begin{aligned}
  \beta'(\phi)(a_{0},\cdots,a_{n+2})
  =& (-1)^{n+1} \phi(b'(a_{0},\cdots,a_{n+2}))\\
  =& (-1)^{n+1} \left(\phi(a_{0}a_{1},\cdots,a_{n+2}) + \cdots + (-1)^{n+1}\phi(a_{0},\cdots,a_{n+1}a_{n+2})\right)\\
  =&      (-1)^{n+1} a_{0}\biggl(a_{1}f(a_{2},\cdots,a_{n+1}) \\
    &- f(a_{1}a_{2},\cdots,a_{n+1}) + \cdots +(-1)^{n} f(a_{1},\cdots,a_{n}a_{n+1})   \\
    & + (-1)^{n+1}f(a_{1},\cdots,a_{n})a_{n+1}\biggr)a_{n+2} \ .
\end{aligned}
\end{equation}
On définit ainsi l'opérateur de cobord $\beta$ sur $C^{n}(A,M)$ de la manière suivante:
\begin{equation}
\begin{aligned}
  \beta(f)(a_{1},\cdots,a_{n+1})= &a_{1}f(a_{2},\cdots,a_{n+1}) \\
   &+ \sum_{i=1}^{n} (-1)^{i}  f(a_{1},\cdots,a_{i}a_{i+1},\cdots, a_{n+1}) \\
  &+(-1)^{n+1}f(a_{1},\cdots,a_{n})a_{n+1} \ .
\end{aligned}
\end{equation}
On notera l'analogie avec l'opérateur $b$ de Hochschild intervenant dans la formule~(\ref{eq:b-Hochschild}). 
D'autre part, cette définition est souvent prise comme définition de base pour la cohomologie de Hochschild.

Nous allons maintenant donner une interprétation des premiers groupes de cohomologie de Hochschild.
Le premier groupe $H¹(A,M)$ s'interprète de la manière suivante. 
Un cocycle de Hochschild, $z\in Z¹(A,M)$ correspond à une dérivation de l'algèbre $A$ à valeurs dans $M$.
Un cobord $\beta m \in B^{1}(A,M)$ (\textit{i.e.} tel que $\beta (m)(a)=am-ma$) correspond à une dérivation intérieure.
Ainsi, un élément du groupe $H¹(A,M)$ décrit une dérivation extérieure (les notions de dérivations sont données dans la section~\ref{sec:derivations}).
Le deuxième groupe de Hochschild $H²(A,M)$ sert à décrire les extensions de l'algèbre $A$ (d'où le nom $\Ext$ pour le foncteur dérivé de Hom).

Il est également intéressant de noter que la différentielle du calcul différentiel universel, introduite en \ref{sec:calculs-différentiel}, définit un cocycle universel de Hochschild en degré $1$ et permet également de définir des cocycles universels de degrés plus élevés.
Précisons cette notion par la proposition suivante:
\begin{prop}
Soit $M$ un $A$-bimodule et $(a_{1},\cdots,a_{n})\mapsto c(a_{1},\cdots,a_{n})$ un n-cocycle de Hochschild à valeurs dans $M$.
Alors il existe un unique homomorphisme de bimodule: $i_{c}:\Ou(A) \to M$ tel que
\begin{align}
  c(a_{1},\cdots,a_{n})&=i_{c}(da_{1}\cdots da_{n}), &
  & \forall a_{i} \in A \ .
\end{align}
De plus $c$ est un cobord \ssi $i_{c}$ a une extension $\tilde{i_{c}}$ en tant qu'homomorphisme de bimodule de $A\otimes \Ou^{n-1}(A)$ dans%
\footnote{La suite exacte (\ref{eq:suite-exacte-omega}) se généralise pour $n\geq 1$ en la suite exacte suivante:%
  \begin{align}%
    \xymatrix@R=2pt@M=6pt{%
      0 \ar[r] &\Ou^{n}(A) \ar^-{\bar{i}}[r] & A  \otimes\Ou^{n-1}(A)  \ar^-{\mu}[r] & \Ou^{n-1}(A) \ar[r] & 0 \ .%
    }%
  \end{align}%
}%
 $M$.\\
\end{prop}
\noindent
Ces résultats et les références associées se trouvent dans~\cite{dubois-violette:japan99}.

\subsubsection{Cohomologie de Hochschild à valeurs dans le dual de l'algèbre}
\label{sec:HH-duale}

Enfin, nous pouvons considérer la situation où le bimodule est $A^{*}$. 
L' espace $A^{*}$ est un $A$-bimodule pour l'action $(a\phi b)(c)=\phi(bac)$ pour tout $a,b,c\in A$.
La cohomologie $H(A,A^{*})$ du complexe $C^{*}(A,A^{*})$ est alors notée $HH^{*}(A)$.
Comme dans le cas de l'homologie de Hochschild, $HH^{*}$ réalise un foncteur de la catégorie des algèbres sur $k$ associatives avec unité vers la catégorie de $k$-modules. 
Ce foncteur peut être étendu aux algèbres avec ou sans unité et la cohomologie de Hochschild $HH^{*}(A)$ d'une algèbre $A$ est alors obtenue en considérant le complexe total associé au bicomplexe $Hom_{k}(CC^{(2)}_{**}(A),k)$ dual du complexe a deux colonnes $CC^{(2)}_{**}(A)$(\ref{eq:HH}) calculant $HH_{*}(A)$. 
Les opérations $b$,$b'$ et $1-\lambda$ se dualisent sur le complexe $Hom_{k}(CC^{(2)}_{**}(A),k)$.
Dans le cas des algèbres avec unité, on peut se ramener à la première colonne de $CC^{(2)}_{**}(A)$ (la deuxième colonne correspondant au bar-complexe, elle a une homologie triviale et peut être éliminée).
On peut alors vérifier que l'espace $C^{n}(A,A^{*})=Hom_{k}(A^{n},A^{*})$ est isomorphe à l'espace $Hom_{k}(A^{n+1},k)$ et que les opérateurs de cobord correspondent.
On identifie deux éléments  $\phi\in C^{n}(A,A^{*})$ et $f\in Hom_{k}(A^{n+1},k) $ de la manière suivante:
\begin{align}
  \phi(a_{1},\dots,a_{n})(a_{0}) = f(a_{0},a_{1},\dots,a_{n}) \ .
\end{align}
Ainsi, les cohomologies définies à partir des complexes $Hom_{k}(A^{n+1},k)$ et $C^{*}(A,A^{*})$ coïncident.
Cette identification est essentielle en cohomologie cyclique pour faire le lien entre différentes approches.

\subsubsection{Quelques résultats sur l'homologie et la cohomologie de Hochschild pour les algèbres de fonctions $C^{\infty}(X)$}
Nous pouvons enfin énoncer un résultat important dû à Connes pour le cas des algèbres de fonctions $C^{\infty}(X)$ sur une variété $X$.
Il fut montré dans \cite{Conn:85,Conn:90,Conn:94} que la cohomologie de Hochschild $HH^{*}_{cont}(C^{\infty}(X))$ des cochaîne continues (pour la famille de seminormes associée à $C^{\infty}(X)$) est isomorphe à l'espace des courants de de~Rham sur la variété $X$.
Ce résultat a également son équivalent en homologie et on montre alors que l'homologie de Hochschild $HH_{cont \ *}(C^{\infty}(X))$, calculée à partir du complexe de Hochschild où le produit tensoriel $\otimes$ est remplacé par le produit tensoriel topologique $\hat{\otimes}$, est isomorphe à l'espace des formes de de~Rham $\Omega_{dR}(X)$.
Ce résultat généralise pour des algèbres de Fréchet le théorème de Hochschild-Kostant-Rosenberg donnant un isomorphisme entre l'homologie de Hochschild $HH_{*}(A)$ d'une algèbre $A$ lisse et les formes de Kähler.
Ce type de résultats fut à l'origine des motivations pour vouloir définir une géométrie différentielle non commutative.
On a en effet des objets propres à l'existence d'une structure différentielle en géométrie ordinaire qui semblent pouvoir être caractérisés de manière complètement algébrique (avec une notion de continuité en plus lorsque l'on travail avec des algèbre topologiques).

\subsection{Homologie cyclique}\label{sec:homologie-cyclique}
L'homologie cyclique réalise un autre foncteur, que l'on note $HC$, de la catégorie des algèbres associatives vers les groupes abéliens.
Nous allons la définir directement à partir d'un complexe gradué qui, comme nous allons le voir, est intimement lié à celui calculant l'homologie de Hochschild à valeurs dans l'algèbre.

Pour une algèbre associative $A$, on introduit le bicomplexe cyclique $CC_{\bullet \bullet}(A)$:
\begin{align}
  \begin{aligned}
  \xymatrix{
    \vdots \ar_-{b}[d] &
    \vdots \ar_-{-b'}[d] &
    \vdots  \ar_-{b}[d] &
    \\    
    A^{\otimes n+1} \ar_-{b}[d] &
    A^{\otimes n+1} \ar_-{1-\lambda}[l] \ar_-{-b'}[d] &
    A^{\otimes n+1} \ar_-{Q}[l] \ar_-{b}[d] &
    \cdots \ar_-{1-\lambda}[l] \\    
    A^{\otimes n} \ar_-{b}[d] &
    A^{\otimes n} \ar_-{1-\lambda}[l] \ar_-{-b'}[d] &
    A^{\otimes n} \ar_-{Q}[l] \ar_-{b}[d] &
    \cdots \ar_-{1-\lambda}[l] \\    
    \vdots \ar_-{b}[d] &
    \vdots  \ar_-{-b'}[d] &
    \vdots \ar_-{b}[d] &
    \\    
    A^{\otimes 2} \ar_-{b}[d] &
    A^{\otimes 2} \ar_-{1-\lambda}[l] \ar_-{-b'}[d] &
    A^{\otimes 2} \ar_-{Q}[l] \ar_-{b}[d] &
    \cdots \ar_-{1-\lambda}[l] \\     
    A  &
    A \ar_-{1-\lambda}[l]  &
    A \ar_-{Q}[l]  &
    \cdots \ar_-{1-\lambda}[l] \\
  }
\end{aligned} 
\label{diag:CC}
\end{align}
où les opérateurs $b$, $b'$ et $\lambda$ sont ceux introduits dans la section précédente et l'opérateur $Q$ est défini de la manière suivante:
\begin{equation}
\begin{aligned}
  Q &: A^{\otimes n+1}\to A^{\otimes n+1} ,&
  Q&=\sum_{p=0}^{n} (\lambda)^{p} \ .
\end{aligned}
\end{equation}
On vérifie que les carrés dans le bicomplexe (\ref{diag:CC}) sont tous anticommutatifs, c'est-à-dire:
\begin{equation}
  \begin{aligned}
    b(1-\lambda) &=(1-\lambda)b' \\
    b' Q &= Q b 
  \end{aligned}  \ .
\end{equation}

\begin{defn}
  L'homologie cyclique $HC(A)$ est l'homologie du complexe total associé au bicomplexe $CC_{\bullet \bullet}(A)$, c'est-à-dire le complexe:
  \begin{equation}
    \xymatrix{
      {\bf D}: & \cdots \ar[r]  & D_{n} \ar^-{\partial}[r]& D_{n-1} \ar[r]& \cdots \ar[r]&  D_{1}\ar^-{\partial}[r]& D_{0}\ar[r]& 0 
    } \ ,
  \end{equation}
où les $D_{n}$ sont les diagonales du bicomplexe $CC_{\bullet \bullet}(A)$ et l'application $\partial$ est la somme des applications $b$, $Q$, $-b'$ et $1-\lambda$.
\end{defn}

\begin{thm}[Suite exacte de Connes]
\label{thm:sequence-Connes}
  Pour une algèbre $A$ quelconque, l'homologie de Hochschild $HH(A)$ et l'homologie cyclique $HC(A)$ sont reliées par la suite exacte longue:
  \begin{equation}
    \begin{aligned}
      \xymatrix{
        \cdots \ar[r] &
        HC_{n}(A) \ar^-{S}[r] &
        HC_{n-2}(A) \ar^-{B}[r] & 
        HH_{n-1}(A) \ar^-{I}[r] & 
        HC_{n-1}(A) \ar^-{S}[r] & 
        HC_{n-3}(A) \ar^-{B}[r]&
        \cdots\\
        &
        \cdots \ar^-{I}[r] & 
        HC_{1}(A) \ar^-{S}[r] & 
        HC_{-1}(A) \ar^-{B}[r] & 
        HH_{0}(A) \ar[r] & 
        0 \ ,
      }
    \end{aligned}
    \label{eq:connes} 
  \end{equation}
  où  $ HC_{-1}(A)=0$.
\end{thm}
Cette suite vient de la suite exacte courte de complexes:
\begin{equation}
  \xymatrix{
    0\ar[r] & D_{n}^{Hochschild}\ar^-{I}[r] & D_{n} \ar^-{S}[r] & D_{n-2} \ar[r] &0 \ ,
} 
\end{equation}
où $D^{Hochschild}$ est le complexe total du bicomplexe~(\ref{eq:HH}) calculant $HH(A)$.
Il correspond aux deux premières colonnes du complexe cyclique et l'application $I$ est l'inclusion.
On peut alors calculer le quotient du complexe cyclique par $D^{Hochschild}$ et on obtient à nouveau le complexe cyclique avec un décalage de $2$ en chaque degré.

Le bicomplexe cyclique mélange le complexe de Hochschild à deux colonnes~(\ref{eq:HH}) et la résolution périodique de période $2$ du module trivial $ A^{\otimes n+1}/(1-\lambda)$ sur l'algèbre du groupe $\gZ/(n+1)\gZ$:
\begin{equation}
  \xymatrix{
    \cdots \ar[r]&
    A^{\otimes n+1} \ar^-{1-\lambda}[r] &
    A^{\otimes n+1}\ar^-{Q}[r]&
    A^{\otimes n+1} \ar^-{1-\lambda}[r]&
    \cdots \ar^-{1-\lambda}[r]&
    A^{\otimes n+1} \ar[r]&
    0
  } \ .
  \label{eq:resol-2-per}
\end{equation}
Ainsi, on a une surjection naturelle $Tot(CC)\to C_{\lambda}$, où $C_{\lambda}$ est le complexe de Connes, $C^{\lambda}_{n}(A)=A^{\otimes n+1}/(1-\lambda)$.
L'opérateur de bord sur ce complexe est donné par l'image par cette surjection de l'opérateur de Hochschild $b$ agissant sur la première colonne du complexe cyclique $CC_{\bullet \bullet}(A)$~(\ref{diag:CC}).
On montre par des arguments de suite spectrale, que cette surjection est un quasi-isomorphisme (cf..~\cite{Loda:92}) et donc $H^{\lambda}(A)\simeq HC(A)$ (il faut que $\gQ$ soit un sous-ensemble du corps de base $k$ de l'algèbre).

\subsubsection{Liens avec le calcul différentiel universel}
Tout comme nous l'avons fait pour l'homologie de Hochschild, il est possible d'établir un lien avec le calcul différentiel universel.
 On définit un opérateur $B: \Omega^{n}(A)\to \Omega^{n+1}(A)$ qui est une représentation au niveau des complexes de l'opérateur $B$ introduit dans (\ref{eq:connes}).
 Cet opérateur sera appelé opérateur de Connes.
 Sous l'isomorphisme $\Omega^{n}(A)\simeq A^{\otimes (n+1)}\oplus A^{\otimes n}$, il correspond à l'opérateur de $A^{\otimes (n+1)}\oplus A^{\otimes n}$ dans $A^{\otimes (n)}\oplus A^{\otimes n-1}$ donné par la matrice:
 \begin{equation}
   \left( 
     \begin{matrix}
       0 & 0\\
       Q & 0
     \end{matrix}
   \right) \ .
 \end{equation}
 Ainsi, $B$ agit sur un élément de la forme $x_{0} dx_{1} \cdots dx_{n}$ de la manière suivante:
\begin{align}
  B(x_{0} dx_{1} \cdots d_{x_{n}}) &= dx_{0} dx_{1} \cdots dx_{n} + (-1)^{n} dx_{n}dx_{0}\cdots dx_{n-1}+\cdots +(-1)^{n^{2}} dx_{1}\cdots dx_{n}dx_{0} \ .
\end{align}
On voit que $B$ est fortement relié à la différentielle $d$ sur l'algèbre différentielle graduée $\Omega(A)$ tout comme l'est l'opérateur de Hochschild $b$ (voir formule~(\ref{eq:b-karoubi})).

En utilisant l'identification entre le complexe (\ref{HH-tot}) et (\ref{eq:HH-tot-Omega}), on voit que le bicomplexe (\ref{diag:CC}) est isomorphe au bicomplexe:
\begin{align}
&  \begin{aligned}
    \xymatrix{ \vdots \ar^-{b}[d]& \vdots \ar^-{b}[d]&
      \vdots \ar^-{b}[d]  \\
      \Omega^{2}(A) \ar^-{b}[d]& \Omega^{1}(A) \ar_-{B}[l] \ar^-{b}[d] &
      \Omega^{0}(A) \ar_-{B}[l]\\
      \Omega^{1}(A) \ar^-{b}[d]& \Omega^{0}(A) \ar_-{B}[l]
      &\\
      \Omega^{0}(A)& &
      \\
    }
\end{aligned}&
& \ .
\label{eq:bicomp-omega}
\end{align}
On vérifie que l'on a bien $bB+Bb=B^{2}=0$ et que l'homologie du complexe total calcule $HC(A)$.

\begin{rem}
  En fait, l'homologie cyclique $HC(A)$  peut être calculée de manière plus générale à l'aide de tout complexe $M$ calculant l'homologie de Hochschild, muni d'un opérateur $B$ de degré +1 satisfaisant $bB+Bb=B^{2}=0$ ($M$ est appelé un \textit{complexe mixte}).
On remplace alors dans le bicomplexe (\ref{eq:bicomp-omega}) les colonnes $\Omega(A)$ par $M$.
Ainsi, pour une algèbre avec unité, on peut remplacer $\Omega(A)$ par $\Ou(A)$ et l'opérateur $B$ devient $(1-\lambda)sQ$, où $s$ est l'homotopie définie dans (\ref{eq:homotopie}).
Nous renvoyons à~\cite{Loda:92,rosenberg,cuntz} pour plus de détails sur ces constructions.
De même, pour les algèbres commutatives, on peut prendre le calcul différentiel de Kähler et la différentielle usuelle pour l'opérateur $B$ (cf.~\cite{Loda:92,rosenberg}).
\end{rem}

Pour une algèbre $A$ donnée, on peut également définir d'autres groupes d'homologie: $HC^{-}(A)$ et $HP(A)$ appelés respectivement homologie cyclique négative et homologie cyclique périodique. 
Ces groupes sont calculés à partir de bicomplexes similaires au bicomplexe cyclique (\ref{diag:CC}) ou (\ref{eq:bicomp-omega}) (cf~\cite{Loda:92,rosenberg,cuntz}) et sont nécessaires pour pouvoir définir le caractère de Chern-Connes qui est un homomorphisme canonique de $K(A)$ dans $HP(A)$.
Le caractère de Chern-Connes généralise le caractère de Chern en géométrie différentielle et se construit naturellement à partir du générateur de $HC(\gC)$.

\subsubsection{Résultats pour les algèbres de fonctions $C^{\infty}(X)$}

Dans le cas d'une algèbre de fonctions $C^{\infty}(X)$, Connes a montré~ \cite{Conn:85,Conn:90,Conn:94} (dans le cas dual) qu'il y a un isomorphisme entre l'homologie cyclique des chaînes continues et la cohomologie de de~Rham:
\begin{equation}
  HC_{cont \ *}(C^{\infty}(X)) \simeq \Omega^{*}(X)/d\Omega^{*}(X) \oplus H^{*-2}_{dR}(X) \oplus H^{*-4}_{dR}(X) \oplus  \dots 
\end{equation}
On voit ainsi que l'homologie cyclique fournit une bonne généralisation de l'homologie de de~Rham en géométrie non commutative et c'est encore une des motivations pour vouloir définir une géométrie différentielle non commutative.
On peut également voir que le caractère de Chern-Connes défini de manière complètement algébrique coïncide dans le cas d'une algèbre de fonctions $C^{\infty}(X)$ avec le caractère de Chern usuel qui est un isomorphisme entre la $K$-théorie topologique de la variété $X$ et les éléments pairs (ou impairs) de sa cohomologie de de~Rham. 

\subsection{Cohomologie cyclique}

La cohomologie cyclique $HC^{*}(A)$ d'une algèbre associative $A$ avec ou sans unité est définie de manière générale à partir du complexe dual au complexe cyclique $CC_{\bullet \bullet}(A)$~(\ref{eq:complexe-cyclique}): c'est l'homologie du complexe total associé au bicomplexe $Hom_{k}(CC_{\bullet \bullet}(A),k)$.
Dans les cas où $\gQ$ est inclus dans le corps $k$ sur lequel est basé l'algèbre, la cohomologie cyclique de $A$ peut se calculer à partir du  \textit{complexe cyclique de Connes} $C^{*}_{\lambda}(A)$ qui est le dual du complexe $C_{*}^{\lambda}(A)$ introduit dans la section~\ref{sec:homologie-cyclique}.
Ce complexe est défini en degré $n$ de la manière suivante:
\begin{equation}
  C^{n}_{\lambda}(A)= \{ f\in Hom_{k}(A^{n+1},k), f(a_{0},\dots,a_{n}) = (-1)^{n}f(a_{n},a_{0},\dots, a_{n+1})   \}  \ .
\label{eq:complexe-cyclique}
\end{equation}
L'opérateur de cobord que l'on notera $\delta$ est le dual de l'opérateur de Hochschild $b$ sur $C_{*}(A,A)$ correspondant à la première colonne de $CC_{\bullet \bullet}(A)$~(\ref{eq:complexe-cyclique}).
Il agit donc sur un élément $\phi\in  C^{n}_{\lambda}(A)$ de la manière suivante:
\begin{multline}
  \delta\phi(a_{0},\dots,a_{n+1}) =  \phi(a_{0}a_{1},a_{2},\dots,a_{n+1}) + \sum_{i=1}^{n} (-1)^{i}\phi(a_{0},\dots,a_{i}a_{i+1},\dots, a_{n+1} )\\
  + (-1)^{n+1} \phi(a_{n+1}a_{0},a_{1},\dots,a_{n}) \ .
\end{multline}

On peut remarquer qu'un élément $f\in C^{*}_{\lambda}(A)$ peut être considéré comme un élément de $C^{*}(A,A^{*})$ satisfaisant la condition de cyclicité:
\begin{equation}
  f(a_{0},\dots,a_{n}) = (-1)^{n}f(a_{n},a_{0},\dots, a_{n+1})    \ .
\end{equation}
Ainsi, $C^{*}_{\lambda}(A)$ est un sous-complexe du complexe  $C^{*}(A,A^{*})$ calculant la cohomologie de Hochschild $HH^{*}(A)$ à valeur dans $A^{*}$ (voir section~\ref{sec:HH-duale}).
De plus, la suite exacte de Connes définie en homologie dans le théorème~\ref{thm:sequence-Connes} a son équivalent en cohomologie où $HH_{*}$ est remplacé par $HH^{*}$, $HC_{*}$ par $HC^{*}$ et le sens des flèches est renversé.

Nous allons maintenant voir que la cohomologie cyclique est intimement liée à la notion de trace.
Il faut pour cela introduire la notion de cycle qui est une version algébrique de la notion d'intégration de formes différentielles sur une variété.
\begin{defn}
  Un \textit{cycle} de dimension $n$ est un triplet $(\Omega,d,\int)$ où $(\Omega,d)$ est
  une algèbre différentielle graduée et $\int: \Omega^{n}\to k$ est une trace graduée fermée sur
  $\Omega$.

  Un \textit{cycle} sur une algèbre associative $A$ est un cycle $(\Omega,d,\int)$ et un homomorphisme $\rho:A\to \Omega^{0}$.
  On le notera également $(\Omega,d,\int,\rho)$
\end{defn}

La propriété importante associée aux cycles d'une algèbre $A$ est la suivante:
\begin{prop}
  A tout cycle de dimension $n$ sur $A$, on peut associer une cochaîne $\tau$ du complexe  $C^{n}_{\lambda}(A)$ définie par:
\begin{equation}
  \tau(a_{0},\dots,a_{n})= \int\rho(a_{0})d\rho(a_{1})\dots\rho(a_{n}) \ .
\end{equation}
Réciproquement, pour toute cochaîne $\tau$ de  $C^{n}_{\lambda}(A)$, il existe un cycle de dimension n $(\Omega,d,\int)$ et un homomorphisme $\rho:A\to\Omega^{0}$ tel que $\tau$ en soit le caractère. 
\end{prop}
La démonstration du fait que le caractère $\tau$ d'un cycle $(\Omega,d,\int,\rho)$ soit une cochaîne se montre en utilisant de manière essentielle les propriétés d'algèbre différentielle graduée de $(\Omega,d)$ et le fait que $\int$ soit une trace graduée fermée.
Pour démontrer la réciproque, il suffit de prendre pour $(\Omega,d)$ le calcul différentiel universel $\Omega(A)$, pour $\rho$ l'application identité et pour $\int$ l'application linéaire $\hat{\tau}$ associée à une cochaîne $\tau$ définie sur $\Omega^{n}(A)$ par:
\begin{equation}
  \hat{\tau}((a_{0}+\lambda_{0})da_{1} \dots da_{n})= \tau(a_{0},a_{1},\dots,a_{n})\ .
\end{equation}
On vérifie que $\int=\hat{\tau}$ est une trace graduée fermée grâce au fait que $\tau$ est une cochaîne cyclique de Hochschild.
Cette proposition est démontrée de manière plus détaillée dans \cite{Conn:85,Conn:94,Conn:90}.
C'est en fait ce type de considérations qui mena à la découverte de la cohomologie cyclique et c'est de cette manière qu'elle fut originellement introduite.

\subsubsection{Remarques finales}
Nous pourrions encore voir d'autres propriétés telles que l'invariance de Morita ou l'invariance par difféotopie de l'homologie et la cohomologie cyclique, ou encore énoncer la version cohomologique des isomorphismes donnés à la fin de la section précédente, mais nous allons nous arrêter là car ces résultats ne nous seront pas utiles directement par la suite.

\section{Calcul différentiel basé sur les dérivations}
\label{sec:calc-diff-der}
Nous avons vu jusqu'à présent des outils permettant de caractériser des algèbres associatives et que la notion de calcul différentiel universel joue un rôle essentiel dans la construction de certains invariants.
Ces invariants généralisent d'une certaine manière les invariants que l'on obtient en géométrie ordinaire tels que la cohomologie de de~Rham.
Nous allons maintenant essayer de définir un équivalent des formes de de~Rham dans le cadre des algèbres associatives.
Il y a pour cela plusieurs points de vue possibles. 
On peut considérer par exemple que l'homologie de Hochschild fournit une généralisation des formes de de~Rham puisque, dans le cas des algèbres commutatives, elle est directement reliée à ces dernières.
Malheureusement, l'homologie de Hochschild n'est généralement pas munie d'une structure d'algèbre (sauf dans le cas commutatif) et donc ne fournit pas un bon candidat pour remplacer les formes différentielles.
Un autre point de vue serait de considérer que tout calcul différentiel sur une algèbre $A$ (\textit{i.e.} toute algèbre différentielle graduée étant isomorphe à $A$ en degré $0$) est une généralisation du complexe de de~Rham.
Ainsi, pour une algèbre donnée $A$, nous pouvons toujours considérer le calcul différentiel universel.
On voit bien que ce point de vue est un peu trop général et on se sert généralement de la structure particulière de l'algèbre $A$ ou du problème (physique ou mathématique) que l'on veut étudier afin de nous guider dans le choix d'un calcul différentiel particulier.

Le calcul différentiel que nous allons voir dans cette section fut introduit par Michel Dubois-Violette dans~\cite{dubois-violette:88} et est appelé calcul différentiel basé sur les dérivations.
C'est une tentative de définition de la notion de formes différentielles ne faisant pas intervenir d'autres objets que l'algèbre elle-même et ses dérivations qui sont l'analogue des champs de vecteurs en géométrie différentielle.

\subsection{Dérivations}\label{sec:derivations}
\begin{defn}[Dérivations]
  Soit $A$  une algèbre associative et $M$ un $A$-bimodule.
Une dérivation de $A$ à valeurs dans $M$ est une application $k$-linéaire $D:A\to M$ satisfaisant la relation:
  \begin{align}
    &    D(ab)=a(Db)+(Da)b &
    &    \forall a,b \in A  \ .
  \end{align}
  L'ensemble des dérivations de $A$ dans $M$ forme un module sur le centre de l'algèbre $Z(A)$ et on le note $\der(A,M)$ ou simplement $\der(A)$ lorsque $M=A$
.
En particulier, une dérivation définit un cocycle de Hochschild de degré $1$.
L'ensemble de toutes les dérivations dans un bimodule $M$ est précisément l'ensemble $Z¹(A,M)$ des cocycles de Hochschild de degré $1$.
\end{defn}

\begin{defn}[Dérivations intérieures]
A tout élément $m\in M$, on associe une dérivation, appelée {\bf dérivation intérieure}, que l'on note $ad_{m}$.
Elle est définie par:
\begin{equation}
  ad_{m}(a)=[m,a]=ma-am \ .
\end{equation}
On note l'ensemble de toutes les dérivations intérieures  $\Int(A,M)$, ou simplement $\Int(A)$, lorsque $M=A$.
L'espace $\Int(A,M)$ coïncide avec $B¹(A,M)$, l'ensemble des cobords de Hochschild de $A$ à valeurs dans $M$ de degré $1$.
\end{defn}

\begin{rem}
On peut également associer à tout élément $a\in A$ une application $ad_{a}:M \to M$ définie pour tout $A$-bimodule $M$ par la formule $ad_{a}(m)=[a,m]$.  Cette application peut alors s'étendre au complexe de Hochschild $C_{n}(A,M)$ de la manière suivante:
  \begin{align}
    ad_{a}(m\otimes a_{1}\cdots a_{n})=\sum_{i=0}^{n} m \cdots \otimes [a,a_{i}]\otimes \cdots \otimes a_{n} \ .
  \end{align}
Cette opération commute avec l'opérateur $b$ de Hochschild et  passe donc en homologie. 
On peut en fait construire une homotopie $h(a): C_{n}(A,M)\to C_{n+1}(A,M)$ définie par:
\begin{align}
  h(a)(m\otimes a_{1}\cdots a_{n} )= \sum_{i=0}^{n}(-1)^{i}m\otimes \cdots a_{i}  \otimes a \otimes a_{i+1}\otimes \cdots \otimes a_{n} \ ,
\end{align}
tel que $bh(a)+h(a)b = -ad_{a}$.
Ainsi, l'application $ad_{a *}: H_{n}(A,M)\to H_{n}(A,M)$ est l'application nulle.
\end{rem}

\begin{rem}
  Pour $M=A$, $\der(A,M)=\der(A)$ a une structure d'algèbre de Lie pour le crochet $[X,Y]=XY-YX$, avec $X,Y \in \der(A)$.
\end{rem}

\begin{rem}
  $\Int(A)$ est un idéal d'algèbre de Lie de $\der(A)$ et également un sous-$Z(A)$-module.
On note $Out(A)= \der(A)/\Int(A)$ l'algèbre de Lie quotient.
C'est une algèbre de Lie et un $Z(A)$ module et ses éléments sont appelés {\bf dérivations extérieures}.
L'espace $Out(A)$ est isomorphe au groupe $HH¹(A)$.
De même on peut définir $Out(A,M)=\der(A,M)/\Int(A,M)$ pour un bimodule $M$ quelconque et c'est un $Z(A)$-module isomorphe à $H¹(A,M)$.
\end{rem}

\subsection{Calcul différentiel basé sur les dérivations}
  Nous allons construire une algèbre différentielle graduée  $\Oder(A)$, appelée calcul différentiel basé sur les dérivations.
  Cette construction s'inspire de la manière dont sont construites les formes différentielles en géométrie ordinaire sur une variété $V$.
En effet, celles-ci sont construites comme les applications différentiables $n$-linéaires antisymétriques sur $\Gamma(V)$, l'algèbre de Lie formée  par les champs de vecteurs sur $V$, et la différentielle $d$ est définie par la formule de Koszul.
On peut montrer que $\Gamma(V)$ est l'algèbre de Lie des dérivations de l'algèbre de fonctions $C^{\infty}(V)$.
  Cette construction peut être rendue complètement algébrique et être ainsi généralisée aux algèbres associatives quelconques en remplaçant l'algèbre de Lie des champs de vecteurs par l'algèbre de Lie $\der(A)$.
Nous allons donner directement cette construction.

On définit le complexe $\uOder(A)$ comme étant l'ensemble des applications $Z(A)$-multi\-linéaires antisymétriques de $\der(A)$ dans $A$.
C'est une algèbre naturellement $\gN$-graduée que l'on peut décomposer de la manière suivante:
\begin{equation}
  \uOder(A)= \bigoplus_{n=0}^{\infty} \uOder^{n}(A) \ ,
\end{equation}
où $\uOder^{0}(A)$ est identifié à $A$ et $\uOder^{n}(A)$ est l'ensemble des applications $n$-$Z(A)$-multi\-linéaires antisymétriques de $\der(A)$ dans $A$.

La structure d'algèbre est donnée par le produit:
\begin{equation}
  (\omega \eta)(X_{1}, \cdots, X_{m+n})= \sum_{\sigma \in S_{m+n}} \frac{(-1)^{|\sigma|}}{m!n!} \omega(X_{\sigma(1)}, \cdots, X_{\sigma(m)} ) \eta(X_{\sigma(m+1)}, \cdots, X_{\sigma(m+n)} ) \ ,
\end{equation}
pour tout  $\omega \in \uOder^{m}(A)$ et   $\eta \in \uOder^{n}(A)$. 
On  définit une différentielle $\hd$ de degré $1$ par la formule de Koszul, en posant pour tous $X_{1}, \dots, X_{n+1} \in \der(A) $ et tout  $\omega \in \uOder^n(A)$:
\begin{multline}
  \hd\omega(X_1, \dots , X_{n+1}) = \sum_{i=1}^{n+1} (-1)^{i+1} X_i  \omega( X_1, \dots \omi{i} \dots, X_{n+1}) \\
  + \sum_{1\leq i < j \leq n+1} (-1)^{i+j} \omega( [X_i, X_j], \dots \omi{i} \dots \omi{j} \dots , X_{n+1}) \ .
  \label{differential}
\end{multline}

 On montre que la différentielle $\hd$ est une application de complexe, \textit{i.e.} $\hd^{2}=0$, en utilisant l'identité de Jacobi sur $\der(A)$.

 \begin{rem}
   Lorsque $A=C^{\infty}(V)$, l'algèbre $\uOder(A)$ coïncide avec l'algèbre des formes de de~Rham sur $V$.
 \end{rem}

 On peut également définir une sous-algèbre différentielle graduée de $\uOder(A)$:
\begin{defn}
  $\Oder(A)$  est la plus petite sous-algèbre différentielle graduée de $\uOder(A)$ contenant $A$.
  De manière équivalente, c'est la sous-algèbre différentielle graduée de $\uOder(A)$ engendrée par $A$.
\end{defn} 
Du fait que  $\Oder(A)$ soit engendrée par $A$, elle est un quotient du calcul différentiel universel $\Omega(A)$ introduit dans la section~\ref{sec:calc-diff-univ}.
Le noyau de ce quotient peut être construit explicitement en utilisant une filtration de $\Omega(A)$.
Cette construction est donnée dans~\cite{masson:95}.

\subsubsection{\texorpdfstring{Propriété universelle de $\Oder^{1}(A)$}{Propriété universelle}}

 Nous avons vu que le calcul différentiel $\Omega(A)$ (resp.  $\Ou(A)$) est un objet universel dans la catégorie des algèbres différentielles graduées et que
$\Omega$ (resp.  $\Ou$) est le  foncteur adjoint à gauche du foncteur d'oubli qui va vers la catégorie des algèbres associatives (resp. algèbres associatives avec unité).
Nous allons voir qu'il est possible de caractériser $\Omega$ autrement.
En effet, pour une algèbre $A$, le calcul différentiel universel du premier ordre $\Omega^{1}(A)$ peut être caractérisé par la propriété universelle suivante dans la catégorie des $A$-bimodules:
\begin{prop}
  Pour toute dérivation $X: A \to M$, vers un $A$-bimodule $M$, il existe une unique application $i_{X}:\Omega^{1}(A)\to M$, telle que le diagramme suivant soit commutatif:
  \begin{align}
  \begin{aligned}
    \xymatrix{
      A \ar^-{X}[dr] \ar[r]& \ar@{-->}[d]\Omega^{1}(A) \\
      & M  }
  \end{aligned} \ .
\end{align}
\end{prop}
La  catégorie des $A$-bimodules admet un produit tensoriel sur l'algèbre $A$.
Nous pouvons ainsi définir, de manière naturelle, le calcul différentiel universel d'ordre $n$ comme $\Omega^{n}(A)=\Omega^{1}(A) \otimes_{A} \cdots \otimes_{A} \Omega^{1}(A)$, avec $n$ facteurs. 
On retrouve ainsi la même algèbre graduée que dans les constructions précédentes en posant $\Omega(A)=\bigoplus_{n\geq 0}\Omega^{n}(A)$.

Cette approche avec les bimodules va nous permettre de rester plus près des notions que l'on côtoie en géométrie ordinaire et de définir d'autres calculs différentiels universels généralisant le calcul différentiel universel de Kähler (ou de de~Rham).
En effet, sur une algèbre commutative, la catégorie des modules à gauche est équivalente à la catégorie des modules à droite qui est équivalente à la catégorie des bimodules.
De la notion de bimodule sur une algèbre commutative, nous pouvons essayer de garder certaines propriétés lorsque l'on passe aux algèbres associatives quelconques.

\subsubsection{bimodules centraux}\label{sec:bimodules-centraux}
Afin de garder la symétrie entre modules à gauche et modules à droite, il est naturel de considérer la catégorie des bimodules centraux définie de la manière suivante: 
\begin{defn}
  Soit $A$ une algèbre et $M$ un $A$-bimodule.
  Alors $M$ est un bimodule central si on a la relation $zm=mz$ pour tout $z\in Z(A)$, le centre de $A$, et pour tout $m\in M$.  
\end{defn}
Un tel bimodule peut être obtenu à partir d'un bimodule $M$ quelconque de deux manières.
On peut soit considérer le bimodule $M_{Z}$ obtenu par quotient de $M$ par son sous-bimodule $[Z(A),M]$, soit le bimodule $M^{Z}$ qui est le sous-bimodule de $M$ constitué des éléments de $M$ qui commutent avec le centre de $A$.
D'après cette définition, nous pouvons associer à toute algèbre un calcul différentiel du premier ordre: $\Omega_{Z}^{1}(A)= \Omega^{1}(A)/[Z(A),\Omega^{1}(A)]$.
Ce calcul différentiel est solution du problème universel suivant:
\begin{prop}
  Pour toute dérivation $X: A \to M$, vers un $A$-bimodule central $M$, il existe une unique application $i_{X}:\Omega_{Z}^{1}(A)\to M$, telle que le diagramme suivant soit commutatif:
  \begin{align}
    \begin{aligned}
      \xymatrix{
        A \ar^-{X}[dr] \ar[r]& \ar@{-->}^-{i_{X}}[d]\Omega_{Z}^{1}(A) \\
        & M \ .}
    \end{aligned}  
  \end{align}
\end{prop}

On peut construire une algèbre différentielle graduée $\Omega_{Z}(A)$ ayant une propriété universelle analogue à celle de $\Omega(A)$.
On la construit en considérant le quotient de $\Omega(A)$ par l'idéal engendré par $[Z(A),\Omega^{1}(A)]$.
La propriété universelle satisfaite par $\Omega_{Z}(A)$ est la suivante:
\begin{prop}
  Tout morphisme $\phi$ de $A$ dans $\cC^{0}=\cO(\cC)$, où $(\cC,\delta)$ est une algèbre différentielle graduée, tel que $\phi(z) \delta \phi(x)= \delta \phi(x) \phi(z)$, pour tout $x\in A$ et $z\in Z(A)$, admet une unique extension par un homomorphisme d'algèbre différentielle graduée $\tilde{\phi}_{Z}: \Omega_{Z}(A) \to \cC$. 
\end{prop}

Il est possible de considérer la catégorie $\mathbf{Alg}_{Z}$ dont les objets sont les algèbres associatives et les flèches sont les morphismes d'algèbres $\phi:A\to B$ tel que $\phi(Z(A))\subset Z(B)$.
Dans \cite{dubois-mich:94,dubois-violette:japan99}, il est défini une notion de bimodule pour une catégorie d'algèbres quelconque. 
D'après leur définitions, les bimodules pour la catégorie $\mathbf{Alg}_{Z}$ correspondent alors aux bimodules centraux.
De même, nous pouvons définir la catégorie des algèbres différentielles graduées centrales que l'on notera $\mathbf{Diff}_{Z}$ en considérant les algèbres différentielles graduées $\cC$ telles que $\cC^{n}$ est un $\cC^{0}$-bimodule central. Les morphismes de cette catégorie sont les morphismes d'algèbre différentielle graduée qui sont également des morphismes de bimodules centraux.
On peut alors voir que $\Omega_{Z}$ réalise un foncteur contravariant de $\mathbf{Alg}_{Z}$ dans $\mathbf{Diff}_{Z}$ et que ce foncteur est l'adjoint à gauche du foncteur d'oubli allant de $\mathbf{Diff}_{Z}$ dans $\mathbf{Alg}_{Z}$.

\subsubsection{\texorpdfstring{Bimodules diagonaux et $\Oder(A)$}{Bimodules diagonaux}}
\label{sec:bimod-diag-Oder}
Lorsque l'algèbre $A$ est commutative, les modules de dimension finie sont des modules projectifs et sont des sous-bimodules de modules libres sur $A$.
On peut essayer de garder cette notion de finitude dans le cas général.
Pour une algèbre $A$, on considère les $A$-bimodules $M$ qui sont isomorphes à un sous-bimodule de $A^{\cI}$, pour un ensemble $\cI$ quelconque.
On appelle ces modules {\it bimodules diagonaux}.
Nous pouvons introduire une notion de dualité entre les bimodules diagonaux sur $A$ et les modules sur son centre $Z(A)$.
En effet, à tout $A$-bimodule $M$, on peut associer le $Z(A)$-bimodule, $M^{\star A}=Hom_{A}^{A}(M,A) $,  l'ensemble des homomorphismes de bimodules de $M$ dans $A$.
Réciproquement, à tout $Z(A)$-module $N$, on associe le $A$-bimodule, $N^{\star A}=Hom_{Z(A)}(N,A)$, l'ensemble des homomorphismes de $Z(A)$-modules de $N$ dans $A$.
On remarque que $N^{\star A}$ est un $A$-bimodule central.
Ainsi, cette relation de dualité peut être restreinte à une dualité entre les $A$-bimodules centraux et les $Z(A)$-modules.
Cette dualité généralise la dualité entre modules à gauche et modules à droite dans le cas des algèbres commutatives.

On peut définir une application canonique:
\begin{equation}
  \begin{aligned}
    c: M &\to M^{\star A \star A}\\
    m &\mapsto \omega_{m}: \biggl(Hom_{A}^{A}(M,A)\to A: \eta \mapsto \eta(m)\biggr)  \ .
  \end{aligned} 
\label{eq:apps-c}
\end{equation}
On a alors la proposition démontrée dans~\cite{dubois-violette:japan99}:
\begin{prop}
Soit $M$ un $A$-bimodule.
$M$ est diagonal \ssi l'homomorphisme canonique $c:M \to M^{\star A \star A}$ est injectif.
\end{prop}
Ainsi, on voit que tout bimodule diagonal est central et que le dual d'un $Z(A)$-module est un $A$-bimodule diagonal.
On a également les propriétés que tout sous-bimodule d'un bimodule diagonal est diagonal et, que le résultat du produit ou du produit tensoriel sur $A$ de deux bimodules diagonaux est un bimodule diagonal.

Nous pouvons maintenant définir un calcul différentiel du premier ordre $\Omega_{diag}^{1}(A)$ de la même manière que nous l'avons fait avec les bimodules centraux.
Ce calcul différentiel n'est autre que l'image par l'application $c$ (\ref{eq:apps-c}) du bimodule $\Omega(A)$ et il satisfait la propriété universelle suivante:
\begin{prop}
  Pour toute dérivation $X: A \to M$, vers un $A$-bimodule $M$ diagonal, il existe une unique application $i_{X}:\Omega_{diag}^{1}(A)\to M$, telle que le diagramme suivant soit commutatif:
  \begin{align}
  \begin{aligned}
    \xymatrix{
      A \ar^-{X}[dr] \ar[r]& \ar@{-->}[d]\Omega_{diag}^{1}(A) \\
      & M \ .}
  \end{aligned} 
\end{align}
\end{prop}

De même, on peut définir une algèbre différentielle graduée $\Omega_{diag}(A)$ qui est  le quotient de $\Omega(A)$ par l'idéal engendré par le noyau de l'application $c$.
L'algèbre différentielle graduée $\Omega_{diag}(A)$ satisfait une  propriété universelle analogue  à celle satisfaite par $\Omega_{Z}(A)$:
\begin{prop}
  Tout morphisme $\phi$ de $A$ dans $\cC^{0}=\cO(\cC)$, où $(\cC,\delta)$ est une algèbre différentielle graduée, tel que $\phi$ induise une structure de $A$-bimodule diagonal sur $\cC^{1}$, admet une unique extension par un homomorphisme d'algèbres différentielles graduées $\tilde{\phi}_{diag}: \Omega_{diag}(A) \to \cC$. 
\end{prop}

\begin{rem}
  Notons que du fait que les bimodules diagonaux ne soient pas des bimodules d'une catégorie d'algèbres particulière au sens de~\cite{dubois-violette:japan99}, nous n'avons pas d'interprétation de $\Omega_{diag}$ en tant que foncteur contravariant, contrairement à $\Omega$, $\Omega_{u}$ et $\Omega_{Z}$.
\end{rem}

Enfin, nous pouvons voir le lien avec le calcul différentiel basé sur les dérivations. 
Tout d'abord, notons qu'il y a une application naturelle de $\Omega_{diag}^{1}(A)$ vers $\Oder(A)$ qui en fait un isomorphisme~\cite{dubois-violette:japan99}:
\begin{equation}
  \begin{aligned}
    \Omega_{diag}^{1}(A)&= \Oder^{1}(A) \ .
  \end{aligned}
\end{equation}
Cela nous permet ainsi de donner une interprétation de $\Oder^{1}(A)$ en tant que calcul différentiel universel du premier ordre diagonal.
Ensuite, $\der(A)$ est naturellement un $Z(A)$-module et est relié par la dualité précédente au calcul différentiel basé sur les dérivations de la manière suivante:
\begin{equation}
  \begin{aligned}
    (\Oder^{1}(A))^{\star A}&\simeq \der(A)\\
    \der(A)^{\star A} &\simeq \uOder^{1}(A)
    \label{eq:dual}
  \end{aligned} \ \ .
\end{equation}
L'isomorphisme $ (\Oder^{1}(A))^{\star A}\simeq \der(A)$ est réalisé par l'opération de Cartan (voir section~\ref{sec:operations-de-cartan}) $\cX\mapsto i_{\cX}$.
Cette construction donne ainsi un statut satisfaisant à  $\Oder^{1}(A)$.
Cependant l'algèbre différentielle graduée $\Oder(A)$ ne semble pas être caractérisée par une propriété universelle.
Pour l'instant cette algèbre différentielle graduée n'a donc pas de statut particulier.
Néanmoins, nous pouvons essayer de comprendre ce qu'elle capture de l'algèbre et dans quelles situations elle peut être utilisée.
Cela sera en particulier le calcul différentiel que nous étudierons dans la section~\ref{sec:alg-des-endom} afin de caractériser un analogue algébrique des fibrés principaux. 

Nous avons vu qu'à toute algèbre associative, nous pouvions associer l'algèbre de Lie $\der(A)$.
Le calcul différentiel basé sur les dérivations utilise de manière essentielle cette structure d'algèbre de Lie et mélange en quelque sorte, de manière minimale, la structure d'algèbre de Lie de $\der(A)$ et la structure d'algèbre associative de $A$.
Il y a des algèbres pour lesquelles il peut ne pas être naturel d'utiliser ce calcul différentiel comme par exemple l'algèbre du plan de Manin engendrée par deux générateurs $x$ et $y$ satisfaisant la relation $xy=qyx$, avec $q$ un nombre  complexe.
Dans cet exemple, nous avons plutôt envie d'étudier, non pas des dérivations au sens usuel du terme, mais plutôt une notion de dérivation $q$-déformée, afin de se rapprocher de la situation commutative.
Cela sera généralement le cas pour les algèbres obtenues par déformation d'algèbres commutatives.
Il est donc clair, que selon le type de non commutativité introduite, il faudra utiliser des outils différents.
Il n'existe malheureusement pas aujourd'hui de manière systématique d'étudier une algèbre non commutative.
C'est un des buts de la géométrie non commutative d'essayer de clarifier ces méthodes et d'en donner une vision plus homogène.

\subsection{Opérations de Cartan}\label{sec:operations-de-cartan}
 Nous allons définir des opérations au sens de H. Cartan~\cite{cartan:51} sur le calcul différentiel universel $\Ou(A)$ ainsi que sur les calculs différentiels $\uOder(A)$ et $\Oder(A)$.

Tout d'abord, nous pouvons rappeler les définitions de base.
Soit $\cG$ une algèbre de Lie et $(\Omega,d)$ une algèbre différentielle graduée.
On dit que l'on a une opération de $\cG$ sur $\Omega$ si il existe une application linéaire $i$ qui à tout élément $X\in \cG$ associe une dérivation $i_{X}$ de degré $-1$ et une application linéaire $L$ qui à tout élément $X\in \cG$ associe une dérivation $L_{X}$ de degré $0$, tel que l'on ait%
\footnote{Nous considérons ici des commutateurs gradués.}:
\begin{equation}
  \begin{aligned}
    L_{X} &= [i_{X},d] & 
    [i_{X},i_{Y}]&=0\\
    [L_{X},L_{Y}]&= L_{[X,Y]} &
    [L_{X},i_{Y}]&=i_{[X,Y]} 
  \end{aligned} \ \ , \label{eq:cartan}
\end{equation}
pour tout $X, Y \in \cG$.
Notons que l'on a pour conséquence de ces relations et du fait que $d^{2}=0$:
\begin{equation}
[L_{X},d] = 0 \ .
\end{equation}

Nous introduisons les terminologies suivantes:
\begin{itemize}
\item Nous appellerons élément \textbf{horizontal}, tout élément $\omega$ de $\Omega$ satisfaisant la relation $i_{X}=0, \forall X\in \cG$.
\item Nous appellerons élément \textbf{invariant}, tout élément $\omega$ de $\Omega$ satisfaisant la relation $L_{X}=0, \forall X\in \cG$.
\item Nous appellerons élément \textbf{basique}, tout élément $\omega$ de $\Omega$ invariant et horizontal.
\end{itemize}
Cela nous permet de caractériser deux sous-algèbres différentielles graduées de $\Omega$ con\-struites à partir d'une opération:
\begin{itemize}
\item L'ensemble des éléments invariants de $\Omega$ forme une sous-algèbre différentielle graduée de $\Omega$ que l'on note $\Omega_{\inv}$.
\item L'ensemble des éléments basiques de $\Omega$ forme une sous-algèbre différentielle graduée de $\Omega$ que l'on note $\Omega_{\bas}$.
\end{itemize}

Nous allons maintenant donner deux exemples d'opérations de Cartan de $\der(A)$ sur $\Ou(A)$ et $\uOder(A)$.

\subsubsection{\texorpdfstring{Opérations sur $\Ou(A)$}{Opérations sur le calcul différentiel universel}}

Considérons l'algèbre de Lie $\der(A)$.
D'après la propriété universelle de $\Omega(A)$, toute dérivation $X\in \der(A)$ se relève en une unique application $i_{X}: \Omega^{1}(A)\to A$, telle que $i_{X} d=X$.
Il est alors possible d'étendre cette application en une opération de Cartan de $\der(A)$ sur l'algèbre différentielle graduée $\Omega(A)$.
Nous avons la proposition suivante:

\begin{prop}
  Soit $X\in \der(A)$.
Alors l'homomorphisme de bimodule $i_{X}:\Omega^{1}\to A$ se prolonge de manière unique en une dérivation graduée de degré $-1$ sur $\Omega(A)$. 
\end{prop}
Étant donnée que 
$\Omega^{n} =\Omega^{1}(A)  \otimes _{A} \cdots \otimes_{A} \Omega^{1}(A)$, l'application $i_{X}$ se prolonge sur $\Omega^{n}$ de la manière suivante:
\begin{equation}
  \begin{aligned}
    i_{x}(a_{0}da_{1}\cdots da_{n}) = \sum_{p=1}^{n}(-1)^{p+1}a_{0}da_{1}\cdots da_{p-1}\cdot X(a_{p}) \cdot da_{p+1}\cdots da_{n} \ .
  \end{aligned}
\end{equation}

L'application linéaire $L_{X}$ est définie par $L_{X}=i_{X}\circ d + d\circ i_{X}$.
On montre aisément que les relations~(\ref{eq:cartan}) sont satisfaites.

 \begin{rem}
   L'application $i_{X}$ que nous venons de définir peut servir~\cite{dubois-violette:88} à associer une filtration $F$ à $\Omega(A)$ en posant:
   \begin{align}
     F^{p}\omega^{n}=\{ \omega \in \Omega(A) / i_{X_{1}} \cdots i_{X_{n-p+1}} \omega=0, X_{i}\in \der(A) \} \ .
   \end{align}
   Cette filtration permet de caractériser~\cite{masson:95} le quotient de $\Omega(A)$ vers $\Oder(A)$.
 \end{rem}
 
 \begin{rem}
   On voit ainsi qu'à toute sous-algèbre de Lie $\cG$ de $\der(A)$, on peut associer une opération de Cartan de $\cG$ sur $\Omega(A)$.
\end{rem}

\subsubsection{\texorpdfstring{Opérations sur $\uOder(A)$}{Opérations sur le calcul différentiel basé sur les dérivations}}
Nous pouvons construire des applications de Cartan $i_{X}$ et $L_{X}$ sur $\uOder(A)$, ainsi que sur $\Oder(A)$, en prenant les images respectives par les applications canoniques $\Omega(A)\to \uOder(A)$ et $\Omega(A)\to \Oder(A)$ des applications%
\footnote{Afin de simplifier les notations, nous notons $i_{X}$ et $L_{X}$ les opérations de Cartan indifféremment sur  $\uOder(A)$, $\Oder(A)$ ou  $\Omega(A)$.}
 $i_{X}$ et $L_{X}$ définies sur $\Omega(A)$.

Cependant, ces applications peuvent être définies plus directement sur $\uOder(A)$ et $\Oder(A)$ de la manière suivante:
\begin{defn}  
Pour toute dérivation $X \in \der(A)$, on définit l'application $i_{X}: \uOder^{n}(A) \rightarrow \uOder^{n-1}(A)$, appelée {\bf produit intérieur} par:
   \begin{align}
 (i_{X}\omega)(X_{1},\cdots,X_{n-1}) &=\omega(X, X_{1},\cdots,X_{n-1}) & &
 \text{pour tout $\omega \in \uOder(A)$}.
\end{align}
\end{defn}

Le produit intérieur $i_{X}$ est bien une dérivation graduée de degré $-1$ et est l'image de l'application $i_{X}$ définie sur $\Omega(A)$.
On définit également la dérivation de degré $0$,  $L_{X}=[i_{X},d]$, que nous appellerons {\bf dérivée de Lie} par analogie avec ce qui se passe en géométrie différentielle.
On vérifie à nouveau que $i_{X}$ et $L_{X}$ définissent bien une opération de Cartan.

\begin{rem}
  On voit que $\Oder(A)$ est stable par l'action de $i_{X}$. Les opérations $i_{X}$ et $L_{X}$ peuvent donc se restreindre à $\Oder(A)$.
\end{rem}

L'algèbre de Lie $\der(A)$ admet un idéal naturel $\Int(A)$, qui est donc une sous-algèbre de Lie de $\der(A)$.
Nous avons donc une opération naturelle de $\Int(A)$ sur $\uOder(A)$, ce qui nous permet d'introduire la notion suivante:
\begin{defn}
  On définit  l'algèbre différentielle graduée  $\Omega_{\Out}(A)$ comme étant la sous-algèbre différentielle graduée de $\uOder(A)$  constituée des éléments basiques de $\uOder(A)$ pour l'opération de $\Int(A)$.
\end{defn}

\section{Conclusion}
Il existe d'autres outils importants, tels que la $K$-théorie et les notions liées aux algèbres d'opérateurs, pouvant servir en géométrie non commutative pour caractériser des algèbres.
Seuls les outils abordés dans ce chapitre nous seront utiles pour la suite.
En effet, le calcul différentiel basé sur les dérivations sera utilisé afin de décrire un analogue des théories de jauge en géométrie non commutative. 
Nous verrons dans quelles mesures ce formalisme peut permettre de remplacer celui des fibrés principaux utilisé pour les théories de jauge ordinaires.


\chapter{Théories de jauge et géométrie non commutative}
\label{cha:theories-de-jauge}

Le but de ce chapitre est d'établir un cadre cohérent pour pouvoir définir des théories de jauge en géométrie non commutative.
Habituellement, nous entendons par théorie de jauge, la construction d'un fibré principal au dessus d'une variété, ainsi que la construction de fibrés associés. La théorie des connexions (ordinaires) est essentielle pour comprendre la géométrie de  ces structures.
Les connexions correspondent à ce qu'on appelle les champs de jauge en physique et les sections de fibrés associés correspondent eux aux champs de matière.

En géométrie non commutative, nous devons remplacer la notion d'espace par la notion d'algèbre.
Ainsi, afin de définir une théorie de jauge en géométrie non commutative, on doit remplacer une variété différentiable $M$ (correspondant à l'espace-temps en physique) par son algèbre de fonctions $C^{\infty}(M)$.
Nous savons que les fibrés vectoriels, qui sont des fibrés associés à un fibré principal au dessus de $M$, sont remplacés par les modules projectifs de type fini.

Dans le cas des algèbres commutatives de la forme $C^{\infty}(M)$, il est donc légitime de se demander quelle notion peut remplacer celle de fibré principal. 
Nous allons voir qu'un bon candidat est l'algèbre d'endomorphismes définie à partir d'un fibré vectoriel au dessus de $M$.
En plus de remplacer la notion de fibré principal, cette algèbre aura l'avantage de pouvoir également remplacer l'algèbre $C^{\infty}(M)$.
En effet, ces deux algèbres sont équivalentes de Morita et donc l'ensemble des modules projectifs de type fini (\textit{p.d.t.f.}) sur l'une est équivalent à l'ensemble des modules projectifs de type fini sur l'autre.
Du point de vue de la physique, ceci veut dire qu'un champ physique (sections d'un fibré vectoriel) peut tout aussi bien être décrit par un élément d'un module \textit{p.d.t.f.} sur $C^{\infty}(M)$ que par un élément d'un module \textit{p.d.t.f.} sur une algèbre d'endomorphismes.
Notons que l'algèbre des endomorphismes d'un fibré vectoriel est généralement non commutative. 
Le fait de décrire les champs physiques dans ce cadre permet par exemple de mettre sur le même pied d'égalité les champs de jauge usuels et les champs de Higgs en introduisant la notion de connexion non commutative.

Nous allons donc, dans un premier temps, nous attacher à donner une notion de connexion non commutative, ou connexion algébrique, généralisant la notion de connexion sur les fibrés (connexions ordinaires).
Dans un deuxième temps, nous donnerons la définition d'une algèbre d'endomorphismes et nous en ferons une étude détaillée. 
A la fin de ce chapitre, nous aurons établi un cadre cohérent pour pouvoir construire une théorie de jauge à partir d'une algèbre d'endomorphismes et des exemples de constructions possibles seront donnés dans le chapitre~\ref{cha:modeles-physiques}.

\section{Connexions}

Nous allons voir différentes approches qui permettent de généraliser la notion de connexion sur les fibrés vectoriels en géométrie et donnent une notion de connexion sur des modules.
Nous donnerons une première définition pour des modules simples (à gauche ou à droite), puis donnerons deux approches possibles des connexions sur les bimodules.

\begin{defn}
  Soit $A$ une algèbre associative avec unité, $\Omega$ un calcul différentiel sur cette algèbre ($\Omega$ ici n'est pas nécessairement le calcul différentiel universel) et $M$ un $A$-module.
Alors, une $\Omega$-connexion sur $M$ est%
  \footnote{ou simplement connexion sur $M$ si il ne peut y avoir de confusion}
 une application linéaire $\nabla: M \to \Omega^{1}\otimes_{A} M$ telle que l'on ait
  \begin{align}
    \nabla(a m ) &=a\nabla(m) + d(a) \otimes_{A} m &
    & \forall a\in A, m \in M \ .
   \end{align}
Du fait que $\Omega^{1}$ soit un $A$-bimodule, l'espace vectoriel $ \Omega^{1}\otimes_{A} M$ est naturellement équipé d'une structure de $A$-module. Nous pouvons ainsi étendre la définition de $\nabla$ au complexe $\Omega\otimes_{A} M$ en posant:
\begin{align}
    \nabla(\omega \otimes_{A} m ) &=\omega \cdot \nabla(m) + d\omega \otimes_{A} m &
    & \forall \omega\in \Omega, m \in M \ . 
\end{align}
\end{defn}

Cette dernière définition nous permet de considérer l'application linéaire $\nabla^{2}=\nabla \circ \nabla$ qui est un endomorphisme de $\Omega \otimes_{A} M$ dont la restriction $\nabla^{2} : M \to \Omega^{2}\otimes_{A} M$ est un homomorphisme de $A$-modules.
Cet homomorphisme est appelé la {\it courbure} de la connexion $\nabla$.
Une connexion pour laquelle la courbure est nulle, {\it i.e.} $\nabla^{2}=0$ est appelée {\it connexion plate}.
De même, la différence de deux connexions $\nabla^{1}$, $\nabla^{2}$ définit un homomorphisme de $A$-modules: $\nabla^{1}- \nabla^{2}: M \to \Omega^{1}\otimes_{A} M$, ce qui donne à l'espace des connexions sur un module $M$, une structure d'espace affine modelée sur $\Hom(M,\Omega^{1}\otimes_{A}M)$.

Notons qu'un $A$-module $M$ n'admet pas nécessairement une connexion. 
Cependant, pour certaines catégories de modules, nous pouvons montrer l'existence d'au moins une connexion.

Nous pouvons par exemple considérer le cas des modules libres.
Tout module libre peut se mettre sous-la forme $A\otimes E$, où $E$ est un espace vectoriel.  
Il existe alors une connexion plate canonique qui est $\nabla= d \otimes I_{E}$. 
  
De même, pour un module projectif $M$, défini par un projecteur $P: A \otimes E \to M$, on peut définir une connexion canonique $\nabla= P ( d\otimes I_{E})$. Ainsi, tout module projectif admet au moins une $\Omega$-connexion.
Dans la situation où $\Omega$ est le calcul différentiel universel $\Ou(A)$, la réciproque est vraie (\textit{c.f.}~\cite{cuntz:95}): un $A$-module admet une $\Ou(A)$-connexion \ssi c'est un module projectif.

\subsubsection{Modules à droite}
De manière analogue, nous pouvons définir la notion de connexion pour les modules à droite. Si $N$ est un $A$-module à droite, une $\Omega$-connexion sur $N$ est une application linéaire $\nabla: N \to N\otimes_{A}\Omega^{1}$ telle que $\nabla(na)=\nabla(n)a+ n\otimes_{A}d(a)$ pour tout $n\in N$ et $a\in A$.

\subsubsection{Connexion duale}
Lorsque l'on a une connexion $\nabla$ sur un module $M$, il est possible de lui associer une connexion $\nabla^{*}$ sur son dual%
\footnote{Le dual d'un $A$-module à gauche $M$ est l'ensemble des homomorphismes de $A$-modules à gauche de $M$ dans $A$.}
 $M^{*}$ en posant:
\begin{align}
  \langle m, \nabla^{*} (n)\rangle &=  d(\langle m,n\rangle ) - \langle\nabla(m),n\rangle &
  &\forall m\in M, n\in M^{*}.
\end{align}

\subsubsection{Connexion hermitienne}
Lorsque l'on a un module $M$ hermitien, c'est-à-dire un module $M$ sur $*$-algèbre $A$ doté d'une structure hermitienne $h: M \times M \to A$, une connexion $\nabla$ est dite hermitienne lorsque l'on a:
\begin{align}
  d(h(m,n))&= h(\nabla m,n) + h(m,\nabla n)&
  &  \forall m,n \in M.
\end{align}

\subsubsection{Transformations de jauge}
De même, nous pouvons introduire une notion similaire à celle de groupe de transformation de jauge sur les fibrés vectoriels.
Soit $M$ un $A$-module à droite. Le groupe $\Aut(M)$ de tous les automorphismes de $M$ agit sur l'espace affine des connexions sur $M$ de la manière suivante:
\begin{align}
  \nabla \to \nabla^{U}= U^{-1} \circ \nabla \circ U \ ,
\end{align}
où $U\in \Aut(M)$ (on a naturellement $\Aut(M)\subset \Aut(M\otimes_{A} \Omega)$).

Si $A$ est une $*$-algèbre et $M$ un module avec structure hermitienne $h$, alors le sous-groupe de $\Aut(M)$ constitué des éléments $U$ qui préservent $h$, c'est-à-dire, tel que 
\begin{align*}
  h(Um,Un)&=h(m,n) && \forall m,n \in M
\end{align*}
sera appelé le \textbf{groupe de jauge}. On le notera $\Aut(M,h)$ et ses éléments seront appelés \textbf{transformations de jauge} de $M$.

\subsection{Connexions sur un bimodule}
Nous avons vu précédemment que la notion de bimodule est déjà présente dans la définition de connexion sur un module simple.
De plus, la question de définir la notion de connexion sur un bimodule se pose dès que l'on travaille avec une algèbre avec involution et que l'on veut étudier des notions de réalité sur les modules.
Il est donc naturel d'essayer d'étendre cette notion aux bimodules.

Pour un bimodule, il est délicat d'appliquer la définition de connexion que nous avons donnée précédemment car nous devons pour cela oublier la structure de bimodule et considérer un bimodule $M$  comme un module simple, à gauche ou à droite.
Cette démarche n'est pas très satisfaisante et il est alors souhaitable d'avoir une notion de connexion se servant de la structure de bimodule. De manière plus précise, nous voudrions qu'une connexion sur un bimodule soit une application linéaire entre bimodules.
C'est, par exemple, absolument nécessaire pour discuter de la notion de connexion linéaire où le bimodule en question est un calcul différentiel.

Une notion de connexion a été proposée dans~\cite{mourad:94,dubois-violette:95} et est la suivante:
\begin{defn}
  Soit $M$ un $A$-bimodule.
  Une $\Omega$-connexion à gauche du $A$-bimodule $M$ est une $\Omega$-connexion $\nabla$ du module $M$ considéré comme $A$-module à gauche tel qu'il existe un homomorphisme de bimodule $\sigma: M \otimes_{A}\Omega^{1}\to \Omega^{1} \otimes_{A} M$ tel que:
  \begin{align}
    \nabla(ma)&=\nabla(m)a + \sigma(m \otimes_{A} da)&
    &\forall a\in A, m\in M \ .
  \end{align}
\end{defn}
Nous pouvons également donner une définition similaire de connexion à droite sur un bimodule.

Avec cette définition, il est possible de définir (voir~\cite{mourad:94}) une notion de produit tensoriel sur des connexions, ainsi que les notions de connexion duale et connexion hermitienne, comme nous l'avons fait pour les connexions sur des modules simples.

\subsubsection{Connexion et opérateurs différentiels}
Il est également possible de voir (\textit{c.f.}~\cite{dubois-violette:95}) les connexions sur les bimodules en terme d'opérateurs différentiels du premier ordre dans les bimodules.
Cette approche est plus proche de ce que nous allons voir par la suite.
Pour plus de précisions, nous renvoyons le lecteur intéressé sur cet article.

Nous allons maintenant voir une autre notion de connexion qui part d'un autre point de vue sur les connexions dans le cadre des fibrés vectoriels sur une variété et qui sera basée sur les dérivations.

\subsection{Connexions basées sur les dérivations}
Nous allons maintenant voir une notion de connexion qui fut proposée dans~\cite{dubo-mich:95}.
En géométrie ordinaire, une connexion peut être vue comme une application qui relève un champ de vecteur sur la variété de base d'un fibré vectoriel en un champ de vecteur sur le fibré.
On appelle souvent cela ``transport parallèle''.
Il est ainsi naturel de considérer la définition suivante pour une connexion algébrique:
\begin{defn}
  Soit $A$ une algèbre et $M$ un $A$-bimodule.
 Une connexion est une application $\nabla$ qui à toute dérivation $X\in \der(A)$ sur $A$ associe une dérivation $\nabla_{X} \in \der(A,M)$ sur $M$.
\end{defn}

Cette définition a l'avantage de pouvoir s'appliquer à toutes les catégories de modules (il suffit de considérer un module à gauche comme un bimodule avec une structure de module à droite triviale).
Nous avons vu précédemment que $\der(A)$ est un $Z(A)$-module et est intimement lié à la notion de bimodule central.
D'autre part, dans la situation classique (géométrique), une connexion est une application linéaire par rapport à la multiplication par une fonction sur la base.
Il est donc naturel de se restreindre aux $A$-bimodules centraux et de considérer plutôt la définition suivante:

\begin{defn}
  Soit $M$ un $A$-bimodule central. Une $\Oder(A)$-connexion sur $M$ est une application linéaire:
  \begin{align}
    \nabla: \der(A)&\to \End_{A}(M)
  \end{align}
  telle que  
\begin{align}
  \nabla_{zX}(m)&=z\nabla_{X}(m)\\
  \nabla_{X}(amb)&=a\nabla_{X}(m)b + X(a)mb+amX(b) \ ,
\end{align}
pour tout $m\in M$, tout $X\in \der(A)$, tout $z\in Z(A)$ et tous $a,b \in A$.
\end{defn}

Lorsque $A$ a une involution, nous pouvons définir une notion de réalité de la manière suivante:
\begin{defn}
  Soit $A$ une algèbre involutive et $M$ un $A$-bimodule central avec involution, alors une $\Oder(A)$-connexion sur $M$ est dite \textbf{réelle} si l'on a:
  \begin{align}
    \nabla_{X}(m^{*})&= (\nabla_{X}(m))^{*} &
    &\forall m\in M, X\in \der_{\gR}(A) \ ,
  \end{align}
  où $\der_{\gR}(A):=\{X\in \der(A)/ (X(a))^{*}=X(a^{*}) , \forall a \in A\}$.
\end{defn}

\begin{defn}
  La \textbf{courbure} d'une connexion $\nabla$ est un homomorphisme de $Z(A)$-modules, définie par l'application linéaire antisymétrique: 
\begin{equation}
  \begin{aligned}
    \der(A)\times \der(A) & \to \Hom_{A}^{A}(M) \\
    (X,Y) &\mapsto R_{X,Y}=\nabla_{X}\nabla_{Y}-\nabla_{Y}\nabla_{X} - \nabla_{[X,Y]} \ .
  \end{aligned}
\end{equation}

La courbure vérifie donc les propriétés suivantes:
\begin{equation}
  \begin{aligned}
    R_{zX,Y}(m)&=zR_{X,Y}(m)\\
    R_{X,Y}(amb)&=aR_{X,Y}(m)b
  \end{aligned} \ \ ,
\end{equation}
pour tout $m\in M$, tous $X,Y \in \der(A)$, tout $z\in Z(A)$ et tous $a,b\in A$.
De ce fait, $R$ est un élément de $\Hom_{Z(A)}(\bigwedge^{2}_{Z(A)}\der(A), \Hom_{A}^{A}(M))$.
D'autre part, $R$ satisfait de manière canonique les identités suivantes:
\begin{equation}
  [\nabla_{X},R_{Y,Z}]+[\nabla_{Y},R_{Z,X}]+[\nabla_{Z},R_{X,Y}]=R_{[X,Y],Z}+[\nabla_{Y},R_{Z,X}]+[\nabla_{Z},R_{X,Y}] \ ,
\end{equation}
pour tous $X,Y,Z \in \der(A)$ et sont appelées \textbf{identités de Bianchi}.
\end{defn}

Un $A$-bimodule central $M$ n'admet pas nécessairement de connexions, mais dans le cas où il en existe au moins une, on remarquera que l'espace des connexions sur $M$ est un espace affine modelé sur $\Hom_{Z(A)}(\der(A), \Hom_{A}^{A}(M,M))$.

\begin{rem}
  Cette notion de connexion peut être reliée à la notion de connexion, plus générale, sur un bimodule qui a été donné dans la section précédente.
Il faut pour cela introduire la famille de  $Z(A)$-modules:
  \begin{align}
    \uOder^{n}(A,M)=\Hom_{Z(A)}(\bigwedge^{n}_{Z(A)}\der(A),M) \ .
  \end{align}
Une connexion $\nabla$ peut alors être vue comme application linéaire de $\nabla: M \to \uOder^{1}(A,M)$ qui satisfait:
  \begin{align}
    \nabla(amb) &= da \otimes_{A}mb + a\nabla(m)b + am \otimes_{A}db &
    & \text{ $\forall a,b  \in A$, $\forall m\in M$} \ ,
  \end{align}
où les isomorphismes canoniques de bimodules entre $\uOder^{1}(A)\otimes_{A} M$,  $\uOder^{1}(A,M)$ et $M \-\otimes_{A}\uOder^{1}(A)$ ont été utilisés. 
Cette approche peut se révéler assez riche et permet notamment dans le cas des connexions linéaires (quand le bimodule est $\uOder(A)$) d'introduire une notion de torsion analogue à celle introduite dans la situation géométrique.
Ce point de vue est largement développé dans~\cite{dubo-mich:95}.
\end{rem}

Notons que pour  une algèbre donnée et un bimodule $M$ sur cette algèbre, il n'existe pas toujours de connexions sur $M$. 
Il est également possible de rencontrer la situation inverse, c'est-à-dire qu'il existe toujours au moins une connexion, cette connexion étant parfois canonique.
Nous pouvons voir quelques exemples.

 \subsubsection{Le cas commutatif}
 
 Dans le cas où $A$ est l'algèbre $C^{\infty}(M)$ des fonctions indéfiniment dérivables sur une variété $M$ para-compacte, de dimension finie, $(\Oder(A),d)$ est simplement le complexe $(\Omega(M),d) $ de de~Rham et $\der(C^{\infty}(M))=\Gamma(TM)$ est l'algèbre de Lie des champs de vecteurs sur $M$.
 Si l'on considère le module $\Gamma(E)$ des sections $C^{\infty}$ d'un fibré vectoriel $E$ au dessus de $M$, une connexion sur $\Gamma(E)$ est alors une connexion au sens usuel (géométrique) du terme.

 \subsubsection{\texorpdfstring{Le cas où $\Out(A)=0$}{Le cas où Out(A)=0}}
 Considérons maintenant une algèbre $A$ non commutative pour laquelle l'algèbre de Lie des dérivations extérieures est nulle,\textit{i.e.} $\Out(A)=0$.
 Une telle algèbre n'a, par définition, que des dérivations intérieures, \textit{i.e.} $\der(A)\simeq\Int(A)$.
On peut alors construire une connexion canonique $\tilde{\nabla}: M \to \Oder^{1}(A,M)$, définie par:
 \begin{equation}
   \begin{aligned}
\tilde{\nabla}_{\ad_{x}}(m)= \ad_{x}(m) \ ,
   \end{aligned}
 \end{equation}
pour tout $x \in A$ et tout $m\in M$.
Cette connexion a une courbure nulle.

La connexion canonique peut être prise comme connexion de référence, puisque toute connexion $\nabla$ sur $M$ peut s'écrire sous la forme:
\begin{align*}
\nabla&= \tilde{\nabla} + \Gamma &
&\text{où}
&\Gamma \in \Hom_{Z(A)}(\Int(A),\Hom_{A}^{A}(M,M)) \ .
\end{align*}

\subsubsection{Le cas où le centre est trivial} 

Dans le cas où le centre de l'algèbre est trivial, \textit{i.e.} $Z(A)=k \cdot \gone$, alors la dérivée de Lie 
\begin{align}
  X \mapsto L_{X}=i_{X}d + d i_{X}
\end{align}
définit une connexion sur les bimodules $\Oder^{n}(A)$ et $\uOder^{n}(A)$.
Ces connexions ont toutes une courbure nulle car la dérivée de Lie est un homomorphisme d'algèbres de Lie.

\section{L'algèbre des endomorphismes}\label{sec:alg-des-endom}
\subsection{Motivations}
L'étude des algèbres d'endomorphismes a été motivée par les travaux effectués dans \cite{dubois-violette:89,dubois-violette:89:II,dubois-violette:90,dubois-violette:90:II} pour des algèbres de matrices et de fonctions à valeurs matricielles.
Ces algèbres sont en fait des cas particuliers d'algèbres d'endomorphismes.

Comme nous allons le voir, les algèbres d'endomorphismes, munies d'un calcul différentiel spécifique, peuvent se substituer aux fibrés principaux utilisés habituellement pour décrire  les champs de jauge en physique.
Elles auront l'avantage d'inclure dans la notion de connexion non seulement les champs de Yang-Mills, mais aussi, des champs scalaires jouant des rôles similaires aux champs de Higgs du modèle standard.
Il avait en fait déjà été remarqué que l'on pouvait retrouver de tels champs scalaires en faisant des réductions dimensionnelles de fibrés sur des variétés ayant des symétries particulières~\cite{forgacs:80,hudson:84,farakos:87}. 
L'avantage d'une approche algébrique à partir d'algèbres d'endomorphismes par rapport aux approches basées sur des fibrés principaux, ou des techniques de réductions dimensionnelles, est que nous ``gommons'' tout l'aspect continu sur les fibres qui n'est à priori pas nécessaire pour décrire la physique.
De cette manière, nous avons une description plus minimale des champs physiques que sont les champs de Yang-Mills et les champs de Higgs.

Néanmoins, pour les algèbres d'endomorphismes, les champs scalaires obtenus ne sont pas exactement ceux du modèle standard et sont en fait plus nombreux.
Notons qu'il est tout de même possible, en considérant des algèbres légèrement différentes et un calcul différentiel non commutatif également différent, d'avoir une description  exacte du Lagrangien classique du modèle standard~\cite{connes:90,conn-lott:90,coqu:90}.

Ces types de modèles sont donc à voir, pour l'instant, comme des ``modèles jouets'' nous donnant un nouveau regard sur les théories de jauge.
Ce regard différent nous permettra peut-être de surmonter un jour les diverses difficultés que l'on rencontre actuellement pour comprendre complètement les théories quantiques des champs de ces modèles ou encore pour trouver un modèle géométrico-algébrique permettant d'unifier les théories de jauge à la gravitation (qui est aussi une théorie de jauge, mais traitée différemment des autres).

Notons également que les algèbres d'endomorphismes sont des cas particuliers d'algèbres construites à partir de fibrés en algèbres (n'étant pas nécessairement des fibrés d'endomorphismes).
Ces algèbres sont intéressantes à étudier du point de vue de la $K$-théorie et peuvent être classées par ce que l'on appelle les classes de Dixmier-Douady qui correspondent à des invariants topologiques de variétés. Les algèbres d'endomorphismes étant Morita équivalentes à des algèbres de fonctions ordinaires, la classe de Dixmier-Douady est nulle dans ce cas.
Ces invariants possèdent une interprétation en théorie des cordes et correspondent aux charges des D-branes\dots

 \subsubsection{Les matrices}
Une des algèbres d'endomorphismes la plus simple que l'on puisse considérer est l'algèbre des endomorphismes d'un espace vectoriel $\gC^{n}$, c'est-à-dire l'algèbre  $M_n(\gC)$ des matrices complexes de taille  $n\times n$.
Cette algèbre n'a que des dérivations intérieures et l'algèbre de Lie  $\der(M_n) = \Int(M_n)$ peut être identifiée à l'algèbre de Lie $\ksl_n:=\ksl(n,\gC)$. 
On montre  que l'on a:
\begin{align}
  \Omega_\der(M_n) &\simeq M_n \otimes \exter \ksl_n^\ast \ ,
\label{soudure}
\end{align}
où $\ksl_n^\ast$ est le dual de  $\ksl_n$. 
On note  $\d'$ la différentielle sur ce complexe.

Dans cette situation, il existe une $1$-forme particulière que l'on note  $\theta$, définie par:
\begin{align}
  i\theta : \der(M_n) & \longrightarrow \ksl_n \subset M_{n}\\
  \ad_{\gamma} & \longmapsto \gamma - \frac{1}{n} \tr (\gamma)\gone \ ,
  \label{eq:theta}
\end{align}
pour tout $\gamma \in M_n$.
Cette  $1$-forme satisfait la relation:
\begin{align*}
 \d' (-i\theta) + (i\theta)^2 = 0
\end{align*}
et pour tout  $\gamma \in M_n = \Omega^0_\der(M_n)$, on a  $\d' \gamma = [i\theta, \gamma]$. 

Cette  $1$-forme  $\theta$ peut aussi être vue comme la $1$-forme canonique permettant d'identifier explicitement les algèbres de Lie $\der(M_n)$ et $\ksl_n$.

\subsubsection{L'algèbre des fonctions à valeurs matricielles}
On peut maintenant compliquer légèrement l'exemple précédent en considérant l'algèbre des fonctions sur une variété $M$ à valeurs matricielles $\kA = C^\infty(M)\otimes M_n$.
Comme nous le verrons dans la section suivante, c'est l'algèbre des endomorphismes d'un fibré vectoriel trivial au dessus de $M$, de fibre $\gC^{n}$.
Le calcul basé sur les dérivations pour cette algèbre a été étudié dans~\cite{dubois-violette:90:II}. 
Le centre de l'algèbre $\kA$ est exactement $C^\infty(M)$ et l'algèbre de Lie des dérivations $\der(\kA)$ se décompose canoniquement en tant que module sur l'algèbre $C^\infty(M)$ de la manière suivante:
\begin{equation}
\label{decder1}
\der(\kA) = [\der(C^\infty(M))\otimes \gone ] \oplus [ C^\infty(M) \otimes \der(M_n) ] \ .
\end{equation}
Cela implique la décomposition canonique du complexe de formes:
\begin{align*}
  \Omega_\der(\kA) = \Omega(M) \otimes \Omega_\der(M_n) \ .
\end{align*}
La différentielle $\hd$ sur $\Omega_\der(\kA)$ est alors la somme $\hd = \d + \d'$ où $\d$ et $\d'$ sont les différentielles qui ont été définies dans les deux exemples précédents.
La  $1$-forme $\theta$ est bien définie dans $\Omega^1_\der(\kA)$ si on l'étend  sur  $\der(\kA)$ par l'application nulle sur les éléments de  $\Gamma(TM)$. 

On peut déjà à ce niveau définir la notion de connexion et voir apparaître les champs de Yang-Mills et des champs scalaires.
Ceci a été fait dans cette situation dans~\cite{dubois-violette:89,dubois-violette:89:II,dubois-violette:90:II}.

Ces modèles sont développés dans le chapitre~\ref{cha:modeles-physiques}.



\subsection{L'algèbre des endomorphismes d'un fibré vectoriel}
\label{sec:algebra-endom-vect}
 Nous allons étudier dans cette section une algèbre particulière qui est l'algèbre des sections du fibré des endomorphismes d'un fibré vectoriel complexe de fibre $\gC^{n}$.
 Nous allons voir que cette algèbre possède beaucoup de similarités avec la notion de fibré principal et qu'elle contient, en un certain sens, la notion de connexion sur un fibré principal et qu'elle en donne une généralisation naturelle. 
 Cela nous permettra de mettre en rapport la notion de connexion non commutative avec les champs de Yang-Mills-Higgs en physique.

 \begin{defn}
   Soit $\cE$ un fibré vectoriel de fibre $\gC^{n}$ associé à un fibré principal $E$ de groupe de structure $SU(n)$.
   Nous supposerons que la variété de base $M$ est une variété différentiable de classe $C^{\infty}$, de dimension finie et para-compacte.
   On note $\End(\cE)$ le fibré des endomorphismes de $\cE$.
C'est également un fibré associé à $E$.
   Ses sections sont à valeurs dans l'algèbre des matrices de taille $n$ par $n$. 
   Ainsi, l'ensemble de ces sections forme une algèbre que l'on notera $\kA$.
   Cette algèbre a  une structure hermitienne naturelle, héritée de celle sur les matrices.
On associe à cette structure hermitienne une involution que l'on notera $a\mapsto a^{\ast}, \forall a\in \kA$.
 \end{defn}

On peut voir~\cite{dubois-violette:98} que les deux calculs différentiels $\Omega_\der(\kA)$ et $\underline{\Omega}_\der(\kA)$ coïncident.
Nous noterons $\hd$ la différentielle sur $\Omega_\der(\kA) =\underline{\Omega}_\der(\kA)$.

\begin{rem}
Nous avons vu dans la section~\ref{sec:operations-de-cartan} que la sous-algèbre de Lie $\Int(\kA)$ opère au sens de H.
Cartan sur  $\Omega_\der(\kA)$.
Ainsi, les formes horizontales pour cette opération sont exactement les formes différentielles sur $M$ à valeurs dans $\End(\cE)$, c'est-à-dire les formes tensorielles à valeurs dans $\End(\cE)$, et les formes basiques sont les formes différentielles ordinaires sur $M$.
Dans ce qui suit, la notion d'horizontalité fera toujours référence à cette opération.
\end{rem}

\subsubsection{Le centre de l'algèbre}

 Le centre de l'algèbre $\kA$, $Z(\kA)$ est exactement l'algèbre de fonctions  $C^{\infty}(M)$  et on identifie $f\in C^{\infty}(M)$ avec $f \gone \in \kA$.
Nous pouvons définir des applications naturelles sur $\kA$:
\begin{align*}
  \tr : &\kA \rightarrow C^\infty(M) &
  &\mbox{et} &
  \det : &\kA \rightarrow  C^\infty(M)
\end{align*}
 qui correspondent aux applications trace et déterminant sur les fibres de $\End(\cE)$.

On a également une application naturelle:
\begin{align}
  \rho : \der(\kA) \rightarrow \der(C^\infty(M)) = \Gamma(TM)
\label{rho}
\end{align}
qui n'est autre que l'application quotient, s'inscrivant dans la suite exacte courte d'algèbres de Lie et de modules sur l'algèbre $C^{\infty}(M)$:
\begin{equation}
\xymatrix@1{ {0} \ar[r] & {\Int(\kA)} \ar[r] & {\der(\kA)} \ar[r]^-{\rho} & {\Out(\kA) \simeq  \Gamma(TM)} \ar[r] & {0}} \ .
\label{sesder}
\end{equation}

Cette suite exacte courte généralise la décomposition que l'on a dans la situation triviale (\ref{decder1}).
Dans le cas où le fibré n'est pas trivial, cette suite ne peut pas être scindée canoniquement et pourra être scindée grâce à une connexion ordinaire.

\subsubsection{Connexions ordinaires}

À toute dérivation $\cX \in \der(\kA)$,  on peut associer un champ de vecteurs $\rho(\cX)=X \in \Gamma(TM)$ sur la base $M$.
D'autre part,  la  $1$-forme $i\theta$ définie précédemment est bien définie ici sur les dérivations intérieures $\Int(\kA)$ par la relation:
\begin{align*}
  i\theta(\ad_\gamma) = \gamma - \frac{1}{n}\tr(\gamma) \gone \ ,
\end{align*}
pour tout $\gamma\in \kA$.

\begin{rem}
Afin de simplifier les notations, nous considérerons que pour toute dérivation intérieure  sous la forme $\ad_\gamma$, l'élément  $\gamma$ est sans trace.
Ainsi, une dérivation intérieure peut être considérée comme une section sans trace du fibré des endomorphismes de $\cE$.
Nous noterons $\kA_0$ l'ensemble des éléments sans trace de $\kA$.  
\end{rem}

Nous pouvons naturellement associer  à une connexion $\nabla^{\cE}$ une connexion $\nabla$ sur $\End\cE$ définie en prenant le produit tensoriel de la connexion $\nabla^\cE$  sur $\cE$ avec la connexion $\nabla^{\cE^\ast}$ sur  $\cE^\ast$, le fibré vectoriel dual\footnote{$\nabla^{\cE^\ast}$ est définie par la relation: 
  \begin{align}
    X \langle \epsilon , e \rangle = \langle \nabla^{\cE^\ast}_X \epsilon , e
    \rangle + \langle \epsilon , \nabla^\cE_X e \rangle
  \end{align}
  pour toutes les sections  $\epsilon$ de  $\cE^\ast$ et $e$ de $\cE$.
}
de $\cE$.
Nous pouvons alors associer à $\nabla^\cE$  une $1$-forme non commutative $\alpha \in \Omega^1_\der(\kA)$ définie par la relation:
\begin{align}
  \alpha(\cX)&= -i\theta(\cX - \nabla_X), &
  \text{où } X &= \rho(\cX) \ .
\end{align}

La $1$-forme $\alpha$ prend ses valeurs dans $\kA_{0}$ et peut donc être considérée comme une extension à toutes les dérivations de la $1$-forme $-i\theta$. Elle satisfait à la relation suivante:
\begin{align}
  \alpha(\ad_\gamma) &= -\gamma &
  & \forall \gamma \in \kA_{0} \ .
\end{align}

On voit déjà sur cet aspect que l'algèbre $\kA$ joue un rôle similaire au fibré principal.
En effet, l'application canonique $\nabla^\cE \mapsto \alpha$ est un isomorphisme d'espaces affines entre l'espace affine des connexions  $SU(n)$ sur le fibré $\cE$ et l'espace affine des $1$-formes non commutatives sans trace et antihermitienne sur $\kA$ satisfaisant  $\alpha(\ad_\gamma) = -\gamma$. 
Ainsi,  les connexions ordinaires peuvent être décrites par des $1$-formes non commutatives.

\subsubsection{Décomposition des dérivations}
Grâce à une connexion $\nabla^\cE$ sur $\cE$ et à la $1$-forme non commutative qui lui est associée, toute dérivation  $\cX \in \der(\kA)$ peut se décomposer de la manière suivante:
\begin{equation}
\label{decder} 
\cX = \nabla_{\rho(\cX)} - \ad_{\alpha(\cX)} \ .
\end{equation}
Cette décomposition n'est pas canonique et n'est définie que par le choix d'une connexion sur $\cE$ et par l'application $C^\infty(M)$-linéaire:
\begin{equation}
  \begin{aligned}
    \Gamma(TM) &\to \der(\kA)\\
    X&\mapsto \nabla_X
  \end{aligned} \ \ .
\end{equation}
On voit ainsi qu'une connexion $\nabla^\cE$ sur $\cE$ permet de scinder la suite exacte courte~(\ref{sesder}) de  modules sur  l'algèbre $C^\infty(M)$ et nous permet de décomposer  toute dérivation $\cX\in\der(\kA)$ en une dérivation ``horizontale'' $\nabla_{\rho(\cX)}$ et une dérivation ``verticale'' (intérieure) $-\ad_{\alpha(\cX)}$.
Ce point sera développé un peu plus loin où nous verrons qu'une telle connexion permet également de scinder d'autres suites exactes courtes étant en rapport avec le fibré principal $E$.
Nous verrons également dans le chapitre  \ref{sec:theories-de-jauge-endo} la suite exacte courte duale de~(\ref{sesder}).

\begin{rem}
Notons que cette situation est similaire à la situation classique (commutative) dans laquelle on peut interpréter une connexion comme une application linéaire qui à un champ de vecteur sur $M$ associe un champ de vecteur sur un fibré principal.
Nous verrons d'ailleurs à la fin de ce chapitre comment construire les classes caractéristiques usuelles à partir de ce type de décompositions.  
\end{rem}

\subsubsection{Différentielle covariante}
Tout comme dans la théorie des fibrés principaux, nous pouvons introduire le concept de dérivée covariante sur les formes non commutatives.
\begin{defn}
  Soit $\omega\in\Oder^{p}(\kA)$ et $\cX_{1}, \dots , \cX_{p+1}\in \der(\kA)$.
  La différentielle covariante de $\omega$ est définie par:
  \begin{align}
    D: \Oder^{p}(\kA) &\longrightarrow (\Oder^{p+1}(\kA))_{|\hor} \\
    \omega & \longmapsto D\omega(\cX_{1}, \dots , \cX_{p}) = \hd\omega(\nabla_{\rho(\cX_{1})}, \dots , \nabla_{\rho(\cX_{p+1})}) \ ,
  \end{align}
  où $ (\Oder^{p+1}(\kA))_{|\hor}$ est le sous-espace horizontal vis-à-vis de l'action de $\Int(\kA)$.
\end{defn}

\subsubsection{Courbure}
\label{sec:courbure}
De même qu'une connexion $\nabla^\cE$ peut être représentée par une $1$-forme $\alpha\in \Oder^{1}(\kA)$, la forme de courbure $R^\cE$ de la connexion $\nabla^\cE$ peut être représentée par une $2$-forme non commutative.
En effet, un petit calcul montre que l'on a la relation suivante: 
\begin{align*}
  R^\cE(\rho(\cX),\rho(\cY)) &= \hd\alpha(\cX, \cY) + [\alpha(\cX), \alpha(\cY) ] = D\alpha(\cX,\cY)&
  & \forall \cX, \cY \in \der(\kA),
\end{align*}
Ainsi, $R^\cE$ est représentée par la  $2$-forme non commutative $D\alpha = \hd\alpha + \alpha^2=\rho^{*}R^{\cE}$ qui est un élément horizontal de $\Omega^2_\der(\kA)$. 
Elle peut également s'exprimer de la manière suivante:
\begin{align*}
  R^\cE(X,Y) &= D\alpha(\nabla_{X},\nabla_{Y}) =  - \alpha([(\nabla_{X},\nabla_{Y}])&
  & \forall X, Y \in \Gamma(TM),
\end{align*}
Cette relation nous permet d'interpréter la courbure associée à une connexion ordinaire comme étant l'obstruction à ce que le sous-espace des dérivations horizontales forme une algèbre de Lie.

On voit ainsi que la $1$-forme $\alpha$ vient se substituer à la $1$-forme de connexion sur le fibré principal $E$ et joue un rôle tout à fait analogue.
Nous verrons plus loin que la $1$-forme $\alpha$ définit en fait une connexion algébrique sur le module $\kA$ et que la $2$-forme  $\hd\alpha + \alpha^2$ est la courbure de cette connexion.

\subsubsection{Transformations de jauge ordinaires}
L'algèbre de Lie des dérivations réelles sur $\kA$ agit naturellement sur l'espace des connexions $SU(n)$ par la dérivée de Lie définie sur $\Omega_\der(\kA)$. 
Si l'on restreint cette action aux dérivations intérieures, la dérivée de Lie correspond alors aux transformations de jauge infinitésimales sur les connexions ordinaires.
Ainsi, on a:
\begin{align*}
  \cL_{\ad_\xi} \alpha = -\hd \xi - [\alpha, \xi]= -D\xi \ ,
\end{align*}
pour tout $\xi \in \cA$, avec $\tr \xi =0$ et $\xi^\ast + \xi = 0$ ($\ad_\xi$ est alors une dérivation intérieure réelle).
Les éléments $\xi$ sont exactement les éléments de l'algèbre de Lie du groupe des transformations de jauge de $\cE$.

\subsection{\texorpdfstring{Relations entre  $\Omega_\der(\kA)$ et les formes sur un fibré principal}{Relations avec les formes sur un fibré principal}}
 \label{sec:omega_der}

Nous allons voir que le calcul différentiel  $\Omega_\der(\kA)$ est intimement relié au calcul différentiel $\Omega(E)$ des formes de de~Rham sur le fibré principal $E$.
Cette analyse nous permettra de mieux comprendre le rapport entre l'algèbre $\kA$ et le fibré principal.
D'autre part, ce point de vue nous sera utile comme outil de calcul et nous servira dans l'étude des connexions symétriques au chapitre suivant.
Nous allons voir que l'algèbre $\kA$ peut être vue comme la sous-algèbre basique d'une algèbre plus grosse ayant une structure particulièrement simple.

Nous rappelons que $E$ est un fibré principal de groupe de structure $SU(n)$ auquel est associé le fibré vectoriel $\cE$.
Nous noterons $C^\infty(E)$ l'algèbre des fonctions sur $E$.

\subsubsection{\texorpdfstring{L'algèbre $\kB$}{L'algèbre B}}
Nous pouvons considérer l'algèbre $\kB = C^\infty(E) \otimes M_n$ des fonctions sur $E$ à valeurs matricielles.
Nous notons alors  $(\Omega_\der(\kB), \hd) = (\Omega(E) \otimes \Omega_\der(M_n), \d + \d')$ son calcul différentiel basé sur les dérivations.

Nous avons une application naturelle $\xi \mapsto \xi^E$ qui associe à tout élément $\xi \in \ksu(n)$ un champ de vecteur vertical sur $E$.
Il est alors facile de voir que l'ensemble $\{ \xi^E + \ad_\xi\ / \ \xi \in \ksu(n) \}$ est une sous-algèbre de Lie de $\der(\kB)$ qui est isomorphe à $\ksu(n)$.
Cette algèbre de Lie définit une opération de Cartan (cf formule~(\ref{sec:omega_der})) de  $\ksu(n)$ sur $\Omega_\der(\kB)$ dont la sous-algèbre basique sera notée $\Omega_{\der,\Bas} (\kB)$.
On montre~\cite{masson:99} alors que le calcul différentiel $\Oder(\kA)$ est isomorphe à la sous-algèbre différentielle graduée $\Omega_{\der,\Bas} (\kB)$.
 
Nous allons ainsi pouvoir relier la $1$-forme $\alpha$ associée à une connexion sur $\cE$ à la $1$-forme de connexion sur le fibré principal.
En effet, une connexion $SU(n)$ sur $E$ est donnée par une $1$-forme $\omega_E$ sur $E$ à valeurs dans  $\ksu(n) \subset M_n$.
Cette connexion permet de définir une connexion $\nabla$ sur $\cE$ qui définit à son tour une $1$-forme $\alpha \in \Omega^1_\der(\kA)$.
D'après le résultat précédent, cette $1$-forme vient d'une $1$-forme $\alpha^E \in \Omega_{\der,\Bas} (\kB)$, qui n'est autre que $\alpha^E= \omega_E - i \theta$, où $\theta \in \Omega^1_\der(M_n)$ est la $1$-forme canonique définie dans la formule~(\ref{eq:theta}).

La basicité de cette $1$-forme est une conséquence des propriétés sur $\omega_E$ et $i\theta$ et en particulier l'équivariance de $\omega_E$.

\subsubsection{Liens entre les dérivations}
Au niveau des dérivations, le lien entre l'algèbre $\kA$ et l'algèbre $\kB$ peut être résumé dans le diagramme commutatif suivant:
\begin{equation}
\label{diagderivations}
\raisebox{.5\depth}{%
  \xymatrix{
    &           &  0 \ar[d]       &     0  \ar[d]         &    \\
    &  0    \ar[d]\ar[r]    & \cZ_{\der}(\kA) \ar[d]\ar[r]   &  \G(TVE)  \ar[d]\ar[r]   & 0  \\
    0 \ar[r] & \Int(\kA) \ar[d]\ar[r]  & \cN_{\der}(\kA) \ar[d]_{\tau}\ar[r]^{\rho_E}   &  \G_{M}(E)  \ar[d]^{\pi_*}\ar[r] & 0  \\
    0 \ar[r]& \Int(\kA) \ar[d]\ar[r]  &\der(\kA) \ar[d]\ar[r]_{\rho}   & \G(TM)  \ar[d]\ar[r]      & 0  \\
    &  0        &     0     &    0           &  
    }}
\end{equation}
Ce diagramme combine les dérivations sur $\kA$, les dérivations sur $\kB$ et les champs de vecteurs sur $E$:
\begin{itemize}
\item 
La rangée du bas est juste la suite exacte courte ordinaire reliant les champs de vecteurs sur $M$, les dérivations générales et les dérivations  intérieures sur $\kA$.
\item
 Dans la colonne du milieu, $\cN_{\der}(\kA) \subset \der(\kB)$ est le sous-ensemble des dérivations de $\kB$ qui préservent la sous-algèbre basique $\kA \subset \kB$ et $\cZ_{\der}(\kA) \subset \der(\kB)$ est le sous-ensemble des dérivations sur $\kB$ qui s'annulent sur $\kA$.
 Ces deux algèbres de Lie ont été définies de manière plus générale dans~\cite{masson:96}.
Cette suite exacte courte est utilisée dans~\cite{masson:99} pour montrer que  $\kA$ est une ``sous-variété non commutative'' quotient de l'algèbre $\kB$.
 L'algèbre de Lie $\cZ_{\der}(\kA)$ est générée en tant que $C^\infty(E)$-module par les éléments particuliers $\xi^E + \ad_\xi$ pour tout $\xi \in \ksu(n)$. 
\item
La colonne de droite ne fait intervenir que des objets géométriques.
L'espace $\G_{M}(E)$ est défini comme étant:
\begin{align}
  \G_{M}(E)= \{ \hX \in \G(E) / \pi_* \hX(p) = \pi_*\hX(p') \ \forall p , p' \in E \text{ tel que }  \pi(p)=\pi(p')     \} \ .
\end{align}
C'est l'algèbre de Lie constituée des champs de vecteurs sur $E$ qui peuvent être envoyés sur des champs de vecteurs sur $M$ en utilisant l'application tangente $\pi_* : T_p E \rightarrow T_{\pi(p)}M$.
On peut également voir cette algèbre de Lie comme l'algèbre de Lie des automorphismes infinitésimaux du fibré principal $E$ ou encore comme les champs de vecteurs sur $E$ invariant sous l'action du groupe de structure de $E$, $SU(n)$.
\end{itemize}
Dans ce qui suit pour tout $\eta = f \otimes \xi \in C^\infty(E) \otimes \ksu(n)$, nous noterons $\eta^E \in \G(TVE)$ le champ de vecteurs $f \cdot \xi^E$ sur $E$.
 En utilisant ces notations, tout élément dans $\cZ_{\der}(\kA)$ peut être mis sous la forme $\eta^E + \ad_\eta$, où $\eta \in C^\infty(E) \otimes \ksu(n)$.
Ainsi, l'application qui envoie tout élément $\eta^{E} \in \G(TVE)$  sur  $\eta^E + \ad_\eta \in \cZ_{\der}(\kA)$ est un isomorphisme.

 \begin{rem}
   Le diagramme~(\ref{diagderivations}) a des similarités avec le diagramme présenté en page 12 de~\cite{dubois-violette:98} qui implique des structures d'algébroïdes de Lie.
\end{rem}

\subsubsection{Rôle d'une connexion ordinaire}
Une connexion $\omega_E$ sur  $E$ permet de scinder trois suites exactes courtes dans le diagramme~(\ref{diagderivations}) et ce de manière compatible.
 Tout d'abord, cette connexion peut être utilisée pour relever un champ de vecteurs $X$ sur $M$ en un champ de vecteurs horizontal $X^h$ sur $E$.
 Cela nous permet de scinder la suite exacte courte suivante: 
\begin{equation}
\xymatrix@1{0 \ar[r] & \G(TVE) \ar[r] & \G_{M}(E) \ar[r] & \G(TM) \ar[r] & 0}
\end{equation}
en tant que suite exacte courte de  $C^\infty(M)$-modules.

Il a été montré dans~\cite{dubois-violette:98} et rappelé dans la section~\ref{sec:algebra-endom-vect}, que la connexion  $\omega_E$ permet de scinder la suite exacte courte  (\ref{sesder}) de  $C^\infty(M)$-modules en utilisant l'application $X \mapsto \nabla_X$.
Maintenant, en utilisant les notations introduites précédemment, nous pouvons scinder la suite exacte courte correspondant à la colonne du milieu en utilisant l'application $\cX \mapsto \cX^E= \rho(\cX)^h - \ad_{\alpha(\cX)^E}$, où $\alpha(\cX)^E$ est l'élément basique dans $\kB$ associé à $\alpha(\cX) \in \kA$ (notons que $\alpha(\cX)^E=\alpha^E(\cX^E)$).

Ainsi, nous pouvons décomposer tout élément de $\cN_{\der}(\kA)$ en quatre parties, en rendant explicite les noyaux des deux suites exactes courtes dans lesquelles cet espace est impliqué.

Toute dérivation  $\kX \in \cN_{\der}(\kA)$ peut être considérée comme une dérivation de $\kB$ et s'écrire sous la forme $\kX= \hat{X} + \ad_\gamma$, où $\hat{X} \in \G(E)$ et $\gamma \in C^\infty(E)\otimes \ksl_n$.
A ce point, nous pouvons voir que $\hat{X}$ est dans $\G_M(E)$ en utilisant la restriction de $\kX$ au centre $C^\infty(M)$ de $\kA$ considérée comme une sous-algèbre de $\kB$.
Une dérivation $\kX$ est un élément de  $\cN_{\der}(\kA)$ \ssi $L_\xi \kX \in \cZ_{\der}(\kA)$ pour tout $\xi \in \ksu(n)$.
 Cela veut dire qu'il doit exister un élément $\eta \in C^\infty(E) \otimes \ksu(n)$ tel que:
 \begin{equation}
   \begin{aligned}
     \lbrack\xi^E, \hat{X}\rbrack &= \eta^E \\
     L_\xi \gamma &= \eta
   \end{aligned} \ .
  \end{equation}
En appliquant la forme de connexion $\omega_E$ sur la première relation et en utilisant l'équivariance de $\omega_E$, on obtient:
\begin{align*}
  L_\xi (\omega_E(\hat{X})) = (L_\xi^{E}+L_{\ad_{\xi}}) (\omega_E(\hat{X})) = \eta \ .
\end{align*}
Introduisons l'élément $Z =- \alpha^E(\kX)= \gamma - \omega_E(\hat{X}) \in C^\infty(E) \otimes \ksl_n$.
 Alors $L_\xi Z = 0$ pour tout $\xi \in \ksu(n)$, ce qui implique que $Z \in \kA_0$, ou $\ad_Z \in \Int(\kA)$.
 Ainsi, la dérivation $\kX$ peut être réécrite comme $\kX = \hat{X} + \ad_\gamma = X^h + \hat{X}^v + \ad_{\omega_E(\hat{X}) + Z}$.
Avec nos notations, on a $\hat{X}^v = \omega_E(\hat{X})^E$ (partie verticale du champ de vecteur $\hat{X}$).
Finalement, nous pouvons réécrire:
\begin{equation}
  \kX = X^h +\ad_Z + \underbrace{\omega_E(\hat{X})^E + \ad_{\omega_E(\hat{X})}}_{ \in \cZ_{\der}(\kA)} = X^h + \omega_E(\hat{X})^E  + \ad_{\omega_E(\hat{X})} +  \underbrace{\ad_{Z}}_{ \in \Int(\kA)}  \ .
\end{equation}

La situation peut être résumée dans le diagramme suivant où les applications permettant de scinder les différentes suites exactes courtes sont écrites explicitement:
\begin{equation}
  \label{eq:diagramme}
  \raisebox{.5\depth-.5\height}{%
    \xymatrix@R=0pc@C=0pc@M=2pt{
      \cN_{\der}(\kA) \ar@{->>}[rrrrrrr]^{\rho_E} \ar@{->>}[dddddd]_{\tau} &&&& \rule{60pt}{0pt} &&& \G_M(E)
      \ar@{->>}[dddddd]^{\pi_*} \\
      &&&\makebox[100pt][r]{$(\pi_*\hX)^h+\omega_E(\hX)^E+\ad_{\omega_E(\hX)}$}& & \hX \ar@{|->}[ll]&&\\
      &\makebox[15pt][l]{$\cX^E = \rho(\cX)^h-\ad_{\alpha(\cX)}$} &&&&& X^h & \\
      &&&&\rule{0pt}{30pt}&&&\\
      &\cX \ar@{|->}[uu] &&&&& X \ar@{|->}[uu] & \\
      &&\nabla_X &&& X \ar@{|->}[lll]&&\\
      \der(\kA) \ar@{->>}[rrrrrrr]_{\rho} &&&&&&& \G(TM)
      }}
\end{equation}

Dans ce qui suit, $\Omega_\der(\kA)$ sera identifié avec la sous-algèbre basique de $\Omega_{\der}(\kB)$ correspondante.

Nous pouvons maintenant étudier les conséquences de cette construction sur une connexion non commutative donnée par une $1$-forme $\omega \in \Omega^1_\der(\kA)$.
Une telle $1$-forme se décompose de la manière suivante:
\begin{align*}
  \omega = a - \phi \in [ \Omega^1(E) \otimes M_n ] \oplus [ C^\infty(E)   \otimes M_n \otimes \ksl_n^\ast ]
\end{align*}
avec la condition de basicité: 
\begin{gather*}
(L_{\xi^E} + L_{\ad_\xi})a = 0  \ \ \ \ \ \ \ \  (L_{\xi^E} + L_{\ad_\xi})\phi = 0 \\
i_{\xi^E} a - i_{\ad_\xi}\phi = 0
\end{gather*}
pour tout $\xi \in \ksu(n)$.
Nous avons décomposé la dérivée de Lie $L$ et le produit intérieur en une partie géométrique et une partie algébrique.
 
À ce niveau, nous pouvons faire des commentaires généraux.
Tout d'abord, la relation d'invariance sur $a$ n'est rien d'autre que la condition d'équivariance pour une connexion ordinaire sur $E$.
Cependant, la deuxième relation empêche $a$ d'être une telle connexion.
Cette relation généralise la relation de verticalité pour une connexion ordinaire et connecte la valeur de $a$ sur un champ de vecteurs vertical aux valeurs de $\phi$.
Pour une connexion ordinaire, $\phi$ doit être remplacé par $i \theta$ et alors la relation de verticalité usuelle est retrouvée.

Dans la deuxième relation, l'invariance de $\phi$ a une interprétation géométrique naturelle.
En effet, $\phi$ peut être vu comme une application $E \rightarrow M_n \otimes \ksl_n^\ast$.
En utilisant les résultats standards de géométrie différentielle, la relation d'invariance sur $\phi$ nous permet d'interpréter $\phi$ comme une section d'un fibré vectoriel au dessus de $M$, associé à $E$, dont les fibres sont isomorphes à $M_n \otimes \ksl_n^\ast$. Nous noterons ce fibré $\End\cE\otimes \End_{0}^{*}\cE $.
Dans cette identification, la dérivée de Lie $L_{\ad_\xi}$ sur $M_n \otimes \ksl_n^\ast$ n'est rien d'autre que l'action infinitésimale de $SU(n)$ sur $M_n \otimes \ksl_n^\ast$ qui intervient dans la construction du fibré vectoriel associé à $E$:  $\End\cE\otimes \End_{0}^{*}\cE $.
Ce type d'identification sera souvent utilisé dans la suite.

\subsubsection{Connexion symétrique}
Considérons maintenant la situation suivante:

Si l'on a une action d'un groupe de Lie compact connexe $G$ sur le fibré principal $E$ qui commute avec l'action naturelle du groupe de structure $H=SU(n)$ sur $E$,
alors, à tout $Y\in \cG$, l'algèbre de Lie de $G$,  nous pouvons associer un champ de vecteurs $Y^E$ sur $E$.
Ce champ de vecteurs induit une action de Cartan de $\cG$ sur $\Omega(E)$.
Cette opération s'étend naturellement en une opération sur $\Omega_\der(\kB) = \Omega(E) \otimes \Omega_\der(M_n)$ où $\cG$ agit seulement sur la partie dépendante de $E$.
Du fait que les actions de $G$ et de $H$ commutent, l'opération de $\cG$ laisse invariante la sous-algèbre basique $\kA$ de $\kB$ et se restreint en une opération sur $\Omega_\der(\kA)$.
Alors l'action originale de $G$ sur $E$ donne lieu à une action de $\cG$ sur $\kA$.
Cette action est celle que l'on utilisera pour caractériser les connexions non commutatives $G$-invariantes sur $\kA$ dans la section~\ref{invar-nonc-conn}.

\subsection{Point de vue local}\label{sec:point-de-vue-local}
Dans cette partie, nous allons étudier plus précisément la relation entre les algèbres $\kA$ et  $\kB$ du point de vue local.

\subsubsection{Objets locaux à partir de $\kA$}
Caractérisons tout d'abord les objets locaux dans $\kA$.
Une telle discussion a été faite dans~\cite{dubois-violette:98,masson:99} et nous en rappelons ici quelques points essentiels.
Restreinte à un ouvert $U$ au-dessus duquel le fibré $\End(\cE)$ se trivialise, l'algèbre $\kA$ est isomorphe à $\kA\loc:=C^{\infty}(U)\otimes M_n$ et nous pouvons associer à tout élément $a\in \kA$, un élément $a\loc\in \kA\loc$.
Ainsi, au-dessus d'une intersection  $U\cap U' \neq \emptyset$ de deux ouverts trivialisants $U$ et $U'$, nous avons des fonctions de transition $g: U\cap U' \to SU(n)$ qui relient $a\loc'$ à $a\loc$ de la manière suivante:
\begin{align*}
  a\loc'=\Ad_{g^{-1}} a\loc \ .
\end{align*}

Nous pouvons aussi associer à toute dérivation $\cX \in \der(\kA)$, une dérivation locale $\cX\loc \in \der(C^{\infty}(U)\otimes M_n)$.
Une telle dérivation peut être décomposée en deux parties :
\begin{align}
  \cX\loc=X_{|U} +\ad_{\gamma\loc} \ ,
\label{eq:der-loc}
\end{align}
où $X=\rho(\cX)$ (voir équation~(\ref{rho})) et $X_{|U}$ est sa restriction à l'ouvert $U$.
Il est possible de donner une expression explicite pour $\gamma\loc$ si l'on considère une connexion sur $\End(\cE)$, à laquelle on associe la $1$-forme non commutative $\alpha$ et la $1$-forme locale $A\loc \in \Omega^1(U)\otimes \ksu(n)$.
Alors on a :
\begin{align}
  \gamma\loc&= A\loc(X_{|U})-\alpha(\cX)\loc \ .
\label{eq:gamma-loc}
\end{align}
Au-dessus d'une intersection $U\cap U' \neq \emptyset$,  $\cX\loc'$ et  $\cX\loc$ sont reliées de la manière suivante:
\begin{align*}
  X_{|U}' &=X_{|U}\\
  \gamma\loc' &= \Ad_{g^{-1}}\gamma\loc + g^{-1} X_{|U}(g) \ .
\end{align*}

Enfin, nous pouvons considérer les $1$-formes locales.
À tout élément $\omega \in \Omega_{\der}^1(\kA)$, on associe un élément $\omega\loc \in \Omega_{\der}^1(C^{\infty}(U)\otimes M_n)$ au-dessus de $U$.
Cette $1$-forme locale se décompose naturellement en deux parties: $\omega\loc= a + \phi\circ i\theta$ où $a \in \Omega^1(U)\otimes M_n$, et $\phi \in C^{\infty}(U)\otimes M_n \otimes \ksl_n^*$.
Alors au-dessus d'une intersection $U\cap U'\neq \emptyset$,  $\omega\loc$ et  $\omega\loc'$ sont reliés de la manière suivante:
\begin{equation}
  \begin{split}
    a' &= \Ad_{g^{-1}} \circ a - \Ad_{g^{-1}} \circ \phi \circ \Ad_{g} \circ g^*\theta^H \\
    \phi' &= \Ad_{g^{-1}} \circ \phi \circ \Ad_{g} \ \ \ ,
  \end{split}
  \label{transition-relation}
\end{equation}
où $\theta^H$ est la forme de Cartan usuelle sur le groupe $H$ et $g^*\theta^H= g^{-1}dg$.

 Les fonctions de transition ressemblent à celles rencontrées pour les théories de jauge usuelles (en géométrie ordinaire).
Dans le cas présent, ces relations sont ``twistées''  par le champ scalaire $\phi$.

\subsubsection{Objets locaux à partir de $\kB$}
Voyons maintenant les relations entre ces objets locaux et les objets locaux que l'on peut obtenir à partir de l'algèbre $\kB = C^{\infty}(E)\otimes M_n$ restreinte à des ouverts trivialisants de $E$.
Tout d'abord, considérons une section locale $s: U \to E$.
Nous pouvons lui associer l'application \textit{pull-back} $s^*: C^{\infty}(E)\to C^{\infty}(U)$ qui s'étend de manière triviale en une application $s^*: \kB \to C^{\infty}(U)\otimes M_n$.
Alors par définition, l'image de $\kB_{\cH-\bas |U}$ par l'application $s^*$ est l'algèbre $\kA\loc$ obtenue par localisation de l'algèbre $\kA$ au-dessus de $U$.

Pour les dérivations, nous avons montré précédemment qu'avec l'aide d'une connexion ordinaire, il est possible d'associer à tout élément $\cX \in \der(\kA)$, un élément $\cX^E \in \cN(\kA) \subset \der(\kB)$, où explicitement $\cX^E= \rho(\cX)^h - \ad_{\alpha(\cX)^E}$.
De manière analogue, en utilisant l'inclusion $ \Omega_{\der}(\kA) \hookrightarrow \Omega_{\der}(\kB)$, on peut associer à tout élément $\omega \in \Omega_{\der}(\kA)$, un élément $\varpi \in \Omega_{\der}(\kB)_{\cH-\bas}$.
Alors nous pouvons comparer les expressions de $\varpi(\cX^E)$ et $\omega(\cX)$.
Au-dessus d'un ouvert $U \subset M$ trivialisant $E$ et en utilisant l'expression locale d'une connexion sur $E$, on a:
\begin{equation*}
  \cX^E_{|s(x)} = s_* X_{|x}- A\loc(X)_{|x}- \ad_{\alpha(\cX)^E_{|s(x)}} \ ,
\end{equation*}
où $x \in U$.
D'après la basicité de la $1$-forme $\varpi$, nous pouvons montrer que 
\begin{align*}
  \omega\loc(\cX\loc)=s^*(\varpi_{|U}(\cX^E_{|U})) = s^*\varpi_{|U}(\cX\loc) \ ,
\end{align*}
et que 
\begin{align*}
  s^*\varpi_{|U}=\omega\loc \ .
\end{align*}
Cela généralise le résultat précédent obtenu pour les éléments des algèbres $\kA$ et $\kB$, et on a de manière plus générale:
\begin{align*}
  s^*\Omega_{\der}(\kB)_{\cH-\bas |U}=\Omega_{\der}(\kA\loc) \ .
\end{align*}
Cette relation montre que l'on peut obtenir des expressions locales $\omega\loc$ soit à partir de la $1$-forme $\omega \in \Omega_{\der}^1(\kA)$ en se restreignant à un ouvert trivialisant de $\kA$, soit à partir de la $1$-forme $\varpi \in \Omega_{\der}^1(\kB)_{\cH-\bas}$ en se restreignant à un ouvert trivialisant de $\kB$.
Cela montre encore une fois le rôle similaire que peuvent jouer l'algèbre $\kA$ et le fibré principal $E$.

Enfin notons que les relations de transition~(\ref{transition-relation}) auraient pu être obtenues en considérant la $1$-forme $\varpi$ au-dessus des intersections $U\cap U' \neq \emptyset$ d'ouverts trivialisants, avec les relations de transition $s'= s \cdot g$ entre sections locales $s: U\to E$ et $s':U'\to E$.

\subsubsection{Objets locaux exprimés grâce à une connexion de référence}

Il a été montré dans~\cite{masson:99} que si l'on choisit une connexion de référence, alors il est possible d'exprimer les $1$-formes locales en terme de tenseurs qui se transforment de manière homogène.

Soit  $\omega\in \Oder(\kA)$ une $1$-forme non commutative et  $\omega\loc \in \Omega(U)\otimes \Oder(M_{n})$ sa forme locale.
Pour obtenir les relations de transition~(\ref{transition-relation}), nous avons décomposé $\omega\loc$ de la manière suivante:
\begin{align}
  \omega\loc= a + \phi\circ i\theta \ ,
\label{ea:decomp-theta-loc}
\end{align}
où $a \in \Omega^1(U)\otimes M_n$ et $\phi \in C^{\infty}(U)\otimes M_n \otimes \ksl_n^*$. 
Cette décomposition est naturelle vis-à-vis de la décomposition (\ref{eq:der-loc}) des dérivations locales.
Comme nous le voyons dans la formule (\ref{transition-relation}), les relations de transition sont relativement compliquées et pour des formes de degré plus élevé, on obtiendra des expressions difficilement manipulables.

Il est donc préférable de décomposer les formes locales autrement, et ce, grâce à une connexion de référence $\nabla^{\cE}$ sur $\cE$.
On note $\alpha$ la $1$-forme non commutative qui lui est associée et $A\loc$ la $1$-forme de connexion locale sur un ouvert $U$.

Alors d'après la formule (\ref{eq:gamma-loc}), la forme locale $\omega\loc$ peut se décomposer de la manière suivante:
\begin{align}
  \omega\loc= \tilde{a} - \tilde{\phi}\circ \alpha\loc \ ,
  \label{eq:decomp-alpha-loc}
\end{align}
où $\tilde{a} \in \Omega^1(U)\otimes M_n$, et $\tilde{\phi} \in C^{\infty}(U)\otimes M_n \otimes \ksl_n^*$.
On peut passer de la décomposition~(\ref{ea:decomp-theta-loc}) à la décomposition~(\ref{eq:decomp-alpha-loc}) en faisant le changement de variables:
\begin{equation*}
\begin{aligned}
  \tilde{a} &= a + \phi\circ A\loc \\
  \tilde{\phi} &= \phi \ .
\end{aligned}
\end{equation*}
Ainsi, les relations de transition lors d'un changement de carte sont les suivantes:
\begin{equation}
  \begin{aligned}
    \tilde{a}' &= \Ad_{g^{-1}} \circ \ \ta \\
    \phi' &= \Ad_{g^{-1}} \circ \ \phi \circ \Ad_{g} \ .
  \end{aligned}
  \label{transition-relation-tilde}
\end{equation}
Ces relations de transition sont homogènes et tous les termes inhomogènes de (\ref{transition-relation}) ont été absorbés par la forme locale $\alpha$.

On reconnaît là les relations de transition pour des formes tensorielles sur un fibré principal.
En effet, dans la décomposition~(\ref{eq:decomp-alpha-loc}), tous les objets peuvent être définis globalement et donc cette décomposition reste vraie pour une $1$-forme globale.
Ainsi, un élément $\omega \in \Oder(\kA)$ peut toujours se mettre sous la forme :
  \begin{align}
    \omega= \tilde{a}\circ \rho - \tilde{\phi}\circ \alpha \ ,
    \label{eq:decomp-alpha}
  \end{align}
  où $\tilde{a} \in \Omega^1(M,\End(E))$ est une forme tensorielle et $\tilde{\phi} \in C^{\infty}(M, F)$ est une section du fibré $F$ associé à $E$ dont les fibres sont isomorphes  à $M_n \otimes \ksl_n^*$.
Nous montrerons au début de la section \ref{sec:theories-de-jauge-endo} que cette décomposition globale peut être obtenue de manière plus directe.

\begin{rem}
  Nous avons obtenu cette décomposition en passant par une décomposition locale, mais nous aurions pu également l'obtenir en passant par l'algèbre $\kB$.
\end{rem}

Il est possible de faire une décomposition similaire à la décomposition~(\ref{eq:decomp-alpha}) pour des formes de degré plus élevé et on obtient alors des relations de transition homogènes pour les formes locales associées (voir~\cite{masson:99}).
Nous verrons dans la section \ref{sec:theories-de-jauge-endo} que cette décomposition à l'aide d'une connexion de référence est également utile lorsque l'on a une opération de Hodge associée à une structure Riemannienne.

\subsection{Morphisme de Chern-Weil}
\label{sec:classe-car}

L'algèbre des endomorphismes peut également se substituer au fibré principal afin de construire les classes caractéristiques usuelles.
Habituellement les classes caractéristiques sont obtenues par le morphisme de Chern-Weil associé à un fibré principal.
Nous allons voir dans cette section que ce morphisme peut être construit directement à partir d'une algèbre d'endomorphismes.

Nous allons nous inspirer de la construction donnée par Lecomte dans~\cite{lecomte:85}, où il est montré que le morphisme de Chern-Weil peut se voir comme étant l'obstruction à scinder la suite exacte courte d'algèbres de Lie associée aux automorphismes infinitésimaux d'un fibré principal.
Nous avons déjà rencontré cette suite dans le diagramme~(\ref{diagderivations}) et pouvons la rappeler ici:
\begin{equation*}
  \xymatrix@1{ {0} \ar[r] &  \G(TVE) \ar[r] &  \G_{M}(E) \ar[r]^-{\pi_*} &  \G(TM) \ar[r] & {0}} \ .
\end{equation*}
Dans \cite{lecomte:85}, il est en fait montré que l'on peut développer une théorie des classes caractéristiques, parallèle  à la construction classique de l'homomorphisme de Chern-Weil d'un fibré principal, dans un cadre purement algébrique relativement à toute suite exacte courte d'algèbres de Lie. Cette construction étant donnée pour des algèbres de Lie, c'est le complexe de Chevalley qui est utilisé de manière essentielle.

Nous allons adapter cette procédure à la suite exacte courte d'algèbres de Lie et  de $Z(\kA)$-modules (\ref{sesder}).
Nous devrons remplacer le complexe de Chevalley utilisé dans~\cite{lecomte:85}  par le calcul différentiel basé sur les dérivations $\uOder(\kA)$ afin de prendre en compte la structure de $Z(\kA)$-module supplémentaire.
Le fait de travailler avec le complexe $\uOder(\kA)$  plutôt qu'avec le complexe de Chevalley nous permettra d'obtenir directement le morphisme de Chern-Weil usuel contrairement à Lecomte~\cite{lecomte:85} qui obtient le morphisme de Chern\--Weil composé avec l'application induite en cohomologie par l'inclusion du complexe de Chevalley dans le complexe de de~Rham, \textit{i.e.} le complexe $\uOder(C^{\infty}(M))$. Cela induit dans~\cite{lecomte:85} des effets non locaux associés à la cohomologie de Chevalley.
Cela revient ainsi à travailler  avec des objets locaux au sens de~\cite{lecomte:85}.
Notons également que la suite~(\ref{sesder}) est reliée à la suite précédente par le diagramme~(\ref{diagderivations}).

Nous pouvons écrire la suite (\ref{sesder}) de la manière suivante:
\begin{equation*}
  \xymatrix@1{ {0} \ar[r] & \kA_{0} \ar[r]^-{ad} & \der(\kA) \ar[r]^-{\rho} & \Gamma(TM) \ar[r] & {0}} \ ,
\end{equation*}
où $\kA_{0}$ est identifiée à $\Int(\kA)$ par l'application $ad$.
Cette suite peut toujours être scindée en tant que suite de $Z(\kA)$-modules à l'aide d'une connexion $\nabla^{\cE}$ sur $\cE$ à laquelle on associe un projecteur $\rho^{*}\nabla: \der(\kA)\to \der(\kA)$, où $\nabla=\nabla^{\cE}\otimes\nabla^{\cE^{*}}$.
La courbure algébrique de ce projecteur (voir section~\ref{sec:courbure}) est la $2$-forme non commutative horizontale $\rho^{*}R^{\cE}$.

Nous pouvons maintenant donner la construction de l'homomorphisme de Chern-Weil à partir de $\kA$.
Remarquons que l'algèbre de Lie $\kA_{0}$ est un $\der(\kA)$-module et un $Z(\kA)$-module.
On peut alors considérer un élément
\begin{align*}
  f\in (S_{Z(\kA)}^{q} \kA_{0}^{\star Z(\kA)})_{\der(\kA)-\inv} \ ,
\end{align*}
où $\star Z(\kA)$ est l'opération agissant sur la catégorie des $Z(\kA)$-modules décrite dans la section~\ref{sec:bimod-diag-Oder}.
Ainsi, $f$ est une application $Z(\kA)$-multilinéaire symétrique de $\otimes_{Z(\kA)}^{q}\kA_{0}$ dans $Z(\kA) \simeq C^{\infty}(M)$ invariante sous l'action de $\der(\kA)$.
La condition d'invariance se traduit par:
\begin{align*}
  L_{\cX}f &= 0&
  & \forall \cX \in \der(\kA)  \ ,
\end{align*}
où la dérivée de Lie est définie de manière habituelle.
Si l'on décompose les dérivations en partie verticale et partie horizontale à l'aide de la connexion $\nabla$ sur $\End \cE$, cette relation devient équivalente à:
\begin{align*}
 &\left\{ \begin{aligned}
    L_{\nabla_{X}}f &= 0 \\
    L_{\ad_{\gamma}}f &=0
\end{aligned}\right. &
&\forall X\in \Gamma(M), \gamma \in \kA_{0} \ .
\end{align*}
On en déduit alors que $f$ est constante (en tant que section sur $M$) et définit un polynôme invariant sur $\ksl(n)$ (on peut s'en convaincre en regardant ces relations localement, c'est-à-dire au-dessus d'un ouvert trivialisant de $\kA$ où $\kA_{0}\simeq C^{\infty}(U)\otimes \ksl(n)$).
La $2$-forme de courbure $R^{\cE}$ permet alors d'associer à une telle application un élément de 
\begin{align*}
  \Omega^{2q}(M) \simeq \uOder^{2q}(Out(\kA),Z(\kA))
\end{align*}
en saturant les arguments de $f$. 
On note cet élément
\begin{align*}
  f_{\nabla}= AS\dot f(R^{\cE}\otimes \dots \otimes R^{\cE})\ ,
\end{align*}
où $AS$ désigne l'antisymétrisation par rapport à $R^{\cE}$.
Par les techniques usuelles, on vérifie que $f_{\nabla}$ est un cobord, \textit{i.e.} $d f_{\nabla}=0$, et que  la classe de cohomologie de $f_{\nabla}$ ne dépend pas de $\nabla$.

Nous avons ainsi construit une application linéaire 
\begin{align*}
  I(\ksl(n)) \simeq (S_{Z(\kA)}\kA_{0}^{\star Z(\kA)})_{\der(\kA)-\inv} &\longrightarrow   H^{\text{pair}}(M)\\
  f &\longmapsto [f_{\nabla}]
\end{align*}
 associant à tout polynôme invariant sur $\ksl(n)$ un élément pair de la cohomologie de de~Rham de la variété $M$.
Ce morphisme s'identifie de manière évidente au morphisme de Chern-Weil usuel. 

Cette construction nous permet ainsi de voir l'homomorphisme de Chern-Weil (ou les classes caractéristiques) comme étant l'obstruction à pouvoir scinder la suite exacte courte de $Z(\kA)$-modules (\ref{sesder}) en tant que suite exacte courte d'algèbres de Lie et fournit une interprétation naturelle des classes caractéristiques dans le cadre des algèbres d'endomorphismes.

Notons que dans le cadre des fibrés principaux, ce morphisme peut être obtenu par des méthodes homologiques à partir du complexe de Weil construit sur l'algèbre de Lie du groupe de structure d'un fibré principal qui est de dimension finie.
Ici nous n'avons pas eu besoin de groupe de structure et avons travaillé directement à partir de l'algèbre de Lie $\Int(\kA)\simeq \kA_{0}$ qui s'identifie à l'algèbre de Lie du groupe de jauge.
Cette algèbre de Lie est de dimension infinie et nous en avons extrait les polynômes invariants par une condition d'invariance vis-à-vis de l'algèbre de Lie $\der(\kA)$ (également de dimension infinie).
La recherche d'une interprétation de cette construction dans un cadre homologique plus général est l'objet de recherches en cours.

Nous terminons par quelques remarques.
Pour une algèbre d'endomorphismes $\kA$,  il serait intéressant de voir les liens possibles entre cette construction des classes caractéristiques et d'autres notions algébriques, telles que la cohomologie basique de $\kA$ (voir \cite{dubois-violette:94}) ou encore son homologie cyclique.
Plusieurs indications mènent à penser qu'un tel lien puisse exister.
En effet, notons que $\kA$ est équivalente de Morita à l'algèbre des fonctions $C^{\infty}(M)$ et donc son homologie cyclique redonne l'homologie de de~Rham de la base $H(M)$ (cf. membre de droite du morphisme de Chern Weil). 
Notons également que la  cohomologie basique de l'algèbre $M_{n}(\gC)$ sur laquelle est modelée $\kA$ coïncide avec les polynômes invariants sur $M_{n}(\gC)$ (cf. membre de gauche du morphisme de Chern Weil). 
Il serait aussi intéressant de voir si cette construction a des liens avec la construction habituelle du morphisme de Chern-Connes (voir~\cite{Loda:92} par exemple) en géométrie non commutative. La notion de morphisme de Chern-Weil doit cependant s'en distinguer étant donné que sa construction n'est pas invariante de Morita (la dimension des fibres est fixée).
Il serait également intéressant d'étudier les liens possibles avec la classe de  Dixmier-Douady~\cite{dixm-doua:63,Dixm:57} construite pour des champs continus de $C^{*}$-algèbres.

Notons tout de même que la construction que nous venons de faire peut s'adapter à toute  algèbre associative $A$, à partir de la suite exacte courte d'algèbres de Lie:
\begin{equation*}
\xymatrix@1{ {0} \ar[r] & \Int(A) \ar[r] & \der(A) \ar[r]^-{\rho} & \Out(A)  \ar[r] & {0}} \ .
\end{equation*}

La notion de classes caractéristiques et de fibré principal en géométrie non commutative est également abordée dans \cite{andr-krie-mich-van:93} pour des algèbres admettant l'action d'un groupe de Lie de dimension finie.

\section{Conclusion}
Nous avons montré dans ce chapitre comment traduire un certain nombre de notions géométriques, définies dans le cadre des fibrés principaux, en des notions purement algébriques en rapport avec l'algèbre des endomorphismes.
Les algèbres d'endomorphismes constituent  ainsi un terrain d'étude intéressant et pourraient être utiles, de part leur ressemblance avec les fibrés principaux, à établir un dictionnaire plus complet entre  géométrie différentielle et géométrie non commutative.
Elles seraient en quelque sorte pour les fibrés ce que sont les $C^{*}$-algèbres commutatives pour les espaces topologiques. 
En effet, c'est en étudiant les $C^{*}$-algèbres commutatives, qui sont équivalentes à des espaces topologiques (voir section~\ref{sec:star-algèbres}), que nous avons pu traduire différentes notions de géométrie en des notions algébriques équivalentes (ex: modules projectifs de type fini \textit{v.s.} fibrés vectoriels) et ainsi établir un pont entre topologie et algèbre.
De manière similaire, nous espérons que l'étude des algèbres d'endomorphismes puisse permettre de traduire différentes notions de géométrie différentielle, en rapport avec la notion de fibré, en des notions algébriques équivalentes et ainsi contribuer à établir un  pont entre géométrie et algèbre.

A l'heure actuelle, une étude plus approfondie des classes caractéristiques pour les algèbres associatives et le lien avec d'autres outils de la géométrie non commutative reste encore à effectuer et semble être une direction de recherche intéressante.


\chapter{Symétries}
\label{cha:symetries}
Dans ce chapitre, nous allons reprendre et résumer certains travaux sur les connexions invariantes en géométrie différentielle.
Comme nous l'avons vu précédemment, il existe un lien assez direct entre le calcul différentiel basé sur les dérivations d'une algèbre d'endomorphismes et l'algèbre des formes différentielles de de Rham sur un fibré  principal.
Les résultats que nous obtiendrons dans cette section, nous seront utiles pour les modèles physiques construits dans le cadre d'algèbres d'endomorphismes.

L'approche géométrique que nous allons suivre possède certaines similarités avec les modèles de théories de jauge sur des algèbres non commutatives. Ces méthodes furent même les premières à fournir une interprétation géométrique des champs scalaires de type Higgs. 
En ce sens, les modèles de géométrie non commutative, que nous verrons au chapitre~\ref{cha:modeles-physiques}, sont des versions algébriques et plus économes de ce type de modèles.
Généralement, lorsque l'on utilise un groupe en physique, nous n'exploitons pas vraiment sa structure de variété. 
En effet,  l'ensemble des champs de vecteurs d'un groupe de Lie forme une algèbre de Lie de dimension infinie.
Or,  on se contente généralement d'exploiter seulement la sous-algèbre de Lie formée des champs de vecteurs invariants à gauche ou à droite qui est une algèbre de Lie de dimension finie.

Par exemple, dans le modèle standard des particules, pour la description des symétries dites ``internes'' des particules, c'est en fait la structure d'algèbre de Lie 
qui est utilisée de manière essentielle plutôt que la structure de groupe. 
Ainsi, les modèles de théories de jauge basés sur des algèbres de matrices capturent la structure algébrique et ``discrète'' de ces constructions géométriques qui possèdent, pour la description des particules, des degrés de liberté continus superflus.

Pour les symétries qui ne sont pas internes, ou faisant partie intégrante des modèles que nous étudions, l'utilisation de groupes de Lie peut toujours s'avérer utile.
Cela est le cas, par exemple, pour les symétries d'espace temps.
Ainsi, si nous voulons étudier un problème ou une situation physique possédant une certaine symétrie, par exemple une symétrie sphérique ou une symétrie de rotation autour d'un axe, nous chercherons alors les solutions aux équations du mouvement invariantes sous l'action de ce groupe.

La construction que nous allons voir joue donc un double rôle dans cette thèse.
Elle nous permettra de présenter les modèles géométriques qui ont précédé les modèles de géométrie non commutative et  de bien voir les différences entre ces deux approches.
D'autre part, cette construction nous fournira des outils pour pouvoir caractériser les degrés de liberté d'une connexion non commutative invariante sous l'action d'un groupe de symétrie et ces résultats seront exploités au chapitre~\ref{chap:theorie-de-born}.
Les résultats présentés dans ce chapitre sont essentiellement ceux obtenus dans \cite{masson-serie:04}.

\section{Connexions invariantes sur un fibré principal}\label{construction_1}

Plusieurs approches ont été proposées pour l'étude des connexions invariantes sous l'action d'un groupe de Lie et elles sont essentiellement de deux types : les approches ``globales'', étudiées dans~\cite{hudson:84,jadczyk:84,harnad:80} et les approches ``locales'', étudiées dans~\cite{forgacs:80,brodbeck:96}. 
Nous reprendrons les approches globales qui permettent de bien appréhender ce qui se passe géométriquement et nous ferons le lien avec les approches locales, souvent plus utiles pour faire des calculs explicites.

\subsection{Réduction de fibrés principaux}\label{reduction}

Considérons un fibré principal $\fiber{H}{E}{M}$ de groupe de structure $H$. Nous adopterons également la notation suivante pour une fibration:
\begin{align*}
  \lfiber[\pi]{H}{E}{M} \ .
\end{align*}
Nous prendrons garde à ce que la première flèche ne soit pas une application mais schématise les fibres du fibré.

Considérons maintenant un groupe de Lie compact $G$ agissant à gauche sur $E$.
Nous noterons cette action $G \circlearrowright E $ et nous supposerons toujours que cette action commute avec celle du groupe $H$. Une manière de garantir cela, est de considérer $G$ comme un sous-groupe du groupe des automorphismes de $E$, noté $\Aut(E)$.
Dans ce cas, $E$ est dit $G$-symétrique.
La projection $\pi$ induit alors une action $G \circlearrowright M$ qui est caractérisée par le diagramme suivant:
\begin{align*}
  \xymatrix{0 \ar[r]& \Int(E) \ar[r] & \Aut(E) \ar[r] & \Out(E) \simeq \Aut(M)
    \ar[r] & 0 \ . \\
    &&&G \ar@{_(->}[ul]\ar[u]&} 
\end{align*}

Il est également possible d'essayer de construire une action de $G$ sur $E$ à partir d'une action de $G$ sur la base $M$ et commutant avec l'action du groupe de structure $H$. Cette question a été étudiée dans~\cite{hussin:94}.
Nous supposerons par la suite que cette action est donnée, et donc, nous partirons toujours d'un fibré principal $G$-symétrique.

Nous demanderons de plus que l'action du groupe $G$ soit simple (voir~\cite{harnad:80,jadczyk:84}), ce qui veut dire que la variété de base $M$ a elle-même une structure de fibré, dont les fibres sont isomorphes à un espace homogène $G/G_{0}$. Ce fibré est obtenu par action du groupe $G$ sur $M$ et donc l'espace de base est isomorphe à $M/G$.
Les groupes d'isotropie de  l'action de $G$ sur $M$ sont tous isomorphes à $G_{0}$.
Nous avons ainsi le diagramme de fibration suivant:
\begin{align*}
  \lfiber{G/G_0}{M}{M/G} \ . 
\end{align*}

Considérons maintenant l'espace $P = \{ x \in M,\ G_x=G_0 \}$, où $G_x$ est le groupe d'isotropie associé à tout point $x \in M$. 
Alors $P$ est un fibré principal de groupe de structure $N(G_0)/G_0$, où $N(G_0)$ est le normalisateur de $G_0$ dans $G$. 
Ce fibré sera noté par: 
\begin{align*}
  \lfiber{N(G_0)/G_0}{P}{M/G} \ .
\end{align*}
Nous pouvons alors considérer le fibré  \fiber{G/G_0}{M}{M/G} comme étant un fibré associé au fibré principal \fiber{N(G_0)/G_0}{P}{M/G} pour l'action naturelle $N(G_0)/G_0 \circlearrowright G/G_0$.
 
Une construction similaire peut être faite sur $E$ sur lequel le groupe $S = G \times H $ agit à droite par l'action:
  \begin{align*}
    G\times H\times E  & \longrightarrow  E \\
      (g,h,p) & \longmapsto g^{-1} p h  \ .
  \end{align*}
Notons tout d'abord que pour tout point $p\in E$, il existe un homomorphisme canonique:
\begin{align*}
  \lambda_p  :  G_{\pi (p)} \longrightarrow   H \ ,
\end{align*}
défini par la relation $g_0 \cdot p = p \cdot \lambda_p (g_0)$ pour tout $g_0 $ dans $G_{\pi (p)} $. 
Le groupe d'isotropie  $S_p$ d'un point  $p$ dans $E$ pour l'action  $E \circlearrowleft S$ est donc $S_p = \{(g_0,\lambda_p(g_0)) / \ g_0 \in G_{\pi (p)} \} $.
Tous ces groupes d'isotropie sont isomorphes et nous noterons $S_0$, correspondant à un point $p_{0}\in E$, l'un d'entre eux.
Ainsi, le fibré $E$ hérite de la structure de fibré suivante:
\begin{align*}
  \lfiber{S/S_0}{E}{M/G} \ .
\end{align*}
On notera que l'action de $S$ sur $E$ est également une action simple.

En procédant comme nous l'avons fait pour l'action de $G$ sur $M$, nous pouvons définir un sous-fibré   $Q:= \{ p \in E,\ S_p=S_0 \}$ de $E$.
Il est donné par le diagramme de fibration suivant:
\begin{align*}
\lfiber{N(S_0)/S_0}{Q}{M/G} \ ,
\end{align*}
où $N(S_0)$ est le normalisateur de $S_0$ dans $S$.
Il est remarquable que sur $Q$ l'application $\lambda_{p |Q}$ est indépendante du point $p \in Q$.
Nous noterons alors cette application $\lambda: G_0 \to H$.

Notons  $\pi_{Q}$ la restriction de la projection $\pi$ à $Q$.
Il est facile de voir que  $\pi_{Q}(Q)  \subset P$ et que le noyau de $\pi_Q$ est isomorphe à $Z_0=Z(\lambda(G_0),H)$,  le centralisateur de  $\lambda(G_0)$ dans $H$.
On peut également montrer que $\pi(Q)$ possède la structure de fibré suivante:
\begin{align*}
\lfiber[\pi_Q]{Z_0}{Q}{\pi(Q)} \ .
\end{align*}

D'après les définitions de base, nous avons:
\begin{align*}
  N(S_0)= \{(g,h) \in S / \  g \in N(G_0), \
  h^{-1} \lambda(g_0) h = \lambda(g^{-1} g_0 g), \forall g_0 \in G_0 \} \ .
\end{align*}
Nous avons également une inclusion naturelle de $Z_0$ dans $ N(S_0)/S_0$ donnée par la composition des deux applications du diagramme suivant:
\begin{align*}
  \xymatrix@1@R=7pt@M=6pt{ {Z_0 \simeq \{ e\} \times Z_0 } \ar@{^{(}->}[r] & {N(S_0)} \ar[r] & {N(S_0)/S_0} } \ .
\end{align*}
De plus, $Z_0$ est un  sous-groupe normal de $ N(S_0)/S_0$ et l'espace quotient $  (N(S_0)/S_0)/Z_0$ est un sous-groupe de $ N(G_0)/G_0$.

Toutes les observations précédentes se résument dans le diagramme suivant:
{
\footnotesize
\begin{equation}
\raisebox{.5\depth-.5\height}{%
\xymatrix@R=7pt@C=10pt@M=6pt{%
Z_0 \ar@{^{(}->}[rrr] \ar@{^{(}->}[ddrr] \ar@{=}[ddd] & & & N(S_0)/S_0 \ar@{>>}[rrr] \ar@{^{(}->}[ddrr] \ar'[dd][ddd] & & & (N(S_0)/S_0) / Z_0 \ar@{^{(}->}[dr] \ar'[dd][ddd] & & \\
& & & & & & & N(G_0)/G_0 \ar@{^{(}->}[dr] \ar'[d][ddd] & \\
& & H \ar@{^{(}->}[rrr] \ar@{=}[ddd] & & & S/S_0 \ar@{>>}[rrr] \ar[ddd] & & & G/G_0 \ar[ddd] \\
Z_0 \ar'[rr][rrr] \ar@{^{(}->}[ddrr] & & & Q \ar@{>>}'[rr][rrr]^-{\pi_Q} \ar@{^{(}->}[ddrr] \ar@{>>}'[dd][ddd] & & & \pi(Q) \ar@{^{(}->}[dr] \ar@{>>}'[dd][ddd] & & \\
& & & & & & & P \ar@{^{(}->}[dr] \ar@{>>}'[d][ddd] & \\
& & H \ar[rrr] & & & E \ar@{>>}[rrr]^-{\pi} \ar@{>>}[ddd] & & & M \ar@{>>}[ddd] \\
& & & M/G \ar@{=}'[rr][rrr] \ar@{=}[ddrr] & & & M/G \ar@{=}[dr] & & \\
& & & & & & & M/G \ar@{=}[dr] & \\
& & & & & M/G \ar@{=}[rrr] & & & M/G \\%
}}\label{diagram}
\end{equation}
}
Nous ferons attention à ce que certaines flèches dans ce diagramme ne représentent pas des applications mais correspondent à des flèches de diagrammes de fibration.
On notera que les flèches horizontales correspondent plutôt à l'action du groupe $H$ (ou de sous-groupes de $H$) et les flèches verticales à des actions des groupes $G$ et $S$.
Nous pouvons également vérifier que le noyau de la projection $\pi(Q) \to M/G $ est isomorphe à $ (N(S_0)/S_0)/Z_0$.

\subsection{Connexions invariantes}\label{inv_con}

L'action $G \circlearrowright E$ induit une action de $G$ sur l'espace $\Omega^1(E)$ des $1$-formes sur $E$. 
Du fait que les actions de $G$ et $H$  commutent, cette action s'étend naturellement en une action sur l'espace affine des connexions sur $E$ inclus dans l'espace $\Omega^1(E) \otimes \cH$, où $\cH$ est l'algèbre de Lie du groupe $H$. 
Pour tout $\omega \in \Omega^1(E) \otimes \cH$ et tout $g \in G$, nous notons cette action par $\omega^g= g^{*} \omega$. 
Nous allons maintenant caractériser une connexion $G$-invariante, qui satisfait $\omega^g=\omega$ pour tout $g \in G$.

Il est utile de faire une décomposition des divers espaces tangents correspondant aux variétés introduites dans la section précédente.
Nous allons faire ces décompositions à l'aide des différentes actions de groupes. 
Tout d'abord, introduisons les notations suivantes pour les algèbres de Lie correspondant aux différents groupes:
\begin{displaymath}
  \begin{array}{|c|c|c|c|c|c|c|c|c|c|} 
    \hline
    \text{Groupe} &    G & H & N(G_0) & G_0 & N(G_0)/G_0 & Z_0 & S=G\times H & S_0 & N(S_0) \\
    \hline
    \text{Algèbre de Lie} &  \cG &  \cH & \cN_{0}  & \cG_0 & \cK & \zo & \cS= \cG \oplus \cH & \so & \cN_{\so} \\
    \hline
  \end{array}
\end{displaymath}
Soit,
\begin{align*}
  \cG &= \cN_0 \subsetplus \cL &
  & \text{et} &
  \cH &= \zo \subsetplus \cM \ ,
\end{align*}
une décomposition d'algèbre de Lie réductive%
\footnote{Une décomposition d'algèbre de Lie $\mathfrak{g}=\mathfrak{h} \subsetplus \mathfrak{l} $ est réductive si  $\mathfrak{h}$ est une sous-algèbre de Lie de $\mathfrak{g}$ et si $\mathfrak{l}$  est un sous-espace réductif, i.e.   $[  \mathfrak{h} , \mathfrak{l} ] \subset \mathfrak{l}$.}
que nous supposerons aussi être une décomposition orthogonale d'espaces vectoriels pour la métrique de Killing.
Il est facile de montrer que nous avons la décomposition orthogonale d'algèbres de Lie:
\begin{align*}
\cN_0= \cG_0 \oplus \cK \ .
\end{align*}
Alors,
\begin{align*} \so =  \{ (X_0, \lambda_* X_0) / X_0 \in \cG_0 \} \ , \end{align*}
 où $\lambda_*: \cG_0 \to \cH$ est l'application  tangente de $\lambda: G_0 \to H$.
Ainsi, $\so$ est isomorphe à $\cG_0$. 

En utilisant cette identification, on montre que la décomposition 
\begin{align*}
  \cN_{\so} =\so \oplus \cK \oplus \zo 
\end{align*}
est une décomposition orthogonale d'algèbres de Lie. 
En fait, tout élément $(X,\xi) \in \cN_{\so}\subset \cG \times \cH$ peut être écrit sous la forme $(X,\xi)=(X_0+X_{\cK},\lambda_*X_0 +\xi_{\zo})$, où $X_0 \in \cG_0, X_{\cK} \in \cK$ et $\xi_{\zo}\in\zo$.

Avec ces décompositions et les actions de groupe, on obtient alors une version infinitésimale du diagramme~(\ref{diagram}):
{\footnotesize
  \begin{align*}
    \raisebox{.5\depth-.5\height}{%
      \xymatrix@R=3pt@C=10pt@M=4pt{       
        {\zo} \ar@{^{(}->}[rrr] \ar@{^{(}->}[ddrr] \ar@{=}[ddd] & & &  {\cK \oplus \zo} \ar@{>>}[rrr] \ar@{^{(}->}[ddrr] \ar@{^{(}->}'[dd][ddd] & & & \cK \ar@{  =}[dr] \ar@{^{(}->}'[dd][ddd] & & \\
        & & & & & & & \cK \ar@{^{(}->}[dr] \ar@{^{(}->}'[d][ddd] & \\
        & & {\zo \subsetplus \cM} \ar@{^{(}->}[rrr] \ar@{=}[ddd] & & & { (\cK \subsetplus \cL) \oplus (\zo \subsetplus \cM)} \ar@{>>}[rrr] \ar@{^{(}->}[ddd] & & & \cK \subsetplus \cL \ar@{^{(}->}[ddd] \\
        {\zo} \ar@{^{(}->}'[rr][rrr] \ar@{^{(}->}[ddrr] & & &{ T_q Q} \ar@{>>}'[rr][rrr]^-{\pi_{Q *}} \ar@{^{(}->}[ddrr] \ar@{>>}'[dd][ddd] & & & {T_x \pi(Q) } \ar@{^{(}->}[dr] \ar@{>>}'[dd][ddd] & & \\
        & & & & & & & {T_x P} \ar@{^{(}->}[dr] \ar@{>>}'[d][ddd] & \\
        & & {\zo \subsetplus \cM} \ar@{^{(}->}[rrr] & & & {T_q E} \ar@{>>}[rrr]^-{\pi_*} \ar@{>>}[ddd] & & & {T_x M} \ar@{>>}[ddd] \\
        & & & {T_{[x]} M/G} \ar@{=}'[rr][rrr] \ar@{=}[ddrr] & & & {T_{[x]} M/G} \ar@{=}[dr] & & \\
        & & & & & & & {T_{[x]} M/G} \ar@{=}[dr] & \\
        & & & & & {T_{[x]} M/G} \ar@{=}[rrr] & & &  {T_{[x]} M/G} \\%
        }}%
    \end{align*}
     }
pour tout point $q\in Q \subset E$ avec $\pi(q)=x$.
On a alors la décomposition:
 \begin{align}
   T_q E= T_q Q \oplus\cL^{Q}_{|q} \oplus \cM^Q_{|q} \ , \label{T_decomposition}
 \end{align}
où $ \cL^Q_{|q}$ (resp. $\cM^Q_{|q}$) est le sous-espace parcouru par les vecteurs  $X_q^E$ en $q\in E$ obtenus à partir des champs de vecteurs fondamentaux $X^E$ sur $E$ associés aux vecteurs $ X\in \cL$ (resp. $X \in \cM$).

Nous sommes maintenant prêts à caractériser une connexion $G$-invariante.
Soit  $\omega \in \Omega^1(E)\otimes \cH$ une $1$-forme de connexion  $G$-invariante. 
Nous pouvons la restreindre à $Q$, sans perdre d'information, puisque ses valeurs sur $E$ peuvent être retrouvées par action du groupe $G$ sur $Q$ en utilisant le fait que $G \cdot Q = E $. 
Alors pour tout $q \in Q$, $\omega_{|q}$  peut être évaluée sur les trois espaces vectoriels $T_q Q$, $\cL^Q_{|q}$ et $\cM^Q_{|q} $:
 \begin{itemize}
 \item
   La restriction à  $\cM^Q_q \subset \cH^Q_q $ est déterminée par la relation $\omega_{|q} (X_q^E) = X$, pour tout $X \in \cM$.
   Il n'y a pas d'autres degrés de liberté.
 \item 
   La restriction à  $T_q Q$ donne une $1$-forme $\mu$ définie par  $\mu (X)= \omega (X)$, pour tout $X \in TQ $. 
   Elle satisfait la propriété d'équivariance suivante:
   \begin{align*}
     &\R^{*}_{(g,h)} \mu = Ad_{h^{-1}} \mu &
     &\forall (g,h) \in N(S_0) \ ,
   \end{align*}
   où $\R_{(g,h)}$ est l'action à droite de  $N(S_0)$ sur $Q$. 
   En tenant compte de cette relation d'équivariance pour un élément $(g_0, \lambda (g_0) ) \in S_0$, on peut montrer que $\mu$ prend ses valeurs dans $\zo$.
   Cette relation d'équivariance restreinte à $Z_0$ fait de $\mu$ une $1$-forme de connexion sur le fibré principal $ \fiber{Z_0}{Q}{\pi (Q)}$.
 \item
   La restriction à $\cL^Q_q $ induit une application:
   \begin{align*}
     \psi_q:    \cL   & \to  \cH   \\
     X& \mapsto \psi_q (X) = \omega_q (X_q^E) \ .
   \end{align*}
   L'application $\psi_{q}$ satisfait la relation d'équivariance suivante:
   \begin{align*}
     &\ Ad_h \circ \psi_q \circ Ad_{g^{-1}} = \psi_{g q h^{-1}} &
     &    \forall (g,h) \in S \ .
   \end{align*}
   Alors pour tout $(g_{0},\lambda(g_{0})) \in S_0$, on a $ Ad_{\lambda(g_0)} \circ \psi_{q} \circ Ad_{g_0^{-1}} =\psi_q $.
   L'application $q \mapsto \psi_q$ de $Q$ dans $\cF$  définit une section du fibré associé à $Q$: $ F^{\cL} = Q \times_{N(S_0)/S_0} \cF_{\cL}$, où la fibre est définie comme étant l'espace vectoriel suivant:
   \begin{align*}
     \cF_{\cL}= \{\ell : \cL\to \cH  , Ad_{\lambda (g_0)} \circ \ell \circ Ad_{g_0^{-1}} = \ell  \} \ .
   \end{align*}
 \end{itemize}

Ainsi, $\omega$ est complètement déterminée par les deux objets $\mu$ et $\psi$ décrits ci-dessus.
Notons que $\mu$ et $\psi$ sont naturellement reliés à  des fibrés principaux construits à partir du fibré $E$: $\psi$ correspondant à une structure ``verticale'' sur le diagramme (\ref{diagram}) et $\mu$ à une structure ``horizontale''.

Il est possible~\cite{jadczyk:84} de décomposer la connexion $\omega$ sur des objets ne faisant référence qu'au fibré $ F^{\cL}$.
Pour cela, il est nécessaire d'introduire une connexion $A$ sur le fibré principal: 
 \begin{align*}
 \lfiber{(N(S_0)/S_0)/Z_0}{\pi (Q)}{M/G}
 \end{align*}
 
 Alors la $1$-forme de connexion $\mu$ peut être mise en bijection avec un couple $(B, \alpha)$ où 
 \begin{itemize}
 \item $B$ est une connexion sur le fibré principal \fiber{N(S_0)/S_0}{Q}{M/G}.
 \item $\alpha$ est une section de fibré vectoriel $F^{\cK}$,  où  $F^{\cK} = Q \times_{N(S_0)/S_0} \cF_{\cK} $ est le fibré associé au fibré principal \fiber{N(S_0)/S_0}{Q}{M/G}. 
L'espace vectoriel $\cF_{\cK}$ est défini de la manière suivante:
 \begin{align*}
 \cF_{\cK}= \{ k : \cK \to \cH  , Ad_{\lambda (g_0)} \circ k \circ  Ad_{g_0^{-1}} =  k  \} \ .
 \end{align*}
 \end{itemize}
 (Notons la similarité entre $F^{\cL}$ et $F^{\cK}$). 

La correspondance entre  $\mu$ et le couple $(B,\alpha)$ est donnée explicitement par les relations:
 \begin{align*}
   \left\{
     \begin{array}{ll}
       B & =  \mu + \pi^* A  - \mu (\pi^* A)^Q  \\
       \alpha_q &= \mu_q |_{\cK_q} 
     \end{array} \right. \ .
 \end{align*}
 En particulier, on a $pr_\cK B = \pi^* A $, où $pr_\cK$ est la projection de $\cK \oplus \zo $ sur $\cK$.
À partir de ce point de vue, il est également possible~\cite{coquereaux:85} de faire un lien avec les théories de type Kaluza-Klein, en voyant la connexion $\omega$ comme faisant partie d'une métrique sur le fibré principal $E$.

 \subsection{Relation avec l'approche de Wang}\label{wang_approach}
 
 Du fait que nous ayons supposé que l'action du groupe $G$ soit simple, l'espace  $M$ est localement isomorphe au produit d'espaces $ M/G \times G/G_0$. 
L'étude des connexions invariantes peut être encore simplifiée en supposant que $M = M/G \times G/G_0$, ce qui revient à restreindre notre étude à des objets localisés autour d'une orbite de $G$ dans $M$. 
Dans cette section, nous supposerons donc que $M=M/G \times G/G_0$. 
Cela nous permettra de faire le lien entre la construction précédente et les approches locales faites dans~\cite{harnad:80,forgacs:80,brodbeck:96}. 
Ce point de vue rejoint également celui de Wang donné dans~\cite{wang:58}, qui aborda le problème des connexions invariantes sur des espaces homogènes.
Nous appellerons cette approche, approche ``locale'' par la suite.

 La structure relativement simple de $M$ va nous permettre de faire une construction similaire à celle faite précédemment, mais en remplaçant l'espace $P$ par l'espace quotient $M/G$. 
Cela simplifiera grandement la structure des fibrés dans les directions correspondant à $G$, ainsi que la décomposition des connexions invariantes.
Nous pourrons même aller un peu plus loin et classifier les fibrés $G$-symétriques.

Tout d'abord, de par la structure de $M = M/G \times G/G_0$,  nous  pouvons inclure $M/G$ dans $M$, en l'identifiant à  $M/G \times \{e G_0\}$ dans $M$.
Alors, les fibrés $G$-symétriques peuvent être classifiés par les couples $([\lambda],\tQ)$, où $[\lambda]$ est la classe de conjugaison d'un homomorphisme $\lambda : G_0 \to H $ pour l'action de  $G$ sur $G_0$ et $\tQ$ est un fibré principal sur $M/G$ de groupe de structure $Z_0= Z(\lambda(G_0),H)$. 
En effet, étant donné un fibré principal $E$ au dessus de  $M = M/G \times G/G_0$, $G$-symétrique, de groupe de structure $H$, nous pouvons construire un couple $([\lambda],\tQ)$  en considérant la restriction $E_{|M/G}$ de $E$ à $M/G$.
Définissons alors $\tQ= \{ p \in E_{|M/G} / \lambda_p=\lambda \}$, pour une application de référence $\lambda=\lambda_{p_0}$.

Réciproquement, nous pouvons associer à tout couple $([\lambda],\tQ)$ un fibré principal $G$-symétrique.
Nous pouvons construire le fibré principal suivant:
 \begin{align*} 
 \lfiber{Z_0 \times G_0}{Q'=\tQ \times G}{M/G \times G/G_0}
 \end{align*}
 défini pour l'action $(z,g_0,\tq,g)\mapsto(\tq \cdot z_0, g \cdot g_0)$. 
Considérons maintenant l'action à gauche de $Z_0 \times G_0$ définie%
\footnote{Remarquons que l'action des sous-groupes $Z_0$ et $G_0$ commutent.}
sur $H$ par:
 \begin{align*}
   \rho:  Z_0 \times G_0 \times H  &\longrightarrow H \\
   (z,g_0,h) &\longmapsto z \cdot \lambda(g_0) \cdot h \ .
 \end{align*}
 Nous pouvons alors construire un fibré associé à $Q'$: $\tE= Q' \times_{(Z_0 \times G_0)} H$, de fibre $H$.
 On peut voir que $\tE$ est un fibré principal $G$-invariant en considérant le diagramme commutatif suivant:
 \begin{align}
\raisebox{.5\depth-.5\height}{%
  \xymatrix{
    {}& {Z_0 \times G_0} \ar[d] &  {Z_0 \times G_0} \ar[d] \\
    {H} \ar[r] & {Q' \times H} \ar[r]^-{pr_1} \ar@{->>}[d]_-{\Psi} & {Q'=\tQ \times G} \ar@{->>}[d] \\
    {H} \ar[r]  & {\tE} \ar[r]^-{\tilde{\pi}} & {M=M/G \times G/G_0}}} \ ,
\label{fibre_associe}
\end{align}
 où $Q'\times H$ est aussi un fibré principal de groupe de structure $H$ et  $G$-invariant. On prendra garde à ce que certaines flèches soient des flèches correspondant à des diagrammes de fibration.

 La composition des applications $E\mapsto ([\lambda],\tQ)$ et $ ([\lambda],\tQ)\mapsto \tE$ fournit une application $E \mapsto \tE$, qui est un isomorphisme de fibrés principaux avec groupe de structure $H$ et $G$-invariant.
Nous pouvons réaliser l'application inverse en considérant l'application qui à tout point $\Psi(\tq,g,h) \in \tE$, associe le point $g \cdot \tq \cdot h \in E$ où $\tq\in \tQ$ est considéré comme un point de $E$.

En utilisant cet isomorphisme, nous pouvons envoyer une connexion $G$-invariante sur $E$ sur une connexion $G$-invariante $\omega $ sur $\tE$.
Du fait que la projection $\Psi$ du fibré principal  $Q'\times H$ de groupe de structure $(Z_0 \times G_0)$ est aussi une application  $G$-équivariante de fibrés $G$-symétriques de groupe de structure $H$, on peut montrer (voir~\cite{harnad:80,brodbeck:96} et la démonstration dans le cas non commutatif traité dans la section~\ref{local_approach}) que $\pso$ peut être écrit sous la forme générique suivante:
 \begin{align}
  \pso_{|(\tq,g,h)} = Ad_{h^{-1}} ( \Lambda_{|\tq} \circ \theta^G_{~|g} + \tom_{|\tq}) +\theta^H_{~|h} \ ,  
  \label{class_decomposition}
\end{align}
où $\tom$ est une $1$-forme de connexion sur $\fiber{Z_0}{\tQ}{M/G}$ et $\theta^G$ et $\theta^H$ sont les formes de Cartan usuelles définies sur les groupes $G$ et $H$ respectivement. L'application $ \Lambda \in C^{\infty}(\tQ)\otimes \cG^* \otimes \cH $  satisfait la propriété d'équivariance:
\begin{align*}
  &\R_{z_0}^* \Lambda = Ad_{z_0} \Lambda &
  & \forall z_0 \in Z_0
\intertext{et les relations:}
  &Ad_{\lambda(g_0)} \circ  \Lambda  \circ Ad_{g_0^{-1}} =  \Lambda  &
  &\forall g_0 \in G_0 \\
  &  \Lambda_{\tq}(X_0) = \lambda_*(X_{0}) &
  &\forall X_0 \in \cG_0 \ \text{et} \ \forall \tq \in \tQ \ .
\end{align*}
Cette application caractérise ainsi une section du  fibré associé à $\tQ$, de fibre $\cG^* \otimes \cH$ et de base $M/G$ et correspond à l'action adjointe de  $Z_0$ sur $\cH$.

\section{Connexions non commutatives invariantes} \label{invar-nonc-conn} 

\subsection{Approche globale}\label{global_approach}

Nous allons maintenant caractériser les connexions non commutatives invariantes sous l'action d'un groupe de Lie $G$ compact.

Soit $\cG$, l'algèbre de Lie de $G$. Une action de $\cG$ sur $\kA$ est une opération de Cartan de $\cG$ sur l'algèbre différentielle graduée $\Omega_\der(\kA)$ et nous considérerons toujours $\cG$ comme une sous-algèbre de Lie de $\der(\kA)$.
\begin{rem}
  Nous procédons ici comme dans la situation géométrique, où nous considérions $G$ comme un sous-groupe du groupe des automorphismes du fibré principal.
\end{rem}

Une connexion non commutative $\cG$-invariante sur le module à droite $\kA$ est une con\-nexion non commutative $\widehat{\nabla}$ qui satisfait la relation suivante:
\begin{align*}
  Y\left( \widehat{\nabla}_\cX a \right) = \widehat{\nabla}_{[Y,\cX]} a +
  \widehat{\nabla}_\cX \left( Ya \right) \ ,
\end{align*}
pour tout $Y\in \cG$, $\cX \in \der(\kA)$ et $a\in \kA$.
Si $\widehat{\nabla}$ est donnée par une $1$-forme non commutative $\omega$, cette condition est équivalente à la condition suivante:
\begin{align*}
  L_Y \omega = 0 \ ,
\end{align*}
pour tout $Y\in\cG$.

Nous allons maintenant nous servir de la caractérisation précédente des connexions ordinaires $G$-symétriques afin de caractériser les connexions non commutatives invariantes pour l'action d'un groupe $G$.
Pour cela, nous allons utiliser le fait que $\kA$ est une sous-algèbre basique de $\kB= C^\infty (E) \otimes M_n$ pour une action de   $\cH=\ksu(n) \subset M_n$ bien choisie, ce qui nous permettra, en particulier, d'utiliser le diagramme~(\ref{diagram}).

Une $1$-forme de connexion non commutative $G$-invariante est  un élément 
\begin{align*}
  \omega \in (\Omega^1_{\der}(\kA))_{\cG-\inv}\simeq \Oder^{1}(\kB)_{\cH-\Bas,
    \cG-\inv}
\end{align*}
et peut se mettre sous la forme suivante (voir section~\ref{sec:omega_der}):
\begin{align*}
\omega = a - \phi \in [\Omega^1(E) \otimes M_n] \oplus [ C^{\infty}(E)\otimes M_n \otimes \der(M_n)^*] \ .
\end{align*}
Alors  $a$ et $\phi$ satisfont aux relations:
\begin{align}
  L_{\xi} ( a-\phi) &= 0 \label{invh} \\
  i_\xi( a-\phi) &= 0 \label{hor} \\
  L_{X} ( a-\phi) &= 0 \label{invg}
\end{align}
pour tout $\xi \in \cH=\ksu(n)$ et tout $X \in \cG$. 

\begin{rem}
  Rappelons que pour une connexion ordinaire, on a $\phi=i\theta$. 
Ainsi, les résultats de la section~\ref{construction_1} pourront facilement être retrouvés.
\end{rem}

Grâce à l'action du groupe  $G$, nous pouvons restreindre $a$ et $\phi$ à $Q \subset E$.
Alors $a$ est complètement déterminé par ses valeurs sur $T_qE$  pour tous les $q\in Q$. 
L'application  $a_q : T_q E \to M_n$  peut être décomposée en plusieurs parties.
Appelons $\mu \in \Omega^1(Q)\otimes M_n$ la restriction de la  $1$-forme $a$ aux sections de $TQ$. 
Par la relation~(\ref{hor}), on a  $\mu_q(\xi_q^E) = \phi_q(\xi)$ pour tout $\xi \in \zo$. 
Cette $1$-forme satisfait la relation d'équivariance:
\begin{align*}
  L_{(X,\xi)}\mu &= ( L_{X} +L_{\xi} )\mu =0   &
 \forall  (X,\xi)& \in \cN_{\so} \subset \cG\times\cH \ .
\end{align*}
(Nous utilisons les notations de la section~\ref{inv_con}. Rappelons également que la dérivée de Lie $L_{\xi}$ contient une partie géométrique et une partie algébrique).
Alors en utilisant l'invariance sous l'action de $\so$ (cette action vient de l'action du groupe $S_{0}$ sur $Q$), on voit que $\mu$ prend ses valeurs dans l'espace vectoriel: 
\begin{align*}
  \wo & := Z( \lambda_* \cG_0, M_n) \ ,
\end{align*}
le centralisateur de $  \lambda_*  \cG_0$ dans $M_n$.
Par conséquent, $\eta_q :=\phi_{q|\zo}$ prend ses valeurs dans $\wo$. 

On constate que  $\wo$ est une sous-algèbre associative de $M_n$ sur laquelle l'algèbre de Lie $\zo$ agit par l'action adjointe.
Il est ainsi naturel de construire l'algèbre différentielle graduée $\Omega_{\zo}(\wo)=\wo \otimes \bigwedge\zo^*$ qui ressemble au calcul différentiel $\Omega_{\der}(M_n)\simeq M_n \otimes \bigwedge\ksl_n^*$. 
Sur $\Omega_{\zo}(\wo)$, la différentielle est définie par une formule de Koszul comme dans la formule~(\ref{differential}), où l'algèbre de Lie  $\zo$  remplace les dérivations de l'algèbre $\wo$. 

Une autre propriété importante de l'algèbre $\wo$ est qu'il y a une application naturelle:
\begin{align*}
  \cN_{ \so} &\to \der\left( C^{\infty}(Q) \otimes  \wo \right)\\
  (X,\xi) &\mapsto  X^Q+\xi^Q + ad_{\xi} \ .
\end{align*}
Notons que $\so$ est envoyée sur $0$ par cette application et celle-ci se factorise en une application de $\cN_{\so}/\so$ dans $\der\left( C^{\infty}(Q) \otimes  \wo \right)$.
Cela nous permet de définir une opération de Cartan de $\cN_{\so}/\so$ sur $\Omega(Q)\otimes \Omega_{\zo} (\wo)$, dont la dérivée de Lie est donnée par: 
\begin{align*}
  L_{(X,\xi)} & = L_{X^Q+\xi^Q} + L_{ad_{\xi}} &
  &\forall (X,\xi) \in \cN_{\so}/\so=\cK\oplus\zo \ .
\end{align*}
En particulier, cette opération induit une opération sur l'algèbre de Lie $\zo$, que l'on notera $L_\xi$, pour tout $\xi \in \zo$.

La différence $\mu-\eta$ est naturellement un élément de degré $1$ dans $\Omega(Q)\otimes \Omega_{\zo}(\wo)$. 
En utilisant les relations~(\ref{invh}),~(\ref{hor})~et~(\ref{invg}), il est alors facile de voir que: 
\begin{align*}
  i_{\xi} (\mu -\eta) &= 0 &
  &\forall \xi \in \zo \\
  L_{(X,\xi)} (\mu -\eta) &= 0 &
  &\forall  ( X,\xi) \in \cN_{\so} \ .
\end{align*}
Cela implique que $\mu -\eta \in \left( \Omega(Q) \otimes \Omega_{\zo}(\wo)\right)_{\zo-\bas}^1$.

Maintenant, introduisons $\alpha\in C^{\infty}(Q)\otimes \wo\otimes \bigwedge \cK^*$ défini par :
\begin{align*}
  \alpha(X) &= \mu(X^Q) & 
  & \forall X\in \cK \ .
\end{align*}
On a $\alpha \in \wo \otimes ( C^{\infty}(Q) \otimes \bigwedge \cK^*)_{\cK-\inv}$. L'action de $\cK$ est induite par l'action de $\cN_{\so}$ sur $C^\infty(Q)$ et par la dérivée de Lie usuelle (induite par la structure d'algèbre de Lie de $\cK$) sur $\bigwedge \cK^*$.

Nous pouvons maintenant introduire l'algèbre différentielle graduée: 
\begin{align*}
  \Omega_{\zo +\cK}(M/G,\wo) := ( \Omega(Q) \otimes \Omega_{\zo} (\wo) \otimes   \bigwedge \cK^*)_{(\zo+\cK)-\bas} \ ,
\end{align*}
équipée de la différentielle naturelle, \textit{i.e.} la somme de la différentielle sur chacune des composantes.
Nous verrons plus loin pourquoi $M/G$ a été introduit dans les notations.

Cette algèbre différentielle graduée nous permet de rassembler les applications $\mu,\eta$ et $\alpha$ en une $1$-forme algébrique et on a:
\begin{align*}
  \mu -\eta-\alpha \in \Omega^1_{\zo+\cK}(M/G,\wo) \ .
\end{align*}
Cette relation nous permet ainsi de caractériser la restriction de $a$ et $\phi$ à $TQ$ de manière simple.

Regardons maintenant les composantes $\cL^Q_{|q}$ et $\cM^Q_{|q}$ de $T_qE$.
En utilisant des arguments similaires à ceux utilisés dans la section~\ref{inv_con}, la restriction de $ \psi := a_{|\cL^Q}$ définit une section du fibré vectoriel associé à \fiber{N(\so)/\so}{Q}{M/G}, dont les fibres sont isomorphes à l'espace vectoriel:
\begin{align*}
  \cF_{\cL}&:=  ( M_n  \otimes \cL^*)_{\so-\inv} =\{\ell: \cL \to M_n / L^{\cL}_{(X,\lambda_* X)} \ell =0 \ \ \forall X \in \cG_0\} \ ,
\end{align*}
où $ (L^{\cL}_{(X, \xi)}\ell)(Y)= - \ell([X,Y])+ [\xi,\ell(Y)]$ pour tout $(X,\xi) \in N(\so)$. 
Dans cette relation, on utilise l'action naturelle de  $\cN_{\so}$ sur l'espace $M_n \otimes \cL^*$, pour laquelle l'espace vectoriel $\cF_\cL \subset M_n \otimes \cL^*$ est invariant.
 $\psi \in C^\infty(Q)\otimes \cF_\cL$ satisfait la relation d'équivariance:
\begin{align*}
  L_{(X,\xi)} \psi&= (L_{X^E+\xi^E}+ L^{\cL}_{(X,\xi)})\psi=0 &
  &\forall (X,\xi) \in \cN_{\so} \ .
\end{align*}
Ainsi, on a $\psi \in \left(C^{\infty}(Q) \otimes   \cF_{\cL}  \right)_{(\zo+\cK)-\inv} $.

De la même manière, la restriction $\zeta := a_{|\cM^Q}=\phi_{|\cM}$ %
définit une section du fibré associé à \fiber{N(\so)/\so}{Q}{M/G} dont les fibres sont isomorphes à l'espace vectoriel: 
  \begin{align*}
    \cF_{\cM} &:=  ( M_n  \otimes \cM^*)_{\so-\inv} =\{m: \cM \to M_n / L^{\cM}_{(X , \lambda_* X)} m =0 \ \ \forall X \in \cG_0\} \ ,
  \end{align*}
où $ (L^{\cM}_{(X,\xi)}m)(Y)= - m([X,Y])+ [\xi,m(Y)]$ pour tout $(X,\xi) \in N(\so)$. 
En effet, on a la relation d'équivariance suivante:
\begin{align*}
L_{(X,\xi)} \zeta&= (L_{X^E+\xi^E}+ L^{\cM}_{(X,\xi)})\zeta=0 &
&\forall  (X,\xi) \in \cN_{\so}
\end{align*}  
et donc $\zeta \in \left(C^{\infty}(Q) \otimes   \cF_{\cM}  \right)_{(\zo+\cK)-\inv} $ \ .
  
D'après la relation:
\begin{align*}
  ( M_n  \otimes \cL^*)_{\so-\inv} \oplus  (  M_n \otimes \cM^* )_{\so-\inv} &=  ( M_n \otimes (\cL^* \oplus \cM^* )_{\so-\inv}=:\cF \ ,
\end{align*}
$\zeta +\psi$ peut être considéré comme une section d'un fibré associé à \fiber{N(S_0)/S_0}{Q}{M/G} dont les fibres sont isomorphes à l'espace vectoriel $\cF= \cF_{L}\oplus \cF_{M}$.

Les résultats précédents peuvent se résumer par l'isomorphisme suivant:
\begin{align*}
  (\Omega^1_{\der}(\kA))_{\cG-\inv} \simeq \Omega^1_{\zo +\cK}(M/G,\wo) \oplus \cP \ ,
\end{align*}
où
\begin{align*}
  \cP &= (C^{\infty}(Q) \otimes \cF)_{(\zo+\cK)-\inv} \ .
\end{align*}
Notons $\cC:=\Omega^0_{\zo+\cK}(M/G,\wo) = (C^{\infty}(Q) \otimes \wo)_{(\zo+\cK)-\inv}$, alors $\Omega_{\zo+\cK}(M/G,\wo)$ est un calcul différentiel associé à $\cC$. 
L'algèbre $\cC$ peut être interprétée comme l'algèbre des sections du fibré associée à \fiber{N(S_0)/S_0}{Q}{M/G} dont les fibres sont isomorphes à l'algèbre $\wo$. 
L'algèbre $\cC$ peut être considérée comme une ``réduction'' de l'algèbre $\kA$ et les éléments de $SU(\cC)$ définissent des transformations de jauge sur l'espace de connexions non commutatives  $\kA$, $\cG$-invariantes. 
Équipée du calcul différentiel $\Omega_{\zo+\cK}(M/G,\wo)$ et du module $\cP$, cette algèbre constitue le bloc élémentaire pour la construction des connexions $\cG$-invariantes sur $\kA$. 

\begin{rem}
  Notons que les objets introduits sont naturellement reliés à des structures de fibré au dessus de $M/G$, tout comme pour les connexions ordinaires invariantes.
  Cependant, notons que dans le cas des connexions non commutatives, nous n'avons pas eu besoin de connexion de référence pour pouvoir redescendre les objets sur $M/G$. 
\end{rem}

\subsection{Approche locale}\label{local_approach}

Comme dans le cas classique, nous pouvons considérer le cas où $M=M/G\times~G/G_0$.
Nous utiliserons les notations introduites dans la section~\ref{wang_approach}. 
L'idée est d'envoyer (par application \textit{pull-back}) une connexion sur $\tE$ vers une connexion sur $Q'\times H$.
Cela permet ainsi d'écrire explicitement la décomposition d'une $1$-forme de connexion, en utilisant en particulier les formes de Cartan sur $G$ et $H$.
Nous allons généraliser cette construction pour les $1$-formes de connexion non commutatives.
Nous devons donc passer à une approche plus algébrique et devons trouver les bonnes algèbres sur lesquelles travailler.
L'idée est d'envoyer les $1$-formes de connexion non commutatives $\cG$-invariantes dans une algèbre différentielle graduée suffisamment ``grosse'', sur laquelle elles prendront une expression relativement simple. 
Nous pouvons résumer la problématique dans le schéma suivant:
\begin{align*}
\raisebox{.5\depth-.5\height}{%
  \xymatrix{
    Q'\times H \ar[d]_{\Psi} &{\xy *[o]=<30pt>\hbox{?}="o"*\frm{o} \endxy}  \ar@{^(->}[r] & \Omega(Q'\times H)\otimes \Omega_{\der}(M_n) \\
    \tE \ar@{~>}[r] & \Omega_{\der}(\kA) \ar@{^(->}[r]^-{\text{basique}}_-{\text{pour $\cH$}} & \Omega(\tE)\otimes \Omega_{\der}(M_n) \ar[u]^{\Psi^*}   
    }} 
\end{align*}
L'algèbre différentielle graduée que nous cherchons doit remplacer $\kA$, de la même manière que l'espace $Q'\times H$ remplaçait $\tE$.
Il est naturel de considérer cette algèbre comme la sous-algèbre basique de $C^{\infty}(Q'\times H)\otimes M_n$ pour l'opération de Cartan de $\cH$. 
Alors le  \textit{pull-back} $\pso$ d'une $1$-forme de connexion $\cG$-invariante $\omega \in \Omega_{\der}(\kA) \subset \Omega(\tE)\otimes \Omega_{\der}(M_n) $ appartient à $ \Omega(Q'\times H)\otimes \Omega_{\der}(M_n)$. 
Du fait que $Z_0$ et $G_0$ n'agissent pas sur la partie $M_n$ de ces algèbres différentielles graduées et que $\Psi^* $ préserve l'invariance sous l'action de $\cG$ ainsi que la basicité vis-à-vis de $\cH$, on a:
\begin{align*}
  \pso \in \left[ \Omega(\tQ) \otimes \Omega(G) \otimes \Omega(H) \otimes \Omega_\der(M_n) \right]_{\substack{G-\inv \hfill \\ H-\bas \hfill\\(Z_0\times G_0)-\bas \hfill}} \ .
\end{align*}

L'avantage de travailler au niveau de l'algèbre différentielle graduée $\Omega(\tQ) \otimes \Omega(G) \otimes \Omega(H) \otimes \Omega_{\der}(M_n)$ est que nous pouvons décomposer facilement les actions des groupes $G$, $G_0$, $Z_0$ et $H$ sur les différents espaces.
Ces actions sont schématisées dans le diagramme suivant:
\begin{align*}
  \xymatrix{
    &&Z_0 \ar@/^/[dll]_{\R^*_{z_0}} \ar@/_/[drr]^{\L^*_{z_0^{-1}}}&&&H \ar@/^/[dl]_{\R^*_{h}} \ar@/_/[dr]^{Ad_{h}\otimes Ad^*_{h^{-1}} }&\\
    \Omega({\tQ}) & \otimes & \Omega(G) & \otimes& \Omega(H) & \otimes&  \left(M_n\otimes \bigwedge \ksl_n^*\right) & \\
    &G \ar@/^/[ur]_{\L^*_{g^{-1}}}&& G_0 \ar@/_/[ul]^{\R^*_{g_0}} \ar@/^/[ur]_{\L^*_{\lambda(g_0)^{-1}}}&&&
    }
\end{align*}
où les symboles $\L$ et $\R$ schématisent une action à gauche et à droite respectivement.
Notons que l'action de $Z_0$ et $G_0$ sur $\Omega(H)$ commutent.

Alors, en utilisant la $H$-basicité et la $G$-invariance, un simple calcul montre que la $1$-forme $\pso$ peut être écrite sous la forme générique:
\begin{align}
  \pso_{|(\tq,g,h)} = Ad_{h^{-1}} \left( \tom_{|\tq} + \Lambda_{\tq}\circ\theta^G_{|g} +  \phi_{\tq}\circ Ad_{h}\circ (\theta^H_{|h}-i\theta)\right) \ , \label{decomposition}
\end{align}
où $\theta^G$ et $\theta^H$ sont les $1$-formes de Cartan sur les groupes $G$ et $H$ et $i\theta$ est la $1$-forme algébrique introduite en section~\ref{sec:alg-des-endom}. 
Il est naturel d'utiliser la $1$-forme $i\theta$ dans cette relation, de manière à rendre explicite l'identification~(\ref{soudure}). 
Dans la formule~(\ref{decomposition}), on a:
\begin{align*}
   \tom &\in \Omega^1(\tQ) &
   \Lambda &\in C^{\infty}(\tQ)\otimes M_n \otimes \cG^*  &
   \phi & \in  C^{\infty}(\tQ)\otimes M_n \otimes \ksl_n^* \ .
 \end{align*}
La $Z_0$-invariance implique que:
 \begin{align*}
   \R^*_{z_0} \tom  &= Ad_{z_0^{-1}} \circ \tom &
   \R^*_{z_0} \Lambda &= Ad_{z_0^{-1}} \circ \Lambda &
   \R^*_{z_0} \phi &= Ad_{z_0^{-1}} \circ \phi \circ Ad_{z_0}  \ ,
 \end{align*}
pour tout $z_0 \in Z_0$. 
Alors $\Lambda$ et $\phi$ peuvent être considérés comme des sections de fibrés associés à \fiber{Z_0}{\tQ}{M/G}.\\
D'autre part, la $G_0$-invariance implique que:
\begin{align}
& Ad_{\lambda(g_0)^{-1}} \tom = \tom  &
& Ad_{\lambda(g_0)^{-1}} \circ \Lambda \circ Ad_{g_0} = \Lambda &
& Ad_{\lambda(g_0)^{-1}} \circ \phi \circ Ad_{\lambda(g_0)} = \phi \ ,
\label{equivariance}
\end{align}
pour tout $g_0 \in G_0 $.\\
La $(G_0\times Z_0)$-horizontalité nous donne les relations suivantes:
\begin{align}
  & \Lambda(X_0) =  \phi(\lambda_* X_0)  & 
  &\forall X_0 \in \cG_0 &
  &\text{et}&
  & \tom(Z^{\tQ}_0)=  \phi(Z_0) & 
  & \forall Z_0 \in \cZ_0 \ .
\label{horizontality}
\end{align}

Les connexions ordinaires $\cG$-invariantes sont retrouvées dans la formule~(\ref{decomposition}) lorsque $\phi = \gone$. 
Dans ce cas, nous avons:
\begin{align*}
  \pso_{|(\tq,g,h)} = Ad_{h^{-1}} \left( \tom_{|\tq} + \Lambda_{\tq}\circ\theta^G_{|g} \right) +  \theta^H_{|h}-i\theta
\end{align*}
qui est à comparer avec la formule~(\ref{class_decomposition}). 
Comme nous l'avons déjà expliqué, le terme supplémentaire $i \theta$ est exactement ce qu'il faut pour pouvoir plonger les $1$-formes de connexion dans le complexe des formes non commutatives.

D'après la relation d'équivariance~(\ref{equivariance}),  $\Lambda$ (resp. $\phi$) est un opérateur d'entrelacement entre la représentation de $G_0$ sur $\cG^{\gC}$ (resp.  $\cH^{\gC}$) avec la représentation de  $G_0$ sur l'algèbre $M_n \simeq \vect_{\gC}(\gone, \cH)$.
Ainsi, par le lemme de Schur, cet opérateur se décompose en une somme directe d'isomorphismes entre représentations irréductibles isomorphes entre $\cG^{\gC}$  (resp.  $\cH^{\gC}$) et $M_n \simeq \vect_{\gC}(\gone, \cH)$. 
Si l'on demande que la connexion soit antihermitienne, on doit alors identifier les isomorphismes qui correspondent à des représentations conjuguées les unes aux autres ou directement regarder les représentations réelles.

\begin{rem}
  Cette caractérisation ``locale'' des connexions non commutatives invariantes
  est équivalente à la caractérisation ``globale''.  Cela se voit en
  décomposant tous les degrés de liberté dans les deux situations et en les
  comparant.
\end{rem}

D'après la décomposition (\ref{decomposition}), il est facile d'écrire une expression locale sur $M$.
Une section $S : M \to \tE$ peut être factorisée par une section locale $s=s_{\tQ}\times s_G$ sur le fibré $Q'=\tQ\times G $ et une section  $s_{H}$ sur le fibré trivial $Q'\times H$.
L'application  
\begin{align*}
  S=\Psi \circ s_H \circ s
\end{align*}
est donnée par le diagramme suivant:
\begin{equation*}
  \begin{aligned}
    \xymatrix{
      {Q' \times H} \ar[r]^-{pr_1} \ar@{->>}[d]_-{\Psi} & {Q'=\tQ \times G} \ar@{->>}[d] \ar@/^/[l]^-{s_{H}} \\
      {\tE} \ar[r]^-{\tilde{\pi}} & {M=M/G \times G/G_0}
      \ar@/_/[u]_-{s=s_{\tQ}\times s_G} \ar@/^/[l]^-{S} }
\end{aligned} \ \  .
\end{equation*}
Alors, une $1$-forme locale de connexion s'écrit 
\begin{align*}
  S^* \omega = s^* \circ s_H^* \circ \Psi^* \tom \ \in \Omega(M)\otimes
  \Omega_\der(M_n) \ .
\end{align*} 

Il est utile d'écrire une section $S$ de la manière suivante:
\begin{align*}
S : & M \longrightarrow  \tE \\
& m \longmapsto \Psi(s_{\tQ}(m),s_G(m),h(m)) \ .
\end{align*}
On obtient finalement:
\begin{align}
S^* \omega =
Ad_{h^{-1}} \left( s_{\tQ}^* \tom +  s_{\tQ}^*\Lambda \circ s_G^*\theta^G +   s_{\tQ}^*\phi \circ Ad_{h}\circ (h^*\theta^H -i\theta)\right) \ .
\label{local_form}
\end{align}

Il est intéressant de regarder comment sont implémentées les transformations de jauge ``passives''.
Dans la présente situation, il est naturel de regarder les transformations de jauge passives qui préservent les symétries de la $1$-forme locale de connexion non commutative $S^*\omega$.
Il existe trois manières de procéder. On peut multiplier:
\begin{itemize}
\item  $s_H$ à droite par un élément $h' \in H$,
\item $s_{\tQ}$ par un élément $z_0 \in  Z_0$, ou encore 
\item $s_G$ par un élément $g_0 \in G_0$.
\end{itemize}
Alors $S$ est modifiée de la manière suivante:
\begin{align*}
  S \longmapsto S'= \Psi \circ (s_{\tQ} \cdot z_0,s_G\cdot  g_0 ,h\cdot h' ) \ .
\end{align*}
Nous pouvons reporter l'action sur $\tQ$ et $G$ sur une action sur $H$ grâce à l'équivariance de l'application $\Psi$:
\begin{align*}
  \Psi \circ (s_{\tQ} \cdot z_0,s_G\cdot  g_0 ,h\cdot h' ) &=  \Psi \circ (s_{\tQ},s_G ,\lambda(g_0^{-1})\cdot z_0^{-1} \cdot h \cdot h' ) \ .
\end{align*}

Ce type de transformations de jauge sera illustré dans les exemples de la prochaine section.

\section{Exemples}

Dans cette section, nous allons appliquer les résultats généraux obtenus sur les connexions invariantes à deux exemples.
Le premier exemple sera une généralisation d'une situation largement étudiée dans le cas classique~\cite{forgacs:80,witten:77}, qui correspond à la situation de la symétrie sphérique pour une théorie de jauge avec groupe de structure $SU(2)$ (voir également~\cite{volkov:98} et ses références pour des exemples d'applications).

Le second exemple sera purement non commutatif et consistera à prendre en considération des symétries sur des algèbres de matrices.

\subsection{Symétrie sphérique}\label{sec:exemple:symetrie-spherique}
Nous allons considérer la situation où  $M=\gR \times \gR^3 \setminus \{0\}$. Le premier facteur est paramétré par la coordonnée de temps $t$ et le second facteur par les coordonnées d'espace $(x,y,z)=\vec{r}$.
Le groupe de symétrie est $G=SU(2)$ et agit sur $\gR^3\setminus\{0\}$ par les matrices de rotations%
\footnote{nous considérons l'espace $\gR^3\setminus\{0\}$ parce que nous voulons une action simple.}. %
Alors $G_0$ est isomorphe à  $U(1)$,  l'espace homogène $G/G_0$ est isomorphe à la 2-sphère $\gS^2$ et $M/G=\gR\times\gR^{+*}$.
Nous allons nous intéresser à des théories de jauge, dont le groupe de structure est $H=SU(2)$ et donc pour la partie non commutative nous avons $M_n= M_2(\gC)$.

Nous pouvons tout d'abord traiter cet exemple par l'approche globale~\ref{global_approach}. 
Notons que de manière générale, tout fibré principal de groupe de structure  $SU(2)$ sur $M$ est trivial du fait que $M=\gR \times\gR^{+*} \times \gS^2 $, où $\gR \times\gR^{+*}$ est un espace contractible et  $\dim( \gS^2)=2$. 
Alors, nous pouvons toujours identifier $E$, au fibré trivial:
\begin{align*}
  E &= M\times SU(2) = M/G \times \gS^2 \times SU(2) \ .
\end{align*}
Nous pouvons ainsi remonter l'action de $G$ sur la base $M$ en une action sur le fibré principal $E$. Pour cela nous devons spécifier une action de $G$ sur le groupe de structure $H$.
Nous pouvons étendre cette action de manière triviale, en considérant l'action $(g, h) \mapsto  h \ \forall g\in G, \forall h\in H$. 
Alors la théorie réduite est une théorie de jauge avec groupe de structure $SU(2)$ sur $M/G$ et la situation n'est donc guère intéressante du point de vue de l'action du groupe $G$.
Une action du groupe $G$ sur $H$, plus intéressante à étudier et naturelle dans notre situation car $G=H$, est la suivante:
\begin{align*}
G\times H &\longrightarrow H\\
(g, h) &\longmapsto g\cdot h \ .
\end{align*}
Avec cette action, on voit que l'application  $\lambda$ est un endomorphisme et que nous pouvons considérer le fibré réduit $Q$ de telle manière que $\lambda=\gone$.
Alors nous avons $Z_0= G_0=U(1)_Z$, où: 
\begin{align*}
U(1)_Z :=\{\exp{(2 \epsilon T_3)}, \epsilon \in \gR\} 
\end{align*}
et $\{T_1,T_2,T_3\}$ sont des générateurs antihermitiens de $\ksu(2)$ qui satisfont:
\begin{align*}
[T_1,T_2]&= T_3 & [T_2,T_3]&= T_1 & [T_3,T_1]&= T_2 & \ .
\end{align*}
Le fibré réduit $Q$ est isomorphe à  $ M/G \times \{N,S\} \times U(1)$, où $N$ et $S$ sont les pôles sud et nord de la  $2$-sphère $\gS^2$.
Sans perdre de généralité, nous pouvons restreindre le fibré $Q$ au point $N$, la relation de symétrie venant de l'action du groupe $\gZ_2$ correspondant à la conjugaison sur les nombres complexes.
Dans la présente situation, le diagramme~(\ref{diagram}) devient:
{
  \footnotesize
  \begin{align*}
    \xymatrix@R=7pt@C=3pt@M=4pt{
      U(1)_Z \ar@{^{(}->}[rrr] \ar@{^{(}->}[drr] \ar@{=}[dd] & & & \gZ_2 \ltimes U(1)_Z \ar@{>>}[rrr] \ar@{^{(}->}[drr] \ar'[d][dd] & & &\gZ_2 \ar@{^{(}->}[drr] \ar'[d][dd] & & \\  
      & & SU(2) \ar@{^{(}->}[rrr] \ar@{=}[dd] & & & \gS^2 \times SU(2) \ar@{>>}[rrr] \ar[dd] & & &\gS^2 \ar[dd] \\
      U(1)_Z \ar'[rr][rrr] \ar@{^{(}->}[drr] & & &M/G \times  \{N,S\}  \times U(1)_Z  \ar@{>>}'[rr][rrr]^-{\pi_Q} \ar@{^{(}->}[drr] \ar@{>>}'[d][dd] & & & \{N,S\}\times M/G \ar@{^{(}->}[drr] \ar@{>>}'[d][dd] & & \\
      & & SU(2) \ar[rrr] & & & M/G\times\gS^2\times SU(2) \ar@{>>}[rrr]^-{\pi} \ar@{>>}[dd] & & & M/G \times \gS^2 \ar@{>>}[dd] \\
      & & & M/G \ar@{=}'[rr][rrr] \ar@{=}[drr] & & & M/G \ar@{=}[drr] & & \\
      & & & & & M/G \ar@{=}[rrr] & & & M/G  } 
  \end{align*}
  }
Ici nous avons $\cL=\cM= \vect_{\gR}(T_1,T_2)$, $\cK=0$ et $\wo=\vect_{\gC}(\gone,T_3)$.
On peut alors facilement voir que $\cF\simeq \cL^{\gC}\oplus \cM^{\gC}$.
Finalement, une connexion non commutative est caractérisée par deux sections $\psi$ et $\zeta$ sur $M/G$ à valeurs dans $ \vect_{\gC}(T_1,T_2)$ et une $1$-forme non commutative $\mu -\eta \in (\Omega(M/G)\otimes \Omega(\wo))^1$. 
Ici $\Omega^1(\wo)$ est simplement $\wo$ du fait que $\zo$ est un espace vectoriel de dimension $1$.
 Si nous regardons seulement les connexions antihermitiennes, alors nous pouvons considérer seulement des espaces vectoriels sur $\gR$, et $\psi$ et $\zeta$ peuvent être interprétés comme des champs scalaires complexes (cela est dû au fait que $\cL=\cM \simeq \gC$). Dans ce cas, on a $\cC= C^{\infty}(M/G)\otimes \wo$ et $SU(\cC) =\{ e^{\chi T_3}=\cos{\frac{\chi}{2}} \, \gone +\sin{\frac{\chi}{2}} \, T_3 , \    \chi\in C^{\infty}(M/G) \}$.

Dans cet exemple particulier, l'approche ``locale'' de la section~\ref{local_approach} est bien adaptée à la structure de l'espace de base $M$. 
Alors, nous allons faire le reste de cette analyse en utilisant ces techniques. Le fibré principal $E$ peut être construit à partir d'un fibré principal de groupe de structure $U(1)$ et des classes de conjugaison $[\lambda]=[\gone]$.

Une connexion invariante est donnée explicitement par la formule (\ref{decomposition}).
Pour simplifier l'analyse, nous allons considérer seulement les connexions antihermitiennes sans trace%
\footnote{Les termes avec trace dans une connexion correspondent aux termes proportionnels à $\gone$ dans l'algèbre $M_n=\vect_{\gC}(\gone, \cH)$ et peuvent être étudiés indépendamment des termes sans trace.}.
En utilisant les notations et résultats de la section~\ref{local_approach}, nous sommes amenés à caractériser la décomposition de la représentation adjointe de $SU(2)$  en représentations irréductibles de $U(1)_Z$. 
La représentation adjointe de $SU(2)$ se décompose en la représentation fondamentale de $U(1)$ sur $\vect_{\gR}T_3$ et la représentation de dimension $2$ sur $\vect_{\gR}(T_1,T_2)$, correspondant à la représentation fondamentale de $SO(2)$. 
La propriété d'invariance~(\ref{equivariance}) donne:
\begin{align*}
  \Lambda(T_1) &=\Lambda_1 T_1 + \Lambda_2 T_2 &
  &&
  \phi(T_1) &=\phi_1 T_1 + \phi_2 T_2 \\
  \Lambda(T_2) &= -\Lambda_2 T_1 + \Lambda_1 T_2 &
  &&
  \phi(T_2) &= -\phi_2 T_1 + \phi_1 T_2
\end{align*}  
et en utilisant (\ref{equivariance}) et (\ref{horizontality}), on a:
 \begin{align*}
& \Lambda(T_3)=\phi(T_3)= \eta T_3 \ ,
\end{align*}
où $\eta$ est une fonction sur $M/G$.

\subsubsection{Choix de jauge (section locale)}
Maintenant, si nous voulons écrire l'expression locale~(\ref{local_form}) de la $1$-forme de connexion, nous devons faire le choix d'une section locale ou choix de jauge. Il y a deux jauges particulièrement intéressantes à considérer.
 
Une première jauge, que nous allons appeler  jauge ``singulière'', peut être définie en considérant la section locale constante: 
\begin{align*}
s_H : Q' &\to Q' \times H \\
q' &\mapsto (q',e) \ .
\end{align*}
Le deuxième choix de jauge naturel à considérer est le suivant:
\begin{align*}
  s_{G} : \gS^2 &\to SU(2) \\
  (\vartheta, \varphi) &\mapsto g=e^{\varphi T_3}e^{\vartheta T_2} \ ,
\end{align*}
où nous avons choisi $(\vartheta,\varphi)$ comme coordonnées locales sur $\gS^2$, qui sont les coordonnées sphériques habituelles.
Dans $SU(2)$,  $\vartheta$, $\varphi$ correspondent à deux des paramètres d'Euler.
Nous avons alors: 
  \begin{multline}
    S^* \omega = a T_3 + (\Lambda_1 T_1 + \Lambda_2 T_2) \d \vartheta + (\Lambda_1 T_2 - \Lambda_2 T_1) \sin\vartheta \d \varphi + \eta  T_3 \cos\vartheta \d \varphi \\
      - \left[  (\phi_1 T_1 +\phi_2 T_2) \theta^1  + (\phi_1 T_2- \phi_2 T_1) \theta^2 + \eta T_3\theta^3 \right]   \ ,
  \label{singular_form}
  \end{multline}
où $a=a_r \d r + a_t \d t \in \Omega^{1}(\gR\times \gR^{+*})$ et $i \theta= T_a \theta^a$.
Cette $1$-forme de connexion généralise ce qu'on appelle habituellement l'ansatz de Witten~\cite{witten:77} qui est retrouvé en prenant $\phi_1=\eta=1$ et $\phi_2=0$ (i.e. $\phi = \gone$). 
Le terme correspondant à la $1$-forme $\cos\vartheta \d \varphi$ est souvent appelé terme de monopole et on remarquera qu'il n'est plus constant dans notre approche, puisqu'il est factorisé par une fonction $\eta$. 
Les singularités apparaissant dans~(\ref{singular_form}) sont dues au fait que l'on essaye d'étendre le système de coordonnée sphérique de manière globale sur $\gS^2$.
Cette extension n'est pas possible dans cette jauge et c'est pour cela que nous l'appelons jauge ``singulière''.
Cependant, nous pouvons introduire une autre jauge dans laquelle l'extension de cette $1$-forme locale à une $1$-forme globale est possible.
Cette jauge est la jauge ``régulière'' ou encore ``radiale'', définie par la section suivante:
\begin{align}
S : M &\longrightarrow \tE \\
(r,\vartheta,\varphi) &\longmapsto \Psi (s_{\tQ}(r), e^{\varphi T_3}e^{\vartheta T_2},e^{-\vartheta T_2}e^{-\varphi T_3}) \ .
\label{section0}
\end{align}
Elle peut être obtenue à partir de la jauge ``singulière'' en faisant une transformation de jauge passive qui consiste à multiplier la section $s_H$ par l'élément $h'= e^{-\vartheta T_2}e^{-\varphi T_3} \in H$. 
Alors, en appliquant la formule~(\ref{local_form}), la $1$-forme locale s'exprime de la manière suivante:
\begin{equation}
  \begin{split}
{S}^* \omega = a T_r 
   & + \Lambda_1 [T_r, \d T_r] - \Lambda_2 \d T_r \\
    &  -\phi_1 [T_r, \hd T_r] +\phi_2 \hd T_r  - \eta T_r \theta^r  \ ,
  \end{split}\label{radial_gauge}
\end{equation}
où 
\begin{align*}
  T_r&= \sin\vartheta \cos\varphi \ T_1 + \sin\vartheta \sin\varphi \ T_2+  \cos\vartheta \ T_3 \\
  \theta^r &= \sin\vartheta \cos\varphi \ \theta^1 + \sin\vartheta \sin\varphi \ \theta^2+ \cos\vartheta \ \theta^3
\end{align*}
et $\hd= \d+ \d'$ est la différentielle non commutative introduite dans la section~\ref{sec:alg-des-endom}. 
L'absence de singularité s'explique simplement par le fait que les angles $(\vartheta, \varphi)$ des coordonnées sphériques n'apparaissent pas explicitement dans l'expression (tout s'exprime en fonction du générateur $T_r$).
Pour illustrer le fait que nous avons obtenu en fait une $1$-forme globale, nous pouvons donner quelques formules en coordonnées euclidiennes.
Introduisons la notation $S^*\omega= a + A^a_i T_a \d x^i -  \phi^a_b T_a \theta^b $. 
Alors la formule~(\ref{radial_gauge}) nous donne :
\begin{align*}
  \begin{aligned}
    A_i^a&=\frac{\Re(\psi -i \phi)}{r} \ P^a_i \ + \ \frac{\Im(\psi -i \phi)}{r} \ g^{ab}\epsilon_{ibc} \hn^c\\
    \phi_b^a&= \Re(\phi) \ P^a_b \ +\ \Im(\phi) \ g^{ac}\epsilon_{bcd}\hn^d \ 
    + \ \eta\ \hn^a \hn_b 
\end{aligned} &&&,
\end{align*}
avec $\hn^a=\frac{x^a}{r}$, $P^a_b=\delta^a_i-\hn^a\hn_i$, $g^{ab}$ la métrique euclidienne et $\epsilon_{abc}$ le tenseur totalement antisymétrique tel que $\epsilon_{123}=1$. 
Nous avons introduit les notations pratiques $\psi=-\Lambda_2+ i \Lambda_1$ et $\phi=\phi_1 + i \phi_2$.

Finalement, nous voudrions montrer que les deux autres transformations de jauge passives (sur $s_G$ et $s_{\tQ}$ mentionnées à la fin de la section~\ref{local_approach}) peuvent être effectuées. 
Elles vont en fait correspondre à des symétries $U(1)$ résiduelles.
En terme des deux champs scalaires complexes $\psi$ et $\phi$, l'équation~(\ref{radial_gauge}) devient:
  \begin{multline*}
    S^* \omega = a T_r 
    +  \Re(\psi -i \phi) \d T_r  + \Im(\psi -i \phi) [T_r, \d T_r] \\
    + \Re(-i \phi) \d' T_r + \Im(-i \phi) [T_r,\d' T_r] - \eta T_r \theta^r \ .
  \end{multline*}
Les transformations de jauge passives ``$U(1)$'' correspondent aux transformations sur $s_G$ et $s_{\tQ}$ données par:
\begin{align*}
  \begin{aligned}
    s_G &\leadsto s_G\cdot e^{\chi_0 T_3}\\
    s_{\tQ} &\leadsto s_{\tQ}\cdot e^{\chi_1 T_3}
\end{aligned}&&&,
\end{align*}
où $\chi_0(r,t)$ et $\chi_1(r,t)$ sont deux fonctions arbitraires de  $r$ et $t$, ce qui correspond aux transformations suivantes sur les champs $\psi$ et $a$ : 
\begin{align}
\begin{aligned}
\psi &\leadsto e^{i(\chi_1+\chi_0)} \psi \\
a &\leadsto a - \eta \d (\chi_1+\chi_0)
\end{aligned} &&& .
\label{gauge-transformation}
\end{align}
Nous pouvons remarquer que les champs scalaires $\phi$ et $\eta$ restent inchangés sous ces transformations de jauge passives.
Nous pouvons également remarquer la similarité avec les transformations de jauge usuelles dans le cas abélien.

\subsubsection{Vraies transformations de jauge}
Il est également possible de regarder à quoi correspondent les transformations de jauge du point de vue de la géométrie non commutative (voir section~\ref{sec:alg-des-endom}). 
Commençons par redéfinir le champs $\phi$ en posant $\phi = 1- \phi'$. 
Alors, une transformation de jauge symétrique, paramétrée par un élément $e^{\chi T_3} \in SU(\cC)$, où $\chi$ est une fonction sur $M/G= \gR\times\gR^{+*}$, nous amène aux transformations suivantes:
\begin{align}
\begin{aligned}
\phi'  &\leadsto e^{-i\chi} \phi' \\
\psi &\leadsto e^{-i \chi} \psi \\
a &\leadsto a + \d \chi 
\end{aligned}&&& .
\end{align}
Notons que ces transformations ressemblent beaucoup plus à des transformations de jauge ordinaires.

\subsection{Un exemple purement non commutatif}
\label{sec:exemple-nc} 
Nous présentons dans cette section un exemple purement non commutatif, dans le sens où l'espace de base $M$ est réduit à un point.
L'algèbre des endomorphismes $\kA$ est alors tout simplement l'algèbre des matrices $M_{n}$ qui fut traitée dans la section~\ref{sec:alg-des-endom}.
Alors quelque soit le groupe $G$ agissant sur cette algèbre, nous avons $G=G_0$ et $\lambda $ est un homomorphisme de $G$ dans $H$.
Le problème se réduit ainsi à caractériser les connexions non commutatives $G$-invariantes sur le module $M_n$.
Pour simplifier l'analyse, nous nous restreindrons comme précédemment aux connexions sans trace.

La procédure générale à suivre est de tout d'abord étudier la représentation $\lambda$ de $G$ dans $M_n$ et ensuite de déterminer comment la représentation  $Ad^{H}\circ\lambda$ se décompose en représentations irréductibles de $G$. 
Ces représentations irréductibles correspondent aux différents degrés de liberté d'une connexion non commutative invariante, c'est-à-dire un champ scalaire pour chaque opérateur d'entrelacement entre représentations équivalentes.

Enfin, afin d'illustrer ce résultat, prenons le cas particulier où $G=SU(2)$.
Il est bien connu que les représentations de $SU(2)$ dans  $M_n$ sont paramétrées par les partitions de $n$ (voir~\cite{dubois-violette:japan99} par exemple).
Par exemple, pour $\kA=M_3(\gC)$, les représentations sont indexées par les partitions ``1+1+1'', ``2+1'' et ``3''  de $3$. 
Elles sont décrites de la manière suivante:
\begin{itemize}
\item la représentation ``1+1+1'' correspond à la somme de $3$ copies de la représentation triviale.
  La représentation $Ad^{H} \circ \lambda$ est décomposée en la somme de $8$ copies de la représentation triviale de  $SU(2)$. 
  Cela donne donc lieu à  64 champs scalaires.
  Ce cas n'est pas très intéressant du fait que le groupe $SU(2)$ n'agit pas sur $M_3(\gC)$.
\item la représentation ``2+1''  correspond à la représentation réductible de $SU(2)$ qui est la somme de la représentation fondamentale et la représentation triviale.
  La représentation $Ad^{H} \circ \lambda$ est décomposée en représentations irréductibles de $SU(2)$ de dimensions $3$, $2$, $2$ et $1$ ($6$ champs scalaires).
\item la représentation ``3'' correspond à la représentation de dimension $3$ de $SU(2)$. 
  Dans ce cas, la représentation $Ad^{H} \circ \lambda$ se décompose en représentations irréductibles de $SU(2)$ de dimensions $3$ et $5$ (2 champs scalaires).
\end{itemize}


\chapter{Modèles physiques}\label{cha:modeles-physiques}
Dans ce chapitre, nous allons voir différents modèles construits dans le cadre de la géométrie non commutative des algèbres d'endomorphismes.
Ce seront essentiellement des modèles de théorie classique des champs généralisant les théories de jauge.
Nous savons que les champs de Yang-Mills ordinaires peuvent être décrits au moyen de connexions sur des fibrés.
La théorie des connexions sur des algèbres non commutatives et le calcul différentiel non commutatif ont été introduits dans le chapitre~\ref{cha:theories-de-jauge} et permettent de généraliser ce type de modèles.
Afin de pouvoir garder une interprétation en terme de théorie des champs ordinaire, les algèbres que l'on considère sont généralement fortement reliées aux algèbres de fonctions sur l'espace-temps.
Comme nous l'avons vu, les algèbres d'endomorphismes peuvent se substituer complètement aux fibrés principaux lorsque l'on étudie les théories de jauge ordinaires.
Le fait de travailler dans ce cadre va nous permettre  d'obtenir des généralisations des modèles ordinaires de manière très économe et de fournir une interprétation ``géométrique'' des champs de Higgs.

Dans un premier temps, nous allons considérer des algèbres d'endomorphismes correspondant à des fibrés triviaux et nous verrons comment interpréter les champs de Higgs comme partie d'une connexion dans les ``directions non commutatives''.
Dans un second temps, nous étudierons ces modèles dans le cas d'algèbres d'endomorphismes non triviales.
Nous verrons qu'il est alors possible de définir la notion de structure Riemannienne sur une algèbre d'endomorphismes et que la notion de connexion Riemannienne dans ce cadre permet de généraliser la théorie d'Einstein de la gravitation. 
Cette généralisation reproduit un mécanisme similaire à celui rencontré dans les théories usuelles de type Kaluza-Klein.
Enfin, nous décrirons l'action de ``Maxwell'' non commutative pouvant être construite à partir d'une telle structure Riemannienne.
On verra que ce modèle correspond du point de vue de la géométrie ordinaire à un modèle de Yang-Mills-Higgs sur un espace courbe où le mécanisme de Higgs usuel est couplé de manière intéressante à la "géométrie'' de l'algèbre d'endomorphismes.

\section{Modèles de Yang-Mills-Higgs et géométrie non commutative}\label{sec:maxw-non-comm}

Ces types de modèles non commutatifs furent les premiers à utiliser la géométrie non commutative pour tenter de comprendre la structure du Lagrangien du modèle standard et furent étudiés dans une série de papiers~\cite{dubois-violette:89,dubois-violette:89:II,dubois-violette:90,dubois-violette:90:II,dubois-violette:91}.
D'autres approches furent également proposées dans \cite{coqu:90,coquereaux:93} ou encore dans \cite{conn-lott:90} puis développées dans \cite{cham-conn:96,connes:96} afin de reproduire le Lagrangien classique du modèle standard et l'action de Einstein-Hilbert au sein d'une écriture extrêmement compacte.
L'idée générale sous-jacente à tous ces modèles est de se donner une algèbre $\cA$ et un calcul différentiel $\Omega$ sur cette algèbre, ainsi qu'un $\cA$-module muni d'une $\Omega$-connexion.

Dans cette section, nous allons voir des exemples d'algèbres $\cA$ de la forme $\cA=C^\infty(\gR^{s+1})\otimes\cA_0$, avec $\cA_{0}$ une algèbre, et nous prendrons comme calcul différentiel celui basé sur les dérivations $\Omega=\Oder(\cA)$ présenté dans la section~\ref{sec:calc-diff-der}.
Les modèles construits sur ce type d'algèbres ont alors une interprétation en terme de théorie de Yang-Mills couplée à des champs de Higgs qui sont la partie d'une connexion dans les ``directions non commutatives''.
Nous allons reprendre la présentation qui en est faite dans \cite{dubois-violette:japan99}.

\subsubsection{Algèbres matricielles}
Nous pouvons tout d'abord faire une description de la situation pour une algèbre matricielle $\cA=M_n(\gC)$. 
Toutes les dérivations de  $M_n(\gC)$ sont intérieures, grâce à quoi l'algèbre de Lie $\der(M_n(\gC))$ est isomorphe à $\ksl_{n}(\gC)$.

Comme il a déjà été mentionné dans la section~\ref{sec:alg-des-endom}, on a:
\begin{align}
  \Omega_\der(M_n(\gC)) &\simeq M_n \otimes \exter \ksl_n^\ast \ .
\end{align}

Il est utile d'introduire une base de matrices $E_k, k\in \{1,2,\dots,n^2-1\}$  de taille $n \times n$,  hermitiennes et sans trace.
Les dérivations intérieures $\partial_k=\ad(iE_k)$ forment une base des dérivations réelles.
On a donc $\der_\gR(M_n(\gC))\simeq \ksu(n)$ et $[\partial_k,\partial_\ell]=C^m_{k\ell}\partial_m$, où $C^m_{k\ell}$ sont les constantes de structures de $\ksu(n)$ (ou $\ksl(n)$). 

On définit les éléments $\theta^k \in \Oder^1(M_n(\gC))$ par la relation $\theta^k(\partial_\ell)=\delta^k_\ell \gone$. Ces éléments engendrent l'algèbre différentielle graduée $\Oder(M_{n})$ qui admet la présentation suivante~\cite{dubois-violette:90}~\cite{dubois-violette:rapallo:90}: 
\begin{align*}
\left\{\begin{aligned}
&  E_kE_\ell = g_{k\ell}\bbbone+(S^m_{k\ell}-\frac{i}{2} C^m_{k\ell})E_m \\
&  E_k\theta^\ell = \theta^\ell E_k \\
&  \theta^k \theta^\ell = -\theta^\ell \theta^k\\
&  dE_k = -C^m_{k\ell}E_m\theta^\ell\\ 
&  d\theta^k = -\frac{1}{2} C^k_{\ell m}\theta^\ell \theta^m
\end{aligned}\right. &&&,
\label{eq:presentation}
\end{align*}
où $g_{k\ell}=g_{\ell k}$,$\;$ $S^m_{k\ell}=S^m_{\ell k}$ sont réels, $g_{k\ell}$ sont les composantes de la métrique de Killing de $\ksu(n)$ et $C^m_{k\ell}=- C^m_{\ell k}$ sont les constantes de structures (réelles) de $\ksu(n)$.
La formule donnant $dE_k$ peut être inversée et l'on a:
\begin{equation}
\theta^k=-\frac{i}{n^2} g^{\ell m}g^{kr} E_\ell E_r dE_m \ ,
\end{equation}
où $g^{k\ell}$ sont les composantes de la matrice inverse de $(g_{k\ell})$.

L'algèbre différentielle graduée $\Oder(M_{n})$ admet une $1$-forme canonique $\theta$  définie par:
\begin{align*}
  i\theta : \der(M_n) & \longrightarrow \ksl_n\\
  \ad_{\gamma} & \longmapsto \gamma - \frac{1}{n} \tr (\gamma)\gone \ ,
\end{align*}
pour tout $\gamma \in M_n$. 
Elle satisfait la relation:
\begin{align*}
 \d' i\theta - (i\theta)^2 = 0 \ .
\end{align*}
Elle sert également à exprimer la différentielle non commutative sur les éléments de l'algèbre $M_{n}$:
\begin{align*}
\d' \gamma &= [i\theta, \gamma] &
& \forall \gamma \in M_n = \Omega^0_\der(M_n) \ .
\end{align*}
En composante, on a $\theta=E_k\theta^k$ ou encore $i\theta= T_{k}\theta^{k}$, où $T_{k}=i E_{k}$ est une base de générateurs antihermitiens de $\ksl(n)$.
Notons également que c'est une forme réelle, \textit{i.e.} $\theta=\theta^\ast$, et indépendante du choix de la base $(E_k)$.

\begin{rem}
  La forme $\omega=d\theta$ permet de définir une structure symplectique sur $M_n(\gC)$ (voir~\cite{dubois-violette:90}).
De plus $\theta$ est invariante, \textit{i.e.} $L_X \theta=0$, et tout élément invariant de $\Oder^1(M_n(\gC))$ est un multiple de $\theta$. Ainsi,  $\theta$ est un \textbf{élément canonique invariant} de $\Oder^1(M_n(\gC))$. 
\end{rem}

\subsubsection{Connexions sur $M_{n}$}

La $\ast$-algèbre $M_n(\gC)$ est une algèbre simple avec seulement une représentation irréductible dans $\gC^n$. 
Un module (à droite) projectif de type fini sur $M_{n}(\gC)$ est toujours de la forme $M_{K,n}(\gC)$, l'ensemble des matrices de taille $K \times n$ sur lesquelles  $M_n(\gC)$ agit à droite.
Alors le groupe des automorphismes sur un tel module, $\Aut(M_{K,n}(\gC))$, est le groupe $GL(K)$ agissant par multiplication à gauche.
Le module $M_{K,n}(\gC)$ est naturellement doté d'une structure hermitienne donnée par :
\begin{align*}
  h(\Phi,\Psi)=\Phi^\ast \Psi \ ,
\end{align*}
où $\Phi^\ast$ est la matrice hermitique conjuguée de $\Phi$. 
Ainsi, le groupe de jauge est le groupe $U(K)$ des matrices de taille $K$ unitaires.

Le module $M_{K,n}(\gC)$ admet une connexion canonique $\Onabla$  donnée par:
\begin{align}
  \Onabla\Phi&=- i\Phi \theta &
  \text{où } & \Phi \in M_{K,n}(\gC) \ . \label{eq:Onabla}
\end{align}
Le fait que cela définisse une connexion vient de la relation:
\begin{equation}
\Onabla(\Phi M) = (\Onabla \Phi)M +\Phi [i\theta,M]
\end{equation}
et de l'expression de $d'M$ pour $M\in M_n(\gC)$. 
Cette connexion est hermitienne et sa courbure est nulle.

Nous pouvons alors écrire toute connexion $\nabla$ sous la forme $\nabla \Phi= \Onabla \Phi + A \Phi $ où $A=A_k \theta^k$ avec $A_k \in M_K(\gC)$ et $A\Phi := A_k \Phi \otimes \theta^k$.
La connexion $\nabla$ est hermitienne \ssi les matrices $A_k$ sont antihermitiennes, c'est-à-dire $A_k^\ast=-A_k$.
La courbure de $\nabla$ est donnée par $\nabla^2 \Phi=F\Phi  = F_{k\ell} \Phi \otimes \theta^k \theta^\ell$ avec 
\begin{equation}
F =\frac{1}{2}([A_k, A_\ell] - C^m_{k\ell} A_m) \theta^k \theta^\ell.
\end{equation}

Alors $\nabla^2=0$ \ssi les matrices $A_k$ forment une représentation de l'algèbre de Lie $\ksl(n)$ dans $\gC^K$.
De plus, deux telles connexions sont dans la même orbite vis-à-vis de l'action du groupe $\Aut(M_{K,n}(\gC))=GL(K)$ \ssi les représentations correspondantes de $\ksl(n)$ sont équivalentes.
Cela implique~\cite{dubois-violette:90} que l'\textit{ensemble des orbites des connexions plates hermitiennes} $(\nabla ^2=0)$ peut être mis en bijection avec l'\textit{ensemble des classes de représentations de} $\ksu(n)$ \textit{sur} $\gC^{K}$ \textit{unitairement équivalentes}.
A titre d'exemple, pour $n=2$, ces orbites sont classées par le nombre de partitions de l'entier $K$, \textit{i.e.} ${\rm card}\{(n_r) \vert \sum_r \ \ n_r.r=K\}$.
On peut remarquer que l'on retrouve ici une relation analogue à celle trouvée dans le deuxième exemple dans la section~\ref{sec:exemple-nc}.
\subsubsection{Fonctions à valeurs matricielles}

Nous pouvons maintenant revenir au cas $\cA=C^\infty(\gR^{s+1})\otimes M_n(\gC)$. 
Soit $x^\mu, \; \mu \in \{0,1,\dots,s\}$, les coordonnées canoniques de $\gR^{s+1}$.
On a  :
\begin{equation}
  \Omega_\der(C^\infty(\gR^{s+1}) \otimes M_n(\gC))=\Omega_\der(C^\infty(\gR^{s+1}))\otimes \Omega_\der(M_n(\gC)) \ .
\end{equation}
Ainsi, la différentielle $\hd$ peut s'écrire comme la somme de deux différentielles $\hd=d+d'$, où $d$ est la différentielle le long de $\gR^{s+1}$ et $d'$ est la différentielle de $\Omega_\der(M_n(\gC))$. 

Nous allons considérer des modules projectifs de type fini de la forme $C^\infty(\gR^{s+1}) \otimes M_{K,n}(\gC)$.
Un tel module est doté d'une structure hermitienne naturelle :
\begin{align*}
  h(\Phi,\Psi)(x)&=\Phi(x)^\ast \Psi(x) &
  &  \forall x\in \gR^{s+1} \ .
\end{align*}

En tant que $C^\infty(\gR^{s+1})$-module, c'est un module libre et cela nous permet de définir, pour tout élément  $\Phi \in C^\infty(\gR^{s+1}) \otimes M_{K,n}(\gC)$, la différentielle $d$ de la manière suivante:
\begin{align*}
  d\Phi(x)=\frac{\partial \Phi}{\partial x^\mu}(x) dx^\mu \ .
\end{align*}

Une connexion sur le $C^\infty(\gR^{s+1}) \otimes M_n(\gC)$--module $C^\infty(\gR^{s+1}) \otimes M_{K,n}(\gC)$ est de la forme
\begin{equation}
\nabla \Phi =d'\Phi -i\Phi \theta + A\Phi
\end{equation}
avec $A=A_\mu dx^\mu + A_k\theta^k$, où $A_\mu$ et $A_k$ sont des fonctions sur $\gR^{s+1}$ à valeurs  dans les matrices de taille $K$ (\textit{i.e.} des éléments de $C^\infty (\gR^{s+1}) \otimes M_K(\gC))$ et où: 
\begin{equation}
A\Phi(x)=A_\mu(x)\Phi(x) dx^\mu +A_k(x)\Phi(x)\theta^k \ .
\end{equation}
Une telle connexion est hermitienne \ssi les matrices $A_\mu(x)$ et $A_k(x)$ sont antihermi\-tiennes, $\forall x \in \gR^{s+1}$.
La courbure de $\nabla$ est donnée par $\nabla^2\Phi=F\Phi$ où
\begin{multline}
  F=\frac{1}{2}(\partial_\mu A_\nu-\partial_\nu A_\mu  +[A_\mu,A_\nu])dx^\mu dx^\nu +\\
  +(\partial_\mu A_k + [A_\mu ,A_k])dx^\mu \theta^k
  +\frac{1}{2}([A_k,A_\ell]-C^m_{k\ell}A_m)\theta^k \theta^\ell \ .
\end{multline}

Une connexion $\nabla$ est plate (\textit{i.e.} $\nabla^2=0$) \ssi chaque terme de la formule précédente s'annule et est donc équivalente de jauge à une connexion pour laquelle:
\begin{align*}
  &  A_\mu=0, &
&  \partial_\mu A_k=0 &
& \text{et}&
&  [A_k,A_\ell]=C^m_{k\ell}A \  .
\end{align*}

Deux connexions plates sont équivalentes de jauge \ssi les représentations correspondantes de $\ksu(n)$ dans $\gC^K$ (données par les matrices $A_\ell$) sont équivalentes.
Ainsi, à nouveau, l'\textit{ensemble des orbites des connexions plates hermitiennes} $(\nabla ^2=0)$ peut être mis en bijection avec l'\textit{ensemble des classes de représentations de} $\ksu(n)$ \textit{sur} $\gC^{K}$ \textit{unitairement équivalentes}.

\subsubsection{Action de Yang-Mills}
Si l'on considère $\gR^{s+1}$ comme l'espace-temps de dimension ($s+1$), alors l'algèbre $C^\infty(\gR^{s+1}) \otimes M_n(\gC)$ peut être interprétée comme l'algèbre des ``fonctions dérivables sur un espace temps non commutatif'' et il est naturel de considérer l'action généralisée de Yang-Mills (sur un espace-temps euclidien):
\begin{multline}
  \Vert F\Vert^2 = \int d^{s+1}x \, {\rm tr}\Bigl\lbrace \frac{1}{4} \sum (\partial_\mu A_\nu-\partial_\nu A_\mu +[A_\mu,A_\nu])^2\\
  +\frac{1}{2} \sum(\partial_\mu A_k + [A_\mu,A_k])^2 + \frac{1}{4}\sum  ([A_k,A_\ell]-C^m_{k\ell} A_m)^2\Bigl\rbrace
\label{eq:Yang-Mills-nc}
\end{multline}
 où $g_{\mu \nu}=\delta_{\mu \nu}$ est la métrique sur l'espace-temps et la base $E_k$ des matrices hermitiennes $n \times n$ est choisie de telle sorte que $g_{k\ell}=\delta_{k\ell}$, \textit{i.e.} ${\tr}(E_kE_\ell)=n\delta_{k\ell}$. 
 
On peut introduire une involution de Hodge sur $\Omega_\der(M_n(\gC))$ et l'analogue de l'intégration sur les éléments de $\Omega^{n^2-1}_\der(M_n(\gC))$ (essentiellement la trace). 
En combinant ces notions avec les notions équivalentes sur $\gR^{s+1}$, on obtient un produit scalaire sur $\Omega_\der(C^\infty(\gR^{s+1}) \otimes M_n(\gC))$ (Voir~\cite{dubois-violette:90,dubois-violette:89} pour plus de détails ainsi que dans la section \ref{sec:struct-riem}).

Cette action de  Yang-Mills non commutative correspond à l'algèbre $C^\infty(\gR^{s+1}) \otimes M_n(\gC)$ et peut être interprétée comme l'action d'une théorie des champs sur un espace-temps de dimension $(s+1)$. 
Ainsi, cette théorie des champs correspond à une théorie de Yang-Mills $U(K)$ avec potentiel de jauge $A_\mu(x)$ couplée de manière minimale avec des champs scalaires $A_k(x)$ à valeurs dans la représentation adjointe. 
Ces champs scalaires interagissent entre eux par un potentiel quartique.

Cette action est positive et s'annule pour les champs de jauge étant sur la même orbite que le potentiel de jauge $A_\mu=0$ et $A_k=0$.
En effet, la condition $\Vert F \Vert^2=0$ est équivalente à $F=0$ et donc les orbites de jauge sur lesquelles cette action s'annule sont classées par les classes de représentations de $\ksu(n)$ dans $\gC^K$ unitairement équivalentes.
Ces différentes orbites peuvent s'interpréter comme les différents vides pour la théorie quantique des champs construite sur ce modèle.
Afin de pouvoir faire un développement perturbatif de cette théorie quantique des champs, nous devons choisir un vide et développer les champs autour de ce vide.
Cela correspond à faire une translation dans l'espace des champs.
Ainsi, les variables $A_\mu, A_k$  sont bien adaptées pour travailler autour du vide spécifié par la représentation triviale de $\ksu(n)$,  $A_k=0$. Mais si l'on choisit un vide spécifié par une représentation $R_k$ de $\ksu(n)$, (\textit{i.e.} $[ R_k, R_\ell]= C^m_{k\ell} R_n$), on doit alors utiliser les variables $A_\mu$ et $ B_k=A_k- R_k$ ($R_{k}$ est identifié à une fonction constante à valeurs matricielles).
Si l'on fait ce changement de variables dans l'action, on observe que les composantes $A_\mu$ deviennent massives et que les champs scalaires $B_k$ ont des masses différentes.
Ainsi, le spectre de masse de la théorie dépend du choix du vide indexé par $R_{k}$. 
Ce mécanisme est tout à fait analogue au mécanisme de Higgs, à ceci près que l'invariance de jauge n'est pas brisée dans le présent modèle.
En effet,  le terme de masse du champ $A_\mu$ n'est pas invariant de jauge mais cela est compensé par le fait que les transformations de jauge sur les champs scalaires $B_k$ sont inhomogènes.
 Ainsi, du point de vue de l'interprétation en théorie des champs, les champs $A_{k}$ peuvent être interprétés comme des champs de Higgs.

Ces modèles furent les premiers à proposer une action classique de type Yang-Mills-Higgs et ils admettent des variantes et des généralisations.
En effet, nous avons choisi de prendre le calcul différentiel basé sur les dérivations, ce qui rend le modèle assez rigide.
Il est possible d'utiliser d'autres types de calculs différentiels $\Omega$ afin d'obtenir d'autres modèles de théories de jauge non commutatives dont l'interprétation en terme de théorie des champs est beaucoup plus proche du modèle standard~\cite{connes:90,conn-lott:90,coqu:90}.
Il existe de plus une manière élégante de combiner la notion de spineur avec la notion de calcul différentiel et de métrique~\cite{Conn:94}.
On peut également donner une notion de réalité~\cite{connes:95} en géométrie non commutative, ainsi qu'une notion de principe de moindre action~\cite{cham-conn:96}.

Le problème majeur de ce type de théories reste la quantification.
En effet, lorsque que l'on quantifie une théorie des champs de type Yang-Mills-Higgs, la structure non commutative est perdue.
Comme il est suggéré dans~\cite{dubois-violette:japan99}, il pourrait être utile d'avoir une symétrie de type B.R.S.~\cite{becc-roue-stor:75} basée sur la géométrie non commutative qui garantirait une interprétation en terme de géométrie non commutative au niveau quantique.
Malheureusement, aucune symétrie de ce type n'a été trouvée à ce jour.

\section{Théories de jauge pour les algèbres d'endomorphismes}
\label{sec:theories-de-jauge-endo}
Dans cette section, nous allons généraliser la construction précédente et construire l'analogue d'une théorie Yang-Mills-Higgs sur un fibré non trivial à partir d'une algèbre d'endomorphismes.
Pour ne pas trop alourdir les notations, lorsque nous étudierons les connexions non commutatives sur les modules de $\kA$, nous considérerons le module libre $\kA$.
Nous allons tout d'abord rappeler quelques notions essentielles à la compréhension de la structure de l'algèbre des endomorphismes qui sont  la décomposition des dérivations et des formes non commutatives et leurs liens avec les notions relatives à la géométrie de la  variété de base $M$.
Cela nous permettra d'étudier les structures Riemanniennes sur l'algèbre des endomorphismes et de voir en quoi ces structures sont similaires aux structures Riemanniennes dans les théories de type Kaluza-Klein.
Nous verrons enfin comment construire l'action de ``Maxwell'' non commutative généralisant l'action (\ref{eq:Yang-Mills-nc}) dans le cas où le module est l'algèbre $\kA$ elle-même, c'est-à-dire un module libre de rang $1$ (d'où le nom de théorie de ``Maxwell'').

\subsection{Décomposition des degrés de liberté}
\label{sec:decomp-deg-lib}

\subsubsection{Dérivations}
Rappelons  que nous avons la suite exacte courte suivante au niveau des dérivations de l'algèbre:
\begin{align*}
  \begin{aligned}
    \xymatrix@R=2pt{
      \kA_{0} \ar@{^(->}^-{ad}[r] & \der(\kA) \ar@{->>}^-{\rho}[r] & \Gamma(TM)\\
      \gamma \ar@{|->}[r] & \ad_{\gamma}   & \\
      & \cX \ar@{|->}[r] & \cX_{|Z(\kA)} }
\end{aligned}
&&&,
\end{align*}
où $ad$ est l'application envoyant un élément $\gamma \in \kA_{0}$ vers $\ad_{\gamma} \in \der(\kA)$ et réalise un isomorphisme entre $\kA_{0}$ et $\Int(\kA)$.

Nous avons vu précédemment qu'une connexion $\nabla$ sur $\End(\cE)$ permet de scinder cette suite de la manière suivante:
\begin{align*}
  \begin{aligned}
    \xymatrix@R=2pt{
      \kA_{0} \ar@{^(->}^-{ad}[r] & \der(\kA) \ar@{->>}^-{\rho}[r] & \Gamma(TM)\\
      -\alpha(\cX) & \cX \ar@{|->}_-{-\alpha}[l]   & \\
      & \nabla_{X} & X \ar@{|->}_-{\nabla}[l] }
\end{aligned}&&&,
\end{align*}
où  $\alpha$ est la $1$-forme non commutative associée à $\nabla$.

Ainsi, toute dérivation $\cX \in \der(\kA)$ peut se décomposer de la manière suivante: 
\begin{align*}
\cX &= \nabla_{X} + \ad_{\gamma}&
&\mathrm{avec}&
&X=\rho(\cX)&
& \mathrm{et} &
&\gamma=-\alpha(\cX) \ .
\end{align*}
 
\subsubsection{Formes}

Il existe également une suite exacte courte naturelle au niveau des $1$-formes non commutatives:
\begin{align*}
  \begin{aligned}
    \xymatrix@R=2pt{
      \Omega^{1}(M,\End(\cE)) \ar@{^(->}^-{\rho^{*}}[r] & \Oder^{1}(\kA) \ar@{->>}^-{ad^{*}}[r] & \Omega_{\Int}^{1}(\kA)\\
      \omega^{M} \ar@{|->}[r] & \omega^{M}\circ \rho   & \\
      & \omega \ar@{|->}[r] & \omega\circ ad }
\end{aligned}&&&,
\end{align*}
où $\Omega_{\Int}^{1}(\kA)\simeq  \kA\otimes_{Z(\kA)} \kA_{0}^{*}$ (nous appellerons formes intérieures ses éléments).

Comme précédemment, une connexion $\nabla$ sur $\End(\cE)$ permet de scinder cette suite de la manière suivante:
\begin{align*}
  \begin{aligned}
    \xymatrix@R=2pt{
      \Omega^{1}(M,\End(\cE)) \ar@{^(->}^-{\rho^{*}}[r] & \Oder^{1}(\kA) \ar@{->>}^-{ad^{*}}[r] & \Omega_{\Int}^{1}(\kA)\\
      \omega\circ\nabla & \omega \ar@{|->}_-{\nabla^{*}}[l]   & \\
      & - \mu\circ \alpha & \mu \ar@{|->}_-{-\alpha^{*}}[l] }
\end{aligned}&&& ,
\end{align*}
où  $\alpha$ est la $1$-forme non commutative associée à $\nabla$.

Ainsi, toute forme $\omega \in \Oder^{1}(\kA)$ peut se décomposer de la manière suivante: 
\begin{align}
\omega &= \rho^{*}\omega^{M} - \alpha^{*}\omega_{\Int}&
&\mathrm{avec}&
&\omega_{\Int}=ad^{*}\omega&
& \mathrm{et} &
&\omega^{M}=\nabla^{*}\omega \ .
\label{eq:decomp-formes}
\end{align}
En effet, on a pour toute dérivation $\cX\in \der(\kA)$:
\begin{align*}
  \omega(\cX)&= \omega(\nabla_{\rho(\cX)}-\ad_{\alpha(\cX)}) \\
  & =\omega^{M}\circ\rho(\cX) -\omega_{\Int}(\alpha(\cX)) \\
  &= \rho^{*}\omega^{M}(\cX) - \alpha^{*}\omega_{\Int}(\cX) \ .
\end{align*}

Pour les formes de degré plus élevé, nous pouvons faire une décomposition analogue.
Nous avons vu que l'algèbre différentielle graduée $\Oder(\kA)$ est engendrée par $\kA$.
Nous pouvons ainsi toujours exprimer une forme de degré $n\geq 1$ comme une combinaison linéaire de produit de $1$-formes se décomposant en une partie tensorielle et une partie intérieure.

Nous allons exprimer cette décomposition de manière locale.
Tout d'abord, rappelons que localement la forme $\alpha$ associée à une connexion $\nabla$ peut se décomposer de la manière suivante:
\begin{align*}
  \alpha\loc &= A -i\theta
  = (A^{r} -i \theta^{r}) E_{r} \ ,
\end{align*}
où les matrices $E_{a}$ forment une base des matrices hermitiennes de $\ksl(n)$.
Alors toute forme non commutative $\omega$ peut se décomposer localement de la manière suivante:
\begin{align*}
  \omega\loc &= \sum_{p,q}\omega_{\mu_{1}\cdots \mu_{p},r_{1}\cdots r_{q}} dx^{\mu_{1}}\wedge \cdots \wedge dx^{\mu_{p}}\alpha^{r_{1}}\cdots \alpha^{r_{q}} \ ,
\end{align*}
où $\omega_{\mu_{1}\cdots \mu_{p},r_{1}\cdots r_{q}} \in \cF(M)\otimes M_{n}$.
On a alors les relations de transitions simples:
\begin{align*}
  \omega'_{\mu_{1}\cdots \mu_{p},r_{1}\cdots r_{q}} G^{r_{1}}_{s_{1}} \cdots G^{r_{q}}_{s_{q}} &=
g^{-1} \omega_{\mu_{1}\cdots \mu_{p},s_{1}\cdots s_{q}} g \ .
\end{align*}
où la matrice $(G^{r}_{s})$ est définie par:
\begin{align*}
  \Ad_{g^{-1}}E_{s} &=  G^{r}_{s} E_{r} \ .
\end{align*}

\begin{rem}
On obtient ces relations en considérant la manière dont se transforment les composantes locales de la forme $\alpha$ lors de changements de carte, \textit{i.e.} de la manière suivante:
  \begin{align*}
    \alpha'^{s}(\cX'\loc) &= G^{s}_{r}\alpha^{r}(\cX\loc) \ .
  \end{align*}
\end{rem}

\begin{rem}
  Il est clair que dans cette décomposition locale, les indices $\mu$ correspondent à des indices de formes tensorielles et que les indices $a$ correspondent aux indices de formes intérieures de la décomposition (\ref{eq:decomp-formes}).
\end{rem}

\subsection{Structure Riemannienne et opérateur de Hodge}
\label{sec:struct-riem}
Nous allons compléter l'étude des algèbres d'endomorphismes faite dans la section~\ref{sec:alg-des-endom} et généraliser les décompositions qui ont été faites localement dans la section~\ref{sec:point-de-vue-local} et la sous-section précédente sur les structures Riemanniennes.
\subsubsection{Métrique sur les dérivations}

\begin{defn}
Nous appellerons (pseudo-)métrique sur $\der(\kA)$ une application $Z(\kA)$-bilinéaire symétrique (\textit{i.e.} un élément de $(S^{2}_{Z(\kA)}\der(\kA))^{\star \kA}$ ):
\begin{align*}
  h: \der(\kA)\otimes_{Z(\kA)}\der(\kA) \longrightarrow Z(\kA)
\end{align*}
non dégénérée, c'est-à-dire telle que l'application:
\begin{align*}
  h^{\flat}:  \der(\kA)&\longrightarrow \Oder^{1}(\kA) \\
  \cX &\longmapsto [\cY \mapsto h(\cX,\cY)]  
\end{align*}
est injective.
\end{defn}

 Cette propriété de non dégénérescence est, dans notre cas, équivalente au fait que $\Oder^{1}(\kA)$ est engendré en tant que bimodule sur $\kA$ par son sous-$Z(A)$-module $(\Im h^{\flat})$, \textit{i.e.}:
\begin{align*}
  \kA \cdot (\Im h^{\flat}) \simeq  (\Im h^{\flat}) \cdot \kA \simeq \Oder^{1}(\kA) \ .
\end{align*}

Nous avons vu dans la section~\ref{sec:bimod-diag-Oder} que $\der(\kA)$ est isomorphe à $(\Oder^{1}(\kA))^{\ast \kA}=\Hom_{\kA}^{\kA}(\Oder^{1}(\kA),\kA) $ (voir eq.(\ref{eq:dual})).
Nous pouvons ainsi naturellement  voir le $Z(\kA)$-module $\der(\kA)$ comme un sous-module de $\Hom_{Z(\kA)}^{\kA}(\Oder^{1}(\kA),\kA)$ en utilisant l'inclusion:
\begin{align*}
 \der(\kA) \simeq   \Hom_{\kA}^{\kA}(\Oder^{1}(\kA),\kA) \subset \Hom_{Z(\kA)}^{\kA}(\Oder^{1}(\kA),\kA) \ .
\end{align*}

Ceci nous mène naturellement à la définition suivante:
\begin{defn}
On définit le $\kA-Z(\kA)$ bimodule $\kA-\der(\kA)$ comme étant le bimodule engendré par $\der(\kA)$ en tant que sous-$\kA-Z(\kA)$-bimodule de $\Hom_{Z(\kA)}^{\kA}(\Oder^{1}(\kA),\kA)$.
On définit de même le $Z(\kA)-\kA$ bimodule $\der(\kA)-\kA$ comme étant le bimodule engendré par $\der(\kA)$ en tant que sous-$Z(\kA)-\kA$-bimodule de $\Hom^{Z(\kA)}_{\kA}(\Oder^{1}(\kA),\kA)$.
\end{defn}

Alors, nous pouvons prolonger une métrique $h$ sur $\der(\kA)$ en un homomorphisme de bimodules:
\begin{align*}
  h: \kA-\der(\kA)\otimes_{Z(\kA)}\der(\kA)-\kA \longrightarrow \kA \ .
\end{align*}
L'application $h^{\flat}$ devient alors un homomorphisme de $\kA-Z(\kA)$ bimodules de $\kA-\der(\kA)$ dans $\Oder^{1}(\kA)$ et la condition de non dégénérescence de $h$ est équivalente au fait que cette application est un isomorphisme, \textit{i.e.}:
\begin{equation*}
\xymatrix{
  \kA-\der(\kA) \ar^-{h^{\flat}}_-{\simeq}[r] & \Oder^{1}(\kA) \ .
}
\end{equation*}

Comme nous l'avons fait pour les $1$-formes non commutatives, nous pouvons associer canoniquement à une métrique $h$ sur $\der(\kA)$ une application $Z(\kA)$-linéaire symétrique $h_{\Int}= ad^{*}h: \kA_{0}\otimes_{Z(\kA)} \kA_{0} \to Z(\kA)$ que nous appellerons métrique intérieure. Elle est définie par restriction aux dérivations intérieures de la manière suivante:
\begin{align*}
   ad^{*}h(\gamma,\delta) &= h(\ad_{\gamma},\ad_{\delta}) &
  & \forall \gamma, \delta \in \kA_{0} \ .
\end{align*}

Nous dirons qu'une métrique intérieure $\mu: \kA_{0}\otimes_{Z(\kA)} \kA_{0} \to Z(\kA)$ est non dégénérée si l'application $\mu^{\flat}:\kA_{0}\to \Omega_{\Int}(\kA): \gamma \mapsto [\eta \mapsto \mu(\gamma,\eta)]$ est injective.
Nous appellerons métriques non dégénérées les métriques sur $\der(\kA)$ dont la métrique intérieure est non dégénérée.

De même, à toute métrique $h^{M}$ sur $\Gamma(M)$, on peut associer une métrique (dégénérée) $\rho^{*}h^{M}$ sur $\der(\kA)$ définie par la relation:
\begin{align*}
  \rho^{*}h^{M}(\cX,\cY) &= h^{M}(\rho(\cX),\rho(\cY)) \ .
\end{align*}

De la même manière que nous l'avons fait sur les $1$-formes non commutatives, si l'on introduit une connexion $\nabla$ sur le fibré $\cE$ et que l'on note $\alpha$ la $1$-forme non commutative correspondante, nous pouvons associer à toute métrique intérieure $\mu$ une métrique $\alpha^{*}\mu$ sur $\der(\kA)$ définie par:
\begin{align*}
  \alpha^{*}\mu(\cX,\cY)  &= \mu(\alpha(\cX),\alpha(\cY)) &
  &\forall \cX, \cY \in \der(\kA)  
\end{align*}
et à toute métrique sur $h$ sur $\der(\kA)$ une métrique $h^{M}=\nabla^{*}h$ sur $\Gamma(M)$ définie par:
\begin{align*}
  \nabla^{*}h(X,Y)&= h(\nabla_{X},\nabla_{Y}) \ .
\end{align*}

Nous serions maintenant tentés de décomposer une métrique quelconque sur $\der(\kA)$ en une métrique sur $\Gamma(M)$ et une métrique intérieure à l'aide d'une connexion arbitraire $\nabla$ sur $\cE$.
Nous allons voir que cela est toujours possible pour les métriques non dégénérées et ce de manière unique.
Nous pouvons résumer la situation dans la propriété suivante:
\begin{prop}
\label{sec:metrique-sur-les-derivations}
Soit $h$ une métrique sur $\der(\kA)$ non dégénérée, \textit{i.e.} telle que la métrique intérieure $h_{\Int}=ad^{*}h$ est non dégénérée, alors il existe une unique connexion $\nabla$ sur $\cE$ telle que:
\begin{align}
  h(\nabla_{X}, \ad_{\gamma})&= 0 &
  & \forall X\in \Gamma(M) ,  \gamma \in \kA \ .
\label{eq:orthogonalite}\end{align}
\end{prop}
\demo
Tout d'abord montrons l'existence d'une telle connexion localement.
Il suffit de considérer la connexion $\nabla$ définie par ses expressions locales: $(\nabla_{X})\loc= X\loc + \ad_{A(X\loc)}$, pour tout $X\in \Gamma(M)$, où $X\loc$ est la dérivation locale associée à $X$ et $A$ est la $1$-forme locale de connexion associée à $\nabla$ définie par la relation suivante:
\begin{align*}
  ad^{*}h\loc(A(X),\gamma) &= -h(X\loc,\ad_{\gamma}) \ .
\end{align*}
Cette relation définit complètement $A$ étant donné que la métrique intérieure $\ad^{*}h$ est non dégénérée.
En effet, on a alors:
\begin{align*}
  A(X\loc):= - h\loc(X\loc,\ad_{E_{a}})h_{\Int}^{ab} E_{b} \ ,
\end{align*}
où $(h_{\Int}^{ab})$ est la matrice inverse de la matrice $(h_{ab})$ définie par:
\begin{align*}
  h_{ab} = \ad^{*}h\loc(E_{a},E_{b}) \ .
\end{align*}
Les matrices $E_{a}$ forment une base des matrices hermitiennes de $\ksl(n)$.

On doit vérifier que ce potentiel de jauge $A$ se transforme bien comme une forme locale de connexion.
Nous devons tout d'abord nous donner les relations de transition de la métrique $h$ au-dessus de l'intersection de deux ouverts trivialisants $U$ et $U'$.
Sur $U\cap U'\neq \emptyset$, on a  $h'\loc(\cX'\loc,\cY'\loc)=h\loc(\cX\loc,\cY\loc)$ pour tous $\cX, \cY \in \der(\kA)$, ainsi:
\begin{align*}
&  h'\loc(X,Y) = h\loc(X,Y) \\
&  h'\loc(\ad_{g^{-1}\gamma g},\ad_{g^{-1}\eta g}) = h\loc(\ad_{\gamma},\ad_{\eta}) \\
&  h'\loc(X,\ad_{g^{-1}\eta g})+h'\loc(\ad_{g^{-1}(X\cdot g)}, \ad_{g^{-1}\eta g}) = h\loc(X, \ad_{\eta}) \ ,
\end{align*}
pour tous $X,Y \in \Gamma(M)$ et tous $\gamma, \eta \in \cF(U\cap U')\otimes M_{n}$.\\
Alors, on a:
\begin{align*}
  A'(X) &= -h'\loc(X,\ad_{E_{a}})h'^{ab}E_{b} \\
  &= - h\loc(X,\ad_{g E_{a} g^{-1}})h'^{ab}E_{b} + h'\loc(\ad_{g^{-1}(X\cdot g)}, \ad_{E_{a}})h'^{ab}E_{b} \\
  &= - h\loc(X, \ad_{ E_{e}}) (G^{-1})^{e}_{a} G^{a}_{c}G^{b}_{d} h^{cd}E_{b} + g^{-1}(X\cdot g)\\
  &= g^{-1}A(X)g +g^{-1}(X\cdot g) \ .
\end{align*}

On vérifie également que la relation d'orthogonalité (\ref{eq:orthogonalite}) est bien satisfaite localement:
\begin{align*}
  h\loc((\nabla_{X})\loc, \ad_{\gamma}) &=  h\loc(X\loc + \ad_{A(X\loc)}, \ad_{\gamma}) \\
  &= h\loc(X\loc, \ad_{\gamma}) + \ad^{*}h\loc(A(X\loc), \gamma) \\
  &= 0 \ .
\end{align*}
Ces égalités se recollent bien entre les différents ouverts trivialisants et la relation~(\ref{eq:orthogonalite}) est bien vérifiée de manière globale.

L'unicité de cette connexion est liée à la non dégénérescence de $h$ et $h_{\Int}$.
En effet, soit deux connexions $\nabla$ et $\nabla'$ vérifiant (\ref{eq:orthogonalite}), alors on a:
\begin{align}
  h(\nabla_{X}-\nabla'_{X}, \ad_{\gamma})&= 0 &
  & \forall X\in \Gamma(M) ,  \gamma \in \kA \ .
\end{align}
Or, nous savons que la différence de deux connexions $\nabla_{X}-\nabla'_{X}=\ad_{\mu(X)}$ est une dérivation intérieure qui s'exprime à l'aide d'une forme tensorielle $\mu \in \Omega(M,\End(\cE))$. 
Ainsi, nous avons:
\begin{align}
  h(\ad_{\mu(X)}, \ad_{\gamma})= h_{\Int}(\mu(X),\gamma) = 0 &
  & \forall X\in \Gamma(M) ,  \gamma \in \kA_{0} 
\end{align}
et étant donné que la métrique intérieure $h_{\Int}$ est non dégénérée, nous avons $\mu=0$, soit $\nabla=\nabla'$.
\hfill $\boxempty$
\\

Soit $h$ une métrique sur $\der(\kA)$ non dégénérée et $h_{\Int}=ad^{*}h$ la métrique intérieure correspondante, alors, d'après la proposition précédente, nous pouvons lui associer une connexion $\nabla$ sur $\cE$ de manière unique.
Cette connexion nous permet de décomposer la métrique $h$ de la manière suivante:
\begin{align}
&\begin{aligned}
  h(\cX,\cY) 
  &=h( \nabla_{\rho(\cX)} - \ad_{\alpha(\cX)} ,  \nabla_{\rho(\cY)} - \ad_{\alpha(\cY)} ) \\
  &=\rho^{*}\nabla^{*}h(\cX,\cY) + \alpha^{*}ad^{*}h(\cX,\cY)\\
  &=\rho^{*}h^{M}(\cX,\cY) + \alpha^{*}h_{\Int}(\cX,\cY)
\end{aligned} &
& \forall \cX,\cY \in \der(\kA)\ ,
\label{eq:metrique}
\end{align}
avec  $h^{M}=\nabla^{*}h$ et $h^{\Int}=ad^{*}h$.
Cette décomposition généralise la décomposition (\ref{eq:decomp-formes}) que l'on a obtenue pour les formes non commutatives.

Ainsi, une métrique non dégénérée sur $\der(A)$ se décompose en deux parties: une partie pouvant s'interpréter comme une métrique ordinaire et une partie purement non commutative caractérisée par un ensemble de champs scalaires (ou sections d'un certain fibré).
Réciproquement, il est possible de ``construire'' une métrique non dégénérée à partir d'une  connexion sur $\cE$, d'une métrique ordinaire sur $M$ et d'une métrique définie sur les dérivations intérieures.

Un candidat naturel pour la métrique $h_{\Int}$ est la métrique de Killing~\cite{masson:99} définie par:
\begin{align*}
  h_{Killing}(\gamma, \delta)&= \frac{1}{n}\tr(\gamma\delta) &
  & \forall \gamma,\delta \in \kA_{0} \ .
\end{align*}
Cette métrique fournit une origine particulière pour les métriques intérieures et peut être caractérisée par le fait qu'elle est l'unique (à la multiplication par un scalaire près) métrique intérieure \textit{invariante} pour l'action:
\begin{align*}
  L_{\ad_{X}}h_{\Int}(Y, Z) &= - h^{\Int}(\ad_{X}Y,Z) - h^{\Int}(X,\ad_{X},Y) \ ,&
  &\forall X,Y,Z\in \kA_{0} \ .
\end{align*}
De ce point de vue, la métrique de Killing joue un rôle analogue à la forme $\theta$ définie sur les dérivations intérieures. En effet, nous avions remarqué dans la section \ref{sec:maxw-non-comm} que la forme $\theta$ est l'unique (à la multiplication par un scalaire près) forme non commutative sur $M_{n}$ invariante vis-à-vis des dérivations (intérieures) de $M_{n}$.

\begin{rem}
  La décomposition~(\ref{eq:metrique}) est tout à fait similaire à celle rencontrée dans les théories de type Kaluza-Klein et est son pendant non commutatif.
  Cette comparaison est la même qu'entre les connexions non commutatives et les connexions symétriques dans les théories de réduction dimensionnelle.
\end{rem}

\subsubsection{Analyse locale}
Nous pouvons donner les expressions locales pour la métrique $h$ dans la base $(\partial_{\mu},\ad_{E_{a}})$ par la matrice suivante:
\begin{align*}
h\loc = \left(  \begin{matrix}
    h_{\mu  \nu} & h_{\mu b} \\
    h_{a \nu} & h_{a b}
  \end{matrix}\right ) \ .
\end{align*}
 Si $h$ est non dégénérée, la matrice $(h_{ab})$ est inversible et nous notons $(h_{\Int}^{ab})$ sa matrice inverse.
Cela nous permet de définir les composantes locales d'une connexion:
\begin{align*}
  A^{a}_{\mu} &= -h_{\Int}^{ab}h_{b \mu} \ .
\end{align*}
Ainsi, toujours dans la base $(\partial_{\mu},\ad_{E_{a}})$:
\begin{align*}
h\loc = \left(  \begin{matrix}
    h_{\mu  \nu} & -A_{\mu}^{a}h_{a b} \\
    -h_{a b} A_{\nu}^{b} & h_{a b}
  \end{matrix}\right ) \ .
\end{align*}
Enfin, dans la base $(\nabla_{\mu},\ad_{E_{a}})$ où $\nabla_{\mu}=\partial_{\mu} + \ad_{A_{\mu}}$, on a:
\begin{align*}
  h\loc = \left(  \begin{matrix}
      h^{M}_{\mu  \nu} & 0 \\
      0 & h_{a b}
    \end{matrix}\right ) \ ,
\end{align*}
où $h^{M}_{\mu \nu} = h_{\mu \nu} - A_{\mu}^{a} A_{\nu}^{b} h_{ab}$.

\subsubsection{Métrique sur les $1$-formes non commutatives}
\newcommand{\ho}{h}
La notion de  métrique peut également être regardée du point de vue dual, c'est-à-dire sur les $1$-formes non commutatives.

\begin{defn}
  On appelle métrique sur $\Oder^{1}(\kA)$ un homomorphisme de bimodules:
\begin{align*}
\ho:  \Oder^{1}(\kA)\otimes_{\kA}\Oder^{1}(\kA) \longrightarrow \kA
\end{align*}
non dégénéré et symétrique (voir définitions suivantes).
\end{defn}

\begin{defn}
  Un homomorphisme $\ho:  \Oder^{1}(\kA)\otimes_{\kA}\Oder^{1}(\kA) \to \kA$ est non dégénéré si l'homomorphisme de $\kA-Z(\kA)$ bimodules:
\begin{align*}
  \ho^{\#}: 
   \Oder^{1}(\kA) &\longrightarrow  \Hom_{Z(\kA)}^{\kA}(\Oder^{1}(\kA),\kA) \\
   \omega &\longmapsto [\eta \mapsto \ho(\omega, \eta)]
\end{align*}
réalise%
\footnote{Nous devons tenir compte du fait que $[Z(\kA),\Oder^{1}(\kA)]=0$ pour les algèbres d'endomorphismes afin que $\ho^{\#}$ soit bien un homomorphisme.}
un isomorphisme:
\begin{equation*}
\xymatrix{
\Oder^{1}(\kA)  \ar^-{\ho^{\#}}_-{\simeq}[r] &  \kA-\der(\kA) 
} \ .
\end{equation*}
Ceci est équivalent à dire que le $\kA-Z(\kA)$ bimodule $\Im (\ho^{\#})$ est engendré par son sous-$Z(\kA)$-module $\der(\kA)$, \textit{i.e.}:
\begin{align*}
  \Im \ho^{\#} \simeq \kA-\der(\kA) \ .
\end{align*}
Cette condition est la version duale de la condition de non dégénérescence pour les métriques sur $\der(\kA)$.
\end{defn}

\begin{defn}
  Un homomorphisme $\ho:  \Oder^{1}(\kA)\otimes_{\kA}\Oder^{1}(\kA) \to \kA$ est dit symétrique si l'on a $ \ho(\omega,\eta)=\ho(\eta, \omega)$ pour tous $\omega,\eta\in \Im g^{\flat}$ où $g$ est une métrique sur $\der(\kA)$.
\end{defn}

Maintenant, rappelons que les formes basiques et les formes horizontales pour l'action de Cartan associé à $\Int(\kA)$ forment des sous-algèbres différentielles graduées de $\Oder(\kA)$ reliées de la manière suivante:
\begin{equation}
  \begin{aligned}
    \xymatrix{
      \Omega(M) \ar@{^(->}^-{\bas}[rr] \ar@{^(->}^-{\inv}[dr] & & \Oder(\kA) \\
      &\Omega(M,\End(\cE)) \ar@{^(->}^-{\hor}[ur] & }
\end{aligned}
\end{equation}
Ainsi, par restriction à $\Omega(M)$, il est naturel d'associer à une métrique $\ho$ sur $\Oder^{1}(\kA)$ une métrique $\rho_{*}\ho$ sur $\Omega(M)$ qui est un homomorphisme de bimodules:
\begin{align*}
\rho_{*}\ho:  \Omega(M)\otimes_{\cF(M)}\Omega(M) \to \cF(M) \ .
\end{align*}
Cet homomorphisme se prolonge naturellement en un homomorphisme de bimodules:
\begin{align*}
\rho_{*}\ho:  \Omega(M,\End(\cE))\otimes_{\cF(M)}\Omega(M,\End(\cE)) \to \kA \ .
\end{align*}
De même, on peut associer naturellement à toute métrique $h_{\Int}$ sur $\Omega_{\Int}(\kA)$ une métrique  $ad_{*}h_{\Int}$ sur $\Oder^{1}(\kA)$ dégénérée sur $\Omega(M)$ définie par:
\begin{align*}
  \ad_{*}h_{\Int}(\omega,\eta) &= h_{\Int}(\ad^{*}\omega,\ad^{*}\eta) \ .
\end{align*}

Si l'on se donne une connexion $\nabla$ sur $\cE$, on peut associer à tout homomorphisme de bimodules $g: \Omega(M,\End(\cE))\otimes_{\cF(M)}\Omega(M,\End(\cE)) \to \kA$ symétrique non dégénérée (\textit{i.e.} une métrique sur les formes tensorielles), une métrique $\nabla_{*}g$ sur $\Oder(\kA)$ définie par:
\begin{align*}
   \nabla_{*}g(\omega,\eta)&:= g(\omega\circ\nabla,\eta\circ\nabla) &
&\forall \omega,\eta \in \Oder(\kA) \ .
\end{align*}

Montrons maintenant une propriété analogue à la propriété~\ref{sec:metrique-sur-les-derivations}.
\begin{prop}
  Soit $\ho$ une métrique sur $\Oder(\kA)$ telle que la métrique $\rho_{*}\ho$ soit non dégénérée, alors il existe une unique connexion $\nabla$ sur $\cE$ tel que:
  \begin{align}
    \ho(\mu\circ\rho,\eta\circ\alpha)&=0 &
    &\forall \mu \in \Omega(M,\End(\cE)), \forall \eta\in\Omega_{\Int}(\kA) \ . 
\label{eq:orthogonalité:2}
\end{align}
\end{prop}
\demo
  Notons $g$ la métrique sur $\Gamma(M)$ inverse de $\rho_{*}\ho$.
  Il suffit de prendre la connexion définie localement par:
  \begin{align*}
    A(X) &= \ho\loc(g\loc^{\flat}(X),i\theta) &
    &\forall X \in \Gamma(U) \ .
  \end{align*}
On vérifie que cela définit bien une connexion en considérant les relations de transitions de la métrique $\ho\loc$.

Nous pouvons alors vérifier la relation d'orthogonalité (\ref{eq:orthogonalité:2}) pour des éléments $\mu\in \Im(g^{\flat})$.
Ainsi, si on prend $\mu=g^{\flat}(X)$, on a:
\begin{align*}
  \ho\loc(g\loc^{\flat}(X),\eta\circ(A-i\theta) ) 
  &= \ho\loc\left(g\loc^{\flat}(X),\eta\circ \ho\loc(g^{\flat}(\cdot), i \theta)\right)  - \ho\loc(g\loc^{\flat}(X),\eta\circ i\theta )\\
  &= \eta \circ \ho\loc(g\loc^{\flat}(X),i\theta )  - \ho\loc(g\loc^{\flat}(X),\eta\circ i\theta )\\
  &= 0 \ ,
\end{align*}
où nous nous sommes servis pour passer à la deuxième ligne du fait que 
\begin{align*}
  \ho\loc(g\loc^{\flat}(X),g\loc^{\flat}(\cdot))= g\loc^{\flat}(X)(\cdot)
\end{align*}
(nous commettons un petit abus de notation ici) et la troisième égalité a un sens du fait que $\ho$ est un homomorphisme de bimodules et que $g\loc^{\flat}(X)$ est à valeurs dans le centre de l'algèbre, ce qui nous permet de faire passer $\eta$ à l'extérieur de la métrique.
L'égalité se prolonge globalement de telle manière que la relation (\ref{eq:orthogonalité:2}) soit satisfaite.

Ainsi, toute métrique $\ho$ non dégénérée peut se décomposer de la manière suivante:
\begin{align*}
  \ho &= \nabla_{*}\ho^{M} + ad_{*} \ho_{\Int} \ ,
\end{align*}
où $\ho^{M} = \rho_{*}h$ est une métrique sur $\Gamma(M)$, $\ho_{\Int}= \alpha_{*} \ho$ est une métrique sur $\Omega_{\Int}(\kA)$, $\nabla$ est la connexion associée à $\ho$ de manière canonique et $\alpha$ est la $1$-forme associée à $\nabla$.

\subsubsection{Analyse locale}
Nous pouvons donner les expressions locales pour la métrique $\ho$ dans la base $(dx^{\mu}, i\theta^{a})$ par la matrice suivante:
\begin{align*}
\ho\loc = \left(  \begin{matrix}
    \ho^{\mu  \nu} & \ho^{\mu b} \\
    \ho^{a \nu} & \ho^{a b}
  \end{matrix}\right ) \ .
\end{align*}
 Si $\ho$ est non dégénérée, la matrice $(\ho^{\mu \nu})$ est inversible et nous notons $(\ho^{M}_{\mu \nu})$ sa matrice inverse.
Cela nous permet de définir les composantes locales d'une connexion:
\begin{align*}
  A^{a}_{\mu} &= \ho^{M}_{\mu\nu} \ho^{a\nu} 
\end{align*}
et on a:
\begin{align*}
  \ho\loc = \left(  \begin{matrix}
      \ho^{\mu  \nu} & \ho^{\mu \nu}A_{\nu}^{b} \\
      A_{\mu}^{a}\ho^{\mu \nu} & \ho^{a b}
    \end{matrix}\right ) \ .
\end{align*}
Ainsi, dans la base $(dx^{\mu},\alpha^{a})$, on a:
\begin{align*}
  \ho\loc = \left(  \begin{matrix}
      \ho^{\mu  \nu} & 0 \\
      0 & \ho^{a b}_{\Int}
    \end{matrix}\right ) \ ,
\end{align*}
où $\ho^{a b}_{\Int} = \ho^{ab} - \ho^{\mu \nu} A_{\mu}^{a} A_{\nu}^{b}$.

\subsubsection{Structure Riemannienne}
\begin{defn}
  Soit $g$ une métrique sur $\der(\kA)$ et $h$ une métrique sur $\Oder(\kA)$.
  Nous dirons que les métriques $g$ et $h$ sont compatibles entre elles et définissent une structure Riemannienne sur $\kA$ si $h^{\#}\circ g^{\flat} = id$.  
  En particulier, on a:
  \begin{align}
    h(g^{\flat}(\cX),g^{\flat}(\cY)) &= g(\cX,\cY) & &\forall \cX , \cY \in
    \der(\kA)\ .
  \end{align}
  Si l'on considère le prolongement de l'application $g^{\flat}$ à $\kA-\der(\kA)$, alors on a $g^{\flat}\circ h^{\#}= id$.  De même, nous avons vu que $g$ se prolonge naturellement en un homomorphisme de bimodules $g:
  \kA-\der(\kA)\otimes_{Z(\kA)}\der(\kA)-\kA \to \kA$. Alors on a:
  \begin{align*}
    g(h^{\#}(\omega),\tilde{h}^{\#}(\eta)) &= h(\omega, \eta) & &\forall
    \omega,\eta \in \Oder(\kA) \ ,
  \end{align*}
  où
  \begin{align*}
    \tilde{h}^{\#}:
    \Oder^{1}(\kA) &\longrightarrow  \Hom^{Z(\kA)}_{\kA}(\Oder^{1}(\kA),\kA) \\
  \eta &\longmapsto [\omega \mapsto \ho(\omega, \eta)] \ .
\end{align*}
\end{defn}

\subsubsection{Expressions locales}
Soit $g$ et $h$ deux métriques sur $\der(\kA)$ et $\Oder^{1}(\kA)$ définissant une structure Riemannienne sur $\kA$.
Alors, dans les bases $(\partial_{\mu},\ad_{E_{a}})$ et $(dx^{\mu}, i\theta^{a})$, elles ont pour composantes locales:
\begin{align*}
&g\loc = \left(  \begin{matrix}
    g_{\mu  \nu} & g_{\mu b} \\
    g_{a \nu} & g_{a b}
  \end{matrix}\right )&
&h\loc = \left(  \begin{matrix}
    h^{\mu  \nu} & h^{\mu b} \\
    h^{a \nu} & h^{a b}
  \end{matrix}\right ) \ .
\end{align*}
Si elles sont non dégénérées, alors on peut définir:
\begin{equation*}
\begin{aligned}
(g_{\Int}^{ab}) &= (g_{ab})^{-1} &
&\text{et} &
(h^{M}_{\mu \nu}) &= (h^{\mu \nu})^{-1} \ ,
\end{aligned}
\end{equation*}
ainsi qu'une connexion dont le potentiel de jauge est donné par la formule suivante:
\begin{align*}
    A^{a}_{\mu} &= -g_{\Int}^{ab}g_{b \mu} = h^{M}_{\mu\nu} h^{a\nu} \ .
\end{align*}
Alors, on a:
\begin{align*}
&g\loc = \left(  \begin{matrix}
    g_{\mu  \nu} & -A_{\mu}^{a}g_{a b} \\
    -g_{a b} A_{\nu}^{b} & g_{a b}
  \end{matrix}\right )&
&h\loc = \left(  \begin{matrix}
      h^{\mu  \nu} & h^{\mu \nu}A_{\nu}^{b} \\
      A_{\mu}^{a} h^{\mu \nu} & h^{a b}
  \end{matrix}\right ) \ .
\end{align*}

Dans la base $(\nabla_{\mu},\ad_{E_{a}})$, $(dx^{\mu},\alpha^{a})$, on a:
\begin{align*}
&  g\loc = \left(  \begin{matrix}
      g^{M}_{\mu  \nu} & 0 \\
      0 & g_{a b}
    \end{matrix}\right ) &
&  h\loc = \left(  \begin{matrix}
      h^{\mu  \nu} & 0 \\
      0 & h^{a b}_{\Int}
    \end{matrix}\right ) \ ,
\end{align*}
où
\begin{align*}
  \begin{aligned}
    g^{M}_{\mu \nu} &= g_{\mu \nu} - A_{\mu}^{a} A_{\nu}^{b} g_{ab} \\
    h^{a b}_{\Int} &= h^{ab} - h^{\mu \nu} A_{\mu}^{a} A_{\nu}^{b}
  \end{aligned} \ \ .
\end{align*}
La condition de compatibilité entre $g$ et $h$ s'exprime par le fait que les matrices définies par $g\loc$ et $h\loc$ sont inverses l'une de l'autre. En particulier, on a:
\begin{equation*}
  \left\{\begin{aligned}
      h^{M}_{\mu \nu} &= g^{M}_{\mu \nu} \\
      h_{\Int}^{ab} &= g_{\Int}^{ab}
    \end{aligned} \right. \ \  .
\end{equation*}

\subsubsection{Dualité de Hodge}

Soit $g$ et $h$ deux métriques sur $\der(\kA)$ et $\Oder(\kA)$ définissant une structure Riemannienne sur $\kA$.
Nous prendrons les mêmes notations que précédemment pour les expressions locales.

Tout d'abord, remarquons que l'on peut étendre la notion de métrique sur les formes de degré plus élevé en posant:
\begin{multline}
  h\loc(dx^{\mu_{1}}\wedge \cdots\wedge dx^{\mu_{p}} \alpha^{a_{1}} \cdots \alpha^{a_{q}}, dx^{\nu_{1}}\wedge \cdots\wedge dx^{\nu_{p}} \alpha^{b_{1}} \cdots \alpha^{b_{q}} ) \\
  = (-1)^{pq} \det\left((h^{\mu_{i}\nu_{j}})_{i,j \in [1,p]}\right) \det\left((h_{\Int}^{a_{i}b_{j}})_{i,j \in [1,q]}\right) \ .
\end{multline}
Cela nous permet ainsi de prolonger la métrique $h$ en un homomorphisme de bimodule:
\begin{align*}
  h: \Oder(\kA) \otimes_{\kA} \Oder(\kA) \to \kA \ .
\end{align*}
Il est alors naturel de définir une opération de Hodge:
\begin{align}
  \star : \Oder^{k}(\kA) \to \Oder^{d+n²-1 -k}(\kA) \ ,
\end{align}
où $d$ est la dimension de la variété $M$ et $n$ le rang des fibres de $\cE$.
Cette opération est définie localement pour une forme $\omega \in \Oder^{r}(\kA)$ donnée par ses expressions locales:
\begin{align*}
  \omega\loc &= \omega_{\mu_{1}\cdots \mu_{p},r_{1}\cdots r_{q}} 
  dx^{\mu_{1}}\wedge \cdots \wedge dx^{\mu_{p}}
  \alpha^{r_{1}}\cdots \alpha^{r_{q}} \ ,
\end{align*}
avec $p+q = r$.
On a alors:
\begin{multline}
  (\star \omega)\loc =\frac{(-1)^{q(d-p)}}{(d-p)!(n²-1-q)!}  \omega_{\mu_{1}\cdots \mu_{p},r_{1}\cdots r_{q}} 
    h^{\mu_{1}\nu_{1}} \cdots h^{\mu_{p} \nu_{p}} 
    \epsilon^{M}_{\nu_{1} \cdots \nu_{p} \nu_{p+1}\cdots \nu_{d}} \\
    \times h_{\Int}^{r_{1} s_{1}} \cdots h_{\Int}^{r_{q} s_{q}} 
    \epsilon_{s_{1} \cdots s_{q} s_{q+1} \cdots s_{n²-1}} 
    dx^{\nu_{p+1}}\wedge \cdots \wedge dx^{\nu_{d}}
    \alpha^{s_{q+1}}\cdots \alpha^{s_{n²-1}} \ ,
\end{multline}
où $\epsilon^{M}$ et $\epsilon$ sont les tenseurs définis de la manière suivante:
 \begin{align*}
   \epsilon^{M}_{\nu_{1} \cdots \nu_{d}} &= \sqrt{\det(g^{M}_{\mu \nu})} \delta_{\nu_{1} \cdots \nu_{d}}^{1 \cdots \  d} \\
   \epsilon_{s_{1} \cdots s_{n²-1}}&= \sqrt{\det(g_{ab})} \delta_{s_{1} \cdots s_{n²-1}}^{1 \cdots \ n²-1}    \ .
 \end{align*}
 Le symbole $\delta_{A_{1} \cdots A_{N}}^{1 \ \cdots \ N}$ est le déterminant de la matrice $(\delta^{I}_{A_{J}})_{I,J\in [1,N]}$ et $\delta$ est le symbole de Krönecker (ce tenseur est souvent appelé tenseur complètement antisymétrique).
On vérifie facilement que cette formule définit bien une forme de degré $(d+n^{2}-1-r)$ et que l'opération de Hodge satisfait:
\begin{align*}
  \star \star &= (-1)^{r(d+n^{2}-1-r)}
\end{align*}
sur les formes de degré $r$.\\
L'opération de Hodge permet alors de donner une formule explicite pour la métrique $h$ agissant sur deux formes $\omega, \eta \in \Oder(\kA)$. En effet, on a:
\begin{align*}
  h(\omega, \eta) &= \star^{-1} ( \omega \star \eta) &
  &\text{si $\omega$ et $\eta$ sont de même degré} \\
  &= 0 &
  &\text{sinon.}
\end{align*}

On peut également introduire une forme volume sur  $\Omega_{\Int}(\kA)$ définie de la manière suivante:
\begin{align*}
   \cV_{\Int} &= \frac{1}{(n²-1)!}\epsilon_{s_{1} \cdots s_{n²-1}} \theta^{s_{1}}\circ ad \cdots \theta^{s_{n²-1}}\circ ad \ .
\end{align*}
Cette forme volume nous permet ``d'intégrer''  une forme $\omega$ le long des fibres (voir~\cite{masson:99}). 
On peut définir la notion d'intégration le long des fibres soit à partir de l'algèbre $\kB$ introduite dans la section \ref{sec:alg-des-endom}, soit à partir de la décomposition locale des formes. Dans les deux cas, cela revient à utiliser l'intégration sur $\Oder(M_{n}(\gC))$ introduite dans~\cite{dubois-violette:90}.
Si l'on reste dans $\Oder(\kA)$, on peut voir qu'une forme $\omega\in\Oder(\kA)$ peut toujours se décomposer de manière unique de la manière suivante: 
\begin{align*}
  \omega &= \rho^{*}a \cdot \alpha^{*}\cV_{\Int} + \eta \ , 
\end{align*}
où la forme  $a\in \Omega(M,\End \cE)$ est définie en posant:
\begin{align*}
  a &=\nabla^{*} \left( \star^{-1}( \star \omega \cdot \alpha^{*}\cV_{\Int} )  \right ) 
\end{align*}
et la forme $\eta\in\Oder(\kA)$ est définie par: 
\begin{align*}
  \eta = \omega - \rho^{*}a \cdot \alpha^{*}\cV_{\Int} \ .
\end{align*}

On définit alors l'intégration le long des fibres en posant:
  \begin{eqnarray*}
  \int_{n.c.}:&   \Oder(\kA) & \longrightarrow \Omega(M) \\
  &  \omega=\rho^{*}a \cdot \alpha^{*}\cV_{\Int} + \eta &\longmapsto \int_{n.c.} \omega = \nabla^{*}  \tr \left(\star^{-1}( \star \omega \cdot \alpha^{*}\cV_{\Int} ) \right) = \tr(a) \ .
\end{eqnarray*}
Cela nous permet de définir un produit scalaire sur $\Oder(\kA)$ défini de la manière suivante:
\begin{align*}
  (\omega,\eta) &= \int_{M}\int_{n.c.} \omega\star\eta \ ,
\end{align*}
où $\int_{M}$ est l'intégration usuelle sur la variété de base $M$.
\subsection{Connexion linéaire sans torsion}

Il est possible d'introduire le concept de connexion linéaire~\cite{dubo-mich:95} sur $\der(\kA)$ et d'associer à une structure Riemannienne l'unique connexion linéaire sans torsion laissant la métrique invariante. Nous appellerons cette connexion la connexion de Levi-Cività.
Tout d'abord, rappelons la définition d'une connexion linéaire.
\begin{defn}
  Une connexion linéaire est une connexion sur le $Z(A)$-module  $\der(\kA)$, c'est-à-dire une application qui à toute dérivation $\cX\in \der(\kA)$ associe une application linéaire $D_{\cX}: \der(\kA) \to \der(\kA)$ satisfaisant:
  \begin{align*}
    D_{\cX}(f \cY) &= \cX(f) \cY + f D_{\cX}(\cY) &
    D_{f\cX}(\cY) &= f D_{\cX}(\cY) &
 \forall \cX,\cY \in \der(\kA),  \forall f\in Z(\kA) \ .
  \end{align*}
\end{defn}

\begin{defn}
  La torsion $T^{D}$ associée à une connexion linéaire $D$ est la $2$-forme non commutative à valeurs dans $\der(\kA)$ ($T^{D} \in  \Oder^{2}(\kA,\der(\kA))$) définie par:
  \begin{align*}
    T^{D}(\cX,\cY) &= D_{\cX}(\cY) - D_{\cY}(\cX) - [\cX,\cY] \ .
  \end{align*}
\end{defn}

\begin{defn}
  Soit $g$ une métrique sur $\der(\kA)$ associée à une structure Riemannienne sur $\kA$.
  La connexion de Levi-Cività associée à cette structure Riemannienne  est l'unique connexion linéaire sans torsion ($T^{D}=0$) telle que:
  \begin{align*}
    \cX(g(\cY,\cZ)) &=  g(D_{\cX}\cY, \cZ) + g(\cY,D_{\cX}\cY)  \ ,
  \end{align*}
c'est-à-dire laissant la métrique invariante.
La connexion de Levi-Cività est alors définie par la relation suivante:
  \begin{multline}
    2 g(D_{\cX}\cY, \cZ) = \cX (g(\cY,\cZ)) + \cY(g(\cX,\cZ)) -\cZ(g(\cX,\cY)) \\
    + g([\cX,\cY],\cZ) + g([\cZ,\cY],\cX) + g([\cZ,\cX],\cY) \ .
  \end{multline}
\end{defn}

\subsubsection{Expressions locales et symboles de Christoffel}
Soit $g$ une métrique sur $\der(\kA)$,  $h$ une métrique sur $\Oder(\kA)$ définissant une structure Riemannienne sur $\kA$ et $D$ la connexion de Levi-Cività associée.
Nous allons caractériser cette connexion dans la base locale de dérivations $(\nabla_{ \mu}, \ad_{E_{a}})$ et définir les symboles de Christoffel associés.
En  reprenant les notations précédentes pour les expressions locales de $g$ et $h$, on trouve les expressions suivantes pour les dérivées covariantes:
\begin{align*}
  D_{\nabla_{\mu}}\nabla_{\nu} &= \frac{1}{2} \ad_{F_{\mu \nu}} + \Gamma_{\mu \nu}^{\sigma} \nabla_{\sigma} + \Gamma_{\mu \nu}^{d} \ad_{E_{d}} \\
  D_{\nabla_{\mu}}\ad_{E_{b}} &= \ad_{\nabla_{\mu}E_{b}} + \Gamma_{\mu b}^{\sigma} \nabla_{\sigma} + \Gamma_{\mu b}^{d} \ad_{E_{d}} \\
  D_{\ad_{E_{a}}} \nabla_{\nu} &= \Gamma_{a \nu}^{\sigma} \nabla_{\sigma} + \Gamma_{a \nu}^{d} \ad_{E_{d}} \\
  D_{\ad_{E_{a}}} \ad_{E_{b}} &= \frac{1}{2} \ad_{[E_{a},E_{b}]}  + \Gamma_{a b}^{\sigma} \nabla_{\sigma} + \Gamma_{a b}^{d} \ad_{E_{d}} \ .
\end{align*}
Les coefficients $\Gamma_{AB}^{C}$ sont appelés coefficients de Christoffel et sont définis  de la manière suivante:
\begin{align*}
  \Gamma_{\mu \nu}^{\sigma} &= \frac{1}{2}h^{\sigma \rho}(\partial_{\mu}g^{M}_{\nu \rho} + \partial_{\nu}g^{M}_{\mu \rho} - \partial_{\rho}g^{M}_{\mu \nu})&
  \Gamma_{\mu \nu}^{d}&=  0 \\
  \Gamma_{\mu b}^{\sigma} &=\frac{1}{2}h^{\sigma \rho} g_{eb} F_{\rho \mu}^{e} &
  \Gamma_{\mu b}^{d}&= \frac{1}{2}h_{\Int}^{dc} \nabla_{\mu} g_{b c}    \\
  \Gamma_{a \nu}^{\sigma} &= \frac{1}{2}h^{\sigma \rho} g_{ea} F_{\rho \nu}^{e} &
  \Gamma_{a  \nu}^{d}&= \frac{1}{2}h_{\Int}^{dc} \nabla_{\nu} g_{a c}   \\
  \Gamma_{a b}^{\sigma} &= -\frac{1}{2} h^{\sigma \rho} \nabla_{\rho} g_{ab}&
  \Gamma_{a b}^{d}&= \frac{1}{2} h_{\Int}^{dc} L_{\ad_{E_{c}}} g_{ab}
\end{align*}
où nous avons utilisé les notations suivantes:
\begin{align*}
&  L_{\ad_{E_{c}}} g_{ab} =  (L_{\ad_{E_{c}}} g_{\Int})(E_{a},E_{b}) 
  = -C_{ca}^{e}g_{eb} -   C_{cb}^{e}g_{ae} \\
&  \nabla_{\mu}E_{a} = [A_{\mu}, E_{a}]  
  = A_{\mu}^{b}C_{ba}^{c} E_{c} \\
 & \nabla_{\mu}g_{ab}=  (L_{\nabla_{\mu}}g_{\Int})(E_{a},E_{b}) 
 = \partial_{\mu}g_{ab} - A_{\mu}^{e}C_{ea}^{f}g_{fb} - A_{\mu}^{e}C_{eb}^{f}g_{af}\\
 &F_{\mu \nu} = [\nabla_{\mu},\nabla_{\nu}] = \partial_{\mu} A_{\nu} - \partial_{\nu} A_{\mu} + [A_{\mu}, A_{\nu}] \ .
\end{align*}

\begin{rem}
  Cette connexion linéaire peut aisément être étendue au bimodule \\
  $\kA-\der(\kA)\otimes_{Z(\kA)} \der(\kA)-\kA$ ainsi qu'à $\Oder^{1}(\kA)\otimes_{\kA}\Oder^{1}(\kA)$. Les dérivées covariantes locales s'exprimeront alors à l'aide des mêmes coefficients de Christoffel.
\end{rem}

Cette connexion linéaire a la même décomposition que la connexion de Levi-Cività définie dans les théories de Kaluza-Klein (\textit{i.e.} sur un fibré principal de groupe de structure $SU(n)$).
Il est possible de calculer la courbure associée à cette connexion, ainsi que la courbure scalaire correspondante.
Cette courbure scalaire nous permet de définir une action d'Einstein-Hilbert pour l'algèbre $\kA$. Cette action se décompose alors en l'action de Einstein-Hilbert standard  associée à la métrique $g^{M}$ sur la base, l'action de Yang-Mills pour la courbure associée à $\nabla$ et une action de type Higgs pour les composantes internes de la métrique qui peuvent s'interpréter comme des champs scalaires.
Nous renvoyons à~\cite{kerner:81,kerner:88,Gior-kern:88} pour plus de détails sur cette construction.

\subsection{Action de Yang-Mills-Higgs sur un fibré non trivial}
Nous allons construire l'équivalent de l'action de Maxwell pour une connexion non commutative sur le module libre $\kA$ de rang $1$.
Cette action peut s'interpréter en terme de théorie de type Yang-Mills-Higgs sur espace courbe. 
Nous verrons qu'elle possède un mécanisme de Higgs se ``couplant'' à la connexion de référence donnée par une structure Riemannienne sur $\kA$. 

Nous allons considérer  une structure Riemannienne sur $\kA$, donnée par une métrique $g$ non dégénérée sur $\der(\kA)$ et une métrique $h$ non dégénérée sur $\Oder^{1}(\kA)$ et une connexion non commutative sur le module libre $\kA$, donnée par un élément $\omega\in \Oder^{1}(\kA)$.

Nous pouvons considérer la connexion ordinaire $\nabla$ sur $\End\cE$ associée à la structure Riemannienne et l'utiliser pour décomposer la forme $\omega$ en une partie tensorielle et une partie intérieure. 
Ainsi, on note:
\begin{align*}
  \omega &= \rho^{*}\ta  + \alpha - \alpha^{*}\varphi \ ,
\end{align*}
où $\ta \in \Omega(M,\End\cE)$ et $\varphi \in \Omega_{\Int}(\kA)$.\\
La forme de courbure associée à $\omega$ est définie par:
\begin{align*}
  \Omega(\cX,\cY) &= \cX(\omega(\cY)) - \cY(\omega(\cX)) -\omega([\cX,\cY])+ [\omega(\cX),\omega(\cY)] \ .
\end{align*}
On a alors en composantes locales:
\begin{equation*}
  \begin{aligned}
    \Omega_{\mu \nu} &=\Omega(\nabla_{\mu}, \nabla_{\nu})&
    & = F_{\mu \nu} + \nabla_{\mu}\ta_{\nu}- \nabla_{\nu}\ta_{\mu}+[\ta_{\mu}, \ta_{\nu}] - \varphi(F_{\mu \nu}) \\
    \Omega_{\mu b} &= \Omega(\nabla_{\mu}, \ad_{E_{b}})&
    & = \nabla_{\mu}\varphi_{b} + [\ta_{\mu}, \varphi_{b}] \\
    \Omega_{a \nu} &= \Omega(\ad_{E_{a}}, \nabla_{\nu})&
    & = -\nabla_{\nu}\varphi_{a} - [\ta_{\nu}, \varphi_{a}] \\
    \Omega_{a b} &=  \Omega(\ad_{E_{a}}, \ad_{E_{b}}) &
    &= [\varphi_{a},\varphi_{b}] -\varphi_{c}C_{ab}^{c} 
  \end{aligned}
\end{equation*}
où
\begin{equation*}
  \begin{aligned}
    \varphi_{a}&=\varphi(E_{a}) &
    \nabla_{\mu}\varphi_{b}&= \partial_{\mu}\varphi_{b}+[A_{\mu},\varphi_{b}]- C_{ab}^{c}A_{\mu}^{a}\varphi_{c}\\
    \ta_{\mu} &= \ta(\partial_{\mu}) &
    \nabla_{\mu}\ta_{\nu} &= \partial_{\mu}\ta_{\nu} + [A_{\mu},\ta_{\nu}]
  \end{aligned}
\end{equation*}
et $A$ est le potentiel de jauge associé à $\nabla$.

L'expression globale de la courbure est alors:
\begin{equation}
  \begin{aligned}
    \Omega &= \hd\omega + \omega \wedge \omega \\
    &= \rho^{*}\left(F_{A+\ta} - \varphi(F_{A})\right)- \alpha^{*}\rho^{*}(\nabla_{A}\varphi+[\ta,\varphi])  +\frac{1}{2} \left([\alpha^{*}\varphi,\alpha^{*}\varphi] - \varphi([\alpha,\alpha]\right) \ ,
  \end{aligned}
\end{equation}
où  $F_{A}$ est la forme tensorielle de courbure  associée à la connexion $\nabla$ et $F_{A+\ta}$ est la forme tensorielle de courbure  associée à la connexion $\nabla + \ad_{\ta}$.
La forme $\nabla_{A}\varphi+[\ta,\varphi]$ est un élément de $\Omega(M,\End\cE)\otimes \Omega_{\Int}(\kA)$ et donc l'application $\alpha^{*}\rho^{*}$ sert à ``remonter'' cet élément dans $\Oder(\kA)$.
Nous rappelons que $\Omega_{\Int}^{1}(\kA)\simeq \kA \otimes \kA_{0}^{*} \supset  \kA \otimes \kA^{*}$ est naturellement un $Z(\kA)$-module, c'est-à-dire un module sur $\cF(M)$. Ainsi, $\varphi\in \Omega_{\Int}^{1}(\kA)\simeq \kA \otimes \kA_{0}^{*}$ peut être considérée comme la section d'un fibré associé à $P$ (le fibré principal correspondant à $\cE$) dont les fibres sont isomorphes à $M_{n}\otimes \ksl_{n}^{*}$. La dérivée covariante $\nabla_{A}$ est donc la connexion sur ce fibré correspondant à la connexion $\nabla$.
 
\subsubsection{Action de ``Maxwell'' non commutative}
\label{sec:action-de-maxwell}
Nous sommes maintenant prêts à écrire l'équivalent de l'action de Maxwell pour la connexion non commutative $\omega$ qui prend la forme suivante:
\begin{align*}
  S[\omega] &=  \Vert\Omega\Vert^{2} =  \int_{M}\int_{n.c.} \Omega \star \Omega \ .
\end{align*}
En se servant de la décomposition en blocs de la métrique, on peut voir que cette action est essentiellement composée de trois termes:
\begin{equation*}
  \begin{aligned}
    S[\omega] &=  \Vert F_{A+\ta} - \varphi(F_{A})\Vert^{2} + \Vert\nabla_{A}\varphi+[\ta,\varphi] \Vert^{2} + \Vert\varphi^{2} - \varphi\Vert^{2}
  \end{aligned}
\end{equation*}
où
\begin{equation*}
  \begin{aligned}
    \Vert F_{A+\ta} - \varphi(F_{A})\Vert^{2} &= \tr\int_{M}  (F_{A+\ta} - \varphi(F_{A}))\star_{M}(F_{A+\ta} - \varphi(F_{A}))\\
    \Vert\nabla_{A}\varphi+[\ta,\varphi] \Vert^{2} &= \int_{M}\int_{n.c.} (\nabla_{A}\varphi+[\ta,\varphi])\star (\nabla_{A}\varphi+[\ta,\varphi])\\
    \Vert\varphi^{2} - \varphi \Vert^{2} &= \int_{M}\int_{n.c.} ((\varphi\circ \alpha)² - \varphi(\alpha²))\star ((\varphi\circ \alpha)² - \varphi(\alpha²)) \ .
  \end{aligned}
\end{equation*}
Nous pouvons donner les expressions locales de chacun de ces termes en se restreignant au-dessus d'un ouvert trivialisant $U$  de $M$:
\begin{equation*}
  \begin{aligned}
    \Vert F_{A+\ta} - \varphi(F_{A})\Vert^{2}\loc &= \tr\int_{U} \frac{1}{2}
    \left(F^{A+\ta}_{\mu \nu} - \varphi(F_{\mu \nu})\right) 
    h^{\mu\rho}h^{\nu\sigma}  
    \left(F^{A+\ta}_{\rho \sigma} - \varphi(F_{\rho \sigma}) \right)\\
        \Vert\nabla_{A}\varphi+[\ta,\varphi] \Vert^{2}\loc &= \tr \int_{U}
    \left( \nabla_{\mu}\varphi_{b} + [\ta_{\mu}, \varphi_{b}] \right) 
    h^{\mu\rho}h_{\Int}^{bc}
   \left ( \nabla_{\rho}\varphi_{c} + [\ta_{\rho}, \varphi_{c}] \right) \\
        \Vert\varphi^{2} - \varphi\Vert^{2}\loc &= \tr \int_{U} \frac{1}{2}
   \left( [\varphi_{a},\varphi_{b}] -\varphi_{c}C_{ab}^{c} \right)
      h_{\Int}^{ae}h_{\Int}^{bf}
    \left( [\varphi_{e},\varphi_{f}] -\varphi_{g}C_{ef}^{g}\right)
  \end{aligned}
\end{equation*}
et on a noté:
\begin{equation*}
  \begin{aligned}
    F^{A+\ta}_{\mu \nu} &= \partial_{\mu}(A_{\mu} +\ta_{\nu})- \partial_{\nu}(A_{\nu}+\ta_{\mu})+[A_{\mu}+\ta_{\mu}, A_{\nu} + \ta_{\nu}] \\
    &=F_{\mu \nu} + \nabla_{\mu}\ta_{\nu}- \nabla_{\nu}\ta_{\mu}+[\ta_{\mu},\ta_{\nu}] \ .
\end{aligned}
\end{equation*}
Cette action généralise naturellement l'action~(\ref{eq:Yang-Mills-nc}) que nous avons obtenue pour une connexion sur le module libre $C^{\infty}(M)\otimes M_{n}(\gC)$ au paragraphe précédent. On retrouve bien cette action en prenant un fibré $\cE$ trivial et un potentiel de jauge $A_{\mu}=0$.
On voit que la principale différence avec  l'action~(\ref{eq:Yang-Mills-nc}) est localement l'introduction d'une connexion de référence venant d'une structure Riemannienne sur $\kA$.
Cette connexion est nécessaire pour décomposer correctement les différentes expressions locales et afin d'obtenir des expressions qui se recollent ``bien'' de manière globale sur $M$.
Notons que de manière globale, des effets dus à la topologie de la variété de base peuvent également intervenir.
En effet, la présence de classes caractéristiques non nulles décrites à partir de l'algèbre des endomorphismes dans la section \ref{sec:classe-car} et pouvant être représentées par la connexion de référence $A$ peut mener à des effets inattendus.

Dans le cas où la signature de la métrique $h$ est positive, c'est-à-dire lorsque $h\loc$ est une métrique euclidienne, la fonctionnelle $S$ est positive et son minimum est atteint lorsque:
\begin{equation*}
  \begin{aligned}
    &    F^{A+\ta}_{\mu \nu} = \varphi(F_{\mu \nu}) \\
    &    [\varphi_{a},\varphi_{b}] = \varphi_{c}C_{ab}^{c}    \\
    &     \nabla_{\mu}\varphi_{b} + [\ta_{\mu}, \varphi_{b}]  =0 \ .
  \end{aligned}
\end{equation*}
On peut exhiber deux solutions particulières qui sont:
\begin{align*}
  & \left\{
      \begin{aligned}
      F^{A+\ta}_{\mu \nu} &= 0\\
      \varphi &= 0
    \end{aligned}
  \right. &
&\mathrm{et}&
  & \left\{
      \begin{aligned}
 &       \ta_{\mu} =  a_{\mu} \gone \ (\mathit{i.e. } \ta_{\mu}^{a}=0) &
&\hfill \mathrm{et}&
        &\partial_{\mu}a_{\mu} - \partial_{\nu}a_{\mu} = 0\\
&        \varphi_{a} = E_{a} &
&\Leftrightarrow&
& \varphi=i\theta \circ \ad \simeq \Id \   (\text{sur $\kA_{0}$}) \ .
\end{aligned}
\right.
  \end{align*}
  On remarque que pour le vide $\varphi=\Id $, le  champ de jauge $A_{\mu}$ n'intervient pas dans la paramétrisation de la famille de solutions décrite.
  Au contraire, pour le vide $\varphi=0$, la courbure du champ de jauge $A+\ta$ doit être nulle donc le champ de jauge $A_{\mu}$  intervient explicitement dans la paramétrisation des différents vides.
  A priori, ici, le champ de jauge $A_{\mu}$ n'a pas de dynamique car, comme nous l'avons vu, celui-ci est donné par la structure Riemannienne sur $\kA$ que nous avons supposée fixée.
  Ainsi, $A$ joue le rôle d'une connexion de référence.
Il est clair que dans la situation triviale le champ de jauge $A_{\mu}=0$ est une bonne connexion de référence et ses fluctuations sont décrites par la forme tensorielle $\ta$. 
On pourrait imaginer des situations où la variété de base est non triviale et où le fibré $\cE$ peut également être non trivial comme par exemple dans l'étude des configurations instantoniques sur $S^{4}$ (le compactifié de $\gR^{4}$).

D'autre part, nous avons vu que dans la situation triviale ($A_{\mu}=0$), du point de vue de la théorie quantique des champs, il y a un mécanisme analogue au mécanisme de Higgs, c'est-à-dire que pour la configuration $\varphi=0$, la masse du champ $\ta_{\mu}$ est nulle, alors que pour la configuration $\varphi=\Id$, le champ $\ta_{\mu}$ acquiert une masse.
Ainsi, ce mécanisme se superpose au mécanisme géométrique faisant intervenir $A$.
Une étude plus systématique du mélange de ces deux mécanismes reste encore à faire et notamment l'étude du mécanisme de Higgs dans la situation où la connexion de référence est non triviale.
Notons enfin qu'il est possible de considérer des variantes de ces mécanismes en prenant des modules projectifs de type fini plus généraux pouvant s'identifier localement aux modules $C^\infty(\gR^{d}) \otimes M_{K,n}(\gC)$.

Il serait également possible de donner une dynamique à la structure Riemannienne sur $\kA$.
Il faudrait pour cela ``brancher'' l'action de Einstein-Hilbert associée à cette structure Riemannienne. 
Le champ de jauge $A$ aurait alors une dynamique et il serait intéressant de comprendre comment celle-ci pourrait influencer le mécanisme de Higgs présenté précédemment.
Le fait que le mécanisme de Higgs soit couplé à une structure Riemannienne est intéressant du fait que ce mécanisme est responsable de la masse des particules élémentaires dans le modèle standard et leur permet donc d'interagir avec le champ gravitationnel.


\chapter{Théorie de Born-Infeld et généralisations}\label{chap:theorie-de-born}

Nous allons dans ce chapitre passer à une étude relativement différente de celles faites dans les chapitres précédents.
Il s'agit en effet de l'étude de la théorie de Born-Infeld qui est une théorie de champs non linéaires généralisant la théorie de Maxwell.
Cette théorie développée dans les années 1920 fut étudiée à plusieurs reprises et dans différents contextes.
Elle fut retrouvée récemment en théorie des cordes comme étant, dans une certaine approximation, l'action effective des états de masse nulle d'hélicité $1$ de la théorie. Ces états correspondent à des champs de jauge pour une théorie de jauge avec groupe de structure $U(1)$, et l'action de Born-Infeld pour ces champs intervient de manière essentielle pour décrire la dynamique d'objets étendus appelés $D$-branes.
Dans ce cadre, lorsque l'on considère $N$ $D$-branes coïncidentes en interaction, il est naturel de remplacer le groupe de structure $U(1)$ par le groupe $U(N)$, l'entier $N$ correspondant aux valeurs que prennent les facteurs de Chan-Patton.
La question de définir une action de Born-Infeld pour des champs de jauge non abéliens se pose alors.
Nous nous proposerons d'utiliser la théorie de ``Maxwell'' non commutative, présentée dans le chapitre précédent, afin de donner une généralisation ``naturelle'' de l'action de Born-Infeld.
Nous allons présenter les travaux effectués dans \cite{serie:03,serie:03:II,serie:04}.
Enfin, nous étudierons les solutions des théories de champs non linéaires obtenues dans ce cadre.

\section{Modèle de Born et Infeld}\label{sec:modele-de-born}
La théorie de Born-Infeld est une théorie non linéaire généralisant la théorie de Maxwell de l'électro-magnétisme.
Il est bien connu que dans la théorie de Maxwell, les charges ponctuelles ont une énergie infinie et que les champs peuvent prendre des valeurs arbitrairement grandes lorsqu'on s'approche de ces charges.
Ainsi, après la découverte de l'électron, les physiciens ont cherché des modèles dans lesquels l'électron peut être représenté par une distribution de charge étendue.
Un des modèles qui eut le plus de succès fut celui proposé par G. Mie~\cite{mie:12} en 1912, un modèle d'électro-statique dans lequel le champ électrique ne peut dépasser une certaine borne et dans lequel toute distribution de charge bornée a une énergie finie.
Par analogie avec la notion de  vitesse maximale $c$ en relativité restreinte, G. Mie a introduit la notion de {\it champ électrique maximal}, noté $E_{0}$.
Pour cela, il a modifié la théorie de Maxwell en introduisant le  Lagrangien\footnote{Ce que l'on appellera par la suite ``Lagrangien'' correspond en fait à ce que l'on appelle habituellement ``densité Lagrangienne'' en physique.} suivant:
\begin{align}
L_{Mie} &= \sqrt{1 - \frac{\vec{E}^{2}}{E_0^2}} \ .
  \label{Mie1}
\end{align}
Il était clair qu'un tel Lagrangien ne pouvait décrire une théorie invariante sous l'action du groupe de Poincaré du fait de l'absence de champ magnétique.
On peut montrer que tout scalaire formé à partir du champ électrique et du champ magnétique, invariant de Lorentz, peut s'exprimer en terme des deux invariants $P= \vec{E}^{2}- \vec{B}^{2}$ et $Q= (\vec{E}\cdot \vec{B})^{2}$.
Ainsi, en 1934, Born proposa~\cite{born:34} une généralisation du Lagrangien de Mie consistant à remplacer le terme ``$\vec{E}^{2}$'' par l'invariant relativiste ``$\vec{E}^{2}- \vec{B}^{2}$'':
\begin{align}
L_{Born}&= \sqrt{1 - \frac{\vec{E}^{2} - \vec{B}^{2}}{\beta^2}} \ ,
  \label{eq:Born}
\end{align}
où $\beta$ est une constante dimensionnée correspondant au champ électrique $E_{0}$ de Mie.

Un peu plus tard,  Born et Infeld ont essayé de rendre cette théorie compatible avec la théorie de la relativité générale d'Einstein.
Il constatèrent que l'on pouvait généraliser la densité\footnote{Nous appelons ici densité, une quantité pouvant être intégrée sans avoir besoin d'une forme volume, c'est à dire en utilisant directement la mesure de Lebesgue sur chaque ouvert de la variété.} $\sqrt{|\det(g_{\mu \nu}) |}$ correspondant à la forme volume usuelle en considérant la racine carrée du déterminant  d'un tenseur $T_{\mu \nu}$ quelconque deux fois covariant.
Ainsi ils ont identifié la partie symétrique d'un tel tenseur au tenseur métrique et la partie anti-symétrique au tenseur représentant la $2$-forme de courbure associée au champ électromagnétique en posant $T_{\mu \nu} = g_{\mu \nu}+  \beta^{-1} F_{\mu \nu}$.
De cette manière, ils obtinrent une action généralisant celle obtenue à partir du Lagrangien de Mie et qui était automatiquement invariante sous l'action du groupe des difféomorphismes et du groupe des transformations de jauge.
L'action de Born-Infeld~\cite{born_infeld:34} est la suivante:
\begin{align} \label{BI}
    S_{BI}[g,F] &= \int_{\Mlor} {\cal{L}_{BI}}(g,F) d^4x = \int_{\Mlor}{L_{BI}(g,F)}
    \sqrt{|g|}d^4x   \\               
    &= \int_{\Mlor} \beta^2 \left( \sqrt{|\det(g_{\mu \nu}) |} -
      \sqrt{|\det(g_{\mu \nu}+\beta ^{-1} \,
        F_{\mu \nu}\,)\,|} \right) d^4x   \label{BI-det} \\
    &= \int_{\Mlor} \beta^2 \left(1-\sqrt{ 1+ \frac{1}{\beta^2} (F,F)
        - \frac{1}{4 \beta^4}(F,{\star F})^2}\right) \sqrt{|g|}d^4x \\
    &= \int_{\Mlor} \beta^2\left(1-\sqrt{1+ \frac{1}{\beta^{2}}
        (\vec{B}^2 - \vec{E}^2) - \frac{1}{\beta^{4}}(\vec{E}.\vec{B})^2}\right)
    \sqrt{|g|}d^4x \ ,
\end{align}
où $\Mlor$ est une variété Lorentzienne de dimension $4$ et $d^4x= dx^0  dx^1 dx^2 dx^3$ représente la mesure de Lebesgue sur un ouvert de $\Mlor$ (nous faisons ici un abus de notation en faisant apparaître explicitement des coordonnées locales sous l'intégrale, le but est de donner une idée de l'expression locale du Lagrangien $L_{BI}$). 
On note encore  $\mathcal{L}_{BI}$ la densité associée au Lagrangien $L_{BI}$.

\begin{rem}
  Lorsque la dimension $d$ de la variété de base $\Mlor$ est quelconque, on appellera, dans la suite, action de Born-Infeld l'action obtenue en remplaçant  la mesure $d^{4}x$ par $d^{d} x$ dans l'écriture (\ref{BI-det}) de l'action de Born-Infeld.
\end{rem}

Si on définit les invariants:
\begin{align}
  P &= \frac{1}{4} F_{\mu \nu} F^{\mu \nu}=\frac{\vec{E}^{2}-\vec{B}^{2}}{2} &
  &{\rm et}&
  S &=\frac{1}{4} F_{\mu \nu} {\tilde{F}}^{\mu \nu}= \vec{E}\cdot\vec{B} , &
  &{\rm avec}& 
  {\tilde{F}}^{\mu \nu} &= \frac{1}{2} \, \, \epsilon^{\mu \nu \lambda \rho } \, F_{\lambda \rho} \ ,
\end{align}
on peut réécrire le Lagrangien de Born-Infeld de la manière suivante:
\begin{equation}
{L}_{BI}  = \beta^2 \, \biggl[ 1 - \sqrt{1 + 2 P - S^2} \biggr] \ .
\label{eq:bi-inv-4d}
\end{equation}

\subsection{Propriétés de l'action de Born-Infeld}

\subsubsection{Propriétés basiques}
\begin{itemize}
\item 
La théorie de Maxwell est retrouvée lorsqu'on prend la limite $\beta \to \infty$:
\begin{equation}\label{limite}
  \begin{split}
    {S_{BI}}
    &=  - \int_{\Mlor} \frac{1}{2} (F,F) \sqrt{|g|}d^4x + o(\frac{1}{\beta ^2})\\
    &= - \frac{1}{2}\int_{\Mlor} F\wedge {\star F} + o(\frac{1}{\beta
      ^2})\\
    &= - \int_{\Mlor}\frac{1}{2}(\vec{B}^2 - \vec{E}^2) \sqrt{|g|}d^4x +
    o(\frac{1}{\beta ^2}) \ .
  \end{split}
\end{equation}
\item
Le champ électrique a une limite supérieure $\beta$ lorsque le champ magnétique est nul:
\begin{equation}\label{Mie}
  {L_{BI}}|_{B=0}
  =\beta^2 \left(1-\sqrt{1- \beta^{-2} \vec{E}^2}\right) \ .
\end{equation}
Pour cette raison, l'énergie d'une particule ponctuelle chargée placée à l'origine est finie et le champ électrique reste borné.
\end{itemize}

\subsubsection{Électrodynamique non linéaire}
Notons que dans le cadre de la théorie de Born-Infeld, le terme ``charge ponctuelle'' ne veut pas forcément dire distribution de charge ponctuelle. 

En effet, il faut interpréter la théorie au sens de l'électrodynamique dans les milieux (Maxwell-Faraday).
Dans cette théorie, on a un champ électrique $\vec{E}$, un champ magnétique $\vec{H}=\vec{H}(E,B)$, un champ de déplacement (ou induction électrique) $\vec{D}=\vec{D}(E,B)$ et une induction magnétique $\vec{B}$, dont les équations du mouvement sont données par les équations de Maxwell-Faraday (en présence de sources macroscopiques):
\begin{equation}
\begin{aligned}
  \vec{\nabla} \cdot \vec{D} &= 4\pi \rho  &  \vec{\nabla} \times \vec{H} -\partial_t \vec{D}    &=  4\pi \vec{j}\\
  \vec{\nabla} \cdot \vec{B} &=  0  &  \vec{\nabla} \times \vec{E} + \partial_t \vec{B}   &=  0 \ .
\end{aligned}
\label{eq:Maxwell-Faraday}
\end{equation}
Dans certaines situations, les relations entre $E,H,D$ et $B$, ainsi que les équations de Maxwell-Faraday peuvent être obtenues à partir d'un principe de moindre action. Il faut alors se donner un Lagrangien $L(E,B)$ dépendant seulement de $E$ et $B$ (des deux invariants $P(E,B)$ et $Q(E,B)$ si l'on veut avoir l'invariance relativiste et l'invariance de jauge) et les champs $D$ et $H$ sont définis comme étant les dérivées fonctionnelles de l'action par rapport à $E$ et $B$.
On peut alors définir une densité de charge effective $  \rho_{eff}=\vec{\nabla}\cdot \vec{E}$.
La densité d'énergie du système est une fonction de $D$ et $B$ et est la transformée de Legendre du Lagrangien:
\begin{align}
  \cH(D,B) &= \vec{E}\cdot\vec{D}- L(E,B) \ .
\label{eq:H-legendre}
\end{align}

\begin{rem}
On peut faire un parallèle avec la mécanique Lagrangienne: le couple $(B,E)$ est à comparer au couple $(q,\dot{q})$ représentant la position et la vitesse d'une particule. $D$ est la variable conjuguée de $E$ et est donc similaire à l'impulsion $p$, variable conjuguée de $q$, tandis que $H$ serait l'analogue d'une force.
On peut alors comparer les équations de Maxwell-Faraday aux équations d'Hamilton-Jacobi.
\end{rem}

\subsubsection{Retour à la théorie de Born-Infeld \dots}

Les équations du mouvement pour la théorie de Born-Infeld peuvent se mettre sous la forme~(\ref{eq:Maxwell-Faraday}) d'équations de Maxwell-Faraday.
Le vecteur déplacement et le champ magnétique sont alors reliés au champ électrique et à l'induction magnétique de la manière suivante:
\begin{equation}
\left\{\begin{aligned}
  \vec{D}&= \frac{\partial L}{\partial \vec{E}} = \frac{\vec{E}+ (E\cdot B)\vec{B}}{\sqrt{1-E^{2}+B^{2}-(E\cdot B)^{2}}}\\
  \vec{H}&= -\frac{\partial L}{\partial \vec{B}} =\frac{\vec{B}- (E\cdot B)\vec{E}}{\sqrt{1-E^{2}+B^{2}-(E\cdot B)^{2}}}
\end{aligned}\right. \ ,
\end{equation}
et la densité d'énergie a la forme suivante:
\begin{equation}
  \cH(B,D)=\sqrt{1+B^{2}+D^{2}+(B\wedge D)^{2}} - 1 \ .
\end{equation}
Nous avons mis la constante $\beta$ égale à $1$, celle-ci peut être rétablie  à tout moment par une redéfinition des champs.

On voit que l'énergie est toujours positive et que les champs $D$ et $B$ n'ont pas de restriction dans les valeurs qu'ils peuvent prendre.
On note également l'invariance de la théorie sous les rotations ``électriques-magnétiques'':
\begin{equation}
  \begin{aligned}
    E+i H &\to e^{i\alpha} (E+i H) \\
    D + i B &\to e^{i\alpha} (D+i B)
  \end{aligned}
\end{equation}
 qui correspond à ce qui est souvent appelé dans la littérature la \textbf{dualité ``électrique-magnétique''} dans la théorie de Born-Infeld.

Enfin d'après (\ref{eq:H-legendre}), le champ électrique et le champ magnétique peuvent s'exprimer en fonction du vecteur déplacement et de l'induction magnétique en dérivant la densité d'énergie par rapport au vecteur déplacement et à l'induction magnétique:
\begin{equation}
\left\{\begin{aligned}
  \vec{E}&= \frac{\partial \cH}{\partial \vec{D}} = \frac{\vec{D} (1+B^{2}) - (B\cdot D)\vec{B}}{\sqrt{1+B^{2}+D^{2}+(B\wedge D)^{2}}}\\
  \vec{H}&= \frac{\partial \cH}{\partial \vec{B}} = \frac{\vec{B} (1+D^{2}) - (B\cdot D)\vec{D}}{\sqrt{1+B^{2}+D^{2}+(B\wedge D)^{2}}}
\end{aligned}\right. \ .
\end{equation}

\begin{rem}
  Il est également possible d'introduire les autres transformations de Legendre de $L(E,B)$ et $\cH(D,B)$. 
  Entre autre, du fait de la symétrie sous l'échange de $B$ et $D$ de $\cH(D,B)$, la transformée de Legendre $\cH(B,D)-B\cdot H$ est une fonction de $D$ et $H$ et a la même forme que le Lagrangien de Born-Infeld $L(E,B)$. 
Pour cette raison, on dit que la théorie de Born-Infeld possède une \textbf{dualité de Legendre}.
\end{rem}

Nous pouvons maintenant voir ce qui se passe pour une charge ponctuelle électrique de charge $e$.
Nous avons donc:
\begin{align}
  \vec{\nabla} \cdot \vec{D} &= 4 \pi e \delta^{3}(r) &
  & \Rightarrow &
  \vec{D} &= \frac{e \hat{r}}{r^{2}} \ ,
\end{align}
où $\hat{r}=\vec{r}/r$.
Il correspond à cette solution le champ électrique et la densité de charge effective $\rho_{eff}=\vec{\nabla}\cdot \vec{E}$ suivants:
  \begin{align}
        \vec{E}&=\frac{e \hat{r}}{\sqrt{e^2+r^4}} &
        &\text{et} &
       \rho_{eff} &= \frac{2 e^{3}}{r (r^{4}+e^{2})^{\frac{3}{2}}} \ .
  \end{align}

Si l'on rétablit la constante $\beta$, on peut introduire un rayon caractéristique $r_{0}$ en posant:  $\beta= \frac{e}{r_{0}^{2}}$ et le champ électrique prend la forme suivante:
\begin{align*}
  \vec{E} &= \frac{\beta \hat{r}}{\sqrt{1+\left(\frac{r}{r_0}\right)^4}} \ .
\end{align*}         

On peut alors calculer l'énergie de cette solution:
\begin{align*}
  E &= \beta^2  \int_0^{\infty} \left( \sqrt{1 +\frac{1}{\left(\frac{r}{r_0}\right)^4} }-1\right) r^2 dr \\
  &= \frac{\Gamma(1/4)^2}{6 \sqrt{\pi}} \  \frac{e^2}{r_0} \\
  &  \simeq 1.23605 \  \frac{e^2}{r_0} \ .
\end{align*}

Si on identifie cette énergie à l'énergie de masse de la particule en posant $E = m_0 c^2$ alors on trouve pour l'électron:
\begin{align*}
  r_0 &= 1.23605 \cdot \frac{e^2}{m_0 c^2} \simeq 3.48 \mathrm{fm} \\
  \beta &\simeq 1.8 \cdot 10^{18}   \mathrm{volt/cm}  \ .
\end{align*}      

Ainsi, on voit que dans la théorie de Born-Infeld, la masse des particules peut-être complètement reliée à sa charge électrique et que les divergences présentes dans la théorie de Maxwell sont éliminées.
Nous avons également vu que l'on pouvait associer à une charge ponctuelle une notion de densité de charge (la densité effective de charge $\rho_{eff}$).

\begin{rem}
Il est également possible de considérer des solutions de type dyon~\cite{schwinger:69} généralisant la solution du monopole électrique de Born-Infeld~\cite{chernitsky:99}. Une telle solution peut être obtenue en faisant une rotation ``électrique-magnétique'' de la solution précédente.
\end{rem}

Les ordres de grandeurs que l'on trouve pour l'électron justifient que l'on utilise la théorie de Maxwell dans le domaine de l'électrodynamique classique. Ainsi, son utilisation ne s'est pas vraiment avérée nécessaire en électrodynamique classique, mis à part pour rendre cohérente la théorie.

D'autre part, nous savons que depuis que  Dirac a introduit son équation pour l'électron, celui-ci est considéré comme une particule ponctuelle.
De manière plus précise, dans le cadre de la théorie quantique des champs, l'électron est considéré comme un  ``quantum'' de champ.
Ainsi, l'intérêt pour les modèles classiques de particules chargées de type Born-Infeld a considérablement diminué.

On pourrait cependant se demander, si un tel Lagrangien ne pourrait pas servir de Lagrangien fondamental pour une théorie quantique et cela fut proposé par plusieurs personnes comme  Dirac~\cite{dirac:62}.
Le problème est que l'on ne sait pas définir une théorie quantique des champs pour un Lagrangien non polynomial (du moins en dimension 4) et de plus, l'introduction d'une telle théorie au niveau quantique ne s'est pas avérée utile.
En effet, la théorie de l'électrodynamique quantique (QED), basée sur la théorie de Maxwell, et ses généralisations aux théories de jauge non abéliennes ont abouti au \textit{modèle standard} comprenant les trois interactions: faible, électro-magnétique et forte. Ce modèle s'est avéré être en très bon accord avec l'expérience.
Ainsi, la théorie de Born-Infeld a longtemps été considérée comme une curiosité pour les physiciens.

\subsection{``Redécouvertes'' de la théorie de Born-Infeld}
\subsubsection{Approche de Boillat}

Il faudra attendre les années 1970 pour voir réapparaître le modèle de Born-Infeld dans  les travaux de G. Boillat~\cite{boillat:70} et Pleba\'nski~\cite{plebanski:70} qui visaient à étudier les propriétés de la propagation des discontinuités des champs électriques et magnétiques dans le cadre de la relativité générale.
En étudiant les Lagrangiens généraux ${\cal{L}} \, (P,S)$, dépendant donc des deux invariants $P$ et $S$, il trouvèrent que l'electrodynamique de Born-Infeld est la seule théorie pour laquelle il y a absence de biréfringence, \textit{i.e.} il y a propagation le long d'un seul cône de lumière et absences d'ondes de choc.
Ainsi, ce fut une manière de particulariser la théorie de Born-Infeld (il y a en fait deux autres Lagrangiens ${\cal{L}} = P/S$ et le Lagrangien de Maxwell qui sont particularisés dans leurs études). 
Il s'est également avéré que le Lagrangien de Born-Infeld a des propriétés de dualité électrique-magnétique et dualité de Legendre (voir~\cite{bialynicki-birula,gibbons:01:II,gibbons:95:II} et remarques précédentes).

\subsubsection{La théorie de Born-Infeld en théorie des cordes}
Il y a eu une renaissance de l'intérêt pour la théorie de Born-Infeld en théorie des cordes à partir de 1985, lorsque l'action de Born-Infeld fut retrouvée~\cite{fradkin:85} comme étant une partie (dans l'approximation de champs constant) de l'action effective, des excitations de masse nulle et spin 1 du champ de cordes.
Ainsi, l'action de Born-Infeld peut être comparée à l'action effective d'Euler-Heisenberg~\cite{euler-heisenberg:35} en électrodynamique quantique.

La dynamique des états sans masse de cordes ouvertes se propageant dans un espace plat peut être décrit, à couplage faible, par une action effective $S(A)$ pour un potentiel de jauge $SU(N)$, $A$.
Cette action admet une expansion en $\alpha'$ dont les termes sont exprimés en terme de puissances de la courbure $F$ associée à $A$ et de ses dérivées covariantes.
Cette action admet plusieurs interprétations différentes en théorie des cordes.
Tout d'abord, elle reproduit, à l'ordre des arbres, l'amplitude de disque calculée directement en théorie des cordes.
Deuxièmement, les équations du mouvement venant du principe variationnel associé à $S$ correspondent à la condition d'annulation de la fonction $\beta$, due à la symétrie conforme du modèle sigma de cordes ouvertes avec des interactions du type ``boucles de Wilson'' $\tr P \exp{i\int_{\partial \Sigma}A}$ sur le bord $\partial \Sigma$ de la feuille d'univers des cordes ouvertes (voir~\cite{abouelsaood:86} pour le cas des cordes bosoniques et dans l'approximation de champs lentement variables).
Il a en fait été montré par Andreev et Tseytlin~\cite{andreev:88}  que l'action effective peut s'identifier avec la fonction de partition du modèle sigma décrit ci-dessus (et de manière préliminaire par Fradkin et Tseytlin~\cite{fradkin:85} dans la limite de champs constants).
Enfin, par $T$-dualité, l'action $S(A)$ décrit également la dynamique à couplage faible de $D$-branes de toutes dimensions.
L'action de Bon-Infeld correspond donc à une action effective à tous les ordres en $\alpha'$ pour un champ de jauge abélien et dans la limite d'un champ lentement variable.
D'autre part, il est naturel dans ce cadre de s'intéresser à des généralisations de la théorie de Born-Infeld pour des théories de jauge non abéliennes, car à priori, ce type d'action devrait intervenir dans l'action effective pour les champs de jauge non abéliens. Nous reviendrons sur ce point un peu plus loin.

\section{Généralisation non abélienne du modèle de Born-Infeld}

\subsection{Rappels sur les théories de jauge et notations}
Une interprétation des champs de jauge en physique consiste à les voir comme étant des $1$-formes de connexion sur un fibré principal de groupe de structure  $G$   sur une variété de dimension $4$ ($\gR^{4}$  la plupart du temps).
Une $1$-forme de connexion est une $1$-forme sur le fibré principal, à valeur dans l'algèbre de Lie $\kg$ du groupe $G$.
Dans un système de coordonnées locales, nous avons:
\begin{equation}
A = A^a_{\mu}\, dx^{\mu} \, L_a
\end{equation}
où $L_a , \, \ \ a = 1,2,...N=\dim(G)$ est une base de la représentation adjointe de $\kg$.
Dans bien des cas, on prend plutôt une connexion dans un fibré associé, et la représentation de l'algèbre de Lie peut alors être différente, en particulier lorsque le champ de jauge est couplé avec des spineurs~\cite{domokos:79,kerner:81,kerner:83}.

Il est toujours possible de plonger l'algèbre de Lie dans une algèbre enveloppante et d'utiliser le produit tensoriel:
\begin{equation}
A = A^a_{\mu} \, d x^{\mu} \otimes \, T_a \, ,
\end{equation}
où $\{T_a\}$ sont les matrices représentant la base $\{L_{a}\}$.

De même, la courbure est une $1$-forme prenant ses valeurs dans l'algèbre enveloppante:
\begin{equation} F = dA + \frac{1}{2}[A, A] = ( \, F^a_{\mu \nu} \, dx^{\mu}
  \wedge d x^{\nu} \, ) \otimes \, T_a \ .
\end{equation}

Nous allons maintenant voir quelques possibilités de généralisation de l'action de Born-Infeld pour des champs de jauge non abéliens.

\subsection{Motivations pour une théorie de Born-Infeld non abélienne}

Une des motivations principales pour trouver une théorie de Born-Infeld non abélienne est donnée par la théorie des cordes où ce type d'actions doit intervenir dans une partie de l'action effective permettant de décrire la dynamique des états de masse nulle et spin $1$.
Ce type d'actions permet également de décrire la dynamique d'objets étendus appelés D-branes correspondants à des conditions au bord de Dirichlet pour les cordes ouvertes.

Une autre motivation pour la théorie de Born-Infeld et ses généralisations non abéliennes, ou d'autres théories du même type possédant des non linéarités, est la possibilité d'avoir des solutions de type soliton ayant le même type de comportement que des solutions obtenues en théorie de Yang-Mills couplées à des champs scalaires ou à la gravitation.
Ces solutions peuvent avoir diverses interprétations physiques, par exemple, elles peuvent servir à décrire certains types de trous noirs chargés en relativité générale ~\cite{bartnik:88}, ou encore peuvent fournir des modèles effectifs pour décrire la dynamique de particules de type ``glueballs''.

En effet, de telles solutions ne peuvent être obtenues à partir de la théorie de Yang Mills.
Une démonstration rigoureuse de ce fait est donnée dans \cite{coleman:77} et nous allons donner ici quelques arguments qui en donnent une vision intuitive.
Pour la théorie de Yang-Mills sur espace plat, la densité associée au Lagrangien est: 
\begin{align*}
  \mathcal{L}_{YM} &= -\frac{1}{4} g_{AB} \, F^A_{\mu \nu} \, F^{B \, \mu \nu} \ ,
\end{align*}

Cette théorie est invariante conforme et donc le tenseur d'énergie impulsion est sans trace:
\begin{equation}
  T^{\mu}_{\, \ \ \mu} = -T_{00} + {\displaystyle{\sum_{i=1}^{3}}} T_{ii} = 0 \ .
  \label{kern-eq.1.19}
\end{equation}
Ainsi, du fait de la positivité de l'énergie, \textit{i.e.} $T_{00} > 0$, la somme de pressions principales est positive: $\sum T_{ii} > 0$, ce qui veut dire que le ``fluide de Yang-Mills'' est soumis à des forces répulsives qui empêchent l'existence de configurations statiques, non singulières, d'énergie finie.

Les diverses possibilités pour pouvoir obtenir de telles solutions sont alors de coupler la théorie de Yang-Mills à d'autres types de champs ou bien d'introduire des non linéarités permettant de briser l'invariance conforme.
Par exemple, en présence de champs scalaires, on peut montrer l'existence de solutions de type soliton, comme la solution de 't Hooft~\cite{'thooft:74} et Prasad-Sommerfield~\cite{prasad:75} pour les monopoles magnétiques.
Lorsque la théorie de Yang-Mills est couplée à la gravitation, il existe des solutions de type sphaleron découvertes par Bartnik et  McKinnon dans \cite{bartnik:88}.
Nous allons par la suite considérer des généralisations de la théorie de Yang-Mills faisant intervenir des non linéarités en s'inspirant de la théorie de Born-Infeld pour l'électrodynamique. Nous verrons que dans certains cas des solutions de type soliton peuvent exister.

\subsection{Généralisation la plus simple en dimension $4$}

Une généralisation de l'action de Born-Infeld en dimension $4$ a été proposée par D. Gal'tsov et R. Kerner dans~\cite{gal'tsov:00}.
L'action qu'ils ont considérée est la suivante:
\begin{align}
  S_{GK}[A] &= \int_{\Mlor} \sqrt{1+ \frac{1}{2\beta} g_{ab} F^{a \, \mu \nu} F^{b}_{\mu \nu} - \left(\frac{1}{4\beta} g_{ab} \tilde{F}^{a \, \lambda \rho} F^{b}_{\lambda \rho}\right)^{2}  } \times \sqrt{|g|} d^{4}x \ .
\end{align}
 C'est une généralisation de l'action de Born-Infeld écrite sous la forme~(\ref{eq:bi-inv-4d}), où l'on a remplacé les deux invariants relativistes $P$ et $S$ de la théorie abélienne par les invariants:
\begin{align*}
  P &= \frac{1}{4} g_{ab} F^{a \, \mu \nu} \, F^{b}_{\mu \nu} \\
  S &=  \frac{1}{4} g_{ab} \, \tilde{F}^{a \, \lambda \rho} \, F^{b}_{\lambda \rho}
  =\frac{1}{8} g_{ab} \, \epsilon^{\mu \nu \lambda \rho} \, F^a_{\mu \nu} F^{b}_{\lambda \rho} \ .
\end{align*}

Dans~\cite{gal'tsov:00}, les solutions aux équations du mouvement ont été étudiées pour $G=SU(2)$ et les auteurs ont montré l'existence de solutions de type sphaleron analogues à celles trouvées par Bartnik et McKinnon dans~\cite{bartnik:88}.

Cette généralisation \textit{ad'hoc} ne permet pas de généraliser l'action de Born-Infeld en dimension $d \neq 4$ du fait que l'invariant $S$ n'est alors pas bien défini.
Nous allons maintenant voir d'autres généralisations de l'action de Born-Infeld basées sur l'écriture du Lagrangien à partir d'un déterminant~(\ref{BI-det}) et qui en fournissent  une généralisation en toute dimension.

\subsection{Critères de généralisation}
\label{sec:crit-de-gener}
La manière la plus simple pour généraliser l'action de Born-Infeld serait à première vue de remplacer les nombres réels par des opérateurs hermitiens, conformément à ce qui est pratiqué en mécanique quantique ou encore en géométrie non commutative.
Ainsi, nous devrions faire les remplacements suivants pour passer du cas abélien au cas non abélien:
\begin{equation}
  \left\{\begin{array}{lcl}
      U(1)& \leftrightsquigarrow & G \\      
      i F_{\mu \nu}  & \leftrightsquigarrow &  F_{\mu \nu}^a \otimes T_a \\      
      g_{\mu \nu}   & \leftrightsquigarrow & g_{\mu \nu} \otimes
      \gone_{d_R} \ ,\\      
    \end{array} \right.
  \label{correspondance}
\end{equation}
où $\gone_{d_R}$ et $ iT_a$ sont des matrices hermitiennes.  

Par analogie avec le cas abélien, nous voulons que le Lagrangien satisfasse aux propriétés suivantes:
\begin{enumerate}
\item Nous devons retrouver la théorie de Yang-Mills usuelle dans la limite $\beta \to \infty$.
\item L'analogue non abélien du champ électrique doit être borné lorsqu'il n'y a pas de champ magnétique.
  Pour satisfaire cette contrainte, nous voudrons que l'expression polynomiale sous la racine commence par les termes: $1-\beta^{-2}(\vec{E}^a)^2 + ...$ lorsque $\vec{B}^a =0$.
\item L'action doit être invariante sous difféomorphismes.
\item L'action doit être réelle.
\end{enumerate}

Nous sommes maintenant confrontés au problème de donner un sens au déterminant intervenant dans l'action de Born-Infeld.
En effet, les entrées de la matrice formée par les composantes du tenseur de courbure $F_{\mu \nu}$ ne sont plus des nombres mais des matrices représentant des éléments d'une algèbre de Lie et donc le déterminant n'a plus de sens.
Il est donc nécessaire de trouver une généralisation à la notion de déterminant pour des matrices dont les entrées ne commutent pas.

\subsection{Prescription de la trace symétrisée}

Une solution proposée par Tseytlin~\cite{tseytlin:97} et qui semble correctement générer l'action effective de théorie des cordes jusqu'à l'ordre $4$ en $F$, consiste à prendre la trace symétrisée de l'action de Born-Infeld classique.
Cela consiste à :
\begin{itemize}
\item 
prendre le développement en puissance de $F$ du Lagrangien de Born-Infeld dans le cas abélien, puis 
\item 
remplacer dans chaque monome le tenseur de courbure abélien par le tenseur de courbure non abélien, puis 
\item 
symétriser chaque monome en $F$ afin de lever l'ambiguïté intervenant sur l'ordre dans lequel doivent être multipliées les matrices, puis
\item 
prendre la trace sur l'expression totale.
\end{itemize}

Le Lagrangien peut être formellement écrit de la manière suivante:
\begin{align}
  S_{sym}[A] &= \str \int_{\Mlor}   \sqrt{ \detm\left(g_{\mu \nu} \gone +iF^a_{\mu \nu} T_a\right) } d^d x\ .
\end{align}
où $d$ est la dimension de la variété $M$.

Après l'opération de symétrisation, tout se passe comme si les objets étaient commutatifs.
Cette action admet également une généralisation supersymétrique qui fut développée par Schaposnik {\it{et al}} (voir~\cite{gonorazky:98,christiansen:98,grandi:00} et les références qui y sont incluses).

\subsection{Prescription de Park (cas euclidien)}

Une autre manière de procéder est de généraliser le déterminant intervenant sous la racine.
Nous avons vu que le déterminant est une manière économe de générer une densité à partir d'un tenseur de rang $2$ (c'est d'ailleurs ce qui mena Born et Infeld à considérer leur Lagrangien). 
D'autre part, dans le cas des théories de jauge abéliennes, la courbure est invariante sous transformations de jauge, mais nous savons que dans le cas d'une théorie de jauge non abélienne, la courbure se transforme de manière homogène.
Habituellement, c'est la trace qui est utilisée pour fabriquer un invariant sous transformations de jauge (par exemple le Lagrangien de Yang-Mills), mais nous allons ici tenter d'utiliser un déterminant.
L'idée que nous allons suivre pour généraliser l'action de Born-Infeld, est donc d'utiliser un déterminant afin de générer un Lagrangien à la fois invariant sous difféomorphismes et sous transformations de jauge.

L'idée de Hagiwara et Park est de considérer un déterminant dans un espace de matrices plus gros obtenu après avoir fait le produit tensoriel des deux espaces de matrices $M_{4}(\gC)$ et $M_{d_{R}}(\gC)$.
Ainsi, il faut remplacer le déterminant sur les matrices $4 \times 4$  (noté par la suite $\detm$), par un  déterminant sur des matrices de taille $4d_{R}$, dont le indices sont indexés par des couples d'indices $(\mu,i)$, où $\mu$ correspond aux indices d'espace-temps et $i$ aux indices matriciel de la représentation $R$ (noté par la suite $\detmg$).

Afin de rester le plus proche de la situation abélienne, une idée naturelle à suivre est de plonger le tenseur de la métrique  $g_{\mu \nu}$ dans la même algèbre enveloppante que la $2$-forme de courbure $F$, en le remplaçant par $g_{\mu \nu} \otimes \gone_{N}$.
Nous pouvons maintenant remplacer le déterminant intervenant sous la racine dans le Lagrangien de Born-Infeld (\ref{BI}) en suivant la procédure (\ref{correspondance}) et en utilisant le déterminant $\detmg$.
Cela nous conduit au terme suivant:
\begin{align*}
  \detmg ( g_{\mu \nu} \otimes \gone_{d_R}+\beta^{-1}\,F^a_{\mu \nu} \otimes iT_a \, ) \ .
\end{align*}
Malheureusement, dans le cas où le groupe de structure n'est pas abélien, cette expression est un nombre complexe.

Une autre possibilité, suggérée par Hagiwara~\cite{hagiwara:81} puis reprise par Park~\cite{park:99}, est de considérer des générateurs anti-hermitiens et de prendre le produit tensoriel avec les composantes de la courbure $F$.
Park  proposa~\cite{park:99} l'action suivante:
\begin{equation}
  S_{Park}[F,g]
  = \int_{\Mlor} \alpha \left( \left|\detmg \left( g_{\mu \nu} \otimes 
\gone_{d_R}+\beta^{-1}\,F^a_{\mu \nu} \otimes T_a \, \right)\right|^{\frac{1}{2 d_R}} 
 -\sqrt{|g|} \right)  d^{d}x\ , 
  \label{park:BI}
\end{equation}
où $\alpha $ et $\beta$ sont des constantes réelles et positives et $d$ est la dimension de $M$.  
La racine d'ordre $2 d_R$ est introduite afin qu'il y ait invariance sous difféomorphismes.
En effet, nous sommes ainsi capables de factoriser l'élément de volume $\sqrt{|g|} d^{d}x$ et l'invariance sous difféomorphismes est plus claire si nous réécrivons le Lagrangien sous la forme suivante:
\begin{align} 
  L_{Park}(F,g) & = \alpha \left(\left| \detmg\left( \gone_{d \times d_R} 
      +  \beta^{-1}\,\hat{F}\,\right)\right|^{\frac{1}{2 d_R}} \, -1\right)
  \label{park:operateur} 
\end{align}  
et
\begin{align}   
  S_{Park}[F,g] &= \int_{\Mlor} L_{Park}(F,g) \sqrt{|g|}d^{d}x \ ,
\end{align}
où
\begin{align} 
  \hat{F} = \  \frac{1}{2} F^a_{\mu \nu} \hat{M}^{\mu \nu} \otimes T_a \, \ \
  \, \ \ \, \ \ \,
  (\hat{M}^{\mu \nu})^{\rho}_{\sigma}  =  \ g^{\rho \rho'}
  \delta^{\mu \nu}_{\rho' \sigma} \ ,
\end{align}
$\hat{F}$ est un endomorphisme de $\Mlor \otimes \gC^{d_R}$ et $M_{\mu \nu}$ sont les générateurs de l'algèbre de Lie du groupe de Lorentz (dans la représentation définissante, \textit{i.e.} de dimension $d$).
Il est également utile d'introduire les notations suivantes:
\begin{align}  
  \hat{F}^a &= \frac{1}{2} F^a_{\mu \nu} \hat{M}^{\mu \nu} \ .
\end{align}

Cette substitution mène à un Lagrangien  qui satisfait les points  1), 3), et  4), donnés dans la section \ref{sec:crit-de-gener}, mais pas le point 2).
Les Lagrangiens obtenus par cette technique font de plus intervenir des invariants d'ordre $3$ dans la courbure $F$, ce qui détruit l'invariance sous conjuguaison de charge, $F \mapsto -F$, et mène à une densité d'énergie mal définie.
Cela n'était en fait pas un problème pour Park, puisqu'il a travaillé sur un espace Euclidien.

\subsection{Une version Minkowskienne}
\label{sec:une-vers-mink}
Nous allons voir une généralisation légèrement différente de l'action de Born-Infeld. 
Cette généralisation fut proposée dans~\cite{serie:03}, qui est une publication faite lors de ma thèse.

Tout d'abord, commençons par reformuler l'action de Born-Infeld (dans le cas abélien) de manière légèrement différente:
\begin{equation}\label{BI:J}
  S_{BI}[F,g] =\int_{\Mlor}  \beta^2 \left( \sqrt{|g|} -
    \left|\detcm \left(\gone_2 \otimes g_{\mu \nu} +\beta^{-1} \, \J \otimes
  i F_{\mu \nu} \,\right)\,\right|^{\frac{1}{4}} \right) d^4x \ ,
\end{equation}
où $\J$ est une matrice complexe $2 \times 2 $ dont le carré est égal à $-\gone_2$.

Le Lagrangien est indépendant du choix de $\J$.
Dans (\ref{BI:J}) (voir aussi (\ref{correspondance})), l'unité imaginaire $i$ peut être considérée comme le générateur anti-hermitien de $\mathfrak{u}(1)$.
Dans la formule (\ref{BI:J}), nous utilisons une notation pratique pour spécifier sur quel espace de matrices est pris le déterminant.

Nous pouvons maintenant appliquer la correspondance (\ref{correspondance}) et cela nous mène à l'action suivante:
\begin{align}\label{notreBI}
  S_{bina}[F,g] &=\int_{\Mlor} \alpha \left( \sqrt{|g|} - \left| \,
  \detcmg\left( \gone_2 \otimes g_{\mu \nu} \otimes \gone_{d_R}
  +\beta ^{-1} \, \J \otimes F_{\mu \nu}^a \otimes T_a \,\right)\,\right|^{\frac{1 }{4 d_R}} 
\right) d^4x  \ .
\end{align}

Cette action satisfait les points 1), 2), 3) et  4), si l'on prend $\J$ comme étant un élément de $SL(2, \gC)$ (et de carré $-\gone_{2}$). 
Le Lagrangien est encore indépendant du choix de $\J$.
En particulier, nous retrouvons le Lagrangien de Born-Infeld si l'on remplace ${T_a}$ par ${i}$ et prenons $d_R=1$.

Nous avons supposé dans (\ref{notreBI}) que $\alpha$ et $\beta$ sont des constantes réelles et positives. 

Il est clair que seule une racine de degré $4 d_R$ mènera à une expression invariante sous difféomorphismes, c'est-à-dire se factorisant sous la forme $\sqrt{g} \times L(g,F)$ où $L(g,F)$ est un scalaire que nous appellerons Lagrangien, et on a:
\begin{align} 
  L_{bina}(g,F) &= \alpha \left( 1 - \left|\detcmg\left(\gone_2 \otimes
  \gone_{4 \times d_R} +\beta
  ^{-1} \, \J \otimes \hat{F} \right) \,\right|^{\frac{1}{4 d_R}}\, \right)  ,  
  \label{notreBI:2}
\end{align}
où $\hat{F}=\frac{1}{2} F_{\mu \nu}^{a} \hat{M}^{\mu \nu} \otimes T_a $.
Enfin, on a\footnote{%
Nous pourrions prendre une racine d'un degré différent ($2$ par exemple), mais nous obtiendrions alors une densité ne pouvant pas servir à fabriquer une mesure d'intégration. Il serait encore possible de multiplier cette densité par la densité $\sqrt{g}$ mise à la puissance nécessaire pour obtenir une densité de poids $1/2$, mais une telle construction introduirait une dissymétrie entre $g$ et $F$.}
:
\begin{align} 
  S_{bina}[g,F] &= \int_{\Mlor} L_{bina}(g,F) \sqrt{|g|}d^4x \ .
\end{align}

Le déterminant défini dans (\ref{notreBI:2}) peut être écrit de différentes manières:
\begin{subequations}\label{determinants}
  \begin{align}
    & \detcmg\left(\gone_2 \otimes \gone + \beta^{-1} \, 
\J \otimes \hat{F}\right)\label{det1}\\
    &= \detcmg\left(  s \otimes \gone + \beta^{-1}\, 
 s \J \otimes \hat{F}\right) \label{det2}\\
    &= \detmg\left(\gone + \beta^{-2} \, \hat{F}^2\right)\label{det3} \ ,
 \end{align}
\end{subequations}
où $s$ et $\J$  sont des éléments de $SL(2, \gC)$, $\J$ satisfait $J^2=-\gone$.
Par exemple, si on choisit  $s=i\sigma_2$, et  $ s \J = -i\sigma_3$  dans  (\ref{det2}), nous avons le déterminant suivant:
\begin{equation}\label{matrice_de_commutation}
  \begin{vmatrix}
    -i \beta^{-1} \hat{F} & \gone \\
    -\gone & i \beta^{-1} \hat{F}
  \end{vmatrix} = |g|^{-2 d_{R}} \begin{vmatrix}
   -i \beta^{-1}  F^a_{\mu \nu} \otimes T_a & g_{\mu \nu} \otimes \gone \\
    - g_{\mu \nu} \otimes \gone & i  \beta^{-1}  F^a_{\mu \nu} \otimes T_ a
  \end{vmatrix} \,
\end{equation}
Ce déterminant est une généralisation directe du déterminant considéré par Schuller dans \cite{schuller:02}, dont l'idée est de considérer la matrice~(\ref{matrice_de_commutation}) dans le cas abélien comme la matrice définissant les relations de commutations entre les coordonnées dans l'espace de phase relativiste d'une particule ponctuelle chargée, couplée de manière minimale au champ de Born-Infeld.
De manière similaire, nous pouvons étendre cette interprétation au cas où les coordonnées de cette particule prennent leurs valeurs dans une algèbre de Lie, \textit{i.e.} en imposant les relations de commutation:
\begin{equation}
  \begin{split}
    [ X_\mu,X_\nu ] &=  - \frac{1}{e \beta^2} F_{\mu \nu}^a \otimes T_a\\
    [ X_\mu,P_\nu ] &= -i g_{\mu \nu} \otimes \gone\\
    [ P_\mu,P_\nu ] &= e F_{\mu \nu}^a \otimes T_a \ ,
  \end{split}
\end{equation}
avec
\begin{equation}  
    X_\mu := X^a_\mu \otimes -i T_a \, , \ \ \, \ \ \, \ \ 
    P_\mu := P^a_\mu \otimes -i T_a \, .
\end{equation}

D'un autre côté, la forme particulière (\ref{det3}) nous permet de vérifier que le Lagrangien est bien réel et en même temps, cela donne une généralisation de l'action de Born-Infeld telle qu'elle est donnée dans~\cite{schuller:02}. 

On peut également remarquer, que lorsque l'on prend $\J = -i \sigma_3$ dans (\ref{det1}), le déterminant peut être écrit comme la valeur absolue d'un nombre complexe:
\begin{equation}
  \begin{vmatrix}
    \gone -i \beta^{-1} \hat{F} & 0 \\
    0 & \gone+ i \beta^{-1} \hat{F}
  \end{vmatrix}
  =
  |\detmg\left(\gone -i  \beta^{-1} \hat{F}\right)|^2 \label{module} \ .
\end{equation}

\subsection{Comparaison avec la trace symétrisée}

Nous allons maintenant faire une comparaison entre cette prescription et celle de la trace symétrisée.
Cela nous donnera en même temps des techniques de calcul pour ce type de Lagrangiens.

Rappelons quelques formules utiles reliant le déterminant d'un opérateur linéaire $M$ et ses traces: 
\begin{equation}\label{trace}
  \begin{split}
    \left(\det(1+M)\right)^\beta 
    & = \exp\left({\beta \,\tr(\, \log(1+M)\,)}\right)\\
    & = \sum_{n=0}^{\infty} \ \sum_{\substack{ \underline{\alpha}=(\alpha_1, \cdots ,
 \alpha_n) \\   \in [S_n]}} (-1)^n \prod_{p=1}^{n} \frac{1}{\alpha_p !} 
\left(- \frac{\beta \, \tr(M^p)}{p} \right)^{\alpha_p} \ ,
  \end{split}
\end{equation}
où $\underline{\alpha} \in [S_n]$ et  $[S_n]$ est l'ensemble des classes d'équivalence du groupe des permutations d'ordre $n$. 
Le multi-indice $\underline{\alpha}$ est donné par un diagramme de  Ferrer-Young  ou de manière équivalente par la relation:
\begin{align}
  \sum_{p=1}^{n}\,  p \, \alpha_p & = n  \ , \  \alpha_p \geqslant 0 \ .
\end{align}
 
En utilisant cette formule de trace, il est possible de développer à tout ordre en $F$, les Lagrangiens définis précédemment.
Pour éviter toute ambiguïté, nous allons noter $\trm$ la trace prise sur les indices d'espace-temps, $ \trg$ la trace dans la représentation matricielle de l'algèbre de Lie $\kg$, et par $\trt$ le produit tensoriel des deux traces précédentes.

Pour simplifier, nous allons absorber le facteur $\beta^{-1}$ dans la définition du champ $F$.
Ainsi, en suivant (\ref{det3}), nous avons :
\begin{equation}\label{traces}
  \begin{split}
    & \left( \detmg\left(1+\hat{F}^2\right) \right)^{\frac{1}{4d_R}}\\
    & =\sum_{n =0}^{\infty} \sum_{\underline{\alpha}=(\alpha_1, \cdots ,
      \alpha_n)} (-1)^{n}
    \prod_{k=1}^{n} \frac{1}{\alpha_k !} 
    \left(\frac{- \trt (\hat{F}^{2k})}{4 d_R \times k} \right)^{\alpha_k}  \\
    & =\sum_{n =0}^{\infty} \sum_{\underline{\alpha}=(\alpha_1, \cdots ,
      \alpha_n)} (-1)^{n} \prod_{k=1}^{n} \frac{1}{\alpha_k !}
    \prod_{\substack{m=1 \\ \alpha_k \neq 0}}^{\alpha_k} \left( - \frac{
      \trm(\hat{F}^{a^m_1} \cdots\hat{F}^{a^m_{2 k }} )}{4 k} \times
    \frac{\trg( T_{a^m_1} \cdots T_{a^m_{2 k}})}{d_R} \right) \ ,
  \end{split}
\end{equation}
où $ \underline{\alpha} \in [S_n] $ satisfait $  \sum_{k=1}^{n}k\alpha_k = n $ . \\
Nous pouvons comparer ce résultat avec celui obtenu par la prescription de la trace symétrisée~\cite{tseytlin:97}:
\begin{multline}
  \frac{1}{d_R}\str  \left( \detm\left(1+i\hat{F}^a T_a\right) \right)^{\frac{1}{2}} = 
  \frac{1}{d_R}\str  \left( \detm\left(1+\hat{F}^a\hat{F}^b T_a T_b\right) \right)^{\frac{1}{4}}\\
  \shoveleft{ \quad    = \frac{1}{d_R}\str \sum_{n =0}^{\infty} \sum_{
      \underline{\alpha}=(\alpha_1, \cdots , \alpha_{n})} (-1)^n \prod_{ k=1
      }^{n} \frac{1}{\alpha_k !}  \left(- \frac{ \trm(\hat{F}^{a_1}
        \cdots\hat{F}^{a_{2 k}} )}{4 k} T_{a_1}
      \cdots T_{ a_{2 k} } \right)^{\alpha_k} }\\
  \shoveleft{ \quad   =\sum_{n =0}^{\infty} \sum_{ \underline{\alpha}=(\alpha_1, \cdots ,
      \alpha_{n})} (-1)^n \Biggl( \prod_{k=1}^{n} \frac{1}{\alpha_k !}
    \prod_{\substack{m=1 \\ \alpha_k \neq 0}}^{\alpha_k} \left(-
      \frac{\trm(\hat{F}^{a^m_1} \cdots\hat{F}^{a^m_{2 k}}) }{4 k } \right)
    \times }\\
  \times   \frac{1}{d_R} \str \left( \prod_{k=1}^{n} \prod_{m=1}^{ \alpha_k}
    T_{a^m_1} \cdots T_{a^m_{2 k}} \right) \Biggr) \ .
\label{Str:2}
\end{multline}

Nous pouvons ainsi faire un développement en série des deux Lagrangiens obtenus et les comparer au Lagrangien de Born-Infeld (cas abélien).

Par exemple à l'ordre $4$ en $F$, nous obtenons pour (\ref{notreBI}):
\begin{multline}
  L_{bina}[F,g] \simeq -\frac{1}{4{d_R}} \trt\hat{F}^2 + \frac{1}{8 {d_R}}
  \trt\hat{F}^4 -\frac{1}{32 {d_R}^2}(\trt\hat{F}^2)^2\\
  \shoveleft{ \qquad \quad     \simeq -\frac{1}{2} (F^a,F^b)K_{ab} +\frac{1}{8}(F^a,F^b)(F^c,F^{d}) (-
    K_{ab}K_{cd} +K_{abcd}+K_{acbd})} \\ 
  +\frac{1}{8} (F^a,{\star
    F}^b)(F^c,{\star F}^d)K_{acbd} \ ,
\end{multline}
tandis que pour la trace symétrisée, on a:
\begin{equation}
  \begin{split}
    L_{Sym}[F,g] & = \frac{1}{{d_R}}\str(\gone - \sqrt{\detm(\gone + i \hat{F})})\\
    & \simeq \frac{1}{d_R} \str ( -\frac{1}{4} \trm \hat{F}^2 + \frac{1}{8}\trm {F}^4 -\frac{1}{32} (\trm \hat{F}^2)^2 )\\
    & \simeq - \frac{1}{2} (F^a,F^b) K_{ab} + \frac{1}{8} \left((F^a,F^b)(F^c,F^d)+(F^a,{\star F}^b)(F^c,{\star F}^d)\right) K_{\{abcd\}} \ ,
  \end{split} 
\end{equation}
avec $ K_{\{abcd\}} =\frac{1}{3} (K_{ab}K_{cd}+K_{ac}K_{bd}+K_{ad}K_{bc} + \frac{1}{4} S^e_{ab} S_{cde} + \frac{1}{4} S^e_{ac} S_{bde} + \frac{1}{4} S^e_{ad} S_{bce})$.

Nous avons adopté les conventions suivantes:
\begin{equation}
  T_a T_b  =- g_{ab} \gone + \frac{1}{2} C_{ab}^{c}
  T_{c}+ \frac{i}{2} S_{ab}^{c} T_{c} \ ,
\end{equation}   
où  $g_{ab}=\frac{c_R}{d_R} \delta_{ab}$ , $S_{cab}=g_{cd} S_{ab}^{d}$ est complètement symétrique et réel, $C_{cab}=g_{cd} C_{ab}^{d}$ complètement anti-symétrique et réel, et
\begin{align}
  K_{a_1 \cdots \ a_n} = \frac{(-1)^{[\frac{n}{2}]}}{d_R}\trg(T_{a_1} \cdots T_{a_n}) \ .
\end{align}

\subsection{\texorpdfstring{Calcul explicite pour G=$SU(2)$}{Calcul explicite pour G=SU(2)}}

Nous allons faire le calcul du Lagrangien généralisé de Born-Infeld $L_{bina}$~(\ref{notreBI:2}) dans le cas d'un groupe de structure $G=SU(2)$.
Les générateurs $T_a=-\frac{i}{2}\sigma_{a}$ sont représentés par les matrices de Pauli usuelles.
Afin de simplifier le calcul, nous allons prendre le facteur $\beta=1/2$ dans (\ref{module}).

Tout d'abord, on peut remarquer que dans (\ref{module}), l'expression $\detmg(\gone -i \beta^{-1} \hat{F})$ est un carré parfait. En effet, si on multiplie l'expression précédente par $ 1 =\detmg ( \gone \otimes -i\sigma_{2}) $, on obtient:
\begin{align}
  \detmg\left(\gone - 2 i \hat{F}\right) &= \detmg \left( \gone \otimes (-i
    \sigma_2) +
    \hat{F}^a \otimes (i\sigma_2 \sigma_a)\right) \\
  &= |g|^{-2} \detmg \left( g_{\mu \nu} \otimes (-i\sigma_2) + F^a_{\mu \nu}
    \otimes (i\sigma_2 \sigma_a)\right) \ .
\end{align}
La matrice sous le déterminant est anti-symétrique et donc son déterminant est un carré parfait.
Cela implique que la plus grande puissance de $F$ dans l'expansion de $\exp(\frac{1}{2} \tr \log(1+ 2 i \hat{F}))$ est $4$, et l'on a: 
\begin{align*}
  \detmg ( \gone + 2 i \hat{F})
  &= \left( \exp\left({\frac{1}{2} \tr \log ( 1 + i \hat{F} )}\right) \right)^2\\
  &= \left( 1 + \Trf{2} - i \Trf{3} - \Trf{4} + \Trff{2}{2} \right)^2\\
  &= \left( 1 + \frac{\Tr{2}}{4} - i \frac{\Tr{3}}{6} - \frac{\Tr{4}}{8}
  +\frac{\Tr{2}^2}{32} \right)^2 \ ,
\end{align*}
où $\Tr{i}= \trt \left( \left(\hat{F} \right)^i\right)$.

En utilisant la formule (\ref{module}), nous avons:
\begin{equation}
  \label{bina:su2}
  L_{bina}= 1-\sqrt[4]{(1+2P-Q^2)^2 + (2K_3)^2} \ ,
\end{equation}
où
\begin{equation}
  \left\{\begin{array}{lclcl}
      2P  & =& \frac{1}{4} \Tr{2} &=& (F^a,F_a)\\
      Q^2 & = &\frac{1}{8}\Tr{4}-\frac{1}{32}\Tr{2}^2 &=& \frac{1}{4}(F^a,{\star F}^b)
      (F^c,{\star F}^d) K_{acbd}\\
      K_3 & =& - \frac{1}{12}\Tr{3} &=& \frac{1}{6}\epsilon_{abc} \trm(\hat{F}^a \hat{F}^b 
\hat{F}^c)
    \end{array} \right. \ .
\end{equation}

Notons que le Lagrangien dépend exclusivement de trois invariants dépendants de $F$, alors qu'il existe $8$ invariants indépendants~\cite{roskies:77,anandan:78} pour $SU(2)$.
\section{Théorie de Born-Infeld non commutative}
\subsection{Rappels sur les champs de jauge en géométrie non commutative et notations}

Il a été développé dans le chapitre  \ref{cha:modeles-physiques} une théorie de Maxwell non commutative basée sur l'algèbre $\cA = C^{\infty}(V) \otimes M_n(\gC)$.
L'idée était d'utiliser non plus un calcul différentiel gradué commutatif, mais une algèbre différentielle graduée (non commutative) afin de décrire les champs de jauge.
Nous avons vu que cela mène naturellement à une théorie de jauge non abélienne couplée à des champs de Higgs.
Nous allons ici rappeler les principales caractéristiques de cette théorie et en donner une présentation minimale afin que le présent chapitre soit autonome. 

Pour l'algèbre $\cA = C^{\infty}(V) \otimes M_n(\gC)$,  les ``champs de vecteurs'' sont générés par les dérivations de $ C^{\infty}(V)$ et les dérivations intérieures de $ M_n(\gC)$. 
L'algèbre différentielle graduée $\Oder(A)$ est générée par les $1$-formes formant une base duale des dérivations.
Pour construire une théorie de jauge, nous allons considérer l'algèbre $\cA$ comme un bimodule sur $\cA$.
Dans ce cadre, un choix de jauge correspond à choisir un unitaire $e$ de $\cA$, satisfaisant $h(e,e)=1$, où $h$ définit une structure hermitienne sur $\cA$.
Alors tout élément de $\cA$ peut être écrit sous la forme $ \, em \, $ avec $m \in \cA$ et une connexion sur $\cA$ est définie par une application: 
\begin{align*}
  \nabla:  \cA & \to  \Omega^{1}(\cA)  &   
   em & \mapsto (\nabla e)\ m  + e \  dm \ .
 \end{align*}
Dans la jauge $e$, une connexion est complètement caractérisée par un élément
$\omega$ de $\Omega^{1}(\cA)$:
\begin{align*}
\nabla e &= e \  \omega \ .
\end{align*}
On peut également décomposer $\omega$ en une partie verticale et une partie horizontale:
\begin{align*}
  \omega &= \omega_h + \omega_v  & & \text{avec}&
  \omega_h &= A  &
  \omega_v &= \theta + \phi \ .
\end{align*}
 $A$ est l'analogue du champ de Yang-Mills,  $ \theta $  est la $1$-forme canonique sur l'algèbre des matrices et joue le rôle d'une origine dans l'espace affine des connexions.
 Elle satisfait l'équation:
  \begin{align*}
    d\theta +\theta \wedge \theta &=0 \ .
  \end{align*}
Enfin, $\phi$ est une forme tensorielle et peut être identifiée à un multiplet de champs scalaires.

Nous pouvons choisir une base locale des dérivations de $\cA$: $\{e_{\mu},e_{a}\}$,  où $e_{\mu}$ sont les dérivations extérieures de $C^{\infty}(V)$ et $e_{a}= ad(\lambda_{a})$, avec $\{\lambda_{a}\}$ une base de matrices anti-hermitiennes de  $M_n(\gC)$, sont les dérivations intérieures de $M_{n}(\gC)$.

La base duale sera notée par $\{ \theta^{\mu}, \theta^{a} \}$.  
En particulier, on a:
\begin{align*}
   A &= A_{\mu} \theta^{\mu}  &
   \theta &= - \lambda_a \theta^a  &
   \phi &= \phi_a \theta^a \ .
 \end{align*}
Si nous choisissons une connexion anti-hermitienne, on a: $ \phi = \phi_a^b \lambda_b \theta^a$. 

La courbure associée à $\omega$ est :
\begin{align*}
  \Omega = d\omega + \omega \wedge \omega 
\end{align*}
et nous pouvons également définir le champ de force: 
\begin{align*}
  F&= dA + A \wedge A \ .
\end{align*}

On a l'identification suivante:
\begin{align*}
  \Omega_{\mu \nu} &= F_{\mu \nu} & \Omega_{\mu a} &= D_{\mu} \phi_a\\
  \Omega_{a \mu } &= - D_{\mu} \phi_a &
  \Omega_{a b} &= [\phi_a,\phi_b] - C_{a b }^{c} \phi_c
\end{align*}
où $ C_{a b }^{c} $ sont les constantes de structure dans la base $\{\lambda_a\}$.

Une transformation de jauge se fait par le choix d'un élément unitaire $U$ de $M_n(\gC)$, satisfaisant $ h(e U, e U ) = 1$. 
Alors dans une jauge $e'=e U$, on a:
\begin{align*}
  \omega' &= U^{-1} \omega U + U^{-1} dU \ .
\end{align*}
La forme $\theta$ est invariante sous ces transformations de jauge, et donc $A$ et $\phi$  se transforment de la manière suivante:
\begin{align*}
  A' &=U^{-1}  A U + U^{-1} dU  &
  \phi' &=U^{-1}  \phi U \ .
\end{align*}

Si l'on prend en compte le fait que toutes les formes apparaissant ici sont à valeurs matricielles, il est assez naturel d'utiliser des invariants provenant de l'application déterminant et trace dans la construction de Lagrangiens.

\subsection{Action de Born-Infeld}

Nous voulons dans cette section, donner une généralisation du Lagrangien de Born-Infeld pour les champs de jauge intervenant dans la géométrie non commutative des algèbres de fonctions à valeurs matricielles. 
De la même manière que la théorie de Born-Infeld généralise la théorie de Maxwell, cette théorie généralisera la théorie de Maxwell non commutative.
Nous avons considéré dans \cite{serie:04} une généralisation de l'action de Born-Infeld pour une connexion non commutative.
Cette généralisation est elle même une généralisation du Lagrangien $L_{bina}$~(\ref{notreBI:2}) et s'obtient en considérant la densité:
\begin{align}
  {\cal L}_{binc} &= \sqrt{\det|g|} - \{|\det ( \gone \otimes g + J \otimes \hat{\Omega} |\}^{1/4n}
  \label{binc}
\end{align}
où $\hat{\Omega} = \Omega_{\alpha \beta}\hat{L}^{\alpha \beta} $, avec $\hat{L}^{\alpha \beta}$ les générateurs de la représentation fondamentale de l'algèbre de Lie de $SO(4+n^2-1)$ et $J$ est un élément de $SL(2,\gC)$ de carré $-\gone$. 

\subsubsection{Limite de Yang-Mills}
Lorsque la partie intérieure de la connexion non commutative $\omega$ correspond à la forme canonique $\theta$, \textit{i.e.} le champ scalaire $\phi=0$, alors on retrouve la densité  $\cL_{bina}$ proposée dans la section \ref{sec:une-vers-mink} pour la connexion ordinaire $A$, c'est-à-dire:
\begin{align*}
  {\cal{L}}_{bina} =\sqrt{g} \,L_{bina} =  \sqrt{|g|} - \left| \, \det\left( \gone_2 \otimes g_{\mu \nu} 
      \otimes \gone_{d_R} +\beta ^{-1} \, \J \otimes F_{\mu \nu}^a \otimes T_a \,\right)\,
  \right|^{\frac{1 }{4 d_R}} \ .
\end{align*}

\subsubsection{Limite $n=1$}
Si l'on considère le cas $n=1$, que cela soit pour le Lagrangien $L_{binc}$ ou pour $L_{bina}$, on retrouve l'action de Born-Infeld (\ref{BI-det}).

\section{Étude numérique de solutions}
Je présente dans cette section des résultats d'études numériques des solutions aux équations du mouvement pour les Lagrangiens $L_{bina}$ et $L_{binc}$ présentés dans les sections précédentes.

\subsection{Solutions statiques à symétrie sphérique}
Nous allons voir dans cette section le résultat d'une étude des solutions statiques à symétrie sphérique pour le Lagrangien $L_{bina}$ dans le cas d'un groupe de structure $SU(2)$.

\subsubsection{Ansatz de 't Hooft}

't Hooft~\cite{'thooft:74} proposa un ansatz pour étudier les solutions statiques, à symétrie sphérique et purement ``magnétique'' des équations de Yang-Mills. 
Cet ansatz est en fait indépendant des équations du mouvement et pourra nous servir pour étudier d'autres types d'équations du mouvement, comme celles obtenues pour les actions de type Born-Infeld.
Cet ansatz est appelé ansatz de  't Hooft-Polyakov et est le suivant:
\begin{equation}
  \begin{split}
    A & = \frac{1-k(r)}{2 } UdU^{-1} \ \ \text{avec} \  U = e^{i \pi T_r} \\
    &=  (1-k(r)) [ T_r, dT_r] =  (1-k(r)) \  (T_{\theta} \sin\theta d\varphi - 
T_{\varphi} d\theta) \\
    &= \frac{1-k(r)}{ r^2} \  ( \vec{r} \wedge  \vec{T} ) \cdot \vec{dx} \ ,
  \end{split}\label{eq:'thooft}
\end{equation}
où nous utilisons les notations habituelles.
Nous pouvons l'exprimer en composantes: 
\begin{equation}
A^a_k = \frac{(1 - k(r))\,}{r^2} \  \epsilon^a{}_{km} \, x^{m} \, ,
\end{equation}
où
\begin{equation*}
\begin{aligned}
  a,b,c... &= 1,2,3 &&;&
  i,j,k... &= 1,2,3 &&;&
  \epsilon^{a}{}_{km} &= \epsilon^{aij} \, g_{ik} \, g_{jm} \, .
\end{aligned}
\end{equation*}
La notion de symétrie sphérique pour les potentiels de jauge en théorie de Yang-Mills fut analysée par P. Forgacs et N.S. Manton dans~\cite{forgacs:80}, ainsi que dans~\cite{bertrand:92}.
L'ansatz de 't Hooft-Polyakov est en fait un cas particulier de l'ansatz de Witten~\cite{witten:77} qui est le plus général possible pour un  potentiel de jauge (ou connexion) à symétrique sphérique (pas nécessairement statique).
L'ansatz de Witten est complètement justifié par les techniques de réduction et est lui même un cas particulier de connexions non commutatives symétriques (voir section~\ref{sec:exemple:symetrie-spherique}).
Les propriétés de l'ansatz de Witten sont également discutées dans~\cite{volkov:98}. 

Nous avons vu que pour ce type d'ansatz, il y a une symétrie $U(1)$ résiduelle et que les champs intervenant dans l'ansatz peuvent s'interpréter dans le cadre d'une théorie de jauge $U(1)$ sur un espace de de~Sitter de dimension $2$, contenant une connexion et un champ scalaire complexe $w$ avec un potentiel de type Higgs.
Dit autrement, les degrés de liberté d'une connexion symétrique sont paramétrés par quatre fonctions réelles $a_0$, $a_1$, $Re(w)$, et $Im(w)$ (nous utilisons les notations introduites dans~\cite{volkov:98}).
La symétrie de jauge nous autorise à faire un choix de jauge $a_1=0$. 

La composante $a_0$ est éliminée si l'on se restreint à ne regarder que les solutions de type magnétique.
Dans le cas statique, les équations du mouvement possèdent une intégrale première (due à la symétrie $U(1)$ globale résiduelle) qui est nulle pour les solutions d'énergie finie à l'infini. 
Cela se traduit par un choix de phase de la fonction $w$ qui permet de retrouver l'ansatz de 't~Hooft~\cite{'thooft:74}.

Dans ce cas, les seules composantes de la courbure $F$ non nulles sont celles de type magnétique:
\begin{align}
  B^a_i &=\frac{1}{e r^2} \left[  \hat{r}_i \hat{r}^a (1-k^2) - r  k' \, P^a_i \, \right] \ ,
\end{align}
où $\hat{r}_i = \frac{x_i}{r}$ et $P^a_i=\delta^a_i - \hat{r}^a \hat{r}_i$ est la projection sur le plan perpendiculaire au vecteur radial.

\subsubsection{Calcul du Lagrangien}

A l'aide de l'ansatz de 't Hooft~(\ref{eq:'thooft}), les trois invariants apparaissant dans le Lagrangien peuvent s'exprimer de la manière suivante:
\begin{equation}
  \begin{split}
    2P &= \frac{1}{ r^4} \left[ (1-k^2)^2 + 2 (r k')^2 \right]\\
    K_3 &=\frac{1}{ r^6} \left( (1-k^2) (r k')^2 \right)\\
    Q^2  &=  0 \ .
  \end{split}
\end{equation}

Alors l'action devient:
\begin{equation}
  S_{bina} =\int \, \left( 1 - \left\{\left( 1 + \frac{(
      1-k^2)^2 + 2 ( r k')^2 }{r ^4} \right)^2 + \frac{4}{r^{12}} (1- k^2)^2 (r k')^4
  \right\}^{1/4} \  \right) r^2  d r \ .
\end{equation}

Pour l'étude des solutions aux équations du mouvement associées à ce Lagrangien, il est utile d'introduire la variable de ``temps'': $\tau = \log(r)$. 
Alors l'action prend la forme suivante: 
\begin{equation}
  S_{bina} =\int   (1 - \sqrt[4]{A} ) \, e^{3 \tau} \, d \tau \ ,
\end{equation}
où
\begin{displaymath}
  \left\{ 
    \begin{array}{ll}
      A & =(1+a^2+2b)^2+4a^2 b^2 = (1+a^2)((1+2b)^2+a^2) \ . \\
      a & =(1-k^2) /r^2\\
      b &=\dot{k}^2 /r^4
    \end{array} \right.
\end{displaymath}

Les équations du mouvement associées sont les suivantes:
\begin{equation}
  A_{k} + A_{\dot{k}} ( \, \frac{3}{4} \frac{\dot{A}}{A}-3 \, ) - \frac{d}{d
  \tau} A_{\dot{k}} = 0 \, ,
\end{equation}
ou de manière équivalente:
\begin{equation}
  \left\{ \begin{array}{ll}
      \dot{k} &= u \\
      \dot{u} &= \gamma(k,u,\tau) u + k (k^2-1) 
    \end{array} \right.
\label{systeme_dynamique}
\end{equation}
avec
\begin{equation}
  \begin{split}
    \gamma (k,u,\tau) &= 1 - 2 \  \frac{ u^2 + 2u k (1- k^2) + (1- k^2 )^2}{ r^4 + (1-
      k^2)^2} \\
    & + \ \frac{
      6 u (1- k^2)\left[ k u^2 + 2u (1-k^2) + k ( 1- k^2)^2 \right] 
      \left[ r^4 + 2 u^2 + (1- k^2)^2\right]}{\left[ r^4 + (1-
      k^2)^2 \right]\left[ (r^4+2 u^2)^2 + (1-k^2)^2(r^4+6 u^2) \right]} \ .
  \end{split}
\end{equation}

Le coefficient $\gamma$ joue le rôle d'une friction (terminologie empruntée à la mécanique) et est similaire à celui trouvé dans~\cite{gal'tsov:00}.
Dans la théorie de Yang-Mills, avec le même ansatz, le facteur correspondant est $\gamma_{YM} = 1$.

Le système (\ref{systeme_dynamique}) n'est pas autonome (\textit{i.e.}, certains coefficients dépendent explicitement de la variable $\tau$) et une étude qualitative du comportement de ses solutions peut être faite en faisant une analyse de points fixes.
L'analyse doit être faite en terme des variables $(\tau,k,u)$ (voir par exemple~\cite{chernavsky:78}).
Il ne peut pas y avoir de points fixes dans la variable temporelle $\tau$ qui ``court'' toujours.
L'idée est donc de trouver des solutions asymptotiques pour la fonction $k$ pour $\tau \rightarrow - \infty$ ($r \rightarrow 0$)
ou $\tau \rightarrow \infty$ ($r \rightarrow \infty$) et de trouver des solutions aux équations du mouvement en faisant un développement asymptotique en puissance de $r$ ou $1/r$.
Pour  $r \rightarrow \infty$, le système admet les points fixes $(k=1,u = 0)$ et $(k=-1,u = 0)$.

\subsubsection{Développement asymptotique}

Bien que les solutions aux équations du mouvement semblent posséder des développements asymptotiques analogues à ceux trouvés dans~\cite{donets:97,dyadichev:00,gal'tsov:00}, une analyse détaillée montre que des solutions du type de Bartnik-McKinnon~\cite{bartnik:88} sont exclues ici.

Nous trouvons  deux développements en puissance de $r$ qui satisfont les équations du mouvement asymptotiquement autour de  $r=0$.
\begin{enumerate}
\item 
  Le premier développement dépend de deux paramètres $k_0$ et  $a$ et commence de la manière suivante:
\begin{align}\label{dev:r=0:1}
  k= k_0+ a r - k_0 \left(\frac{ 5 a^2 }{6 g}+ \frac{g}{12
    a^2}\right) r^2 + \frac{a^8 (52-70 g) - 9 a^4 g^3 + (g-1) g^4}{108 a^5 g^2} r^3+
  O(r^4) \ ,
\end{align}
où $g=1-k_0^2$,  $a \neq 0$ et $ g \neq 0$.
Ce développement présente certaines similarités avec celles obtenues dans~\cite{donets:97,dyadichev:00} qui dépendent du même paramètre $k_0$.
\item
Le second développement dépend d'un paramètre $b$ et commence de la manière suivante:
\begin{align}\label{dev:r=0:2}
  k=\pm \left(  1- b r^2 + \frac{3 b^2 + 92 b^4 + 608 b^6}{10+200 b^2 + 1600
  b^4}\,r^4 + O(r^6) \right)
\end{align}
\end{enumerate}

Autour de $r=\infty$, nous pouvons trouver un développement en puissance de $r^{-1}$, dépendant d'un paramètre $c$, et qui commence de la manière suivante:
\begin{align}\label{dev:inf}
  k=\pm \left( 1- \frac{c}{r} + \frac{3 c^2}{4 r^2} + O(\frac{1}{r^3}) \right) \ .
\end{align}

Nous remarquons que ce développement est le même jusqu'à l'ordre $7$ que celui obtenu pour les solutions aux équations de Yang-Mills.
Cela nous permet d'interpréter les intégrales à l'infini en terme de charge magnétique, énergie, \dots

Maintenant que nous connaissons les solutions asymptotiques aux équations du mouvement, nous allons chercher des solutions globales raccordant ces différents développements.
Ce type de solutions ne peut être obtenu de manière analytique à cause de la complexité des équations du mouvement. Nous allons donc faire une recherche numérique de ces solutions.
 
\subsubsection{Solutions numériques}

La recherche de solutions numériques est basée sur les mêmes techniques que celles développées dans~\cite{donets:97,dyadichev:00,gal'tsov:00}.
Grâce aux développements (\ref{dev:r=0:1}) et  (\ref{dev:inf}), nous évaluons les conditions initiales à utiliser pour l'intégration numérique des équations (\ref{systeme_dynamique}).
Les trois paramètres intervenant dans les différents développements  (deux en $r=0$ et un en  $r= \infty$) sont reliés par deux équations de contraintes.  
Les solutions peuvent donc être indexées par un paramètre réel et nous prendrons $c$ (cf eq.(\ref{dev:inf})), avec $c>0$ ou bien $\tau_c=\log(c)$.

On peut assigner à chaque solution un entier $n$, $n-1$ étant le nombre de fois que la fonction $u$ passe par zéro ou le nombre de tours des solutions dans le plan $(k,u)$.
Quelques solutions sont tracées sur la figure Fig.~\ref{fig:some_plots}.

Quand le paramètre $\tau_c$ varie de $- \infty$ à $+ \infty$, on observe que l'entier $n$ croît de $1$ à l'$\infty$.
Ainsi, à chaque fois que l'entier $n$ augmente de $1$, on peut faire correspondre certaines  valeurs caractéristiques du paramètre $\tau_c$:
\figuresPlots

Les solutions, que nous avons obtenues numériquement, prennent les valeurs $k=1$ à $r = \infty$ et $k=k_0$, avec $-1<k_{0}<1$ en $r=0$.
Ce comportement est radicalement différent des solutions de type sphaleron obtenues par Bartnik-McKinnon dans~\cite{bartnik:88} ou d'autres solutions du même type obtenues dans~\cite{gal'tsov:00}.

Pour ces solutions, les deux paramètres $k_0$ et $a$ de (\ref{dev:r=0:1}) sont des fonctions du paramètre $\tau_c$.
Nous avons évalué l'énergie de ces solutions pour des valeurs du paramètre $k_0$ pour $\tau_c$ variant de $-10$ à $20$. 
L'énergie $E$ est représentée en tant que fonction de $\tau_c$ dans la figure Fig.~\ref{fig:E}.
On observe la présence de minimums locaux sur cette courbe.
Ces minimums locaux semblent survenir aux valeurs critiques définies précédemment du paramètre $\tau_c$.
L'énergie des solutions tend vers la valeur $E_{\tau_c=\infty}= E_{n=\infty}  = 1.23605...$ lorsque $\tau_{c}\to \infty$, qui correspond à l'énergie d'une charge magnétique ponctuelle de type  Born-Infeld~\cite{gal'tsov:00}.

\figuresE

Sur la figure Fig.~\ref{fig:ko}, est tracée la dépendance du paramètre $k_0$ en $\tau_c$ pour ces solutions. 
On peut observer des singularités dans les dérivées de cette courbe aux valeurs critiques du paramètre $\tau_{c}$.

\figuresK

Nous remarquons que toutes les solutions tendent vers les solutions du vide de la théorie de Yang-Mills lorsque $r\to \infty$, ce qui veut dire que loin de l'origine les non-linéarités du modèle sont négligeables.
Au contraire, près de l'origine $r=0$, nous obtenons des solutions non triviales, de type monopole magnétique, paramétrées par la constante $k_{0}$.
\newpage
\subsection{Solutions pour le secteur scalaire}
Dans cette section, je présente les résultats d'une étude des solutions aux équations du mouvement pour certains ansatzs dans le Lagrangien $L_{binc}$ et dans le cas d'une algèbre de matrice $M_{2}(\gC)$.

\subsubsection{Réduction du Lagrangien pour les champs scalaires}

Nous travaillons toujours dans le cas où l'algèbre est $ C^{\infty}(\gR^4) \otimes M_2(\gC)$.
Afin d'avoir une idée du type de solutions que l'on peut obtenir pour la partie champ scalaire, nous allons supposer que les champs de Yang-Mills sont nuls et nous allons prendre l'ansatz pour la partie scalaire $ \varphi \in C^{\infty}( \gR_4 ) $ :
\begin{align*}
 \phi &= \varphi \ \theta \ .
\end{align*}

Dans ce cas, le déterminant intervenant dans l'équation~(\ref{binc}) est:
\begin{align}
  \begin{vmatrix}
    \hat{g}_{\mu \nu} & i D\phi\\
    -iD\phi & \hat{g}_{ab}+i H 
  \end{vmatrix} \ ,
  \label{matrice}
\end{align}
où 
 \begin{align*}
   H &=  \left\{\Omega_{ab} \right\}_{{a,b = 1,2,3}} &
   D\phi  &= \left\{ D_{\mu}\phi_a\right\}_{\substack{\mu=0,1,2,3 \\ a=1,2,3 \ \,  }} &
    \hat{g}_{\mu \nu}&=  g_{\mu \nu}\otimes  \gone_2
   \ .
 \end{align*}
Grâce au lemme de Schur, nous pouvons réduire ce déterminant à celui de la matrice suivante:
\begin{equation*}
  \begin{vmatrix}
  \gone_3   + i H - D_{\mu}{\phi}D^{\mu}\phi
    \end{vmatrix} \ .
\label{matrice2}
\end{equation*}
Pour les mêmes raisons que dans la section précédente, ce déterminant est un carré parfait, et sa racine peut s'exprimer à l'aide d'une somme finie de produits de traces de la matrice $M= i H - D\phi D\phi$ (nous noterons $  D_{\mu}{\phi}D^{\mu}\phi=D\phi D\phi =(D\phi)^2$).
On peut ainsi calculer le Lagrangien de Born-Infeld et on obtient:
\begin{align*}
  L_{binc}&=1- \left\{ \left(1+ 3 \beta^{-2} (D\varphi)^2\right)^2 + 
    16 \beta^{-2} \varphi^2(\varphi-m)^2 \right\}^{\frac{1}{4}}
  \sqrt{1+4 \beta^{-2} \varphi^2(\varphi-m)^2 }  \ .
\end{align*}
où $\beta$ est le paramètre de Born-Infeld et $m$ est un paramètre de masse pour le champ scalaire.
Ces paramètres seront mis égaux à $1$ dans la suite.

\subsubsection{Le cas statique}

Dans cette section, nous montrerons que l'on ne peut pas obtenir de configuration statique non triviale pour ce champ scalaire.
Les arguments que nous allons utiliser sont une adaptation de ceux utilisés dans le théorème de Derrick~\cite{derrick:64}.
L'idée est d'utiliser des dilatations du champ $\varphi (r) \to \varphi_{\lambda}(r)=\varphi (\lambda r) $ de manière à générer une courbe à un paramètre dans l'espace des champs passant par une solution.
Alors le principe variationnel le long de cette courbe est: $\partial S[\varphi_{\lambda}] / \partial \lambda = 0$ à $\lambda=1$,\textit{ i.e.} : 
\begin{align}
\int 4 \pi  r^2 dr \left ( \frac{\partial L}{\partial \varphi'} \varphi' - 3
 L \right ) &= 0 \ .
\label{eq:derrick}
\end{align} 
On peut voir que la fonction sous l'intégrale peut s'écrire de la manière suivante:
\begin{align*}
  f(\varphi, \varphi') &=1/3 \left ( \frac{\partial L}{\partial
      \varphi'} \varphi' - 3 L \right ) \notag \\
  &=  \frac{\sqrt{A}}{B^{3/4}} \left[(1+3p)(1+2p)+ 16 s^2 \right] -1 \ ,
  \label{eq:fonction} 
\end{align*} 
où
\begin{align*}
s &={\varphi} (\varphi-1) &
p &= \varphi'^2 \\
A &= 1+4 s^2 &
B&= (1+3p)^2+16 s^2  \ .
\end{align*}
On voit que la fonction $f$ est toujours positive du fait de la relation suivante:
\begin{align*}
  ((1+2p)(1+3p)+ 16 s^2)^4 \geq ( (1+3p)^2+ 16s^2)^3    \Rightarrow  f \geq 0 \ .
\end{align*}
Ainsi, la condition (\ref{eq:derrick}) est satisfaite \ssi $f=0$ pour tout $r>0$.  
L'équation $f=0$ n'admet que les solutions triviales,  $\varphi' = 0$ et $\varphi = 0 \text{ ou } 1$. 
Cela nous mène à la conclusion, comme dans le cas considéré par Derrick~\cite{derrick:64}, qu'il n'y a pas de solution non triviale dans ce modèle.

\subsubsection{Solutions dépendantes du temps}

Il est possible de faire une analyse des solutions homogènes dépendantes du temps.
Nous considérons donc l'ansatz  $F_{\mu \nu} = 0$, $\hat{\phi}_a = \phi \hat{T}_a$ dans le Lagrangien de Born-Infeld non commutatif. Ce dernier prend alors la forme suivante:
  \begin{align*}
    L_{binc}=1- \left\{ 1+ 6 (D\phi)^2 + 9
      (D\phi)^4 + 16 \phi^2(\phi-1)^2 \right\}^{\frac{1}{4}} \sqrt{1+4 \phi^2(\phi-1)^2 } \ .
  \end{align*}

Si l'on prend le sous-ansatz $\varphi = \varphi(t)$, alors les équations du mouvement sont:
\begin{align*}
  &\dot{\varphi}=u\\
  &(1+4X) g(X,Y) \dot{u} + 4 s s' h(X,Y) = 0 \ ,
\end{align*}
où
\begin{align*}
  & s=\varphi (\varphi -1) \ , \ s'=2\varphi-1\\
  &  X=s^2 \ ,  \ Y=u^2\\
  & g(X,Y)= 16 X (1-9 Y ) + (1-3 Y)^2 \\
  & h(X,Y)= ((1-3Y)^2+16 X)(1-Y+8X) -6(1+4 X)(1-3Y)Y  \ .
\end{align*}

Il y a des points dans l'espace des phases où $\dot{u}$ n'est pas bien défini, ce sont les points où le polynôme $g$ s'annule (4 courbes dans la figures Fig~(\ref{fig:char})).
Cependant, l'indétermination peut être levée aux points où la fonction $4 s s' h(X,Y)$ s'annule (\textit{c.f.} courbes gris clair) en même temps que la fonction $g(X,Y)$.
Ces points sont au nombre de $16$ et l'indétermination est levée pour $14$ d'entre eux (voir sur la figure Fig.~(\ref{fig:char}) les points accentués en gras ainsi que les flèches qui indiquent les directions possibles du vecteur tangent une fois l'indétermination levée).

On remarque que dans une certaine zone de l'espace des phases, les trajectoires sont périodiques et définies pour tous les temps.
Si l'on prend des conditions initiales à l'extérieur de cette zone, on ne peut intégrer le système que sur des temps finis, c'est-à-dire que les solutions $\phi(t)$ trouvées ont leur dérivée seconde qui diverge en un intervalle de temps fini puisqu'elles sont amenées à "heurter" une des courbes où $g=0$.

Il semble cependant que certaines trajectoires peuvent traverser les courbes de singularité $g=0$ aux points où l'indétermination est levée. Cela donne un ensemble de trajectoires "limites" qui passent par ces points et qui peuvent éventuellement être prolongées sur l'intervalle de temps  $\gR$.

\Figurechar
\Figuretraj

\newpage
\subsection{Vers une application en cosmologie}

Nous pouvons essayer de coupler le modèle précédent avec champ scalaire à la métrique de Friedmann-Robertson-Walker afin de voir dans quelle mesure ce type de Lagrangien peut être utile en cosmologie.
Nous nous contenterons ici de faire une analyse des points fixes.

\subsubsection{Analyse des points fixes pour le champ scalaire libre couplé à la gravitation}
Avant de considérer l'action de Born-Infeld, nous allons rappeler l'analyse qui peut être faite pour le cas du champ libre.
Les notations utilisées pour les constantes sont:
\begin{align*}
  \frac{8\pi}{m_{p}^{2}}&=8\pi G = \kappa^{2} \ .
\end{align*}

Nous allons étudier le comportement d'une théorie de gravitation pour un univers homogène et isotrope avec un champ scalaire dont
la densité d'énergie est:
\begin{align*}
  \rho&=b(\frac{1}{2}\dot{\phi}^{2} +\frac{1}{2}m^{2} \phi^{2} ) + \rho_{0} 
\end{align*}
et la pression:
\begin{align*}
  p&=b(\frac{1}{2}\dot{\phi}^{2} - \frac{1}{2}m^{2} \phi^{2} ) - \rho_{0} \ .
\end{align*}

Alors, les équations du mouvement sont:
\begin{equation*}
  \begin{aligned}
    &  H^{2}+ \frac{k}{a^{2}}= \frac{\kappa^2}{3} \rho & \text{(Friedmann)}\\
    & \ddot{\phi}+3H\dot{\phi}+m^{2} \phi=0 & \text{(équation du mouvement pour $\phi$)}
\end{aligned}
\end{equation*}

Voici deux autres équations non indépendantes des deux précédentes mais utiles:
\begin{align*}
&  \dot{H}= -\frac{b \kappa^{2}}{2} \dot{\phi}^{2}+\frac{k}{a^{2}} \\
&\dot{H}+H^{2}=\frac{b \kappa^{2}}{3}(\frac{1}{2} m^{2} \phi^{2}-\dot{\phi}^{2}) + \frac{\kappa^2}{2} \rho_{0} \ .
\end{align*}

Nous pouvons alors réécrire le système d'équations du mouvement comme un système d'ordre 3 pour les variables $(\phi,u,H)$:
\begin{align*}
  \dot{\phi}&=u\\
  \dot{u}&= -m^{2}\phi -3 H u \\
  \dot{H}&= \frac{b\kappa^{2}}{3}(\frac{1}{2}m^{2}\phi^{2}-u^{2}) + \frac{\kappa^{2}}{3}\rho_{0}-H^{2}
\end{align*}
et considérer l'équation de Friedmann comme une contrainte.

Le point fixe de ce système est 
\begin{align*}
  u&=0\\
  \phi&=0\\
  H ^{2}&= \frac{\kappa^2}{3}\rho_{0} \ , 
\end{align*}
mais il faut faire attention à ce que ces conditions soient compatibles avec l'équation de Friedmann.
Dans les cas où ce point fixe existe, il est intéressant de linéariser le système d'équations autour de ce point.
On pose alors $H=H_{0}+\delta H$ et en ne gardant que les termes du premier ordre en $\phi,u,\delta H$,  on obtient le système linéarisé:
\begin{align*}
\left(  \begin{matrix} \dot{\phi} \\ \dot{u} \\ \dot{H}  \end{matrix} \right) =
\left(  \begin{matrix}
    0 & 1 & 0 \\
    -m^{2} & -3H_{0} & 0 \\
    0 &0& -2 H_{0}
  \end{matrix}\right)
\left(  \begin{matrix}    \phi\\u\\H \end{matrix}\right)
\end{align*}

On remarque que les variables $(\phi,u)$ se découplent de $\delta H$.
Pour  $(\phi,u)$, on trouve donc les valeurs propres:
\begin{align*}
  \lambda_{\pm}&=-\frac{3H_{0}}{2} \pm \frac{1}{2}\sqrt{9H_{0}^{2}-4m^{2}} \ .
\end{align*}

Il faut analyser les différents régimes possibles selon la valeur de $H_{0},m^{2}$.
Tout d'abord, nous allons considérer $H_{0}>0$ (univers en expansion). Il faut ensuite distinguer les cas $m^{2}>0$ (point stable) et $m^{2}<0$ (point instable).
\begin{itemize}
\item Dans le cas $m^{2}>0$,
  \begin{itemize}
  \item pour  $9H_{0}^{2}-4m^{2}<0$,
    \begin{align*}
      \lambda_{{\pm}} &=-\frac{3H_{0}}{2} \pm \frac{i}{2}\sqrt{4m^{2}-9H_{0}^{2}} &
    \end{align*}
\item pour $9H_{0}^{2}-4m^{2}>0$,
  \begin{align*}
      \lambda_{\pm}&=-\frac{3H_{0}}{2} \pm \frac{1}{2}\sqrt{9H_{0}^{2}-4m^{2}}   & < 0
  \end{align*}
  \end{itemize}
\item Dans le cas $m^{2}<0$, on a
 \begin{align*}
      \lambda_{+}&=\frac{3H_{0}}{2} ( \sqrt{1-\frac{4m^{2}}{9H_{0}^{2}}}-1)   & > 0 \\
      \lambda_{-}&=-\frac{3H_{0}}{2}(1 - \sqrt{1-\frac{4m^{2}}{9H_{0}^{2}}})  & < 0
   \end{align*}
qui correspond à un point en selle, donc instable.
\end{itemize}

\subsubsection{Champ scalaire de type Born-Infeld couplé à la gravitation}

Nous allons analyser les points critiques pour une théorie de champ scalaire ayant pour Lagrangien:
\begin{align*}
L_{\phi}&=
 1-{\left( {\left( 1 -\frac{3\,{\dot{\phi} }^2}{{\beta }^2} \right) }^2 + \frac{16\,{\phi }^2\, {\left( -\gamma  + \phi  \right) }^2}{{\beta}^2} \right) }^{\frac{1}{4}}\left(1 + \frac{4\,{\phi }^2\,{\left( -\gamma  + \phi  \right) }^2}{{\beta}^2}\right)^{\frac{1}{2}}
\end{align*}
couplé de manière minimale à la métrique de Friedmann-Robertson-Walker.

La densité d'énergie associée à ce champ scalaire a la forme suivante:
\begin{align*}
\rho&=
\frac{{\sqrt{1 + \frac{4\,{\phi }^2\,{\left( -\gamma  + \phi  \right)}^2}{{\beta}^2}}}\,\left( 1 - \frac{3\,{\dot{\phi} }^2}{{\beta }^2} + \frac{16\,{\phi }^2\,{\left( -\gamma  + \phi  \right) }^2}{{\beta }^2} \right) }{{\left( {\left( 1 - \frac{3\,{\dot{\phi} }^2}{{\beta }^2} \right) }^2 +\frac{16\,{\phi }^2\,{\left( -\gamma  + \phi  \right) }^2}{{\beta}^2} \right) }^{\frac{3}{4}}}-1 \ .
\end{align*}

Autour des points fixes, la théorie a un comportement similaire à la théorie libre pour laquelle la densité d'énergie est:
\begin{align*}
  \rho&=b(\frac{1}{2}\dot{\phi}^{2} +\frac{1}{2}m^{2} \phi^{2} ) + \rho_{0} \ .
\end{align*}

Si on linéarise les équations du mouvement obtenues du Lagrangien de Born-Infeld, on obtient alors des équations identiques à celles du champ libre avec des valeurs de paramètres $\rho_{0}$ et $m^{2}$ bien spécifiques.

Les deux paramètres du modèle auquel sont reliés $\rho_{0}$ et $m^{2}$ sont les paramètres $\beta$ et $\gamma$ (masse du champ scalaire).
Les trois points fixes du Lagrangien considérés sont $(\phi=0,u=0)$, $(\phi=,\gamma,u=0)$ et $(\phi=\frac{\gamma}{2},u=0)$.

Pour les deux premiers, on trouve:
\begin{align*}
 m^{2}&=4\gamma^{2} \\ \rho_{0}&=0  \ ,
\end{align*}
ce qui correspond à un point stable avec valeurs propres $\lambda_{{\pm}}= \pm i 2\gamma$.

Pour le point $(\phi=\frac{\gamma}{2},u=0)$, on trouve:
 \begin{align*}
  m^{2}&=-2\gamma^{2} \frac{(  1+\frac{\gamma^{4}}   {2\beta^{2}}  )}   {(1+\frac{\gamma^{4}}{4\beta^{2}})} \\
  \rho_{0} &= \left(1+\frac{\gamma^{4}}{\beta^{2}}\right)^{\frac{1}{4}} \sqrt{1+\frac{\gamma^{4}}{4\beta^{2}}}  \ - \,1 \ ,
 \end{align*}
ce qui correspond à un point instable avec valeurs propres $\lambda_{{\pm}}= \pm i 2\gamma$.
Les valeurs propres peuvent être calculées de manière exacte en fonction des paramètres $\gamma$ et $\beta$ et nous pouvons les donner sous la forme de développement en puissances de $\beta$:
\begin{align*}
  \lambda_{+}&=\sqrt{2}\gamma - \frac{3}{4}\sqrt{\frac{\gamma^{4}}{2\beta^{2}}} + o(\frac{1}{\beta^{2}})\\
  \lambda_{-}&=-\sqrt{2}\gamma - \frac{3}{4}\sqrt{\frac{\gamma^{4}}{2\beta^{2}}} + o(\frac{1}{\beta^{2}}) \ . 
\end{align*}


\chapter*{Conclusion et perspectives}
\addcontentsline{toc}{chapter}{\protect\numberline{}Conclusion et perspectives}
\chaptermark{Conclusion et perspectives}

Le thème central de cette thèse a été l'étude des algèbres d'endomorphismes et leurs diverses applications.

Il a été montré comment ces algèbres pouvaient servir à reformuler les théories de jauge ordinaires, ce qui a permis de fournir une nouvelle grille de lecture d'un ensemble de phénomènes décrits habituellement dans un cadre géométrique.

Outre le fait de fournir un nouveau point de vue sur les théories de jauge en plongeant différents modèles dans un cadre algébrique, nous avons vu qu'il est possible de considérer une généralisation des théories de jauge pour des connexions non commutatives permettant ainsi de relier entre eux des degrés de libertés qui apparaissaient auparavant comme isolés.
Plus précisément, nous avons vu qu'un champ de jauge et un champ de Higgs, habituellement représentés par une connexion ordinaire et une section d'un fibré vectoriel, pouvaient être vus comme les différentes parties d'une connexion non commutative.
De ce point de vue, le fait de travailler avec une algèbre d'endomorphismes plutôt qu'avec un fibré principal permet d'unifier certains concepts de théories classiques des champs.
Les Lagrangiens de type Yang-Mills-Higgs ainsi obtenus possèdent une structure relativement rigide due à la nature géométrique et algébrique des modèles introduits.
Néanmoins, la caractérisation des connexions ordinaires et non commutatives invariantes sous l'action d'un groupe de Lie compact effectuée en détail dans le chapitre \ref{cha:symetries} peut permettre d'introduire plus de diversité dans les modèles obtenus.

Grâce à cette nouvelle structure d'algèbre, certains objets deviennent plus facilement manipulables et/ou caractérisés de manière plus directe.
Rappelons par exemple que pour une algèbre d'endomorphismes donnée, le groupe de jauge est un sous-groupe du groupe des automorphismes intérieurs, et l'algèbre de Lie du groupe de jauge correspond aux dérivations intérieures de l'algèbre. 
Nous avons également vu que le morphisme de Weil, servant à obtenir les classes caractéristiques en géométrie ordinaire, peut être obtenu par une construction complètement algébrique et relativement naturelle à partir de la suite exacte courte reliant les dérivations intérieures aux dérivations de l'algèbre.
Enfin notons que le fait d'avoir une présentation algébrique des théories de jauge permet de se rapprocher des techniques utilisées en théorie quantique des champs.
Nous pouvons donner comme exemple la symétrie B.R.S.T. qui est un outil essentiel en théorie quantique des champs pour l'étude des anomalies ainsi que pour prouver la renormalisabilité d'une théorie de jauge. Cette symétrie est reliée à la géométrie de l'espace affine des connexions et au groupe de jauge, ce dernier étant caractérisé de manière très simple dans ce nouveau contexte. 
Notons également que des techniques homologiques sont utilisées dans ce cadre et il serait intéressant de voir si la construction du morphisme de Weil que nous avons obtenue dans la section~\ref{sec:classe-car} peut s'inscrire dans une construction homologique de même type.

\vspace{.5cm}

Nous avons vu qu'il est possible de définir une structure Riemannienne ainsi que le concept de connexion de Levi-Cività sur une algèbre d'endomorphismes.
En particulier, une telle structure Riemannienne permet d'introduire naturellement une connexion (ordinaire) de référence.
La construction de cette connexion, faite dans la section~\ref{sec:struct-riem}, s'apparente aux constructions de type Kaluza-Klein habituellement effectuées sur les fibrés principaux.
L'introduction d'une métrique s'avère également indispensable pour pouvoir formuler un principe de moindre action pour une connexion non commutative.
Nous avons formulé un modèle de ``Maxwell non commutatif'' dans la section~ \ref{sec:action-de-maxwell} et cela nous a mené à un nouveau type de modèle de théorie des champs couplant de manière originale la métrique (au sens généralisé) et une connexion non commutative.
On a alors une approche intéressante des théories de jauge avec brisure de symétrie sur un fibré non trivial.
La connexion de référence donnée par la structure Riemannienne est décrite par un champ de fond ``$A$'' et une possibilité s'offre de considérer,  pour cette connexion, un représentant d'une classe caractéristique.

Une autre possibilité est de considérer l'action de Einstein-Hilbert pour la structure Riemannienne introduite sur l'algèbre des endomorphismes.
On obtient alors un modèle dans lequel la connexion de référence $A$ et la forme tensorielle $\ta$ ont une dynamique différente.

\vspace{.5cm}

Il serait intéressant d'étudier le rapport entre les algèbres d'endomorphismes munies d'une structure Riemannienne et les algèbres obtenues à partir de fibrés en algèbres de Clifford, également construites de manière naturelle à partir d'un fibré vectoriel (le fibré tangent) et d'une métrique sur ce fibré. 
Une telle approche pourrait permettre d'introduire les spineurs dans notre modèle et éventuellement de traiter la gravitation dans un cadre algébrique similaire à celui développé pour les algèbres d'endomorphismes.
Il serait également intéressant d'adapter et de généraliser les techniques développées pour les algèbres d'endomorphismes à d'autres types d'algèbres non commutatives telles que les algèbres obtenues à partir de fibrés en algèbres d'opérateurs. Ces algèbres furent introduites par Dixmier et Douady dans~\cite{Dixm:57,dixm-doua:63,Dixm:64}.

Nous pouvons conclure par le fait que le présent travail a permis d'une certaine manière de consolider le pont entre géométrie ordinaire et algèbre.
En effet, les algèbres d'endomorphismes fournissent une base pour pouvoir formuler des notions géométriques en rapport avec la notion de fibré principal dans un langage algébrique tout comme les $C^{*}$-algèbres permettent de formuler des notions de topologie dans un langage algébrique.
Enfin nous pouvons espérer que certaines constructions que nous avons obtenues dans ce cadre, comme la construction du morphisme de Weil, pourront fournir de nouvelles méthodes de calcul en géométrie non commutative et qu'elles pourront s'adapter à d'autres types d'algèbres comme des algèbres obtenues par déformation ``à la Moyal'' \cite{moyal:49} (voir également \cite{grac-vari-figu:01}), ou encore les algèbres des ``sphères non commutatives'' étudiées dans~\cite{conn-duboi:03}. 
Bien d'autres exemples d'algèbres non commutatives pourraient encore être citées\dots

Nous avons vu que du point de vue de la physique, la géométrie non commutative sert pour l'instant à formuler essentiellement des théories classiques des champs.
Cette situation peut paraître paradoxale étant donné qu'une des motivations principales pour la géométrie non commutative a été la mécanique quantique.
On peut donc penser qu'une compréhension plus profonde de ces structures est certainement nécessaire afin de pouvoir formuler et décrire des phénomènes quantiques dans le cadre de la géométrie non commutative. 

\vspace{.5cm}

Un autre aspect de cette thèse est l'étude des généralisations de la théorie de Born-Infeld aux théories de jauge non commutatives et non abéliennes.
Une proposition de généralisation a été faite en étendant le déterminant de  l'action de Born-Infeld, habituellement pris sur un espace de matrices représentant des tenseurs $2$ fois covariants, à un espace de matrices de taille plus élevée obtenu en considérant le produit tensoriel de l'espace de matrices représentant des tenseurs $2$ fois covariants avec l'espace des matrices $M_{n}$.
Nous avons ainsi obtenu un généralisation de l'action de Born-Infeld pour des connexions non commutatives. 
Pour les connexions représentant une connexion ordinaire, on obtient alors une action de Born-Infeld pour des champs de jauge non abéliens.
Cette généralisation a été comparée à d'autres généralisations telle que  celle appelée ``prescription de la trace symétrisée'' utilisée en théorie des cordes pour reproduire la partie non dérivative de l'action effective jusqu'à l'ordre $4$.
L'action de Born-Infeld généralisée a été étudiée plus en détail dans le cas d'un groupe de structure $SU(2)$  pour des connexions non abéliennes et non commutatives.
Lors de cette étude,  l'accent a été mis sur la recherche de solutions de type solitonique possédant des symétries particulières (symétrie sphérique) afin d'essayer de caractériser l'effet des non linéarités propres au modèle de Born-Infeld.
L'aspect dynamique des solutions a été étudié dans le secteur scalaire (partie ``non commutative'' d'une connexion).

Enfin notons que dans le cadre de la théorie des cordes, on voit apparaître la notion de théorie des champs sur espace non commutatif, la non commutativité des coordonnées étant mesurée par un champ de fond ``$B_{\mu\nu}$'' représentant une $2$-forme.
On peut alors associer à son champ $H=dB$ une classe de Dixmier-Douady. Il serait intéressant de voir si les algèbres de sections de fibrés en algèbres, généralisant les algèbres d'endomorphismes et correspondant à ces classes, pourraient être utiles dans l'étude de ces théories.
Une telle approche pourrait permettre d'établir de nouveaux outils et techniques de calcul afin d'apporter une lumière nouvelle sur les différentes structures algébriques intervenant dans ce contexte.


\appendix
\chapter{Quelques Lagrangiens pour champs scalaires}
Dans cette annexe, est expliqué le calcul du Lagrangien résultant de différents ansatz sur le Lagrangien de Born-Infeld non commutatif $L_{binc}$ (\textit{c.f.} formule~(\ref{binc})) pour une théorie de jauge non-commutative construite sur l'algèbre $C^{\infty}(M)\otimes M(2,\gC)$.

\section{Notations}
Nous adopterons les notations suivantes:
\begin{align*}
  T_a &= -i \sigma_a\\
  T_a T_b &= - \delta_{a b} + \sum_{c} \epsilon_{abc} T_c\\
  [T_a, T_b] &= C_{ab}^{c} = 2 \epsilon_{abc} T_c
\end{align*}
où $T_{a}$ sont les générateurs antihermitiens de $\ksl(2)$.
On définit également les quantités suivantes:
\begin{align*}
  K^{(n)}_{a_1 \dots a_n} = \frac{(-1)^{[\frac{n}{2}]}}{2} \tr(T_{a_1} \dots T_{a_n})
\end{align*}
soit
\begin{align*}
K_{ab} &=\delta_{ab}\\
K_{abc} &= \epsilon_{abc} \\
K_{abcd} &= \delta_{ab}\delta_{bc}- \delta_{ac}\delta_{bd} + \delta_{ad}
\delta_{bc}\\
\vdots
\end{align*}

\section{Calcul du Lagrangien}

On doit calculer le déterminant de la matrice suivante:
\begin{align}
  \begin{vmatrix}
    1 & i D\hat{\phi}\\
    -iD\hat{\phi} & 1+i \hat{F} 
  \end{vmatrix}
\end{align}
où 
\begin{align*}
\hat{F} &= \left\{ [\hat{\phi}_a, \hat{\phi}_b] - C_{ab}^c \hat{\phi}_c \right\}_{{a,b = 1,2,3}}\\
\hat{\phi}_a &= \phi_a^b T_b
\end{align*}
est la partie non-commutative de la courbure, et
\begin{align*}
D\hat{\phi} & = \left\{ D_{\mu}\hat{\phi}_a\right\}_{\substack{a=1,2,3\\\mu=0,1,2,3}}
\end{align*}
la partie mixte $\hat{F}_{\mu a}$.
La matrice dans (\ref{matrice}) est la matrice $g+i F$ intervenant dans la généralisation de l'action de Born-Infeld avec la partie ordinaire de la courbure $\hat{F}_{\mu \nu} =0 $. C'est une matrice $(7 \times 2)^2$. On peut la réduire par la formule de Schur à une matrice $(3 \times 2)^2$:
\begin{equation}
  \begin{vmatrix}
    1+ i \hat{F} - D_{\mu}{\phi}D^{\mu}\hat{\phi}
    \end{vmatrix}
\label{matrice:2}
\end{equation}
 On remarque alors que le déterminant de la matrice précédente est un carré parfait. Ceci est dû à la décomposition en produit tensoriel de notre matrice $6$ par $6$ et à l'algèbre de matrice $M_{2}(\gC)$ (on ramène le déterminant de cette matrice au déterminant d'une matrice antisymétrique en multipliant (\ref{matrice:2}) par $1 \otimes T_2 $). Bien que cela soit un carré parfait, on ne peut réduire la taille de la matrice, il faut en fait utiliser la formule des traces pour un déterminant, en renormalisant chaque trace par un facteur $1/2$ et en s'arrêtant aux puissances $3$ de la matrice (au lieu de $6$ pour une matrice $6\times 6$ quelconque).
La formule à appliquer est la suivante:
\newcommand{\bv}{\boxempty}
\begin{align*}
\sqrt{\det(1+iF)}&= 1 +  \bv  &-& \bv\bv             &+& \bv\bv\bv \\
    &  & +&  \substack{\bv\\ \bv} &-&  \substack{\bv\bv\\ \bv \hfill}  \\
  &  &  &                   &+&   \substack{\bv\\ \bv \\ \bv}  \\
&= 1 +t_1 &-& t_2 &+&t_3\\
& &+&\frac{1}{2}(t_1)^2 &-&t_1 t_2 \\
&&& &+&\frac{1}{3!}(t_1)^3
\end{align*}
\begin{align*}
  t_1 &=\frac{1}{2} \tr(M)\\
  t_2 &=\frac{1}{2} \tr(\frac{M^2}{2})\\
  t_3 &= \frac{1}{2} \tr(\frac{M^3}{3})
\end{align*}
 et 
$$  M =i \hat{F} - D\hat{\phi} D\hat{\phi} $$
On est maintenant paré pour attaquer le calcul du déterminant.
La forme générale étant donnée ci-dessus, il est intéressant de la considérer dans certains cas particuliers afin d'obtenir une expression plus simple.

\subsubsection{Ansatz $\hat{\phi}_a=\phi T_a$}
Je mène le calcul en détail pour cet ansatz, la méthode sera la même pour les autres ansatz. 
Pour celui-ci, on a:
\begin{align*}
  \hat{\phi}_a&=\phi T_a\\
  \hat{F}_{ab}&=\phi (\phi-1) [T_a, T_b] = 2 s \epsilon_{abc} T_c \\
  D\hat{\phi}_a D\hat{\phi}_b &= (D\phi)^2 T_a T_b = a T_a T_b
\end{align*}
où on a posé $a=(D\phi)^2$ et $s=\phi(\phi-1)$.
On doit maintenant calculer les traces de $M$, $ M^2 $ et $M^3$:
\begin{align*}
  \frac{1}{2} \tr M &= - a \sum_a \frac{1}{2} \tr(T_a T_a) &=& 3 a \\
 \frac{1}{2} \tr M^2 & = \cdots & =& -24 s^2 - 24i as + 9 a^2 \\
 \frac{1}{2} \tr M^3 &=  \cdots &=& 48 i s^3 - 3 \times 48 a s^2 - 3 \times 36 i a^2 s + 27 a^3
\end{align*}
 Soit:
\begin{align*}
  t_1 &=  3 a \\
  t_2 &= -12 s^2 - 12 i a s + \frac{9}{2} a^2 \\
  t_3 &= 16 i s^3 - 48 a s^2 - 36 i a^2 s + 9 a^3
\end{align*}
ainsi on a:
\begin{multline*}
\sqrt{\det(1+i F)} = 1 + t_1 - t_2 + t_3 + \frac{1}{2} t_1^2 - t_1 t_2 + \frac{1}{6} t_1^3 \\
\shoveleft \qquad = 1 + 3 a + 12 s^2 +12 i a s - \frac{9}{2} a^2 + 16 i s^3 - 48 a s^2 - 36 i a^2 s + 9 a^3 +\\
  + \frac{9}{2} a^2 + 36 a s^2 + 36 i a^2 s - \frac{27}{2} a^3 + \frac{27}{6} a^3 \\
\shoveleft \qquad = 1+ 3a + 12 s^2 - 12 a s^2 + 12 i a s + 16 i s^3 \\
\end{multline*}
Alors:
\begin{align*}
  |\det(1+i F)| &= \sqrt{\det(1+i F)\det(1-i F)}\\
&= (1+ 3a + 12 s^2 - 12 a s^2)^2 + 16 s^2 ( 2 a + 4 s^2 )^2 \\
&=(1+4 s^2)^2 (1 + 6 a + 9 a^2 + 16 s^2)
\end{align*}
et donc:
\newcommand*\widefbox[1]{\fbox{\hspace{1em}#1\hspace{1em}}}     
\begin{empheq}[box=\widefbox]{align*}
    L=1- \left\{ 1+ 6 (D\phi)^2 + 9
      (D\phi)^4 + 16 \phi^2(\phi-1)^2 \right\}^{\frac{1}{4}} \sqrt{1+4 \phi^2(\phi-1)^2 } 
\end{empheq}

Afin que le Lagrangien commence par $-\frac{1}{2} (D\phi)^2$, il faut faire un changement de normalisation $\phi \rightarrow\frac{1}{\sqrt{3}} \phi$, ce qui donne comme Lagrangien:
  \begin{empheq}[box=\widefbox]{align*}
    L=1- \left\{ 1+ 2 (D\phi)^2 + (D\phi)^4 + \frac{16}{9} \phi^2(\phi-\sqrt{3})^2 \right\}^{\frac{1}{4}} \sqrt{1+\frac{4}{9} \phi^2(\phi-\sqrt{3})^2 } 
  \end{empheq}

On peut encore compliquer un petit peu cet ansatz en choisissant un champ $\phi$ différent pour chaque générateur, soit
$$ \hat{\phi}_a = \phi_a T_a $$
{\bf Attention} il n'y a pas de sommation implicite sur les indices dans cette dernière expression.

\subsubsection{Ansatz $ \hat{\phi}_a = \phi_a T_a $ (pas de sommation implicite) }
Dans cette sous-section, je n'utilise pas de sommation implicite sur les indices matriciels, sauf pour les indices d'espace-temps, lorsque l'on rencontre un terme $(D\phi_a D\phi_b)$, il faut comprendre $(D_{\mu}\phi_a D^{\mu}\phi_b)$.
Il est utile d'introduire un vecteur "champ magnétique":
\begin{equation*}
\vec{B} = 2 \left( \begin{matrix} \phi_2 \phi_3 - \phi_1 \\ \phi_1 \phi_3 - \phi_2 \\ \phi_1 \phi_2 - \phi_3 \end{matrix} \right)
\end{equation*}
On trouve pour les traces de la matrice $M$ les résultats suivants:
\begin{multline*}
  \frac{1}{2} \tr M = \sum_a D\phi_a D\phi_a \\
 \shoveleft { \frac{1}{2} \tr M^2 = - 2 \sum_a B_a^2 + \sum_{a,b} ( D\phi_a D\phi_b)^2 - 2i \sum_{a,b,c} (\epsilon_{abc})^2 (D\phi_a D\phi_b)B_c } \\
\shoveleft{ \frac{1}{2} \tr M^3 = i \sum_{abc} (\epsilon_{abc})^2 B_a B_b B_c + 6 \sum_e B_e^2 (D\phi_e)^2 -3 \sum_{a,b} \left( B_a B_b (D\phi_a D\phi_b) + B_a^2 (D\phi_b)^2 \right) } \\
-3i \sum_{a,b,c,f}  (\epsilon_{abc})^2 B_a (D\phi_b D\phi_f)(D\phi_c D\phi_f)\\ + \sum_{a,b,c} (D\phi_a D\phi_b)(D\phi_b D\phi_c)(D\phi_c D\phi_a)
\end{multline*}
A partir de ces expressions, on peut calculer le déterminant par la même méthode que dans la sous-section précédente. Il y aura cependant moins de simplifications (quelques-unes tout de même, puisque $t_1$,$ t_2$,$ t_3$ contiennent des fractions et que celles-ci doivent disparaître dans l'expression finale du déterminant qui ne doit contenir que des coefficients entiers!).
On peut également retrouver le résultat précédent en posant $\phi_1=\phi_2=\phi_3 = \phi$.
Un autre moyen de vérifier le calcul est de ne prendre qu'un seul $\phi$  non nul, par exemple  $\phi_1 = \phi$ et $\phi_2=\phi_3=0$.
On trouve alors le résultat très simple:
\begin{align*}
  L= 1 -\sqrt{1+ (D\phi)^2} \sqrt{1+4 \phi^2}
\end{align*}
Ce résultat coïncide également avec les résultats trouvés pour les ansatz suivants dans la même limite.
Passons aux autres ansatz.

\subsubsection{Ansatz $\hat{\phi}_a=\hat{\phi} \delta_{a 1}$ , $\hat{\phi}= \phi^b T_b$ }

\begin{align*}
  \hat{F}_{ab} = -2 \epsilon_{ab1} \hat{\phi}
\end{align*}
 Ainsi on trouve :
\begin{align*}
\frac{1}{2} \tr M &= a\\
\frac{1}{2} \tr M^2 &= -8 s + a^2\\
\frac{1}{2} \tr{M^3} &= a^3
\end{align*}
où $a = D\phi_a D\phi_a $ et $ s = \phi_a \phi_a$.
Le calcul du Lagrangien donne alors:
\begin{align*}
  L= 1 - \sqrt{1+a} \sqrt{1+4s} \ ,
\end{align*}
soit:
  \begin{empheq}[box=\widefbox]{align*}
    L= 1 - \sqrt{1+D\vec{\phi}} \sqrt{1+4 \vec{\phi}^2}
  \end{empheq}

\subsubsection{Ansatz $\hat{\phi}_a= \phi_a T_1$}

\begin{align*}
\hat{F}_{ab} &= - 2 \epsilon_{abc} \phi_c T_1 \\
-D\hat{\phi}_a D\hat{\phi}_{b}&= D\phi_a D\phi_b 1
\end{align*}
\begin{align*}
\frac{1}{2} \tr M &= \left(D\vec{\phi}\right)^2\\
\frac{1}{2} \tr M^2 &= -8 \vec{\phi}^2 +  \left(\left(D\vec{\phi}\right)^2\right)^2\\
\frac{1}{2} \tr{M^3} &= 12 \left( \phi_a \phi_b\left(D\phi_a D\phi_b\right)- \vec{\phi}^2\left(D\vec{\phi}\right)^2\right) + \left( \left(D\vec{\phi}\right)^2 \right)^3
\end{align*}
où $\left(D\vec{\phi}\right)^2=D\phi_a D\phi_a$ et $\vec{\phi}^2=\phi_a  \phi_a$.

Ce qui donne:
  \begin{empheq}[box=\widefbox]{align*}
    L=1-\sqrt{1+\left(D\vec{\phi}\right)^2+ 4 \vec{\phi}^2 + 4 \phi_a \phi_b (D\phi_a D\phi_b)}
   \end{empheq}
  
  \begin{rem}
    Les vides statiques et constants dans l'espace sont donnés par l'équation $\hat{F}_{ab} = 0 $. 
    Ce sont les mêmes vides que la théorie de Maxwell non-commutative construite sur la même algèbre, c'est-à-dire les configurations où $\hat{\phi}$ est une représentation de $\ksu(2)$, \textit{i.e.} $\hat{\phi}=0$ ou $\hat{\phi}_a =T_a$.
    On retrouve bien ces deux vides dans les ansatz précédents, qui fournissent des directions particulières de fluctuations autour de ces vides.
\end{rem}

\begin{rem}
  La signature de la métrique est ici $(-,+,+,+)$ pour la partie espace-temps.
  Cela est important si l'on veut avoir le bon signe devant le terme cinétique. 
  Le signe global du Lagrangien étant donné par la signature de la métrique dans les directions matricielles, il faut en fait que la partie spatiale de la métrique ait le même signe que la métrique sur les indices de matrices.
\end{rem}


\chapter{Algèbre Homologique}\label{cha:algebre-homologique}
\section{Modules}\label{sec:modules}
Les modules d'algèbres nous seront utiles car ils représentent l'analogue non commutatif des fibrés vectoriels.
Je rappelle ici quelques définitions essentielles. Tous ces résultats, ainsi que les démonstrations peuvent être trouvées dans~\cite{vermani}. 
\begin{defn}[Module]
  Soit $\cA$ une algèbre unifère sur $\gC$.  Un groupe abélien $\cM$ est appelé {\bf $\cA$-module à gauche}  si pour tout élément $a \in \cA$ et  $m \in \cM$, il existe un élément uniquement déterminé $a m$, tel que:
  \begin{align}
    & (a+b)m=am+bm &
    &\forall a,b\in \cA, m \in \cM \\
    &  (ab)m=a(bm) &
    &\forall a,b\in \cA, m \in \cM \\
    &a (m+n)=am+an &
    &\forall a\in \cA, m,n \in \cM \\
    & 1m=m &
    &\forall m \in \cM
  \end{align}  
  Nous appellerons {\bf $\cA$-module à droite} un module à gauche sur $\cA^{op}$, l'algèbre opposée de $\cA$. 
  Par convention, lorsque nous disons {\bf$\cA$-module}, nous sous-entendons $\cA$-module à gauche.
\end{defn}

\begin{defn}[Module libre]
  Un $\cA$-module $\cF$ est un {\bf module libre} sur la base $X\neq \varnothing$ ($X$ est un ensemble), si il existe une application $\alpha: X\to \cF$ telle que pour toute application $\beta:X\to \cM$, où $\cM$ est un $\cA$-module, il existe un unique $\cA$-homomorphisme $f: \cF\to \cM$ tel que le diagramme suivant soit commutatif:
  \begin{align}
    \xymatrix{
      X \ar^-{\alpha}[r]\ar_-{\beta}[rd]& \cF  \ar^-{f}@{-->}[d]  \\
      &\cM
  }
\end{align}
{\it i.e.} , $\beta=f \alpha$.
\end{defn}

\begin{thm}
  Pour tout ensemble non vide $X$, il existe un $\cA$-module libre ayant $X$ pour base.
\end{thm}
Ainsi tout $\cA$-module est l'image par un homomorphisme d'un $\cA$-module libre. 

\begin{defn}[Module projectif]
  Un $\cA$-module $\cP$ est un module projectif si pour tout diagramme:
  \begin{align}
    \xymatrix{
      & \cP \ar^-{f}[d]  &\\
      \cM \ar^-{\alpha}[r] & \cN \ar[r] &0
    }
  \end{align}
  de $\cA$-modules où la rangée est exacte ($\alpha$ est un épimorphisme), il existe un homomorphisme $g: \cP \to \cM$ tel que $\alpha g=f$.
\end{defn}

\begin{defn}[Résolution (Projective)]
  Soit $\cM$ un $\cA$-module. 
Un complexe:
\begin{equation}
    \xymatrix{
      {\bf P_{\cM}}: & \cdots \ar[r] & P_{n} \ar^-{d_{n}}[r]& P_{n-1} \ar^-{d_{n-1}}[r]& \cdots \ar^-{d_{1}}[r]& P_{0}\ar[r]& 0
    }
  \end{equation}
  où les $P_{n}$ sont des $\cA$-modules (projectifs) est appelé {\bf résolution (projective)} du module $\cM$ si la suite
  \begin{align}
    \xymatrix{
      {\bf P}: & \cdots \ar[r] & P_{n} \ar^-{d_{n}}[r]& P_{n-1} \ar^-{d_{n-1}}[r]& \cdots \ar^-{d_{1}}[r]& P_{0} \ar^-{\epsilon}[r]& \cM \ar[r]& 0
    }
  \end{align}
est une suite exacte, autrement dit si $H(P)=0$. On dit que $\cM$ augmente le complexe $P_{\cM}$, et on a
$H(P_{\cM}) \simeq \cM$.
\end{defn}

\begin{thm}
  Tout $\cA$-module admet une résolution projective.
\end{thm}

\begin{thm}
  Soit $P$ une résolution projective de $\cM$ et $Q$ une résolution de $\cN$. Alors tout homomorphisme $f: \cM \to \cN$ peut s'étendre en un morphisme de complexes $\tilde{f}: P \to Q$. Deux extensions $\tilde{f}$ et $\tilde{\tilde{f}}$ de $f$ sont homotopes. 
 \end{thm}

 \begin{defn}[Foncteur $\Tor$]
   Soit $\cM$ un module à droite sur $\cA$ et $\cN$ un module à gauche sur $\cA$. $\Tor^{\cA}(\cM,\cN)$ est le groupe abélien obtenu en calculant l'homologie du complexe $P_{\cM}\otimes \cN$, où $P_{\cM}$ est une résolution projective  de $\cM$, ou bien l'homologie du complexe $\cM \otimes P_{\cN}$, où $P_{\cN}$ est une résolution projective de $\cN$.
\end{defn}

$\Tor^{\cA}(\cdot,\cdot)$ est un bifoncteur de ${Mod}-\cA \times \cA-{Mod}$ dans ${Ab}$, la catégorie des Groupes abéliens et est le foncteur dérivé de $\cdot\otimes \cdot$.


\bibliographystyle{alpha-perso}
\bibliography{biblio_articles,biblio_livres}
\end{document}